\documentclass[twoside,12pt]{article}
\usepackage{epsfig}

\usepackage{amsmath,amssymb,bm}
\usepackage{graphicx,graphics,color}
\usepackage{xcolor}
\usepackage[colorlinks=true,linkcolor=blue,citecolor=blue]{hyperref}
\usepackage{subfigure}
\usepackage{mathrsfs}
\usepackage{physics}

\def\Journal#1#2#3#4{{#1} {#2} (#4) #3 }

\def\NPA{{\em Nucl. Phys.} A}

\def\NPB{{\em Nucl. Phys.} B}

\def\PLB{{\em Phys. Lett.} B}

\def\PRL{\em Phys. Rev. Lett.}
\def\PREV{\em Phys. Rev.}

\def\PRA{{\em Phys. Rev.} A}
\def\PRD{{\em Phys. Rev.} D}
\def\PRC{{\em Phys. Rev.} C}

\def\RMP{{\em Rev. Mod. Phys.}}

\newcommand{\be}{\begin{equation}}
\newcommand{\ee}{\end{equation}}
\newcommand{\bea}{\begin{eqnarray}}
\newcommand{\eea}{\end{eqnarray}}

\usepackage{slashed} 
\begin{document}

\title{ \vspace{1cm} 
Nucleon form factors and parton distributions \\ in nonlocal chiral effective theory}
\author{P. Wang,$^{1,2}$ Fangcheng He,$^3$ Chueng-Ryong\ Ji,$^4$ W. Melnitchouk $^5$ \\
\\
$^1$Institute of High Energy Physics, CAS, P. O. Box 918(4), \\ Beijing 100049, China\\
$^2$College of Physics Sciences, University of Chinese Academy of Sciences,\\
Beijing 100049, China \\
\author{Fangcheng He}
$^3$Key Laboratory of Theoretical Physics, Institute of Theoretical Physics, CAS, \\
Beijing 100190, China \\
$^4$Department of Physics, North Carolina State University,\\
 Raleigh, North Carolina 27695, USA\\
$^5$Jefferson Lab, Newport News, Virginia 23606, USA\\}

\maketitle

\vspace{-12.5cm}
~~~~~~~~~~~~~~~~~~~~~~~~~~~~~~~~~~~~~~~~~~~~~~~~
~~~~~~~~~~~~~~~~~~~~~~~~~~~~~~~~~~~~~~~~~~~~~~~~ JLAB-THY-22-3645

\vspace{12.5cm}

\begin{abstract} 
We present a review of recent applications of nonlocal chiral effective theory to hadron structure studies. Starting from a nonlocal meson--baryon effective chiral Lagrangian, we show how the introduction of a correlation function representing the finite extent of hadrons regularizes the meson loop integrals and introduces momentum dependence in vertex form factors in a gauge invariant manner. We apply the framework to the calculation of nucleon electromagnetic form factors, unpolarized and polarized parton distributions, as well as transverse momentum dependent distributions and generalized parton distributions. Assuming that the nonlocal behavior is a general property of all interactions, we also discuss the application to the lepton anomalous magnetic moment in nonlocal~QED.
\end{abstract}

\clearpage
\tableofcontents
\clearpage

\section{Motivation}

Protons and neutrons, or collectively nucleons, are the basic building blocks of nuclei and of all nuclear matter in the universe.
Their properties and structure are therefore crucial for understanding a wide range of nuclear phenomena in applications ranging from nuclear structure to nuclear astrophysics.  
Ultimately, the structure of the nucleon must be derived from first principles in quantum chromodynamics (QCD), the theory of the strong nuclear interactions, although in practice this remains a formidable challenge.
Tremendous progress has been made over the last few decades in unravelling the nucleon structure through both experimental and theoretical efforts.
In the following, we briefly summarize the historical developments in the study of nucleon structure, both empirically and theoretically via lattice QCD and other nonperturbative approaches.

\subsection{\it Phenomenological landscape}

Following in the spirit of Rutherford scattering that first established the granularity of atomic nuclei, nucleon structure investigations using high energy electron scattering have led to many fundamental discoveries, beginning with the 1955 observation of the finite size of the proton by Hofstadter \cite{Hofstadter}.
Over the subsequent decades elastic lepton-nucleon scattering measurements have enabled the mapping of a detailed picture of the electric ($G_E$) and magnetic ($G_M$) form factors of the proton and neutron over a wide range of kinematics.
In particular, after the first elastic scattering experiments were performed at the High Energy Physics Laboratory (HEPL) at Stanford in the 1950s, measurements were later carried out at the Stanford Mark III accelerator \cite{Janssens}, the Cambridge Electron Accelerator \cite{Price}, the Stanford Linear Accelerator Center (SLAC) \cite{Rock, Lung, Arnold}, Bonn \cite{Berger, Bruins}, DESY \cite{Bartel, Galster}, Mainz \cite{Simon, Bernauer2, Anklin2}, NIKHEF \cite{Anklin1}, MIT-Bates \cite{Markowitz}, and at Jefferson Lab \cite{Christy, Qattan, Lachniet}.
These covered an increasingly large range of four momentum transfers squared, up to $Q^2 \approx 9$~GeV$^2$ and 30~GeV$^2$ for the proton electric and magnetic form factors, respectively \cite{Sill, Andivahis}, at SLAC.

The accumulated body of elastic electron-proton and electron-deuteron scattering data suggests that at small $Q^2$, the $Q^2$ dependence of the proton's electric ($G_E^p$) and magnetic ($G_M^p$) and the neutron's magnetic ($G_M^n$) form factors are well approximated by a dipole form,
\begin{equation}
G_E^p \approx \frac{G_M^p}{\mu_p} \approx \frac{G_M^n}{\mu_n} \approx G_D,
\end{equation}
where $\mu_N$ ($N=p,n$) is the nucleon magnetic moment, and $G_D$ is a dipole form factor,
\begin{equation}
G_D = \frac{1}{\big( 1+Q^2/Q_0^2 \big)^2}\, ,
\end{equation}
with mass parameter $Q_0^2 = 0.71$~GeV$^2$.
For the neutron electric form factor $G_E^n$, electric charge conservation requires this to vanish at $Q^2=0$, but clear deviations above zero have been observed experimentally for $Q^2$ up to several GeV$^2$, albeit within relatively large uncertainties.

In addition to the cross section measurements which have mostly constrained the magnetic form factors, both the recoil polarization method and the asymmetry measurements using polarized targets have been used to extract the proton and neutron electric form factors.
The recoil polarization technique was used for the first time at MIT-Bates to measure the proton form factor ratio $G_E^p/G_M^p$ \cite{Milbrath1, Milbrath2}, and since then a number of other recoil polarization  and beam-target asymmetry measurements were carried out at MIT-Bates \cite{Crawford, Geis, Thompson, Gao}, MAMI \cite{Pospischil, Ostrick, Glazier, Rohe, Schlimme} and Jefferson Lab \cite{Paolone, Jones2, Gayou2, Puckett2, Warren, Riordan, Anderson}.
The data indicate a slow decrease of $\mu_p G_E^p/G_M^p$ from unity at low $Q^2$, with a dramatic fall-off at larger $Q^2 \gtrsim 1$~GeV$^2$.
These results are significantly different from the data extracted using the Rosenbluth separation method from the cross section measurements, for which $\mu_p G_E^p/G_M^p$ remains close to 1 for all $Q^2$, but with much larger uncertainties compared with the polarization data.
For the neutron form factors, since there are no free neutron targets, measurements of $G_E^n$ and $G_M^n$ are more difficult, and typically utilize deuterium or (polarized) $^3$He targets.
The data from both the double polarization experiments and cross section experiments suggest that $G_M^n/(\mu_n G_D)$ is close to unity.

From the form factor measurements at low $Q^2$, the electromagnetic radii of the proton and neutron were also extracted in a number of analyses \cite{PDG, Xiong, Bezginov, Epstein, Lee}. 
Among them, the proton charge radius has recently attracted the most attention.
Historically, the consistency of the extracted proton charge radius based on the scattering experiments at Orsay \cite{Lehmann}, Stanford \cite{Hand}, Saskatoon \cite{Murphy}, Mainz \cite{Simon2, Bernauer} and Jefferson Lab \cite{Zhan2} and on the various reanalyses of the world data \cite{Sick, Sick2, Hill} has been steadily improved, with the average extracted $\langle r_E^p \rangle \approx 0.88$~fm.
More recently, the proton radius was obtained from precise measurements of the Lamb shift energies in the hydrogen atom \cite{Mohr, Mohr2, Mohr3} and in muonic hydrogen \cite{Pohl, Antognini}. 
The values extracted from measurements in electronic hydrogen are consistent with the world-averaged electron scattering results. 
However, the precise value of the charge radius extracted from the muonic hydrogen data is $\langle r_E^p \rangle \approx 0.84$~fm, which is smaller than the mean value from the electron scattering experiments -- a result referred to as the ``proton radius puzzle.''
The puzzle has recently motivated various lepton scattering experiments at very low momentum transfers \cite{Gasparian, Mihovilovic, Gilman}. 
In particular, values of the proton charge radius $\langle r_E^p \rangle = 0.831$~fm and 0.833~fm obtained from the electron-proton scattering experiment and from the measurement of the electric hydrogen Lamb shift were recently reported \cite{Xiong, Bezginov}, which are both consistent with the previous muonic hydrogen experiments.

The elastic scattering measurements have also stimulated considerable activity over the past two decades in the determination of the flavor separated form factors of the up, down and strange quarks in the nucleon. 
The strange quark contribution in particular is of interest because it is purely a sea quark effect, whereas the $u$ and $d$ quark contributions involve both valence and sea components which cannot be easily disentangled.
In addition to the proton and neutron electromagnetic form factors, which, under the assumption of charge symmetry, provide two sets of constraints on the quark flavors, the neutral-weak vector form factors measured in parity-violating electron scattering (PVES) on the proton has been used as an additional constraint with which one can then solve for each of the $u$, $d$ and $s$ quark contributions to the form factors.
While PVES measurements are challenging, a number of experiments have been successfully performed, starting with SAMPLE at Bates \cite{Spayde, Spayde2} and A4 at Mainz \cite{Maas, Maas2}, followed by the high precision G0 \cite{Armstrong} and HAPPEX \cite{Acha, Aniol, Aniol2} experiments at Jefferson Lab.
The conclusion from analyses of these experiments have been that the strange electric and magnetic form factors $G_{E,M}^s$ have rather small magnitude, with some indication that $G_M^s < 0$ is favored \cite{Young, Baunack, Jimenez}.

As well as the electromagnetic properties of the nucleon, its axial form factors are also of great importance, since they are related to the quark spin. 
The isovector axial charge of nucleon the $a_3 \equiv g_A$ is determined very precisely from $\beta$-decay experiments \cite{PDG}, while the flavor singlet axial charge $\Delta\Sigma$ is related to the nucleon spin carried by quarks. 
In 1987, the EMC measurement of the spin-dependent $g_1$ structure function of the proton led to the surprising conclusion that the sum of quark spins constituted a very small fraction of the spin of the proton \cite{Ashman}. 
Subsequent polarized DIS experiments with increasing precision have been performed at SLAC \cite{Anthony, Abe2, Anthony2, Anthony5}, HERMES \cite{Ackerstaff, Airapetian0, Airapetian1}, SMC \cite{Adeva, Adeva2}, COMPASS \cite{Alexakhin, Alekseev} and Jefferson Lab \cite{Prok2, Fersch, Parno, Posik, Solvignon, Armstrong2}, providing a rich picture of the proton's spin structure.
Data from these and other polarized high energy scattering processes, such as jet and $W$ boson production in polarized $pp$ collisions at RHIC \cite{Adamczyk, Adare, Adam}, have been utilized in global QCD analyses of spin-dependent parton distribution functions (PDFs) by a number of groups \cite{Florian2, Hirai, Blumlein, Leader3, Khanpour, Noceraet, Sato2}. 
The latest results from the JAM Collaboration's simultaneous analysis \cite{Ethier} of helicity PDFs and fragmentation functions give a fraction 
    $\Delta\Sigma = 0.36(9)$ 
of the proton's spin carried by quarks and antiquarks at a scale of $Q^2=1$~GeV$^2$.

The momentum dependence of the axial form factor has been determined in both neutrino scattering \cite{Fanourakis, Ahrens2, Barish, Millerkl, Baker, Kitagaki2, Armenise} and in pion electroproduction \cite{Liesenfeld, Bloom, Brauel, Guerra2, Joos}. 
There exists a systematic small difference between these experiments, and the discrepancy may be explained with chiral perturbation theory \cite{Bernard}. 
As for the electromagnetic form factors, the axial form factor $G_A$ can also be parameterized as a dipole form with mass parameter 
    $M_A^2 = 1.10^{+0.13}_{-0.15}$~GeV$^2$ 
obtained by the MiniBooNE Collaboration \cite{Aguilar-Arevalo, Golan}.
The corresponding axial radius of the nucleon, 
   $\langle r_A \rangle \approx 0.64$~fm,
is smaller than the electric and magnetic proton radii.

The nucleon form factors can also be obtained from moments of generalized parton distributions (GPDs).
Generally, GPDs are functions of 3 variables: the parton momentum fraction, $x$, the momentum transfer squared in the process, $t$, and the longitudinal momentum transfer, referred to as skewness, $\xi$.
GPDs thus present snapshots of the three-dimensional structure of the nucleon, since they contain both longitudinal and transverse momentum dependence in the proton.
At leading twist, there are four chiral-even (helicity conserving) GPDs: $H$, $E$, $\widetilde{H}$, and $\widetilde{E}$, and four chiral-odd (helicity flipping) GPDs: $H_T$, $E_T$, $\widetilde{H}_T$, and $\widetilde{E}_T$.
In the forward limit, $H$, $\widetilde{H}$ and $H_T$ reduce to the unpolarized, polarized and transversity PDFs.

The nucleon GPDs are accessible via hard exclusive processes, such as deeply-virtual Compton scattering (DVCS) or deeply-virtual meson production (DVMP), as well as more involved processes such as double DVCS and timelike Compton scattering (TCS).
During the last twenty years, both unpolarized and polarized experiments have been performed at Jefferson Lab Hall A \cite{Camacho, Defurne}, HERMES \cite{Airapetian, Airapetian6, Airapetian7, Airapetian8}, H1 \cite{Adloff, Aaron}, ZEUS \cite{Chekanov}, and CLAS \cite{Hadjidakis, Morrow, Bedlinskiy} over a wide range of kinematics.
The 12~GeV program at Jefferson Lab will feature measurements of the complete set of polarization configurations in Halls A, B and C \cite{Gao2, Kubarovsky, Armstrong3}.
Future experiments are also planned at COMPASS \cite{d'Hose, Silva, Kouznetsov} and J-PARC \cite{Sawada, Kroll}, which can provide further constraints on GPDs, as well as at the Electron-Ion Collider (EIC) \cite{Accardi} and the proposed Large Hadron electron Collider (LHeC) \cite{Fernandez}, where exclusive processes are among the main goals of their experimental programs.

The description of nucleon structure in terms of PDFs that depend only on the momentum faction $x$ can be generalized to also take into account the parton transverse momentum, $k_T$.
At leading twist there are eight transverse momentum dependent distributions (TMDs) for the nucleon, and after integration over $k_T$ three of these survive and reduce to the usual collinear unpolarized, helicity and transversity PDFs.
The TMDs can be extracted from processes such as semi-inclusive deep-inelastic scattering (DIS), where the TMD distributions enter as convolutions with TMD fragmentation functions, as well as the Drell-Yan process and $Z^0/W^{\pm}$ boson production in hadronic collisions.
Semi-inclusive DIS experiments were carried out by HERMES \cite{Airapetian10}, COMPASS \cite{Alexakhin2, Adolph, Adolph2} and Jefferson Lab \cite{Qian, Zhangy, Huang, Yan}, while the Drell-Yan process was studied at Fermilab \cite{Zhuly} and RHIC. 
The transverse single spin asymmetry was recently measured in $Z^0/W^{\pm}$ production from $pp$ collisions by STAR \cite{Adamczyk2}.
A unified framework for accommodating the entire set of quantum correlation functions, including PDFs, TMDs and GPDs, is the generalized transverse momentum dependent parton distributions \cite{Meissner, Lorce, Lorce2}, although little is known about these functions phenomenologically.

The experimental studies thus far have revealed a number of intriguing and at times surprising results, which have challenged and deepened our understanding of the rich partonic structure of the nucleon.
Among these, one of the best known is the flavor asymmetry in the light antiquark sea, $\bar{d} > \bar{u}$, which DIS experiments from CERN, SLAC and Jefferson Lab, and Drell-Yan lepton-pair production data from Fermilab, have now painstakingly mapped out~\cite{Arneodo, Ackerstaff2, Baldit, Hawker, Towell}.
Similarly, the analysis of neutrino DIS data from CERN and Fermilab suggest a nonzero strange $s-\bar{s}$ asymmetry, albeit with rather large uncertainty \cite{Zeller, Bentz, Barone, Bazarko, Mason}.
Less well determined, but no less fundamental, is the question of the strange quark polarization, which at one time was thought to be related to the proton spin puzzle, but which recent inclusive and semi-inclusive DIS data analyses give as rather small, $\Delta s^+ = -0.03(10)$ \cite{Ethier}. 
In the TMD sector, for time reversal odd functions such as the Sivers and Boer-Mulders functions, the sign is predicted to change between semi-inclusive DIS and Drell-Yan reactions, and it will be crucial to verify this experimentally.
With more high precision data expected from the 12~GeV Jefferson Lab program and the future EIC, we can look forward to significant progress made toward the resolution of these various puzzles, and a better understanding of the nucleon substructure.

\subsection{\it Nonperturbative models}

Although QCD is the fundamental theory of the strong interactions, the direct application of QCD to hadronic physics is extremely difficult due to its nonperturbative nature at low energies.
In practice, therefore, approximations often need to be made in the form of phenomenological models, such as constituent quark models, chiral quark models, quark-diquark models, as well as approaches based on Dyson-Schwinger equations.

One of the oldest and well known models is the constituent quark model (CQM), which attempts to describe the properties of light hadrons as composite systems of $u$, $d$ and $s$ valence or constituent quarks, into which all other degrees of freedom are absorbed \cite{Lilf}.
In most calculations based on the CQM, different quark wave functions \cite{Chung, Gross, Gross2, Gross3} or potentials between valence quarks \cite{Isgur, Capstick} are assumed {\it a priori}. 
In the hypercentral CQM, a spin independent three-quark interaction inspired by lattice QCD was introduced \cite{Ferraris, Sanctis, Sanctis2}. 
Though the meson degrees of freedom are not explicit in the CQM, their effects are included in the potential associated with Goldstone boson exchange \cite{Glozman, Glozman2, Boffi}. 
In the chiral SU(3) quark model, apart from the confining and one-gluon exchange potentials, the potential due to the exchange of the nonet pseudoscalar mesons and the nonet scalar mesons were also included by Zhang {\it et al.} \cite{Zhang}, and further extended to include the potential from vector meson exchange by Dai {\it et al.} \cite{Dai}.

Meson degrees of freedom are explicitly included in the cloudy bag model (CBM) \cite{Miller2, Thomas, Lu}, where meson cloud contributions are accounted for through loop diagrams.
Similarly to the CBM, explicit meson degrees of freedom are also included in the perturbative quark model, which is based on an effective chiral Lagrangian describing valence quarks in baryons as relativistic fermions moving in a static potential \cite{Oset, Faessler, Lyubovitskij, Faessler1, Obukhovsky}. 
The quark-diquark model was also proposed, where, in contrast to the CQM, two quarks in baryons were assumed to form a tight diquark state \cite{Carroll, Sanctis3, Ferretti}.
The CQM, CBM, quark-diquark model, and similar variants have also been extended to the light-front formalism \cite{Frank, Schlumpf, Miller, Pace, Cloet, Chakrabarti, Zhang2}.
The light-front quark model based on the variational principle has been explored for meson phenomenology~\cite{CJ1999, CJ2015}.
Note that with the quark wave functions, the loop integrals in the quark models are convergent.
In contrast, the nonlocal quark-meson coupling model deals with ultraviolet divergences by placing the meson and quark fields at different locations in the effective Lagrangian \cite{Ivanov, Faessler2, Faessler3}, and in this case the vertex function will result in a regulator in the loop diagrams.
Meson fluctuation effects can also be obtained from the Nambu-Jona-Lasinio model starting from the four-fermion interaction, where mesons are described as bound states with the Bethe-Salpeter amplitude \cite{Klevansky, Weigel, Serrano, Courtoy}.

Beyond the above quark models, other approaches include the Dyson-Schwinger equations, which form a set of coupled integral equations for QCD's Green functions that grants access to nonperturbative phenomena \cite{Roberts, Nicmorus, Eichmann, Bednar}, and vector meson dominance, where the photon couples to the baryon through an intermediate vector meson \cite{Iachello, Gari, Williams, Bijker, Faessler4}.
The chiral quark soliton model, where the nucleon appears as a chiral soliton of a static background pion field in the limit of a large number of colors $N_c$, has been applied in various calculations \cite{Christov, Penttinen, Schweitzer, Ossmann, Wakamatsu, Goeke}.
Most recently, the AdS/QCD approach has been developed to the study the hadron mass spectrum, PDFs, meson and nucleon form factors, and structure functions \cite{Abidin, Brodsky3, Hashimoto, Traini}.

\subsection{\it Lattice QCD}

Among various theoretical methods, lattice QCD is the most rigorous approach that is based directly on the fundamental QCD theory.
For technical reasons, lattice simulations are performed in Euclidean spacetime rather than in Minkowski spacetime, and at finite values of the lattice spacing $a$ and lattice size~$L$.
To compare with experimental data, the lattice simulated results must be extrapolated to the continuum ($a \to 0$) and infinite volume ($L \to \infty$) limits.
Furthermore, if the calculations are performed using larger than physical quark masses, the lattice data also need to be extrapolated to the physical mass point ($m_q \to m_q^{{\rm phys}}$).

In the literature, the nucleon electromagnetic form factors have been studied by various lattice groups for a number of years.
Early simulations utilized the quenched approximation, in which disconnected diagrams were neglected, and at large pion masses due to limitations in computing power \cite{Wilcox, Zanotti, Gockeler, Sasaki}. 
Later, form factors were evaluated within full dynamical, unquenched QCD on the lattice \cite{Alexandrou}.
The unquenching effects were found to be relatively small, since the lowest pion mass in the simulation was 380~MeV, for which the loop effects are small.
Dynamical calculations were further carried out by several lattice groups with clover improved Wilson \cite{Gockeler2}, twisted mass \cite{Alexandrou2} and domain wall \cite{Lin, Syritsyn} actions.
With increased computational speed, lower pion masses were accessible \cite{Bhattacharya, Green, Ishikawa}, and currently the electromagnetic form factors of nucleon can be calculated on the lattice dynamically at the physical pion mass \cite{Alexandrou3, Shintani, Jang}.

The progress in lattice QCD has made it possible to study the individual up, down and strange quark contributions to the nucleon electromagnetic form factors.
The $s$ quark contribution is of special interest, and here the disconnected diagrams are crucial.
As for the electromagnetic form factor case, advances over the past 5 years have made it possible to reduce the lowest pion mass simulations from several hundred MeV to the physical mass \cite{Babich, Sufian, Alexandrou4}.
In Refs.~\cite{Sufian, Alexandrou4}, the light sea quark contributions to the nucleon form factors were also calculated.
In their simulations, no differences were found between the contributions from $\bar{u}$ and $\bar{d}$ quarks, and it is very challenging to obtain a $\bar{d}-\bar{u}$ asymmetry on the lattice.

The axial form factor of the nucleon has also been evaluated on the lattice \cite{Jang3, Alexandrou5}.
In recent years, with the advances in both algorithms and supercomputing hardware, the lattice data for $g_A$ at the physical pion mass have become in excellent agreement with the phenomenological value \cite{Alexandrou5, Green0, Horsley, Bali, Gupta, Chang}.
The accurate axial charge $g_A=1.271(13)$ \cite{Chang} from lattice simulation was recently reported, which is very close to the experimental data.
Many other physical quantities related to nucleon structure, such as the nucleon scalar charge \cite{Bali, Gupta, Green1, Alexandrou10}, tensor charge \cite{Bali, Bhattacharya2, Green2, Yoki, Abdel-Rehim}, $\sigma$ term \cite{Bali2, Alvarez-Ruso, Alexandrou9}, quark spin $\Delta\Sigma$ \cite{Alexandrou8, Lin3}, neutron electric dipole moment \cite{Gupta2, Izubuchi}, among others, have also been evaluated by various lattice groups.

Although considerable progress has been made with lattice QCD simulations, exploring hadron structure with lattice QCD has until recently been limited to only the first few moments. 
Higher moments, such as $\langle x^2 \rangle$, have not been updated using dynamical fermions for more than a decade \cite{Lin2, Gockeler3}. 
The $n$-th form factors (Mellin moments) can be obtained from the PDFs by integrating over the momentum fraction $x$.
Since PDFs are defined in Minkowski space on the light-front, it has not been possible to simulate PDFs directly on a Euclidean lattice. 
In recent years, quasi-PDFs have been proposed to compute on the lattice, within the large momentum effective theory (LaMET) \cite{Ji}. 
Alternatively, the pseudo-PDF has also been suggested, which considers the ratio of equal-time matrix elements of the Wilson line between quarks with the rest-frame density matrix element, and is parametrized in terms of the Ioffe time \cite{Radyushkin, Radyushkin2, Ioffe, Braun}.
Currently, many lattice groups are engaged in computing unpolarized PDFs \cite{Lin4, Alexandrou11, Chen2}, polarized PDFs \cite{Chen, Alexandrou12, Green3} and the transversity PDF \cite{Chen, Alexandrou12, Alexandrou13} using either the quasi-PDF or pseudo-PDF approach, although the efforts are still in their early stages.

\subsection{\it Chiral effective theory}

Chiral perturbation theory ($\chi$PT) is an effective field theory (EFT) which provides another systematic method to study hadron physics. 
It allows the description of low energy properties and processes within a systematic perturbative approach, emphasising the chiral symmetry aspects of QCD.
In the 1970s the concept of EFT emerged \cite{Weinberg, Weinberg2} with the development of $\chi$PT as a model-independent way of describing the application of QCD at low energies \cite{Gasser, Gasser2}. 
Compared with phenomenological quark models, EFT works on the hadronic degrees of freedom, without specifying the nature of the internal quark structure and dynamics, which instead is parametrized by low energy constants.
The chiral EFT can be formulated in either the heavy baryon \cite{Jenkins, Jenkins2, Bernard2,Wang4} or relativistic \cite{Ellis, Becher, Gegelia, Lutz} framework, and has been widely applied in the study of hadron properties \cite{Ellis2, Kubis, Zhusl, Bernard3, Bauer, Wein}.
Historically, most formulations of $\chi$PT are based on dimensional or infrared regularization \cite{Becher, Ellis2, Kubis, Borasoy} and the extended on-mass-shell renormalization scheme \cite{Bauer, Fuchs, Schindler}.
Although $\chi$PT has been a fairly successful approach, for the nucleon form factors it is only valid at relatively small $Q^2$ values, $Q^2 \lesssim 0.1$~GeV$^2$ \cite{Fuchs2}.
The range can be extended up to $Q^2 \lesssim 0.4$~GeV$^2$ by explicitly including vector meson degrees of freedom into the theory \cite{Kubis, Kubis2}.

As an alternative regularization method, finite range regularization (FRR) has been argued to achieve better convergence than dimensional regularization in the calculation of many hadronic observables \cite{Young, Young2, Leinweber}.
Inspired by models that account for the finite size of the nucleon as the source of the pion cloud, EFT with FRR has had many successful applications, ranging from extrapolating lattice data for the vector meson mass to magnetic moments and charge radii, electromagnetic and strange form factors, as well as moments of PDFs and GPDs \cite{Wang, Wang2, Hall, Li, Shanahan}. 
The finite range can be parametrized in terms of various functional forms, such as dipole or Gaussian, and characterized by a mass parameter, $\Lambda$.
For a heavy baryon formulation the nonrelativistic regulator is in 3-dimensional momentum space and hence is not covariant.
In a relativistic formulation, covariance can be implemented by making the regulator a four-dimensional function or by using Pauli-Villars regularization, which ensures gauge invariance \cite{XGWang, XGWang2}.

While the introduction of a regulator function in FRR is to some extent {\it ad hoc}, one can make the formulation more systematic by generating the regulator function through a nonlocal generalization of the effective Lagrangian.
By imposing local gauge invariance on the nonlocal Lagrangian, a covariant regulator is generated automatically, without reference to any specific functional form.
The renormalized charge of the nucleon is preserved with the additional diagrams obtained by expanding the gauge link. 
The nonlocal interaction generates both the regulator that renders loop integral convergent and the $Q^2$ dependence of the form factors at tree level.
The nonlocal chiral EFT has been applied to electromagnetic and strange form factors of the nucleon \cite{He, He2}, to PDFs \cite{He6} and TMDs such as the Sivers function \cite{Salamu, Salamu2, He3}, and most recently to sea quark contributions to nucleon GPDs \cite{He5}.
In addition, assuming the nonlocal behavior to be a general property of all interactions, an example is the application to the lepton anomalous magnetic moments \cite{He4}.

\subsection{\it Outline}

After motivating the subject and the scope of this review, in Sec.~\ref{Sec.2} we begin the technical discussion by introducing the basic elements of chiral perturbation theory, including both the relativistic and heavy baryon formalisms, as well as different regularization methods.
The nonlocal chiral effective theory of baryons and pseudoscalar mesons is presented in Sec.~\ref{Sec.3}, where we start from the nonlocal Lagrangian and derive the corresponding Feynman rules, before discussing also nonlocal methods at the quark level.
Applications of the nonlocal EFT are presented in the following three sections.
First, nucleon form factors are discussed in Sec.~\ref{Sec.4}, with the individual $u$, $d$ and $s$ quark contributions featured, as well as the form factors of octet baryons.
Next, contributions to parton distributions of the proton from meson loops are presented in Sec.~\ref{Sec.5}, including the unpolarized light antiquark ($\bar{d}-\bar{u}$) and strange-antistrange ($s-\bar{s}$) asymmetries, and the spin dependent strange quark distribution ($\Delta s$).
The extension of the convolution framework to transverse momentum dependent PDFs, such as the Sivers function, and GPDs is outlined in Sec.~\ref{Sec.6}, where we focus in particular on the sea quark contributions. 
Finally, Sec.~\ref{Sec.7} summarizes the presentation and notes future applications of the nonlocal chiral EFT.
In the Appendix, Sec.~\ref{Sec.8}, we extend the nonlocal EFT to nonlocal quantum electrodynamics (QED) and discuss the anomalous magnetic moments of leptons.

\clearpage
\section{Chiral effective theory}
\label{Sec.2}

\subsection{\it Chiral effective Lagrangian \label{sec:Lagrangian}}

The QCD Lagrangian obtained by the gauge principle can be written as 
\begin{equation}
{\cal L}_{{\rm QCD}} = \sum_f \bar{q}_f^c\, (i{\gamma^\mu D_\mu} - m_f)\, q_f^c
- \frac14 {\cal G}_{\mu\nu}^a\, {\cal G}^{\mu\nu}_a,
\end{equation}
where $f (=u,d,s,c,b,t)$ is the quark flavor index, $c$ denotes the color index, and 
    $D_\mu = \partial_\mu - \frac12 i g \lambda^a{\cal A}_\mu^a$ 
is the covariant derivative.
The eight Gell-Mann matrices are represented by $\lambda^a (a=1,...,8)$, and the gluon field strength tensor ${\cal G}_{\mu\nu}^a$ is expressed as
\begin{equation} 
{\cal G}_{\mu\nu}^a 
= \partial_\mu {\cal A}_\nu^a 
- \partial_\nu {\cal A}_\mu^a 
+ g f_{abc}\, {\cal A}_\mu^b {\cal A}_\nu^c,
\end{equation} 
where $f_{abc}$ are the SU(3) structure constants.
The masses of the three light quarks $u$, $d$, and $s$ are much smaller than those of the heavier quarks.
Their masses are also small in comparison with the masses of typical light hadrons, such as the $\rho$ meson (770~MeV) or the proton (938~MeV).
In the limit where the masses of the light quarks go to zero, the left-handed and right-handed quark fields are decoupled from each other in the QCD Lagrangian.
In other words, besides the local SU(3) color symmetry, the QCD Lagrangian exhibits a further global SU(3)$_L$ $\times$ SU(3)$_R$ symmetry.

In order to exhibit the global symmetry of the QCD Lagrangian for $u$, $d$, and $s$ quarks in the chiral limit ($\chi$ lim, $m_f \to 0$), the right-handed and left-handed projection operators are introduced as
\begin{equation}
P_R = \frac12 (1+\gamma_5) = P_R^\dag, ~~~ 
P_L = \frac12 (1-\gamma_5) = P_L^\dag,
\end{equation} 
where the subscripts $R$ and $L$ refer to right-handed and left-handed, respectively. 
With the defined projection operators, the right-handed and left-handed quark fields are obtained as
\begin{equation}
q_R = P_R\, q, ~~~ q_L = P_L\, q.
\end{equation}
The QCD Lagrangian in the chiral limit for the light quarks is written as
\begin{equation}
{\cal L}_{{\rm QCD}}^{\rm \chi\, lim}
= \sum_{f=u,d,s} \bar{q}_f\, i{\gamma^\mu D_\mu}\, q_f
- \frac14 {\cal G}_{\mu\nu}^a\, {\cal G}^{\mu\nu}_a,
\end{equation}
where the color indices are now omitted.
It is obvious that ${\cal L}_{{\rm QCD}}^{\chi\, lim}$ is invariant under the following SU(3)$_R$ $\times$ SU(3)$_L$ transformation
\begin{subequations}
\begin{eqnarray} 
\left( \begin{array}{c} u_L \\ d_L \\s_L \end{array}\right)' 
= {\rm exp} \left( -\frac{i}{2}\sum^8_{a=1}\theta_a^L\, \lambda^a \right)
\left( \begin{array}{c} u_L \\ d_L \\s_L \end{array}\right),
\\ 
\left( \begin{array}{c} u_R \\ d_R \\s_R \end{array}\right)'
= {\rm exp} \left( -\frac{i}{2}\sum^8_{a=1}\theta_a^R\, \lambda^a \right)
\left( \begin{array}{c} u_R \\ d_R \\s_R \end{array}\right),
\end{eqnarray}
\end{subequations}
where $\theta_a^L$ and $\theta_a^R$ are independent numbers.

On the other hand, the chiral SU(3)$_L$ $\times$ SU(3)$_R$ symmetry is not realized in the low-energy spectrum.
Applying one of the axial generators $Q_A^a$ ($= Q_R^a-Q_L^a$) to an arbitrary state of a given multiplet of well-defined parity, one would obtain a degenerate state of opposite parity, which contradicts experiment.
In other words, the ground state is not invariant under the same symmetry group of the Lagrangian, and the symmetry is spontaneously broken. 
The light pseudoscalar octet mesons qualify as candidates for these Goldstone bosons.
The finite masses of the physical multiplet are interpreted as a consequence of the explicit breaking of the chiral symmetry due to the finite values of the $u$, $d$, and $s$ quark masses in the QCD Lagrangian.

Because of the absence of analytical tools to derive {\it ab initio} descriptions of low-energy properties and processes directly from QCD, EFTs based on hadronic degrees of freedom were proposed, with the same symmetries as QCD. 
The starting point is the hypothesis that a perturbative description is possible in terms of the most general effective Lagrangian containing all possible terms compatible with the assumed symmetry principles \cite{Weinberg, Weinberg2}.
In our case, the relevant group is
    $G=$ SU(3) $\times$ SU(3) 
      = $\{(L,R)| L \in$ SU(3), $R \in$ SU(3)\}. 
The pseudoscalar fields can be arranged into the SU(3) matrix $U(x) = \exp(i\phi(x)/f)$, where $f$ is the pion decay constant, and the matrix $\phi(x)$ is given by
\begin{eqnarray}\label{eq:meson} \nonumber
\phi(x) = \sum^8_{a=1}\lambda_a \phi_a(x) 
&=& \left(
\begin{array}{ccc}
\phi_3 + \frac{1}{\sqrt3}\phi_8 &
\phi_1 - i\phi_2 &
\phi_4 - i\phi_5 \\
\phi_1 + i\phi_2 & 
-\phi_3 + \frac{1}{\sqrt3}\phi_8 &
 \phi_6 - i\phi_7 \\
\phi_4 + i\phi_5 & 
\phi_6 + i\phi_7 &
-\frac{2}{\sqrt3}\phi_8 
\end{array}
\right) \\
&=& \left( 
\begin{array}{ccc}
\pi^0 + \frac{1}{\sqrt3}\eta &
\sqrt2 \pi^+ & 
\sqrt2 K^+ \\
\sqrt2 \pi^- &
-\pi^0 + \frac{1}{\sqrt3}\eta &
\sqrt2 K^0 \\
\sqrt2 K^- & 
\sqrt2\, \overline{K}^0 &
-\frac{2}{\sqrt3}\eta
\end{array}
\right)
\end{eqnarray} 
in terms of the pion, kaon, and $\eta$ fields.
Under the SU(3)$_L\, \times$ SU(3)$_R$ group, the matrix $U$ transforms as
    $U \to U' =  R\, U L^\dag$.
The lowest-order effective Lagrangian for the pseudoscalar mesons which has the global SU(3)$_L\, \times$ SU(3)$_R$ symmetry is given by
\begin{equation}
{\cal L}_M^{(2)}
= \frac{f^2}{4}{\rm Tr} \big[ \partial_\mu U \partial^\mu U \big].
\end{equation}
The quark mass term of QCD which explicitly breaks the chiral symmetry can be written as
\begin{equation}
{\cal L}^{\rm QCD}_{\rm mass} = - \bar{q}_R M q_L - \bar{q}_L M^\dag q_R, ~~~~ M=\left( \begin{array}{ccc} m_u & 0 & 0 \\
0 & m_d & 0 \\ 0 & 0 & m_s \end{array} \right).
\end{equation}
Although the mass matrix $M$ is a constant matrix and does not transform along with the quark fields, the Lagrangian ${\cal L}^{\rm QCD}_{\rm mass}$ would be invariant if $M$ transformed as $M \to M'= RML^\dag$.
One may construct the general effective Lagrangian ${\cal L}(U,M)$ which is invariant if $M$ has the same transformation property as $U$.
The lowest order symmetry breaking term in the effective theory is then written as
\begin{equation}
{\cal L}_{\rm mass}^{(2)}
= \frac{f^2 B_0}{2} {\rm Tr}\big[  MU^\dag + U M^\dag \big],
\end{equation}
where $B_0$ is related to the chiral quark condensate by
    $B_0 = -\langle\bar{q}q\rangle / (3f^2)$.

The effective Lagrangian which has the global SU(3)$_L\, \times$ SU(3)$_R$ symmetry can be generalized to have a local symmetry by introducing external fields \cite{Gasser, Gasser2, Scherer}.
The matrix $U$ in this case transforms as
    $U \to U' = V_R\, U\, V_L^\dag$,
where $V_R(x)$ and $V_L(x)$ are independent spacetime dependent SU(3) matrices.
As in the case of gauge theories, we define external fields $r_\mu^a(x)$ and $l_\mu^a(x)$ as combinations of external vector and axial-vector fields, $v_\mu^a(x)$ and $a_\mu^a(x)$,
\begin{subequations}
\begin{eqnarray}
r_\mu^a(x) &=& v_\mu^a(x) + a_\mu^a(x), \\
l_\mu(x)   &=& v_\mu^a(x) - a_\mu^a(x),
\end{eqnarray}
\end{subequations}
which transform under SU(3)$_L\, \times$ SU(3)$_R$ as 
\begin{subequations}
\begin{eqnarray}
r_\mu\ \to\ r_\mu'
&=& V_R\, r_\mu V_R^\dag + i V_R\,\partial_\mu V_R^\dag, 
\\
l_\mu\ \, \to\ \, l_\mu' 
&=& V_L\, l_\mu\, V_L^\dag + i V_L\,\partial_\mu V_L^\dag,
\end{eqnarray}
\end{subequations}
where we use the shorthand notation
$r_\mu \equiv \frac12 (\lambda^a r_\mu^a)$ and 
$l_\mu \equiv \frac12 (\lambda^a l_\mu^a)$.
Similarly for the mass term, the scalar and pseudoscalar fields $s^a$ and $p^a$ can be introduced, and transform according to
\begin{subequations}
\begin{eqnarray}
(s+ip)\ \to\ (s+ip)'&=& V_R\, (s+ip)\, V_L^\dag,\\
(s-ip)\ \to\ (s-ip)'&=& V_L\, (s-ip)\, V_R^\dag,
\end{eqnarray}
\end{subequations}
where $s \equiv \lambda^a s^a$ and $p \equiv \lambda^a p^a$.
We should note that the singlet fields can also be included with $a=0$, with the corresponding Gell-Mann matrix $\lambda^0 =$ diag(1,1,1).

For any operator $A$ (such as $U$, for instance) transforming as $A \to V_RAV_L^\dag$, the covariant derivative $D_\mu A$ is defined as 
    $D_\mu A = \partial_\mu A - i r_\mu A + i A\, l_\mu$,
and transforms as
\begin{eqnarray}\nonumber
(D_\mu A)\, \to\, (D_\mu A)'
&=& V_R\, (D_\mu A)\, V_L^\dag .
\end{eqnarray}
The general effective Lagrangian for mesons at lowest order which is locally SU(3)$_L\, \times$ SU(3)$_R$ invariant can therefore be written as
\begin{equation}
L_M^{(2)}
= \frac{f^2}{2} {\rm Tr}\big[ D_\mu U (D^\mu U)^\dag \big]
+ \frac{f^2}{4} {\rm Tr}\big[ \chi U^\dag + U\chi^\dag \big],
\end{equation}
where we define $\chi \equiv 2 B_0 (s+ip)$.
The nonzero vacuum expectation values of $s^0$, $s^3$ and $s^8$ result in nonzero quark masses, which generate the pseudoscalar meson masses.

The general chiral effective Lagrangian can be constructed order by order in terms of small external momentum or pion mass.
At ${\cal O}(q^4)$, for example, this is given by \cite{Gasser2, Scherer}
\begin{eqnarray}\nonumber 
{\cal L}_M^{(4)}
&=& L_1 \big\{ {\rm Tr}\big[D_\mu U (D^\mu U)^\dag\big] \big\}^2 
 +  L_2 {\rm Tr}\big[D_\mu U (D_\nu U)^\dag\big] {\rm Tr}\big[D^\mu U (D^\nu U)^\dag\big] 
\\ \nonumber
&+& L_3 {\rm Tr}\big[D_\mu U (D^\mu U)^\dag D_\nu U (D^\nu U)^\dag\big]
 +  L_4 {\rm Tr}\big[D_\mu U (D^\mu U)^\dag\big] {\rm Tr}\big[\chi U^\dag + U\chi^\dag\big]
\\ \nonumber
&+& L_5  {\rm Tr}\big[D_\mu U (D^\mu U)^\dag (\chi U^\dag + U\chi^\dag)\big]
 +  L_6 \big\{ {\rm Tr}\big[\chi U^\dag + U\chi^\dag\big] \big\}^2 
\\ \nonumber
&+& L_7 \big\{ {\rm Tr}\big[\chi U^\dag - U\chi^\dag\big] \big\}^2
 +  L_8 {\rm Tr}\big[\chi U^\dag \chi U^\dag+U\chi^\dag U\chi^\dag\big] 
\\ \nonumber
&-& iL_9 {\rm Tr}\big[f_{\mu\nu}^R D^\mu U (D^\nu U)^\dag + f_{\mu\nu}^L (D^\mu U)^\dag D^\nu U\big]
 +  L_{10} {\rm Tr}\big[U f_{\mu\nu}^L U^\dag f^{\mu\nu}_R\big] \\
&+& L_{11} {\rm Tr}\big[f_{\mu\nu}^R f^{\mu\nu}_R + f_{\mu\nu}^L f^{\mu\nu}_L\big]
 +  L_{12} {\rm Tr}\big[\chi \chi^\dag\big],
\end{eqnarray}
where the coefficients $L_i$ are the low energy constants.
The field strength tensors $f_{\mu\nu}^R$ and $f_{\mu\nu}^L$ are defined as
\begin{subequations}
\begin{eqnarray}
f_{\mu\nu}^R 
&=& \partial_\mu r_\nu - \partial_\nu r_\mu - i [r_\mu, r_\nu], 
\\
f_{\mu\nu}^L 
&=& \partial_\mu\, l_\nu - \partial_\nu\, l_\mu - i [\,l_\mu, \,l_\nu].
\end{eqnarray}
\end{subequations}
With dimensional regularization, the divergences from loop diagrams cancel with the bare low energy constants $L_i$, resulting in the renormalized constants $L_i^r(\mu)$.
The sum of the scale-dependent constants $L_i^r(\mu)$ and the finite loop contributions leads to scale independent results for physical observables.
The number of free parameters increases rapidly when the order of the effective Lagrangian increases from ${\cal O}(q^2)$ to ${\cal O}(q^4)$, and many more low energy constants will appear at higher orders.
For the higher order Lagrangian, see, for example, Ref.~\cite{Scherer} and references therein.

To construct the effective Lagrangian for baryons, an appropriate transformation law for the baryon fields is needed.
There is some degree of freedom in choosing how the baryon fields transform under \mbox{SU(3)$_L\, \times$ SU(3)$_R$}.
The specific representation of baryons does not matter, however, since the pseudoscalar meson matrix $U$ can be used to transform from one representation to another \cite{Georgi}.
Defining the square root matrix $u$ by $u^2 \equiv U$, this transforms as
\begin{equation}
u\, \to\, u' = \sqrt{RUL^\dag} \equiv RuK^\dag \equiv KuL^\dag ,
\end{equation} 
where $K$ depends on $U$, $L$ and $R$,
\begin{equation}
K= (\sqrt{RUL^\dag})^\dag R\sqrt{U}=\sqrt{RUL^\dag}L(\sqrt{U})^\dag.
\end{equation}  
The octet baryons can be arranged in the traceless $3 \times 3$ matrix $B$ as
\begin{equation}
B = \sum_{a=1}^8 \lambda^a B^a = \left( \begin{array}{ccc} \frac{1}{\sqrt{2}}\Sigma^0 + \frac{1}{\sqrt{6}}\Lambda & \Sigma^+ & p \\ \Sigma^- & -\frac{1}{\sqrt{2}}\Sigma^0 + \frac{1}{\sqrt{6}}\Lambda & n \\ \Xi^- & \Xi^0 &\ -\frac{2}{\sqrt{6}}\Lambda \end{array}\right),
\end{equation}
which transforms as $B \to B' = K B K^\dag$.
One can construct the locally \mbox{SU(3)$_L\, \times$ SU(3)$_R$} invariant effective Lagrangian for baryons with external fields in a similar way as for mesons.
The covariant derivative for baryons, $D_\mu B$, is written as 
\begin{equation}
D_\mu B = \partial_\mu B + \left[\Gamma_\mu, B \right], 
\end{equation}
where $\Gamma_\mu$ is defined as
\begin{equation}
\Gamma_\mu
= \frac12
  \big[ u^\dag(\partial_\mu -i r_\mu)u + u(\partial_\mu - i l_\mu)u^\dag \big],
\end{equation}
and transforms as
\begin{eqnarray}\nonumber
D_\mu B\ \to\ (D_\mu B)'
&=& \partial_\mu (KBK^\dag) 
+ \frac12 
  \Big[ 
    K u^\dag R^\dag
    (\partial_\mu - i R r_\mu R^\dag + R \partial_\mu R^\dag) R u K^\dag
\\ \nonumber
& & \hspace*{2.7cm}
 +\, K u\, L^\dag 
    (\partial_\mu - i L\, l_\mu\, L^\dag + L\partial_\mu L^\dag)\, L u^\dag K^\dag, K B K^\dag 
  \Big] 
\\
&=& K \big\{ (\partial_\mu B + [\Gamma_\mu, B] \big\} K^\dag\,
 =\,K (D_\mu B) K^\dag,
\end{eqnarray}
where for any unitary operator $O$ ($=L$, $R$, $u$ or $K$), one has the identity $\partial_\mu O O^\dag = -O\partial_\mu O^\dag$.
At ${\cal O}(q)$ there exists an axial-vector operator $u_\mu$ given by
\begin{equation}
u_\mu = i\left [ u^\dag (\partial_\mu - i r_\mu )u - u ( \partial_\mu -i l_\mu)u^\dag \right ],
\end{equation}
which transforms under SU(3)$_L\, \times$ SU(3)$_R$ as
\begin{eqnarray} 
u_\mu\ \to\ u_\mu' 
&=& K u_\mu K^\dag.
\end{eqnarray}
The locally chiral invariant Lagrangian can be constructed from the covariant derivative and $u_\mu$ field, and in the SU(3) case at the lowest order is written as 
\begin{equation} \label{eq:LOCT}
{\cal L}^{(1)}_{BM} 
= {\rm Tr} \big[ \bar{B}(i\slashed D - M_B)B \big]
- \frac{D}{2} {\rm Tr} \big[ \bar{B}\gamma^\mu \gamma_5 \{ u_\mu, B \} \big] 
- \frac{F}{2} {\rm Tr} \big[ \bar{B}\gamma^\mu \gamma_5 [ u_\mu, B ] \big],
\end{equation}
where $M_B$ is the octet baryon mass, and $D$ and $F$ are the meson-octet baryon-octet baryon coupling constants.

The decuplet baryon fields $T^{ijk}_\mu$, which include the $\Delta$, $\Sigma^*$, $\Xi^*$ and $\Omega^-$ baryons, are represented by a symmetric tensors with components,
\begin{eqnarray}
T^{111} = \Delta^{++}, ~~ 
T^{112} = \Delta^{+}, ~~ 
T^{122} = \Delta^{0}, ~~ 
T^{222} = \Delta^{-}, 
\nonumber \\
T^{113} = \tfrac{1}{\sqrt3}\,\Sigma^*, ~~ 
T^{123} = \tfrac{1}{\sqrt6}\,\Sigma^0, ~~
T^{223} = \tfrac{1}{\sqrt3}\,\Sigma^-, ~~~
\nonumber \\
T^{133} = \tfrac{1}{\sqrt3}\,\Xi^{*-}, ~~ 
T^{233} = \tfrac{1}{\sqrt3}\,\Xi^{*-}, ~~~~~~~~~~~~
\\
T^{333} = \Omega^-. ~~~~~~~~~~~~~~~~~~~~~~~~
\nonumber
\end{eqnarray}
The decuplet fields transform as tensors under SU(3)$_L\, \times$ SU(3)$_R$,
\begin{equation}
(T_\mu^{ijk})' = K_{il} K_{jm} K_{kn} T_\mu^{lmn}.
\end{equation}
The covariant derivative of the decuplet field is defined by 
\begin{equation}
D_\mu T_\nu^{ijk} = \partial_\mu T_\nu^{ijk} + \Gamma_\mu^{il} T_\nu^{ljk} + \Gamma_\mu^{jl} T_\nu^{ilk} + \Gamma_\mu^{kl} T_\nu^{ijl},
\end{equation}
and also transforms as
\begin{eqnarray} \nonumber
D_\mu T_\nu^{ijk}\ \to\
(D_\mu T_\nu^{ijk})' 
= K_{il} K_{jm} K_{kn} \left( \partial_\mu T_\nu^{lmn} + \Gamma_\mu^{lp} T_\nu^{pmn} + \Gamma_\mu^{mp} T_\nu^{lpn} + \Gamma_\mu^{np} T_\nu^{lmp} \right) = K_{il} K_{jm} K_{kn} D_\mu T_\nu^{lmn}. 
\end{eqnarray}

The lowest order chiral invariant Lagrangian for the decuplet baryons can be written as
\begin{equation}\label{eq:LDEC}
{\cal L}_{TM}^{(1)} 
= \overline{T}_\mu^{ijk}
  \big( i\gamma^{\mu\nu\alpha} D_\alpha - M_T \gamma^{\mu\nu} \big)
  T_\nu^{ijk}
- \frac{{\cal H}}{2}\,\overline{T}_\mu^{ijk}\gamma^\alpha \gamma_5
  (u_\alpha)^{kl} T^\mu_{ijl} 
- \frac{{\cal C}}{2}\,
  \big[ \epsilon^{ijk}\overline{T}_\mu^{ilm}\Theta^{\mu\nu}(u_\nu)_{lj}B^{mk} + {\rm H.c.} 
  \big],
\end{equation}
where ${\cal C}$ and ${\cal H}$ are the meson-octet-decuplet baryon and meson-decuplet-decuplet baryon coupling constants, respectively.
The tensor operator $\Theta^{\mu\nu}$ in Eq.~(\ref{eq:LDEC}) is defined as
\begin{equation}
\Theta^{\mu\nu} = g^{\mu\nu} - \Big( Z + \frac12 \Big) \gamma^\mu \gamma^\nu,
\end{equation}
where $Z$ is the off-shell parameter. 
In practice, the lowest-order Lagrangians are mostly used in calculations, and the high order ones can be found in Refs.~\cite{Gasser3, Fettes}.
With the effective Lagrangian, one can proceed to calculate physical observables or processes at tree or loop level.

\subsection{\it Heavy baryon chiral perturbation theory \label{subsec:green}}

In addition to the weak decay constant $f$, the baryonic effective Lagrangian introduces another energy scale into the problem --- the nucleon mass, which does not vanish in the chiral limit.
A method has been devised that separates an external nucleon four-momentum into a large piece, of the order of the nucleon mass, and a small residual component.
The approach is similar to the nonrelativistic reduction of Foldy and Wouthuysen, which provides a systematic procedure to diagonalize a relativistic Hamiltonian to any desired order in $1/M_N$ \cite{Foldy}.

The heavy baryon formulation of $\chi$PT involves an expansion in terms of $q/(4\pi f)$ and $q/M_N$, where $q$ is the (small) external momentum \cite{Jenkins}.
We can consider the nucleon in the SU(2) case, as an example, where the relativistic Lagrangian is given by
\begin{equation}
{\cal L}_{N\pi}^{(1)}
= \overline{\Psi} \left( 
    i\slashed {D} - M_N + \frac{g_A}{2}\gamma^\mu \gamma_5 u_\mu 
  \right) \Psi, 
\end{equation}
where $\Psi$ is the nucleon field, and $g_A = D + F$ is the axial-vector charge of the nucleon. 
The covariant derivative $D_\mu$ and axial vector $u_\mu$ are defined as above, except the Gell-Mann matrices $\lambda^a$ are replaced by the Pauli matrices, $\tau^a$.
The equation of motion for the nucleon field is written as 
\begin{equation} \label{eom}
 \left( i\slashed {D} - M_N + \frac{g_A}{2}\gamma^\mu \gamma_5 u_\mu \right) \Psi = 0.
\end{equation}
For a four vector $v_\mu$ with the properties $v^2 = 1$ and $v_0 \leq 1$, projection operators can be defined as
\be
P_{v+} = \frac{1+ \slashed v}{2}, ~~~~~ 
P_{v-} = \frac{1- \slashed v}{2},
\ee
satisfying the relations
\be
P_{v+} + P_{v-} = 1, ~~~~~~ P_{v\pm}^2 = P_{v\pm}, ~~~~~~ P_{v\pm} P_{v\mp} = 0.
\ee
The nucleon field $\Psi$ can be separated into two velocity dependent fields, ${\cal N}_v$ and ${\cal H}_v$,
\be \label{hl}
\Psi(x) = e^{-iM_N v\cdot x} \big[\, {\cal N}_v(x) + {\cal H}_v(x)\, \big],
\ee
where ${\cal N}_v$ and ${\cal H}_v$ are written in terms of the projections as
\be \label{HL}
{\cal N}_v = e^{iM_Nv\cdot x}P_{v+} \Psi, ~~~~~~~
{\cal H}_v = e^{iM_Nv\cdot x}P_{v-} \Psi,
\ee
and satisfy the properties
\be
P_{v+}{\cal N}_v = {\cal N}_v, ~~~~~
P_{v-}{\cal H}_v = {\cal H}_v, ~~~~~
P_{v+}{\cal H}_v = P_{v-}{\cal N}_v = 0.
\ee
Inserting Eq.~(\ref{hl}) into Eq.~(\ref{eom}), we obtain
\be \label{eq.NHrelation}
\left( i\slashed D + \frac{g_A}{2}\gamma^\mu \gamma_5 u_\mu 
\right) {\cal N}_v
+ 
\left( i\slashed {D} - 2M_N + \frac{g_A}{2}\gamma^\mu \gamma_5 u_\mu 
\right) {\cal H}_v = 0.
\ee
Furthermore, multiplying $P_{v+}$ and $P_{v-}$ on both sides of Eq.~(\ref{eq.NHrelation}) and using $\gamma$ matrix algebra, we have the relations
\begin{subequations}
\label{eq.NHequations}
\bea 
\left(iv\cdot D + \frac{g_A}{2}\gamma_\mu \gamma_5 u_{\perp}^\mu \right) {\cal N}_v
+ \left( i\slashed D_{\perp} + \frac{g_A}{2}v\cdot u \gamma_5 \right) 
{\cal H}_v &=& 0, \\
\left(i\slashed D_{\perp} - \frac{g_A}{2}v\cdot u \gamma_5 \right) {\cal N}_v
+ \left( -2M_N -iv\cdot D + \frac{g_A}{2}\gamma_\mu \gamma_5 u_{\perp}^\mu \right) 
{\cal H}_v &=& 0,
\eea
\end{subequations}
where for any vector $a_\mu$ we define
    $a^\mu_\perp \equiv a^\mu - v\cdot a\, v^\mu$.
One can solve Eqs.~(\ref{eq.NHequations}) for ${\cal H}_v$, which yields
\be
{\cal H}_v 
= \left( 2M_N + iv\cdot D - \frac{g_A}{2}\gamma_\mu \gamma_5 u_{\perp}^\mu
  \right)^{-1}
\left( i\slashed D_{\perp} - \frac{g_A}{2}v\cdot u \gamma_5 \right) 
{\cal N}_v.
\ee
Since ${\cal H}_v$ is formally suppressed by powers of $1/M_N$ relative to ${\cal N}_v$, the fields ${\cal N}_v$ and ${\cal H}_v$ are often called light and heavy components of the field $\Psi$, respectively.
The equation of motion for the light component ${\cal N}_v$ is
\bea \nonumber
\left(iv\cdot D + \frac{g_A}{2}\gamma_\mu \gamma_5 u_{\perp}^\mu \right) {\cal N}_v\!
&+&\! \left( i\slashed D_{\perp} + \frac{g_A}{2}v\cdot u \gamma_5 \right)
\\ 
\hspace*{4cm} && \times 
\left( 2M_N + iv\cdot D - \frac{g_A}{2}\gamma_\mu \gamma_5 u_{\perp}^\mu \right)^{-1}
\left(i\slashed D_{\perp} - \frac{g_A}{2}v\cdot u \gamma_5 \right) {\cal N}_v\ =\ 0,
\eea
and the corresponding effective Lagrangian for the light component can be written as
\bea \nonumber
{\cal L}_{{\rm eff}} 
&=& \overline{\cal N}_v \left(iv\cdot D + \frac{g_A}{2}\gamma_\mu \gamma_5 u_{\perp}^\mu \right) 
{\cal N}_v
\\ 
&+& \overline{\cal N}_v \left( i\slashed D_{\perp} + \frac{g_A}{2}v\cdot u \gamma_5 \right) 
\left( 2M_N + iv\cdot D - \frac{g_A}{2}\gamma_\mu \gamma_5 u_\perp^\mu \right)^{-1}
\left(i\slashed D_{\perp} - \frac{g_A}{2}v\cdot u \gamma_5 \right)
{\cal N}_v,
\eea
where the first term is the leading order Lagrangian.

To obtain the heavy baryon Lagrangian we can define the spin matrix $S_v^\mu$ according to \cite{Jenkins,Scherer}
\be
S_v^\mu = \frac{i}{2} \gamma_5 \sigma^{\mu\nu} v_\nu = -\frac12 (\gamma^\mu \slashed v - v^\mu),
\ee
in terms of which we can write combinations $\overline{\cal N}_v \Gamma {\cal N}_v$ for various $\Gamma$ as
\bea \nonumber
\overline{\cal N}_v \gamma_5\, {\cal N}_v 
= 0,
~~~~~~~ 
\overline{\cal N}_v \gamma^\mu\, {\cal N}_v 
= v^\mu \overline{\cal N}_v\, {\cal N}_v,
~~~~~~~
\overline{\cal N}_v \gamma^\mu \gamma_5\, {\cal N}_v 
= 2\,\overline{\cal N}_v\, S_v^\mu\, {\cal N}_v, 
\\
\overline{\cal N}_v \sigma_{\mu\nu}\, {\cal N}_v 
= 2 \epsilon_{\mu\nu\rho\sigma} v^\rho \overline{\cal N}_v S_v^\sigma {\cal N}_v,
~~~~~~~~
\overline{\cal N}_v\, \sigma^{\mu\nu}\gamma_5\, {\cal N}_v 
= 2i \big(v^\mu \overline{\cal N}_v S_v^\nu {\cal N}_v 
        - v^\nu \overline{\cal N}_v S_v^\mu {\cal N}_v \big).
\eea
The leading order effective Lagrangian in the heavy baryon formalism can then be written as
\be
{\cal L}_{v,N\pi}^{(1)}
= \overline{\cal N}_v \big[ i v\cdot D + g_A S_v \cdot u \big]{\cal N}_v.
\ee
In the SU(3) case, the lowest order Lagrangian generalizes to
\bea \nonumber
{\cal L}^{(1)}_{v,BM} 
&=& i\, {\rm Tr} \bar{B}_v(v \cdot D) B_v 
 +  D\, {\rm Tr} \bar{B}_v S_v^\mu \{ u_\mu, B_v \}
 +  F\, {\rm Tr} \bar{B}_v S_v^\mu [ u_\mu, B_v ]
\\
&-& i\, \overline{T}_{v,ijk}^\mu (v\cdot D) T_\mu^{v,ijk}
 +  \Delta \overline{T}_{v,ijk}^\mu T_\mu^{v,ijk}
 +  \frac{{\cal C}}{2}
    \left[ \epsilon^{ijk}\overline{T}_\mu^{ilm}\Theta^{\mu\nu}(u_\nu)_{lj}B^{mk} + {\rm H.c.}
    \right]
 + {\cal H}\overline{T}_\mu^{ijk}S^\alpha_v (u_\alpha)^{kl} T^\mu_{ijl},
\nonumber\\
&&
\eea
where $\Delta$ is the decuplet--octet baryon mass difference, 
    $\Delta= M_T - M_B$.

The free propagator of the octet baryon is written as
\be
G_{v,B}^0 (k) = \frac{iP_{v+}}{v\cdot k + i\epsilon},
\ee
which is the Fourier transformation of the Green function $G_{v,B}^0(x,x')$,  satisfying the equation
\be
v\cdot \partial\, G_{v,B}^0(x,x') = -\delta^4(x-x') P_{v+}.
\ee
From Eq.~(\ref{HL}), the derivative $i\partial_\mu$ acting on the light component ${\cal N}_v$ produces a residual momentum 
    $k_\mu = p_\mu - M_B\, v_\mu$. 
The actual choice of $v_\mu$ is a matter of convenience and makes the residual momentum $k_\mu$ small, since $v\cdot k \ll M_B$. 
Similarly, the free propagator for the decuplet baryon in the heavy baryon formalism is given by
\be
G_{v,T}^0 (k)
 = \frac{i \big[v^\mu v^\nu - g^{\mu\nu} -(4/3)S^\mu_v S^\nu_v] P_{v+}}
        {v\cdot k - \Delta + i\epsilon}.
\ee

Both meson and heavy baryon $\chi$PT can be formulated with a power counting scheme, when the external momentum of mesons, the momentum of the external sources, and the residual momentum of the baryon are all small compared with the scale of baryon mass or $4\pi f$.
Given the number of independent loop momenta $N_L$ in a Feynman diagram, the number of meson propagators $I_M$, the number of baryon propagators $I_B$, the number of mesonic vertices $N_M^{2n}$ generated from ${\cal L}_M^{(2n)}$, the number of baryon-meson vertices $N_{BM}^{n}$ generated from the Lagrangian ${\cal L}_{BM}^{(n)}$, the power or chiral dimension, $D$, of the diagram is given by
\be
D = 4 N_L - 2 I_M - I_B + \sum_{n=1}^{\infty} 2 n N_M^{2n} + \sum_{n=1}^{\infty} n N_{BM}^{n}.
\ee
Having outlined the formulation of the $\chi$PT in the heavy baryon case, in the following we generalize the discussion to the baryon $\chi$PT in the relativistic case.

\subsection{\it Infrared regularization}
\label{ssec:IR}

One qualitative difference between the meson and baryon sectors of $\chi$PT is that the baryons remain massive in the chiral limit, while the pesudoscalar meson masses vanish.
In the relativistic case, an additional mass scale $M_B$ therefore appears and consequently covariant $\chi$PT does not naturally satisfy the power counting scheme.
Many early applications of $\chi$PT in the baryonic sector were performed in the framework of the heavy baryon formalism.
It would be desirable to have a method which, on the one hand, is in the framework of the relativistic formalism, while, on the other hand, avoiding the shortcomings of the absence of power counting scheme.
The so-called infrared regularization method was proposed for this purpose, which decomposes one-loop diagrams into singular and regular parts \cite{Becher}.
The singular parts are shown to satisfy the power counting, while the regular parts are absorbed into local counter terms in the Lagrangian.

\begin{figure}[tb] 
\begin{center}
\begin{minipage}[t]{8 cm}
\epsfig{file=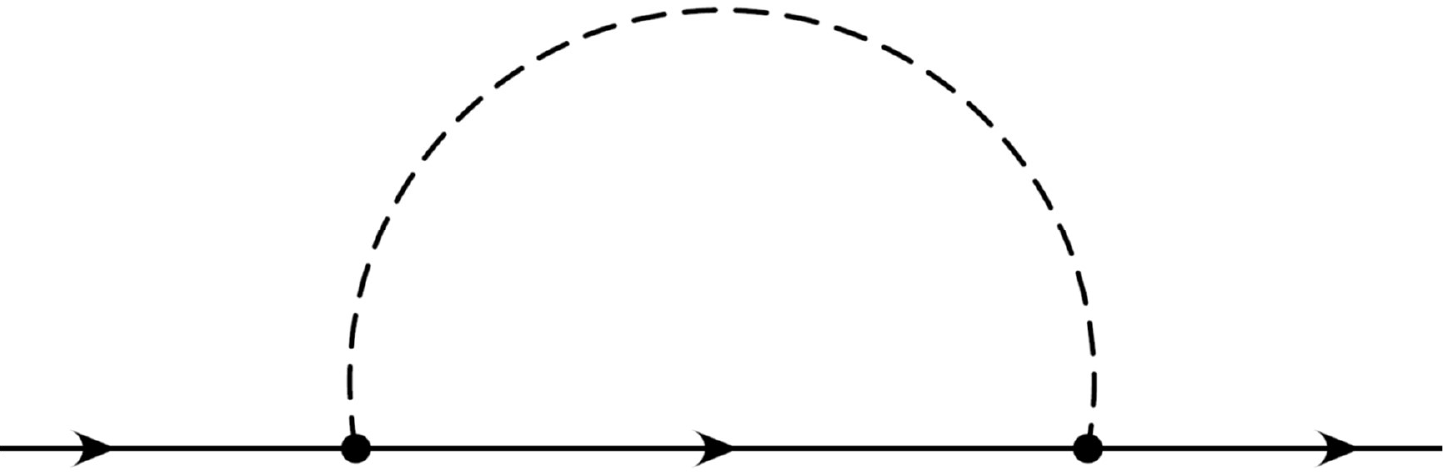,scale=0.5}
\end{minipage}
\begin{minipage}[t]{16.5 cm}
\caption{Nucleon (solid line) self-energy diagram with a pion loop (dashed line).}\label{fig1}
\end{minipage}
\end{center}
\end{figure}

As in Refs.~\cite{Becher, Scherer}, we can take the nucleon self-energy diagram in Fig.~\ref{fig1} as an example, where the momentum of the pion and external nucleon are $k$ and $p$, respectively.
We focus on the corresponding scalar loop integral,
\be
H(p^2) = -i \int\!\frac{\dd^d{k}}{(2\pi)^d} \frac{1}{(k^2 - m_\pi^2 + i \epsilon)\big((p-k)^2 - M_N^2 + i\epsilon\big)}.
\ee
For low values of the pion momentum $k$, the integrand is order of ${\cal O}(q^{-3})$ (note that the off-shellness $p^2 - M_N^2 = {\cal O}(q)$). 
The naive order of this integral for this part is ${\cal O}(q^{d-3})$. 
For large values of the momentum~$k$, such as $k \gg M_N$, we expect the power counting to fail.
Using the Feynman parameterization, $H(p^2)$ can be written as
\be \label{eq:Hdef}
H(p^2) = -i \int_0^1 \dd{z} \int \frac{\dd^d{k}}{(2\pi)^d} \frac{1}{(k^2 - A(z) + i\epsilon)^2},
\ee
where
\be
A(z) = z^2 p^2 - z(p^2 - M_N^2 + m_\pi^2) + m_\pi^2.
\ee
With dimensional regularization, after the $d$-dimension integration the function $H(p^2)$ becomes
\be \label{eq:HinA}
H(p^2) = \frac{\Gamma(2 - d/2)}{(4\pi)^{d/2}} \int_0^1 \dd{z} \left [ A(z) -i\epsilon \right ]^{d/2 - 2}.
\ee
When the external momentum of the nucleon is at threshold, 
    $p_{\rm thr} = M_N + m_\pi$, 
$A(z)$ reduces to the simple form $A(z) = (z p_{\rm thr} - m_\pi)^2$.
Direct integration over $z$ from 0 to 1 leads to mixing of the total contributions, which destroys the power counting rule.
It is preferable, therefore, to separate the contributions into two parts: one with the expected ${\cal O}(q^{d-3})$ behavior, and the other failing to have the expected order,
\be
H(p_{\rm thr}^2) 
= \frac{\Gamma(2 - d/2)}{(4\pi)^{d/2}} 
  \int_0^{z_0} \dd{z} \left [ A(z) -i\epsilon\right ]^{d/2 - 2}
+ \frac{\Gamma(2 - d/2)}{(4\pi)^{d/2}}
  \int_{z_0}^1 \dd{z} \left [ A(z) -i\epsilon\right ]^{d/2 - 2}.
\ee
Note that the simple integrand $A(z)$ is independent of the nucleon mass when $z=0$ or $\infty$, or when $z = m_\pi/p_{\rm thr}$.
We can choose, therefore, $z_0 = m_\pi/p_{\rm thr}$ (or $\infty$,) and define an infrared singular part,
\be
I = \frac{\Gamma(2 - d/2)}{(4\pi)^{d/2}} 
    \int_0^{z_0} \dd{z} 
    \big[ m_\pi - z\,p_{\rm thr} -i\epsilon \big]^{d - 4}
\ee
and an infrared regular part,
\be
R = H - I = \frac{\Gamma(2 - d/2)}{(4\pi)^{d/2}}
    \int_{z_0}^1 \dd{z} 
    \big[ z\, p_{\rm thr} - m_\pi -i\epsilon \big]^{d - 4},
\ee
the latter which can be absorbed in the low energy constants.
Through analytic continuation, for arbitrary $d$ one can express $I$ and $R$ as 
\be \label{infrared}
I = \frac{\Gamma(2-d/2)}{(4\pi)^{d/2}(d-3)} 
    \frac{m_\pi^{d-3}}{p_{\rm thr}},
\ee
and
\be
R = \frac{\Gamma(2-d/2)}{(4\pi)^{d/2}(d-3)} 
    \frac{M_N^{d-3}}{p_{\rm thr}}.
\ee
In the chiral limit $m_\pi \to 0$, the original integral is infrared singular for small $d$.

For a general external momentum which is not at threshold, we introduce two dimensionless variables,
\be
\alpha = \frac{m_\pi}{M_N}, ~~~~~~ \Omega = \frac{p^2-M_N^2-m_\pi^2}{2M_Nm_\pi},
\ee
in terms of which $A(z)$ can be rewritten as
\be
A(z) = M_N^2 \big[ z^2 -2\alpha \Omega z(1-z) + \alpha^2 (1-z)^2 \big].
\ee
After performing the momentum integration, we have
\be \label{HP}
H(p^2) = \kappa (d) \int_0^1 \dd{z} 
         \left[ B(z) -i\epsilon \right]^{d/2 - 2},
\ee
where 
\be
\kappa(d)= \frac{\Gamma(2-d/2)}{(4\pi)^{d/2}} M_N^{d-4}, ~~~{\rm and}~~~
B(z) = \frac{A(z)}{M_N^2}.
\ee
The infrared singularity originates from small values of $z$, where the function $B(z)$ goes to zero as $m_\pi \to 0$. 
To isolate the divergent part, one scales the integration variable $z = \alpha x$ so that the upper limit $z = 1$ in Eq.~(\ref{HP}) corresponds to $x = 1/\alpha$,
\be\label{eq:H-by-C}
H(p^2) = \kappa(d)\, \alpha^{d-3} \int_0^{1/\alpha} \dd{x} 
         \left[ C(x) -i\epsilon \right]^{d/2 - 2},
\ee
where 
\be
C(x) = (\alpha^2 + 2\alpha\Omega + 1)\, x^2 - 2(\alpha + \Omega)\, x + 1.
\ee
On the one hand, when $x$ is zero or $\infty$, $C(x)$ does not depend on $M_N$.
On the other hand, changing the upper limit of $x$ from $1/\alpha$ to $\infty$ does not change the behavior of infrared singularity.
The infrared singular part can therefore be written as
\be
I(p^2) 
= \kappa(d)\, \alpha^{d-3} \int_0^{\infty} \dd{x} 
  \left[ C(x) -i\epsilon \right]^{d/2 - 2}
= \kappa(d)\, \int_0^\infty \dd{z} 
  \left[ B(z) -i\epsilon \right]^{d/2 - 2}.
\ee
The infrared regular part is then given by
\be
R(p^2) = H(p^2) - I(p^2)
       = -\kappa(d)\, \int_1^\infty \dd{z}
         \left[ B(z) -i\epsilon \right]^{d/2 - 2}.
\ee
When $p$ is at threshold, we have $\Omega=1$ and the above expression for $I$ becomes 
\be
I(p^2_{\rm thr}) = \kappa(d)\, \alpha^{d-3} \int_0^\infty \dd{x} 
\left[ \big((\alpha+1)x-1\big)^2 -i\epsilon \right]^{d/2-2},
\ee
which converges for $d<3$.
Through partial integration and analytic continuation, for arbitrary $d$ we then have
\be \label{infrared2}
I(p^2_{\rm thr}) 
= \kappa(d) \alpha^{d-3} \frac{1}{(d-3)(\alpha+1)}
= \frac{\Gamma(2-d/2)}{(4\pi)^{d/2}(d-3)}
  \frac{m_\pi^{d-3}}{p_{\rm thr}},
\ee
which is the same expression as that in Eq.~(\ref{infrared}).

The basic idea of infrared regularization is to replace the general integral $H$ by its infrared singular part $I$ (changing the upper limit from 1 to $\infty$ after using the Feynman parameterization) and drop the regular part $R$. 
In the low-energy region, $H$ and $I$ have the same nonanalytic properties, whereas the contribution of $R$, which is an infinite series in the momentum, can be included by adjusting the coefficients of the most general effective Lagrangian.

Both the heavy baryon $\chi$PT and covariant $\chi$PT in infrared regularization satisfy the power counting scheme. 
To illustrate this we compare the results for the scalar loop integral, $I(p^2_{\rm thr})$.
Writing the four-momentum of the nucleon as $p=M_N v + r$, the nucleon propagator can be expanded in terms of a series in $r^2/(2M_N v\cdot r)$.
The original loop integral can then be rewritten as
\be \label{expand}
I(p^2) 
= \sum_j H_v^j(p^2)
= \frac{-i}{2M_N}\sum_j \int \frac{\dd^d{k}}{(2\pi)^d}
\frac{1}{(k^2-m_\pi^2+i\epsilon)(v\cdot(r-k)+i\epsilon)}
\left[\frac{-(r-k)^2}{2M_Nv\cdot(r-k)+i\epsilon}\right]^j,
\ee
where we have assumed that $r$ is small so that the original $H(p^2)$ can be approximated by its infrared part $I(p^2)$.
For simplicity, we discuss the threshold result for $p^2 = p^2_{\rm thr}$, or $r=m_\pi v$. 
The first term in Eq.~(\ref{expand}) with $j=0$ is the result in the heavy baryon limit, and is expressed as
\be
H_v^0(p^2_{\rm thr}) = \frac{-i}{2M_N}\int \frac{\dd^d{k}}{(2\pi)^d} \frac{1}{(k^2-2m_\pi v\cdot k+i\epsilon)(v\cdot k +i\epsilon)} = \frac{\Gamma(2-d/2)}{(4\pi)^{d/2}(d-3)}
\frac{m_\pi^{d-3}}{M_N}.
\ee
The higher order terms are related by a simple recursion relation,
\be
H_v^{j}(p^2_{\rm thr}) 
= -\frac{m_\pi}{M_N} H_v^{j-1}(p^2_{\rm thr}),~~~~~[j \geq 1]
\ee
which implies that
\be
H_v^{j}(p^2_{\rm thr}) 
= (-1)^j \left(\frac{m_\pi}{M_N}\right)^j H_v^0(p^2_{\rm thr}).
\ee
Consequently, we have
\be
I(p^2_{\rm thr}) 
= \sum_j H_v^j(p^2_{\rm thr})
= \frac{\Gamma(2-d/2)}{(4\pi)^{d/2}(d-3)}\frac{m_\pi^{d-3}}{p_{\rm thr}},
\ee
which is identical to the expression in Eq.~(\ref{infrared2}).
This example shows that the result with infrared regularization is related to an infinite sum of the heavy baryon results. 
Note also that the leading nonanalytic terms in the heavy baryon formalism and the relativistic baryon formalism with dimensional regularization and infrared regularization are all the same.

\subsection{\it EOMS regularization}

In addition to the infrared regularization scheme, another subtraction scheme which is commonly used in the literature is the extended on-mass-shell scheme (EOMS).
The basic idea of this renormalization scheme consists of providing a rule determining which terms of a given diagram should be subtracted in order to satisfy a ``naive'' power counting, and in this section we follow the discussion in Ref.~\cite{Fuchs}.
Considering the same integral $H(p^2)$ as in Sec.~\ref{ssec:IR}, Eqs.~(\ref{eq:Hdef}), (\ref{eq:HinA}), and (\ref{eq:H-by-C}),
after performing the $d$-dimension integral with $d\to 4$, we have 
\be \label{HEOM}
H(p^2) 
= -2\bar{\lambda} 
+ \frac{1}{16\pi^2} 
- \frac{1}{8\pi^2} 
  \frac{\alpha\sqrt{1-\Omega^2}}{(1+2\alpha\Omega+\alpha^2)}\,\arccos(-\Omega) 
- \frac{1}{8\pi^2} 
  \frac{\alpha(\alpha+\Omega)}{(1+2\alpha\Omega+\alpha^2)}\,\ln\alpha,
\ee
where
\be
\bar{\lambda} 
= \frac{M_N^{d-4}}{(4\pi)^2}
\bigg\{ \frac{1}{d-4} - \frac12 \big[\ln (4\pi) +\Gamma'(1)+1\big]\bigg\}.
\ee
Note here that $\Omega$ is of ${\cal O}(m_\pi^0)$ for $p^2 \neq M_N^2$ (it is of ${\cal O}({m_\pi})$ for $p^2 = M_N^2$) and $\alpha$ is of ${\cal O}({m_\pi})$.
We expect $H(p^2)$ to be of ${\cal O}(m_\pi)$ as $d \to 4$.
The first two terms of Eq.~(\ref{HEOM}) obviously violate the power counting.
The dimensionally regularized integral contains a part which, for non-integer $d$, is proportional to non-integer powers of $m_\pi$, but does not violate the power counting. 
On the other hand, the remaining piece of the integral may be expanded in non-negative powers of $m_\pi$ for arbitrary $d$, as indicated in the first two terms discussed above. 
It is this contribution that is responsible for the violation of power counting.

Subtracting the first and second terms from Eq.~(\ref{HEOM}), we can write for the renormalized integral,
\be \label{HREOMS}
H_R(p^2)
= -\frac{1}{8\pi^2}
  \frac{\alpha\sqrt{1-\Omega^2}}{(1+2\alpha\Omega+\alpha^2)}\,\arccos(-\Omega) 
- \frac{1}{8\pi^2}
  \frac{\alpha(\alpha+\Omega)}{(1+2\alpha\Omega+\alpha^2)}\,\ln\alpha.
\ee

It is instructive to compare Eq.~(\ref{HREOMS}) with the result of using infrared regularization, where the integral $H(p^2)$ is divided into an infrared part $I(p^2)$ and a remainder $R(p^2)$ \cite{Becher},
\be
I(p^2) 
= v\bar{\lambda} 
- \frac{1}{8\pi^2}
  \frac{\alpha\sqrt{1-\Omega^2}}{(1+2\alpha\Omega+\alpha^2)}\,
  \arccos\Big(\frac{-\alpha-\Omega}{\sqrt{1+2\alpha\Omega+\alpha^2}}\Big)
- \frac{1}{16\pi^2}
  \frac{\alpha(\alpha+\Omega)}{(1+2\alpha\Omega+\alpha^2)}
  \big( 2\ln\alpha - 1 \big),
\ee
and
\be
R(p^2)
= -(2+v)\bar{\lambda}
+ \frac{1}{8\pi^2}
  \frac{\alpha\sqrt{1-\Omega^2}}{(1+2\alpha\Omega+\alpha^2)}\,
  \arcsin\Big(\frac{\alpha\sqrt{1-\Omega^2}}{\sqrt{1+2\alpha\Omega+\alpha^2}}\Big)
+ \frac{1}{16\pi^2}
  \frac{1+\alpha\Omega}{(1+2\alpha\Omega+\alpha^2)},
\ee
where $v = -(p^2-M_N^2+m_\pi^2)/p^2$. 
The infrared singular and regular parts need further renormalization.
The subtracted parts can both be expanded in powers of $m_\pi$ and absorbed in the most general effective Lagrangian.
It is obvious that final results of infrared regularization and EOMS regularization give the same nonanalytic term which is proportional to $\ln\alpha$.

After integration it will be clear which terms should be subtracted to keep the power counting scheme valid.
However, it is useful to identify those terms which we subtract from a given integral without explicitly calculating the integral beforehand.
For simplicity, we restrict ourselves for the moment to the chiral limit, $m_\pi = 0$, in which case $H(p^2)$ can be written as
\bea
H(p^2)
&=& \frac{\Gamma(2-d/2)M_N^{d-4}}{(4\pi)^{d/2}}
\int_0^1 \dd{z} \big[z^2 - z(1-z)\delta-i\epsilon\big]^{d/2-2}
\nonumber \\
&=& \frac{M_N^{d-4}}{(4\pi)^{d/2}} 
\bigg[ \frac{\Gamma(2-d/2)}{d-3}F(1,2-d/2;4-d;-\delta)
\nonumber \\
&& \hspace*{1.5cm}
+\, (-\delta)^{d-3}\, \Gamma(d/2-1)\, \Gamma(3-d)\,  F(d/2-1,d-2;d-2;\delta)
\bigg],
\eea
where $\delta=(p^2-M_N^2)/M_N^2$ and $F(a,b;c;z)$ is the hypergeometric function \cite{Fuchs}.
Using the expansion
\be
F(a,b;c;z) = 1 + \frac{ab}{c}z + \frac{a(a+1)b(b+1)}{2c(c+1)}z^2 + \cdots,
\ee
as $d \to 4$ we have 
\be \label{M=0}
H(p^2)
= \frac{M_N^{d-4}}{(4\pi)^d/2}
\left[ \frac{\Gamma(2-d/2)}{d-3}
- \delta\ln(-\delta)
+ \delta^2\ln(-\delta)
+ \cdots 
\right],
\ee
where the ellipsis $\cdots$ represents terms that are at least of ${\cal O}(q^3)$.
The subtraction terms can therefore be chosen as
\bea \label{sub}
&&\sum_{l=0}^{N} \frac{(p^2-M_N^2)^l}{l!}\left[\left(\frac{p_\mu}{2p^2}\frac{\partial}{\partial p_\mu}\right)^l\frac{1}{(k^2+i\epsilon)\big((p-k)^2-M_N^2+i\epsilon\big)}\right]_{p^2=M_N^2} \nonumber \\
&=& \frac{1}{(k^2+i\epsilon)(k^2-2k\cdot p+i\epsilon)}{\bigg|}_{p^2=M_N^2} \nonumber \\
&+& (p^2-M_N^2) \left[ \frac{1}{2M_N^2}\frac{1}{(k^2-2k\cdot p + i\epsilon)^2}
-\frac{1}{2M_N^2}\frac{1}{(k^2+i\epsilon)\big((p-k)^2-M_N^2+i\epsilon\big)} \right. \nonumber \\
&& \left. \hspace*{2cm}-\, \frac{1}{(k^2+i\epsilon)(k^2-2k\cdot p+i\epsilon)^2}\right]_{p^2=M_N^2}
+ \cdots,
\eea
where $N$ is determined by the requirement that the terms have no infrared singularity.
These terms are always analytic in the small parameter and do not contain infrared singularities.
In the above example we only need to subtract the first term with $N=0$, while all higher-order terms contain infrared singularities.
For example, the infrared behavior of the first term is $k^3/k^3$ and this integrand has no infrared singularity; however, the last integrand in the second term generates a behavior $k^3/k^4$ and so is infrared singular.
The integral of the first term of Eq.~(\ref{sub}) is given by the first term of Eq.~(\ref{M=0}).
Overall, we find the renormalized integral to be
\be
H_R(p^2) 
= \frac{M_N^{d-4}}{(4\pi)^{d/2}}
\bigg[ 
  \Big( 1-\frac{p^2}{M_N^2} \Big) \ln\Big( 1-\frac{p^2}{M_N^2} \Big)
+ \Big( 1-\frac{p^2}{M_N^2} \Big)^2 \ln\Big(1-\frac{p^2}{M_N^2}\Big)
+ \cdots
\bigg].
\ee

For finite $m_\pi$, 
before explicitly calculating the integral we expand the integrand in terms of $m_\pi^2$, as in Eq.~(\ref{sub}), and subtract those terms which have no infrared singularities.
In the present case we only need to subtract the first term to satisfy the power counting,
\be 
H_{\rm subtr}
= -i \int \frac{d^d k}{(2\pi)^d} 
\frac{1}{(k^2 +i\epsilon)(k^2 - 2p\cdot k+i\epsilon)}\Big|_{p^2=M_N^2}.
\ee
This integral indeed gives the first two terms of Eq.~(\ref{HEOM}),
    $H_{\rm subtr}= -2\bar{\lambda} + 1/(4\pi)^2 + {\cal O}(d-4)$.
The final result will be the renormalized one which satisfies power counting, while in infrared regularization both $I$ and $R$ need further renormalization.
We stress that in covariant baryon $\chi$PT, when we say that an expression is of ${\cal O}(q^i)$, we refer to the minimal power is $q^i$.
This is in contrast to heavy baryon $\chi$PT or meson $\chi$PT, where the expression exclusively consists of terms of ${\cal O}(q^i)$. 
The results of infrared and EOMS regularizations have the same nonanalytic behavior and the leading nonanalytic term is the same as that in the heavy baryon formalism.
The analytic terms in both methods can be expanded in small quantities consisting of an infinite number of terms in the expansion.
These analytic terms can be different in the IR and EOMS schemes.

\subsection{\it Finite-range regularization}

In the preceding two subsections we have discussed the infrared and EOMS regularizations, which make the final results satisfy the conventional power counting.
The subtracted terms can be absorbed in the counter terms in the most general Lagrangian resulting in the renormalized low energy constants. 
These two regularization schemes lead to similar numerical results as with the standard dimensional regularization with proper choice of the renormalized low energy constants.
Numerical calculations of nucleon form factors indicate that the experimental data can only be described well at low momentum transfer, say $Q^2 < 0.1$~GeV$^2$ in infrared, EOMS, and dimensional regularization \cite{Bauer, Fuchs2}.

The situation is the same for the extrapolation of lattice data, where the quark (pion) masses are larger than the physical masses.
When $m_\pi$ is large, the power counting scheme is not valid and high order terms of $m_\pi$ have important contributions. 
Some form of resummation of the chiral expansion is therefore necessary in order to make the chiral perturbation theory applicable beyond the traditional power counting region.
The resummation of the chiral expansion induced through the introduction of a finite range cutoff in the momentum integrals of meson-loop diagrams has been shown \cite{Young2, Leinweber} to be a very good and effective resummation method.

Taking the nucleon mass as an example, in the usual effective field theory the nucleon mass can be expanded in terms of the pion mass up to ${\cal O}(m_\pi^4)$ as
\be
M_N = a_0 + a_2 m_\pi^2 + a_4 m_\pi^4 + \sigma_{\pi N} + \sigma_{\pi\Delta} + \cdots,
\ee
where $\sigma_{\pi N}$ and $\sigma_{\pi \Delta}$ are the self-energies from the $\pi$ loops with intermediate nucleon and $\Delta$, respectively. 
The coefficients $a_i$ $(i=0,2,4)$ are the low energy constants in the effective Lagrangian.
In the heavy baryon limit, the expressions for $\sigma_{\pi N}$ and $\sigma_{\pi\Delta}$ are given by
\be
\sigma_{\pi N} = -\frac{3(D+F)^2}{16\pi^2 f^2} \int_0^\infty \dd{k} \frac{\bm{k}^4}{{\bm k}^2 + m_\pi^2},
\ee
and
\be
\sigma_{\pi \Delta} = -\frac{{\cal C}^2}{6\pi^2 f^2} \int_0^\infty \dd{k} \frac{\bm{k}^4}{\omega(\bm{k})\big(\Delta+\omega(\bm{k})\big)},
\ee
where $\omega(\bm{k})=\sqrt{{\bm k}^2 + m_\pi^2}$, and $\Delta$ is the mass difference between $\Delta$ baryon and nucleon, with the integration measure \mbox{d$k \equiv$ d$|\bm{k}|$}. 
At large momentum these integrals have cubic divergences which need to be regulated, while their infrared behavior gives the leading nonanalytic correction to the nucleon mass.

With dimensional regularization, the nucleon mass $M_N$ can be written as \cite{Tiburzi}
\be
M_N
= a_0 + a_2 m_\pi^2 + a_4 m_\pi^4 
- \frac{3}{32\pi f^2}(D+F)^2 m_\pi^3 
- \frac{{\cal C}^2}{12\pi^2 f^2} {\cal F}(m_\pi,\Delta,\mu).
\ee
For $m_\pi < \Delta$, the function ${\cal F}(m_\pi,\Delta,\mu)$ is given by
\be\label{eq:calF}
{\cal F}(m_\pi,\Delta,\mu) 
= \left( m_\pi^2 - \Delta^2 \right) 
  \left[ 
    \sqrt{\Delta^2 - m_\pi^2}\,
    \ln \left( \frac{\Delta-\sqrt{\Delta^2-m_\pi^2}}{\Delta+\sqrt{\Delta^2-m_\pi^2}} \right) 
- \Delta \ln \frac{m_\pi^2}{\mu^2} 
  \right]
- \frac12 \Delta\, m_\pi^2 \ln \frac{m_\pi^2}{\mu^2}.
\ee
For $m_\pi > \Delta$, the first logarithm in Eq.~(\ref{eq:calF}) becomes an arctangent.
However, this formula also cannot describe lattice data at large pion masses.
To fit lattice data at large $m_\pi$, high order analytic terms from Lagrangian, such as $a_6 m_\pi^6$, are necessary to counter the large, nonanalytic contributions arising from terms such as the order $m_\pi^4 \ln m_\pi$ term.
Physically, when the pion mass is large the loop contribution should be small and vanish as $m_\pi \to \infty$.

In the finite range regularization (FRR), a $\Lambda$ dependent functional cutoff $u(\bm{k})$ is introduced in order to render the integral ultraviolet finite, 
\be\label{eq:sigma_piN_Lam}
\sigma_{\pi N}^\Lambda 
= -\frac{3(D+F)^2}{16\pi^2f^2} \int_0^\infty \dd{k} \frac{\bm{k}^4
    u^2(\bm{k})}{\bm{k}^2+m_\pi^2}.
\ee
For illustration, we choose a dipole regulator,
    $u(\bm{k}) = \Lambda^4/(\bm{k}^2+\Lambda^2)^2$,
which gives for the self-energy,
\bea \label{massNFRR}
\sigma_{\pi N}^\Lambda 
&=& -\frac{3(D+F)^2}{32\pi f^2} 
\frac{\Lambda^5(m_\pi^2 + 4 m_\pi\Lambda+\Lambda^2)}{16(m_\pi+\Lambda)^4} 
\nonumber\\
&=& -\frac{3(D+F)^2}{32\pi f^2}\left(\frac{\Lambda^3}{16} -\frac{5\Lambda}{16}m_\pi^2 + m_\pi^3 -\frac{35}{16\Lambda} m_\pi^4 + \cdots \right).
\eea
The leading nonanalytic term in the integral in Eq.~(\ref{eq:sigma_piN_Lam}) is of ${\cal O}(m_\pi^3)$.
The FRR gives the same ${\cal O}(m_\pi^3)$ leading nonanalytic term as for DR, namely, $-3(D+F)^2/(32\pi f^2)\, m_\pi^3$.

The first two terms in the parentheses of Eq.~(\ref{massNFRR}) which are $\Lambda$ dependent can be absorbed into the $a_0$ and $a_2$ terms, resulting in the renormalized low energy constants $c_0$ and $c_2$ \cite{Leinweber, Young3}.
With the introduction of the regulator, therefore, the loop contribution not only has the same LNA term as in DR, but also has $\Lambda$ dependent higher order terms.
With these higher order terms, the total loop contribution will vary smoothly with the pion mass, and will vanish when the pion mass is sufficiently large.
In this case the analytic terms have good convergence, and very high order terms are not needed to fit the lattice data at large $m_\pi$.
Similarly, the contribution from the $\Delta$ intermediate states in the heavy baryon limit can be written as
\be
\sigma_{\pi\Delta}^\Lambda = -\frac{{\cal C}^2}{6\pi^2 f^2} \int_0^\infty \dd{k} \frac{\bm{k}^4 u^(\bm{k})}{\omega(\bm{k})(\Delta+\omega(\bm{k}))}.
\ee
After the direct integration and Tylor expansion, one funds the same leading nonanalytic term as in DR, namely $({\cal C}^2 m_\pi^4)/(16\pi^2 f^2\Delta) \ln m_\pi$.
Again, higher order terms make the total loop contribution from the intermediate $\Delta$ states decreases smoothly with increasing pion mass.

\begin{figure}[t] 
\begin{center}
\epsfig{file=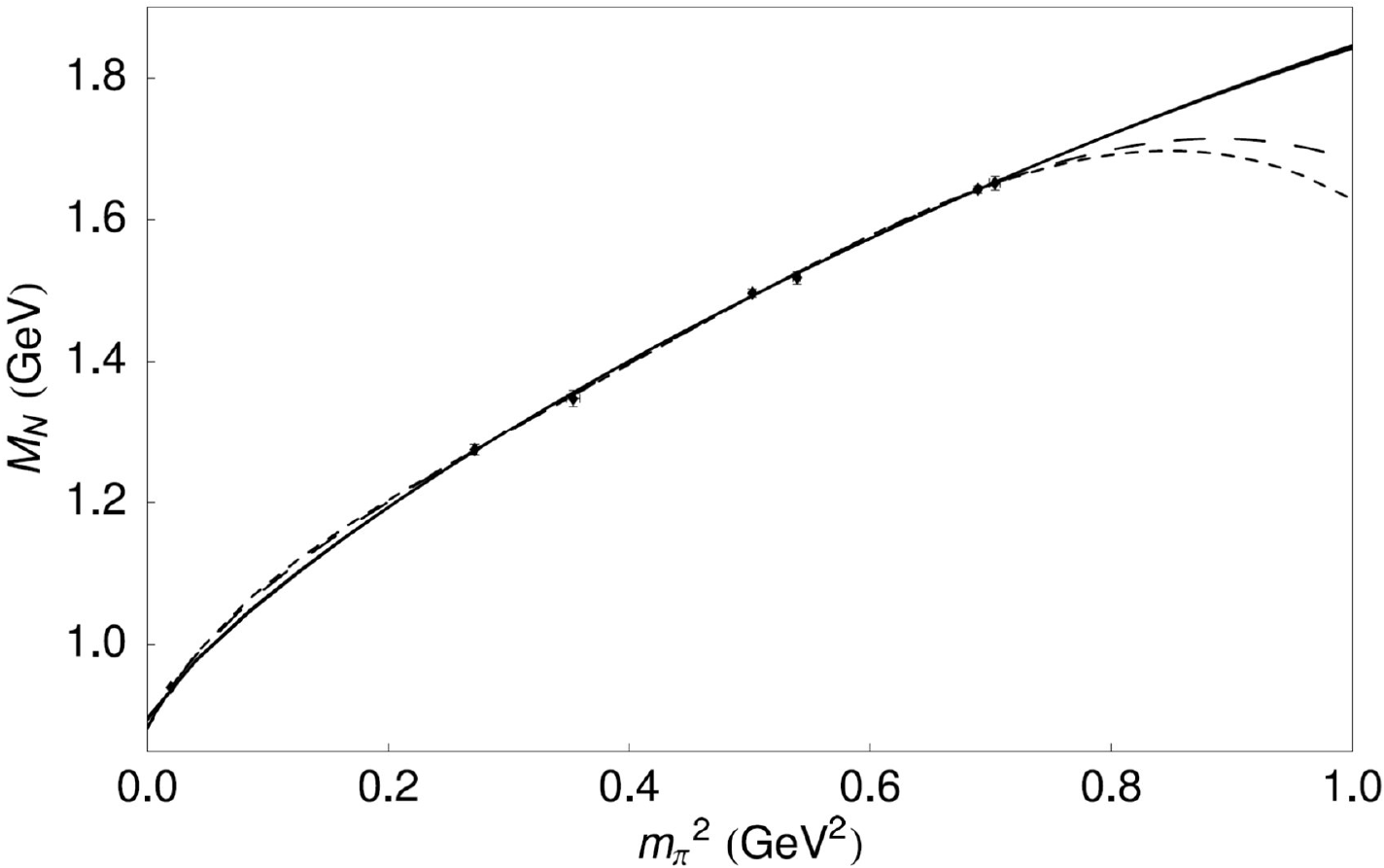,scale=0.75}
\caption{Variation of the nucleon mass $M_N$ with pion mass for various regularization schemes. The short-dash curve corresponds to the simple dimensional regularisation (DR) scheme, and the long-dash curve to the more sophisticated dimensionally regulated (BP) approach. The solid line represents four finite-range regularization schemes, with four kinds of regulators, which are indistinguishable. (Figure from Ref.~\cite{Young3}.)}\label{fig2}
\end{center}
\end{figure}

In Fig.~\ref{fig2}, the pion mass dependence of the nucleon mass is illustrated for various regularization schemes~\cite{Young3}.
The simple dimensional regularization scheme (short-dashed curve) is compared with a sophisticated dimensionally regulated approach (long-dashed curve).
For the FRR, four different functional forms for the ultraviolet vertex regulator are chosen, namely,
    sharp cutoff (SC) \mbox{$u({\bm k}) = \theta(\Lambda-|\bm{k}|)$},
    monopole (MON) $u({\bm k}) = \Lambda^2/(\Lambda^2+\bm{k}^2)$, 
    dipole (DIP) $u({\bm k}) = \Lambda^4/(\Lambda^2+\bm{k}^2)^2$, and
    Gaussian (GAU) $u({\bm k}) = \exp(-\bm{k}^2/\Lambda^2)$.
The results for the four finite range regulators (solid curves) are indistinguishable.
$\chi$PT with FRR can naturally describe lattice data and the linear dependence of lattice data at large $m_\pi$ indicates the smaller analytic contributions beyond the  $a_4$ term.
The good convergent behavior of the FRR result can also be seen from Table~\ref{tab:FRR} \cite{Young3}.
For FRR, the low energy constant $a_4$ is small compared with the $a_0$ and $a_2$ terms, and there is no need to include $a_6$ term.
In contrast, for the DR and branch point (BP) approaches, the $a_4$ is larger than $a_0$ and $a_2$, so that a higher order, $\sim a_6 m_\pi^6$ term is necessary to cancel the large contribution from the loop integral.

\begin{table}[h]
\begin{center}
\caption{Low energy constants $a_i$ in different regularization schemes and parameters $\Lambda$ in FRR. All quantities are in units of appropriate powers of
GeV. (From Ref.~\cite{Young3}.)}
\label{tab:FRR}
\begin{tabular}{lllccc}
\hline \\[-2mm]
~Regulator  & ~$a_0$ & ~~~$a_2$  &  ~$a_4$     &  ~$a_6$   &  $\Lambda$ \\[2mm]
\hline \\[-2mm]
~DR         & 0.882  & ~3.82     &  ~~6.65     &  $-4.24$  &  ---   \\[2mm]
~BP         & 0.825  & ~4.37     &  ~~9.72     &  $-2.77$  &  ---   \\[2mm]
~FRR\,(SC)  & 1.03   & ~1.12     &  $-0.292$   &    ---    & 0.418   \\[2mm]
~FRR\,(MON) & 1.56   & ~0.884    &  $-0.204$   &    ---    & 0.496   \\[2mm]
~FRR\,(DIP) & 1.20   & ~0.972    &  $-0.229$   &    ---    & 0.785   \\[2mm]
~FRR\,(GAU) & 1.12   & ~1.01     &  $-0.247$   &    ---    & 0.616   \\[2mm]
\hline
\end{tabular}
\end{center}
\end{table}

\clearpage
\section{Nonlocal chiral effective theory}
\label{Sec.3}

\subsection{\it Nonlocal effective Lagrangian}

As discussed in the previous section, the advantages of the FRR approach are that there are no explicit ultraviolet divergences in the loop integrals, and good convergent behavior is achieved when applying the effective field theory beyond the power counting region.
This is seen when either relativistic or nonrelativistic regulators are used, in the covariant and heavy baryon formulations, respectively.
A disadvantage, however, is that this prescription violates local gauge symmetry, which is an inevitable feature of introducing hadronic form factors to represent the extended structure of hadrons.

The problem of preserving gauge invariance in theories with hadronic form factors can be formally alleviated by introducing nonlocal interactions into the gauge invariant local Lagrangian, which allows one to consistently generate a covariant regulator.
Physically, nonlocal interactions are more realistic than the local effective interactions, since hadrons are not point particles. 
The non-pointlike nature of the hadrons can be taken into account by defining the baryon field at spacetime point $x^\mu$ and displacing the meson or photon by a distance $a^\mu$ to spacetime point $x^\mu+a^\mu$, with a correlation function, $F(a)$, parametrizing the nonlocal interaction.

The presence of gauge links in the nonlocal Lagrangian connecting different spacetime coordinates generates additional diagrams which are needed to ensure the local gauge invariance of the theory. 
This guarantees that the proton and neutron charges, for example, are unaffected by meson loops, or that contributions to the strangeness in the nucleon from diagrams with intermediate state kaons and hyperons sum to zero.
These basic features of the theory are not guaranteed for a local Lagrangian,  
but arise automatically in the nonlocal theory in which the Ward identities and charge conservation are necessarily satisfied.

Starting from the local effective Lagrangian, we can expand Eqs.~(\ref{eq:LOCT}) and (\ref{eq:LDEC}) and introduce the minimal substitution for the electromagnetic field $\mathscr{A_\mu}$, in which case the local Lagrangian density can be rewritten in the form
\begin{eqnarray}
{\cal L}^{\rm (local)}(x)
&=&\bar B(x)(i \gamma^\mu \mathscr{D}_{\mu} - M_B) B(x)
 + \frac{C_{B\phi}}{f}
   \big[\, \bar{p}(x) \gamma^\mu \gamma^5 B(x)\,
	  \mathscr{D}_{\mu} \phi(x) + {\rm H.c.}
   \big]
\nonumber \\
&+&\overline{T}_\mu(x)
   (i \gamma^{\mu\nu\alpha} \mathscr{D}_{\alpha} - M_T \gamma^{\mu\nu})\,
   T_\nu(x)
 + \frac{C_{T\phi}}{f}
   \big[\, \bar{p}(x) \Theta^{\mu\nu} T_\nu(x)\,
	 \mathscr{D}_{\mu} \phi(x) + {\rm H.c.}
   \big]
\nonumber \\
&+&\frac{iC_{\phi\phi^\dag}}{2 f^2}\,
   \bar p(x) \gamma^\mu p(x)
   \big[ \phi(x) (\mathscr{D}_{\mu} \phi)^\dag(x)
	- \mathscr{D}_{\mu} \phi(x) \phi^\dag(x)
   \big]
\nonumber \\
&+& \mathscr{D}_{\mu} \phi(x)
    (\mathscr{D}_{\mu} \phi)^\dag(x)\, + \cdots ,
\label{eq:Llocal}
\end{eqnarray}
where for the interaction part we show only those terms that are relevant for the calculation of proton form factors and PDFs, and we keep the dependence on the spacetime coordinate $x^\mu$ explicitly. 
The covariant derivatives are defined as 
\begin{subequations}
\begin{eqnarray}
\mathscr{D}_{\mu} B(x)
&=& \left[ \partial_\mu - i e^q_B\, \mathscr{A_\mu}(x)
    \right] B(x),
\\
\mathscr{D}_{\mu} T^\nu(x)
&=& \left[ \partial_\mu - i e^q_T\, \mathscr{A_\mu}(x)
    \right] T^\nu(x),
\\
\mathscr{D}_{\mu} \phi(x)
&=& \left[ \partial_\mu - i e^q_\phi\, \mathscr{A_\mu}(x)
    \right] \phi(x),
\end{eqnarray}
\end{subequations}
where $e^q_B$, $e^q_T$ and $e^q_\phi$ are the quark flavor charges of
the octet baryon $B$, decuplet baryon $T$ and meson $\phi$,
respectively.
For example, for the proton one has the charges
$e^u_p = 2 e^d_p = 2$,
$e^s_p = 0$,
while for the $\Sigma^+$ hyperon
$e^u_{\Sigma^+} = 2 e^s_{\Sigma^+} = 2$,
$e^d_{\Sigma^+} = 0$,
and similarly for other baryons.
For the mesons, the flavor charges for the $\pi^+$ are
$e^u_{\pi^+} = -e^d_{\pi^+} = 1$
but $e^q_{\pi^0} = 0$ for all $q$,
while for the $K^+$ these are
$e^u_{K^+} = -e^s_{K^+} = 1$,
$e^d_{K^+} = 0$,
and similarly for the charge conjugate states.
The coefficients $C_{B\phi}$, $C_{T\phi}$ and $C_{\phi\phi^\dag}$ in Eq.~(\ref{eq:Llocal}) depend on the
coupling constants $D$, $F$ and ${\cal C}$, and are given explicitly
in Table~\ref{tab:C}.

\begin{table}
\begin{center}
\caption{Coupling constants $C_{B\phi}$, $C_{T\phi}$ and
	$C_{\phi\phi^\dag}$ for the $p B \phi$, $p T \phi$
	and $p p \phi \phi^\dag$ interactions, respectively,
	for the various allowed flavor channels.}
\begin{tabular}{c|ccccc}
\hline
\\[-2mm]
\hspace*{0.5cm}$(B\phi)$\hspace*{0.5cm}
& \hspace*{0.5cm}$(p \pi^0)$\hspace*{0.4cm}
& \hspace*{0.4cm}$(n \pi^+)$\hspace*{0.4cm}
& \hspace*{0.4cm}$(\Sigma^+ K^0)$\hspace*{0.4cm}
& \hspace*{0.4cm}$(\Sigma^0 K^+)$\hspace*{0.4cm}
& \hspace*{0.4cm}$(\Lambda  K^+)$\hspace*{0.4cm}
\\[2mm]
\hspace*{0.5cm}$C_{B\phi}$\hspace*{0.5cm}
& \hspace*{0.4cm}$\frac12 (D+F)$\hspace*{0.4cm}
& \hspace*{0.4cm}$\frac{1}{\sqrt2} (D+F)$\hspace*{0.4cm}
& \hspace*{0.4cm}$\frac{1}{\sqrt2} (D-F)$\hspace*{0.4cm}
& \hspace*{0.4cm}$\frac12 (D-F)$\hspace*{0.4cm}
& \hspace*{0.4cm}$-\frac{1}{\sqrt{12}} (D+3F)$\hspace*{0.4cm}
\\[2mm]
\hline
\\[-2mm]
($T\phi$)
& $(\Delta^0 \pi^+)$
& $(\Delta^+ \pi^0)$
& $(\Delta^{++} \pi^-)$
& $(\Sigma^{*+} K^0)$
& $(\Sigma^{*0} K^+)$
\\[2mm]
$C_{T\phi}$
& $-\frac{1}{\sqrt 6}\, {\cal C}$
& $-\frac{1}{\sqrt 3}\, {\cal C}$
& $\frac{1}{\sqrt 2}\, {\cal C}$
& $\frac{1}{\sqrt 6}\, {\cal C}$
& $-\frac{1}{\sqrt{12}}\, {\cal C}$
\\[2mm]
\hline
\\[-2mm]
$(\phi\phi^\dag)$
& $(\pi^+\pi^-)$
& $(K^0 \overline{K}^0)$
& $(K^+ K^-)$
&
&
\\[2mm]
$C_{\phi\phi^\dag}$
& $\frac{1}{2}$
& $\frac{1}{2}$
& 1
&
&
\\[2mm]
\hline 
\end{tabular}
\label{tab:C}
\end{center}
\end{table}

Using the methods described in Refs.~\cite{Faessler3,He,He2,Terning}, the nonlocal version of the local Lagrangian in Eq.~(\ref{eq:Llocal}) can be written as
\begin{eqnarray}
&&{\cal L}^{\rm (nonloc)}(x)
= \bar B(x) (i\gamma^\mu \mathscr{D}_{\mu} - M_B) B(x)
+ \overline{T}_\mu(x)
  (i\gamma^{\mu\nu\alpha} \mathscr{D}_{\alpha} - M_T \gamma^{\mu\nu})
  T_\nu(x)
\nonumber\\
&&\ +\
  \bar{p}(x)
  \left[
    \frac{C_{B\phi}}{f} \gamma^\mu \gamma^5 B(x)\,
  + \frac{C_{T\phi}}{f} \Theta^{\mu\nu} T_\nu(x)
  \right] 
  \mathscr{D}_{\mu} 
  \left(
    \int\!\dd^4{a}\, {\cal G}_\phi^q(x,x+a)\, F(a)\, \phi(x+a)
  \right) + {\rm H.c.}
\nonumber\\
&&\ +\
  \frac{i C_{\phi\phi^\dag}}{2 f^2}
  \bar{p}(x) \gamma^\mu p(x)\,
  \Bigg \{                                 
  \int\!\dd^4{a}\, {\cal G}_\phi^q(x,x+a)\, F(a)\, \phi(x+a)\
  \mathscr{D}_\mu 
  \left( \int\!\dd^4{b}\, {\cal G}_\phi^q(x+b,x)\, F(b)\, \phi^\dag(x+b)
  \right)
\nonumber\\
&&\hspace*{3.35cm} -\
  \mathscr{D}_{\mu} 
  \left( \int\!\dd^4{a}\, {\cal G}_\phi^q(x,x+a)\, F(a)\, \phi(x+a)
  \right) 
  \int\!\dd^4{b}\, {\cal G}_\phi^q(x+b,x)\, F(b)\, \phi^\dag(x+b) 
  \Bigg \}
\nonumber \\
&&\ +\
  \mathscr{D}_{\mu} \phi(x) (\mathscr{D}_{\mu} \phi)^\dag(x)
 + \cdots,
\label{eq:nonlocal}
\end{eqnarray}
where
\be \label{eq:partial}
\mathscr{D}_\mu \Psi(x)
= \left[ \partial_\mu 
        - ie_\Psi^q \int\!\dd^4{a}\, \mathscr{A}_\mu(x-a)\, F(a)
  \right] \Psi(x)
\ee
for a field $\Psi$.
The gauge link ${\cal G}_\phi^q$ is introduced to preserve local gauge invariance,
\begin{equation}
{\cal G}_\phi^q(x,y)
= \exp\left[ 
        -i e^q_\phi \int_x^y \dd{z}^\mu 
                    \int \dd^4{a}\, \mathscr{A}_\mu(z-a)\, F(a)
      \right],
\label{eq:link}
\end{equation}
where the function $F(a)$ is the correlation function in coordinate space.
One can verify that the nonlocal Lagrangian in Eq.~(\ref{eq:nonlocal}) is invariant under the gauge transformations
\begin{subequations}
\begin{eqnarray}
B(x) &\to&
    B'(x) = B(x) \exp\left[i e^q_B\, \theta(x)\right],
\\
T_\mu(x) &\to&
    T_\mu'(x) = T_\mu(x) \exp\left[i e^q_T\, \theta(x)\right],
\\
\phi(x)  &\to&
    \phi'(x)  = \phi(x)  \exp\left[i e^q_\phi\, \theta(x)\right]
\end{eqnarray}
for the matter fields, and
\begin{equation}
\hspace*{-0.5cm}
{\mathscr{A}^\mu}(x)\ \to\
    \mathscr{A}'^\mu(x) = {\mathscr{A}^\mu}(x) + {\partial^\mu} \theta'(x)
\end{equation}
\end{subequations}
for the electromagnetic field, where 
    $\theta(x) = \int \dd^4{a}\, \theta'(x+a)\, F(a)$ is an arbitrary
function of the spacetime coordinate~$x$.

The nonlocal Lagrangian density in Eq.~(\ref{eq:nonlocal}) can be further
decomposed by expanding the gauge link (\ref{eq:link}) in powers of
the charge $e_\phi^q$,
\begin{eqnarray}
{\cal G}^q_\phi(x+b,x+a)
&=& \exp \Big[ -i e^q_\phi\, (a-b)^\mu
	      \int_0^1 \dd{t}\, \mathscr{A}_\mu\big(x+at+b\,\bar t\,\big)
	 \Big]
\nonumber \\
&=& 1\ +\ \delta {\cal G}^q_\phi\
       +\ \cdots,
\label{eq:linkexpand}
\end{eqnarray}
where the $\delta {\cal G}^q_\phi$ is 
\begin{eqnarray}
\delta {\cal G}^q_\phi
&=& -\ i e^q_\phi\, (a-b)^\mu
       \int_0^1 \dd{t}\, \mathscr{A}_\mu\big(x+at+b\,\bar t\,\big)
\label{eq:deltaG}
\end{eqnarray}
and we have used a change of variables
	$z^\mu \to x^\mu + a^\mu\, t + b^\mu\, \bar t$,
with $\bar t \equiv 1-t$.
This allows the Lagrangian ${\cal L}^{\rm (nonloc)}$ to be written as a sum of free and interacting parts, where to lowest order the latter consists of purely hadronic (${\cal L}^{\rm (nonloc)}_{\rm had}$), electromagnetic (${\cal L}^{\rm (nonloc)}_{\rm em}$), and gauge link (${\cal L}^{\rm (nonloc)}_{\rm link}$) components.
The higher order terms in Eq.~(\ref{eq:linkexpand}) contribute to higher order electromagnetic corrections, which are in practice negligible.
The higher order terms can also be related to other processes, such as those involving two or more photons emitted in the final state.

The hadronic and electromagnetic interaction parts of the nonlocal
Lagrangian arise from the
covariant derivatives in Eq.~(\ref{eq:nonlocal}),
and are given by
\begin{eqnarray}
{\cal L}^{\rm (nonloc)}_{\rm had}(x)
&=& \bar{p}(x)
    \left[ \frac{C_{B\phi}}{f}\, \gamma^\mu \gamma^5 B(x)
	 + \frac{C_{T\phi}}{f}\, \Theta^{\mu\nu} T_\nu(x)
    \right]
    \!\int\! \dd^4{a}\, F(a)\, \partial_\mu \phi(x+a)
	+ {\rm H.c.}
\nonumber\\
&+& \frac{iC_{\phi\phi^\dag}}{2f^2}\,
    \bar{p}(x) \gamma^\mu p(x)
    \int\!\dd^4{a}\!\int\!\dd^4{b}\, F(a)\, F(b)
\nonumber \\
& & \hspace*{4.8cm} \times
    \left[ \phi(x+a) \partial_\mu \phi^\dag(x+b)
	 - \partial_\mu \phi(x+a) \phi^\dag(x+b)
    \right],~~~
\label{eq:Lnonloc_had}
\end{eqnarray}
and
\begin{eqnarray}
{\cal L}^{\rm (nonloc)}_{\rm em}(x)
&=& e^q_B\, \int \dd^4{a}\, \bar{B}(x) \gamma^\mu B(x)\, \mathscr{A}_\mu(x+a)F(a)\
+\ e^q_T\, \int \dd^4{a}\, \overline{T}_\mu(x) \gamma^{\mu\nu\alpha} T_\nu(x)\,
	    \mathscr{A}_\alpha(x+a)F(a)
\nonumber\\
&+& i e^q_\phi \int \dd^4{a}\, 
    \left[ \partial^\mu \phi(x) \phi^\dag(x)
         - \phi(x) \partial^\mu \phi^\dag(x)
    \right] \mathscr{A}_\mu(x+a)F(a)
\nonumber\\
&-& i e^q_\phi\, \bar{p}(x)
    \left[
      \frac{C_{B\phi}}{f}\, \gamma^\mu \gamma^5 B(x)
    + \frac{C_{T\phi}}{f}\, \Theta^{\mu\nu} T_\nu(x)
    \right]
\nonumber\\
& & \hspace*{3cm} \times
    \int\!\dd^4{a} \int \dd^4{b}\, F(a)\, F(b)\,
    \phi(x+a)\, \mathscr{A}_\mu(x+b) + {\rm H.c.}
\nonumber\\
&-& \frac{e^q_\phi C_{\phi\phi^\dag}}{2f^2}\,
    \bar{p}(x) \gamma^\mu p(x)
    \int\!\dd^4{a} \int\!\dd^4{b} \int\!\dd^4{c}\, F(a)\, F(b)\, F(c)\, \phi(x+a) \phi^\dag(x+b)\, \mathscr{A}_\mu(x+c), 
\nonumber\\
& &
\label{eq:Lnonloc_em}
\end{eqnarray}
respectively.
For the $\delta {\cal G}^q_\phi$ term in Eq.~(\ref{eq:deltaG}),
which explicitly depends on the gauge link, the nonlocal interaction
with the external gauge field yields the additional contribution to
the Lagrangian density,
\begin{eqnarray}
\hspace*{-0.5cm}
{\cal L}^{\rm (nonloc)}_{\rm link}(x)
&=& -i e^q_\phi\, \bar{p}(x)
    \left[ \frac{C_{B\phi}}{f}\, \gamma^\rho \gamma^5 B(x)
	 + \frac{C_{T\phi}}{f}\, \Theta^{\rho\nu} T_\nu(x)
    \right]
\nonumber\\
& & \hspace*{0.5cm} \times
    \int_0^1 \dd{t} \int\!\dd^4{a} \int\!\dd^4{b} \, F(a)\, F(b)\, a^\mu\,
    \partial_\rho \phi(x+a) \mathscr{A}_\mu(x+at+b\,\bar t\,)
    + {\rm H.c.}
\nonumber\\
&+& \frac{e^q_\phi C_{\phi\phi^\dag}}{2f^2}\,
    \bar{p}(x) \gamma^\rho p(x)
    \int_0^1 \dd{t} \int\!\dd^4{a}\!\int\!\dd^4{b}\int\!\dd^4{c}\,
	F(a)\, F(b)\, F(c)\, (a-b)^\mu\,
\nonumber\\
& & \hspace*{0.5cm} \times
    \left[ \phi(x+a)\, \partial_\rho \phi^\dag(x+b)
	 - \partial_\rho \phi(x+a) \phi^\dag(x+b)
    \right] \mathscr{A}_\mu\big(x+at+b\,\bar t+c\big).
\label{eq:Lnonloc_link}
\end{eqnarray}
Note that compared with traditional power counting schemes in chiral perturbation theory that use dimensional regularization, the introduction of the regulator function $F(a)$ in the nonlocal interactions (\ref{eq:Lnonloc_had})--(\ref{eq:Lnonloc_link}) leads to the generation of higher order terms with coefficients that in general will depend on the regulator mass parameter $\Lambda$.
This is analogous to a resummation of the standard chiral perturbation theory, which goes beyond the usual power counting regime.

\subsection{\it Feynman rules for nonlocal Lagrangian}

With the nonlocal effective Lagrangian, one can calculate hadron properties such as the nucleon electromagnetic form factors, strange form factors, unpolarized and polarized parton distributions, generalized parton distribution functions, amongst others.
In this subsection we will present the electromagnetic currents and Feynman rules for the nonlocal Lagrangian which will be needed for the calculation of these quantities.

For the nonlocal theory the contribution from quark $q$ to the current has two parts: the usual electromagnetic current, $J_{q, \rm em}^\mu$, obtained with minimal substitution from Eq.~(\ref{eq:Lnonloc_em}),
\begin{eqnarray}
\hspace*{-0.5cm}
J_{q, \rm em}^\mu(x)
&\equiv& \frac{\delta \int\!\dd^4{y}\,
		{\cal L}_{\rm em}^{\rm (nonloc)}(y)}
	      {\delta \mathscr{A_\mu}(x)}
\nonumber\\
&=& \int\!\dd^4{a}\, F(a) \left(e^q_B\,\bar{B}(x-a) \gamma^\mu B(x-a)
 +  e^q_T\, \overline{T}_\alpha(x-a) \gamma^{\alpha\nu\mu} T_\nu(x-a) \right) \nonumber \\
&+& ie^q_\phi\int\!\dd^4{a}\, F(a) 
    \left[ \partial^\mu \phi(x-a) \phi^\dag(x-a)
         - \phi(x-a) \partial^\mu \phi^\dag(x-a)
    \right]
\nonumber\\
&-& i e^q_\phi\,
    \int\!\dd^4{a}\int\!\dd^4{b}\, F(a)\, F(b)
\nonumber \\
& & \hspace*{1cm} \times\
    \bar{p}(x-b)
    \left[ \frac{C_{B\phi}}{f}\, \gamma^\mu \gamma^5 B(x-b)\
        +\,\frac{C_{T\phi}}{f}\, \Theta^{\mu\nu} T_\nu(x-b)
    \right]
    \phi(x+a-b)
    +\, {\rm H.c.}
\nonumber\\
&-& \frac{e^q_\phi C_{\phi\phi^\dag}}{2f^2}
    \int\!\dd^4{a}\int\!\dd^4{b}\int\!\dd^4{c}\, F(a)\, F(b)\, F(c)\,
 \bar{p}(x-c) \gamma^\mu p(x-c)\, \phi(x+a-c) \phi^\dag(x+b-c),
\nonumber\\
& &
\label{eq:Jem}
\end{eqnarray}
and an additional term obtained from the gauge link,
\begin{eqnarray}
\hspace*{-0.4cm}
\delta J_q^\mu(x)
&\equiv& \frac{\delta \int\!\dd^4{y}\, {\cal L}^{\rm (nonloc)}_{\rm link}(y)}
              {\delta \mathscr{A_\mu}(x)}
\nonumber\\
&=&\ -i e^q_\phi
    \int_0^1\!\dd{t}\int\!\dd^4{a}\int\!\dd^4{b}\, F(a)\, F(b)\,
    a^\mu\, \bar{p}(x-at-b)
    \left[ \frac{C_{B\phi}}{f}\, \gamma^\rho \gamma^5 B(x-at-b)\,
    \right.
\nonumber\\
&& \left.\hspace*{6.2cm}
+\, \frac{C_{T\phi}}{f}\, \Theta^{\rho\nu} T_\nu(x-at-b)
    \right] \partial_\rho \phi\big(x+a\bar t-b\big) + {\rm H.c.}
\nonumber\\
&+& \frac{e^q_\phi C_{\phi\phi^\dag}}{2f^2}
    \int_0^1\!\dd{t}\int\!\dd^4{a}\,\int\!\dd^4{b}\,\int\!\dd^4{c}\,
    F(a)\, F(b)\, F(c)\, (a-b)^\mu\,
    \bar{p}\big(x-at-b\bar t-c\big) \gamma^\rho
\nonumber\\
&& \times\ p\big(x-at-b\bar t-c\big) \Big[ \phi\big(x+(a-b)\bar t-c\big)
	  \partial_\rho \phi^\dag\big(x-(a-b)t-c\big)
\nonumber\\
& & \hspace*{3.5cm}
        -\, \partial_\rho \phi\big(x+(a-b)\bar t-c\big)
	  \phi^\dag\big(x-(a-b)t-c\big)
    \Big],
\label{eq:Jlink}
\end{eqnarray}
respectively.
As for the nonlocal Lagrangian, the nonlocal currents in Eqs.~(\ref{eq:Jem}) and (\ref{eq:Jlink}) include the extra regulator function, $F(a)$.
The local limit can be obtained by taking $F(a)$ to be a $\delta$-function, $F(a) \to \delta^{(4)}(a)$, which is equivalent to taking the form factor in momentum space to be unity.
From the above equations, the currents for the $u$, $d$ and $s$ quarks can be written explicitly in terms of the the mesonic and baryonic currents.
Here, as an example, we only show the currents related to octet baryons which have contributions to the proton form factors and PDFs.
For the $u$ quark, the current can be written explicitly as
\begin{eqnarray}
J_u^\mu
&=& \int\!\dd^4{a}\, F(a)
\Big[ \, 2 \bar p(x-a) \gamma^\mu p(x-a) + \bar n(x-a) \gamma^\mu n(x-a)
+ \bar\Lambda (x-a) \gamma^\mu \Lambda(x-a) 
\nonumber\\
& & \hspace*{2cm}
+\, 2 \overline{\Sigma}^+(x-a) \gamma^\mu \Sigma^+(x-a)
+ {\overline\Sigma}^0 (x-a) \gamma^\mu \Sigma^0(x-a)
\Big]
\nonumber\\
&-& \int\!\dd^4{a}\int\!\dd^4{b}\, F(a)\, F(b)\,
\bar p(x-b) \gamma^\mu \gamma^5 
\bigg[\frac{i(D+F)}{\sqrt 2f} n(x-b) \pi^+(x+a-b)
\nonumber\\
& & \hspace*{6.2cm}
-\, \frac{i(D+3F)}{\sqrt{12}f}\, \Lambda(x-b) K^+(x+a-b)\,
\nonumber\\
& & \hspace*{6.2cm}
+\ \frac{i(D-F)}{2f}\, \Sigma^0(x-b) K^+(x+a-b) + \text{H.c.} 
\bigg] 
\nonumber\\
&-& \frac{1}{2f^2}\,
\int\!\dd^4{a} \int\!\dd^4{b} \int\!\dd^4{c}\, F(a)\,F(b)\,F(c)
\nonumber\\
& & \hspace*{1.6cm}
\times\,
\bar p(x-c) \gamma^\mu p(x-c)
\Big[ \pi^+(x+a-c)\, \pi^-(x+b-c) + 2 K^+(x+a-c)\, K^-(x+b-c) 
\Big]
\nonumber\\
&-& \int_0^1\dd{t} \int\!\dd^4{a} \int\!\dd^4{b}\,
F(a)\, F(b)\, 
a^\mu\, \bar p(x-at-b) \gamma^\rho \gamma^5 
\bigg[ 
\frac{i(D+F)}{\sqrt 2f} n(x-at-b) \partial_\rho \pi^+\big(x+a\bar t-b\big)
\nonumber\\
& & \hspace*{7.2cm}
-\, \frac{i(D+3F)}{\sqrt{12}f} 
    \Lambda(x-at-b)\,  \partial_\rho K^+\big(x+a\bar t-b\big)
\nonumber\\
& & \hspace*{7.2cm}
+\, \frac{i(D-F)}{2f}\,
    \Sigma^0(x-at-b)\, \partial_\rho K^+\big(x+a\bar t-b\big)
+ \text{H.c.} 
\bigg] 
\nonumber\\
&+& \frac{1}{2f^2} 
\int_0^1 \dd{t} \int\!\dd^4{a} \int\!\dd^4{b} \int\!\dd^4{c}\,
F(a)\,F(b)\,F(c)\, (a-b)^\mu\, 
\bar p\big(x-at-b\bar t-c\big) \gamma^\rho p\big(x-at-b\bar t-c\big)
\nonumber\\
& & \hspace*{4.7cm}
\times
\Big[ 
\pi^+\big(x+(a-b)\bar t-c\big)\, 
\overleftrightarrow{\partial_\rho} \pi^-\big(x-(a-b)t-c\big)
\nonumber\\
& & \hspace*{5cm}
+\, 2 K^+\big(x+(a-b)\bar t-c\big)\, 
\overleftrightarrow{\partial_\rho} K^-\big(x-(a-b)t-c\big)
\Big],
\label{eq:ju}
\end{eqnarray}
where 
$\overleftrightarrow{\partial_{\rho\!}} = \overrightarrow{\partial_\rho} - \overleftarrow{\partial_\rho}$. 
For $d$ and $s$ quarks, the currents have the same structure as $J_u^\mu$, except for different coefficients.
Omitting for convenience the integral factors and arguments in Eq.~(\ref{eq:ju}), the $d$ and $s$ quark currents can be obtained by replacing the corresponding expressions by
\begin{eqnarray}
J^\mu_d
&\to & \bar p \gamma^\mu p
 + 2 \bar n \gamma^\mu n
 + 2 \overline{\Sigma}^- \gamma^\mu \Sigma^-
 + \overline{\Sigma}^0 \gamma^\mu \Sigma^0
 + \bar\Lambda \gamma^\mu \Lambda
\nonumber\\
&+&
\bar p \gamma^\mu \gamma^5 
\bigg[
  \frac{i(D+F)}{\sqrt2 f}\, n\, \pi^+
- \frac{i(D-F)}{\sqrt2 f}\, \Sigma^+\, K^0 
\bigg]
\nonumber\\
&+&
\frac{1}{2f^2}\,
\bar p \gamma^\mu p\,
\Big[ \pi^+ \pi^- -\, \overline{K}^0 K^0
\Big]
\nonumber\\
&+& a^\mu\, \bar p \gamma^\rho \gamma^5 
\bigg[ 
  \frac{i(D+F)}{\sqrt2 f}\, n\, \partial_\rho\pi^+
- \frac{i(D-F)}{\sqrt2 f}\,\Sigma^+\, \partial_\rho K^0
\bigg]
\nonumber\\
&-& \frac{1}{2f^2}\,
(a-b)^\mu\, \bar p \gamma^\rho p\,
\Big[ 
  \big( \pi^+ \partial_\rho \pi^- - \pi^- \partial_\rho \pi^+ \big)
- \big( \overline{K}^0 \partial_\rho K^0 - K^0 \partial_\rho \overline{K}^0 \big)
\Big],
\\
& &
\nonumber
\label{eq:jd}
\end{eqnarray}
and
\begin{eqnarray}
\label{eq:js}
J^\mu_s
&\to &\overline{\Sigma}^+ \gamma^\mu \Sigma^+
 + \overline{\Sigma}^0 \gamma^\mu \Sigma^0
 + \bar\Lambda \gamma^\mu \Lambda
\nonumber\\
&+& \bar p \gamma^\mu \gamma^5
\bigg[
  \frac{i(D-F)}{\sqrt2 f}\, \Sigma^+\, K^0
+ \frac{i(D-F)}{2f}\, \Sigma^0\, K^+
- \frac{i(D+3F)}{\sqrt{12} f}\, \Lambda\, K^+
\bigg]
\nonumber\\
&+& \frac{1}{2f^2}\,
\bar p \gamma^\mu p\,
\Big[ 2 K^+ K^- + \overline{K}^0 K^0
\Big]						
\nonumber\\
&+& a^\mu\, \bar p \gamma^\rho \gamma^5 
\bigg[
  \frac{i(D-F)}{\sqrt2 f}\, \Sigma^+\, \partial_\rho K^0
+ \frac{i(D-F)}{2f}\, \Sigma^0\, \partial_\rho K^+
- \frac{i(D+3F)}{\sqrt{12} f}\, \Lambda\, \partial_\rho K^+ 
\bigg]
\nonumber\\
&-& \frac{1}{2f^2} (a-b)^\mu\, \bar p \gamma^\rho p\,
\Big[
  2 \big( K^+ \partial_\rho K^- - K^- \partial_\rho K^+ \big)
  + \big( \overline{K}^0 \partial_\rho K^0 - K^0 \partial_\rho \overline{K}^0 \big)
\Big].
\end{eqnarray}
The terms involving the doubly-strange baryons $\Xi^{0,-}$ and $\Xi^{*0,-}$ are not present because they cannot couple to the proton initial states.
The terms in each current proportional to $a^\mu$ and $(a-b)^\mu$ are the additional contributions generated from the gauge link, which guarantees charge conservation.

\begin{figure}[tb] 
\begin{center}
\epsfig{file=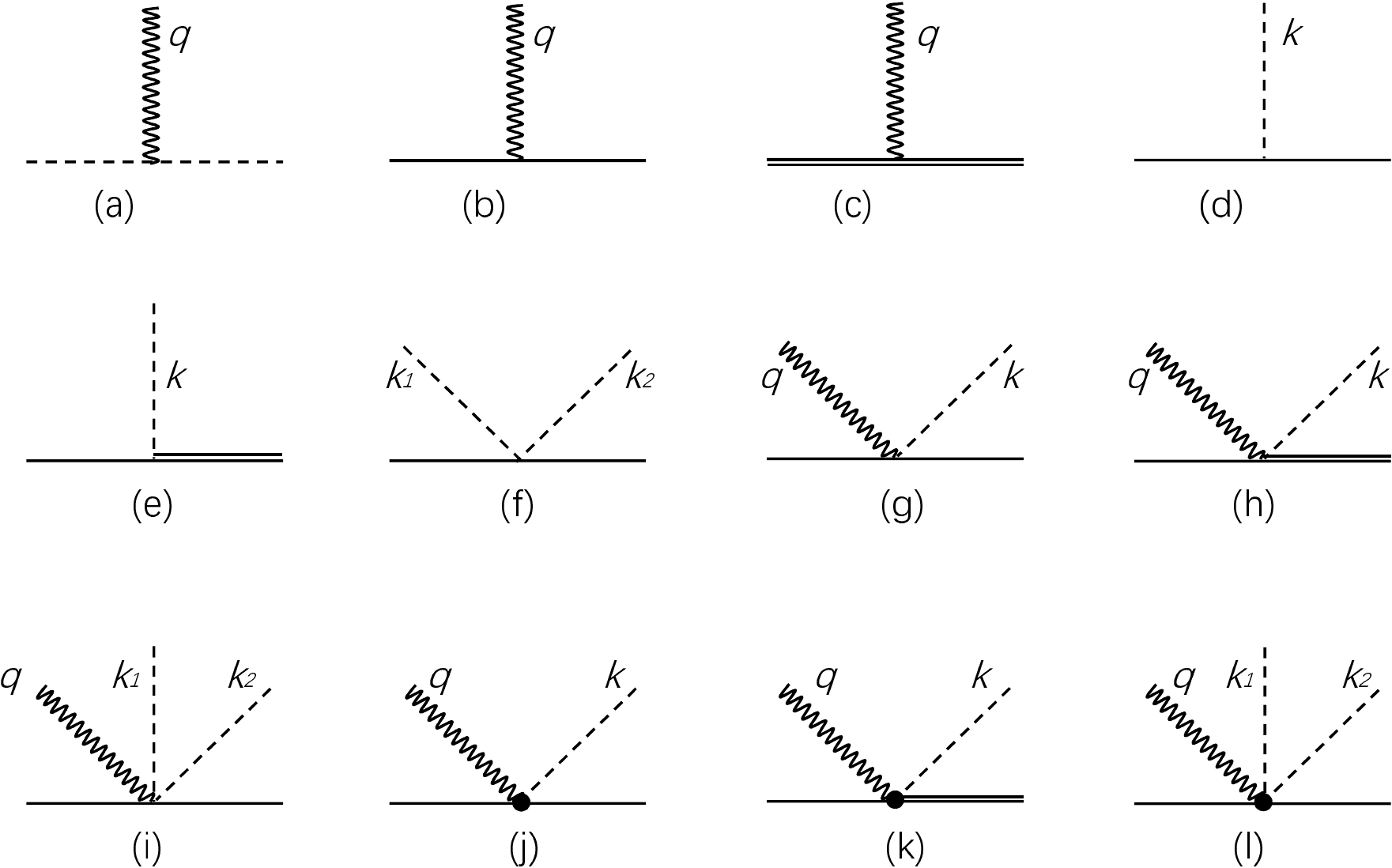,scale=0.6}
\caption{Interaction vertices for the nonlocal Lagrangian, for interactions between octet baryon (solid lines), decuplet baryon (double-solid lines), meson (dashed lines) and photon (wavy lines) fields. The four-momentum of the photon is denoted by $q$, while that of the meson by $k$ or $k_1$ and $k_2$. Vertices generated from the additional interactions associated with the gauge link [diagrams (j), (k) and (l)] are indicated by black circles.}
\label{fig:Feynman}
\end{center}
\end{figure}

With the nonlocal Lagrangian and currents, one can obtain the corresponding Feynman rules.
The vertices for the baryon--meson, baryon--photon, and baryon--meson--photon interactions which will be used in the calculation of the nucleon form factors and PDFs are illustrated in Fig.~\ref{fig:Feynman}, including contributions from the additional interactions associated with the gauge link.
The vertices for the photon coupling to the meson $\phi$, octet baryon $B$, and decuplet baryon $T$ in Figs.~\ref{fig:Feynman}(a), \ref{fig:Feynman}(b) and \ref{fig:Feynman}(c), respectively, are given by
\be\label{eq:vertex_abc}
V^\mu_{\rm (a)} = -i e_\phi^q (k_1^\mu +k_2^\mu) \widetilde{F}(q), ~~~~~~~~~ 
V^\mu_{\rm (b)} = -i e_B^q \gamma^\mu \widetilde{F}(q), ~~~~~~~~~ 
V^{\mu\nu\alpha}_{\rm (c)} = -i e_T^q \gamma^{\mu\nu\alpha}\widetilde{F}(q),
\ee
where $\widetilde{F}(q)$ is the Fourier transformation of the correlation function $F(a)$ and provides the momentum dependence of the form factors at tree level, with $q^\mu$ the four-momentum of the photon and $k^\mu$, $k_1^\mu$, and $k_2^\mu$ are the four-momenta of the meson fields.

For the baryon--meson vertices in Figs.~\ref{fig:Feynman}(d), \ref{fig:Feynman}(e) and \ref{fig:Feynman}(f), we write the vertex functions as
\be
V_{\rm (d)} = \frac{C_{B\phi}}{f}\slashed{k}\gamma_5\, \widetilde{F}(k), 
~~~~~
V_{\rm (e)}^\mu = \frac{C_{T\phi}}{f} 
\Big[ k^\mu - \big( Z+\tfrac12 \big) \gamma^\mu \slashed{k} \Big]
\widetilde{F}(k),
~~~~~ 
V_{\rm (f)} = \frac{C_{\phi\phi^\dag}}{2f^2}\, (\slashed{k}_1+\slashed{k}_2)\, \widetilde{F}(k_1)\widetilde{F}(k_2),
\ee  
where the regulator $\widetilde{F}(k)$ associated with each meson field with momentum $k^\mu$ makes the meson-loop integral convergent.
Similarly, for the standard baryon--meson--photon interactions in Figs.~\ref{fig:Feynman}(g), \ref{fig:Feynman}(h) and \ref{fig:Feynman}(i), using minimal substitution we obtain the vertices
\bea
V_{\rm (g)}^\mu
&=& -\frac{e_\phi^q\, C_{B\phi}}{f}
\gamma^\mu\gamma_5 \widetilde{F}(k) \widetilde{F}(q),
    \nonumber\\
V_{\rm (h)}^{\mu\nu}
&=& -\frac{e_\phi^q\, C_{T\phi}}{f}
\Big[ g^{\mu\nu} - \big(Z+\tfrac12\big) \gamma^\mu \gamma^\nu \Big] \widetilde{F}(k)\widetilde{F}(q), 
    \\
V_{\rm (i)}^\mu
&=& -\frac{e_\phi^q\, C_{\phi\phi^\dag}}{2f^2} 
\gamma^\mu \widetilde{F}(k_1)\widetilde{F}(k_2)\widetilde{F}(q),
    \nonumber
\eea 
where here the vertices depend on both the momenta $q$ and $k$.

For the additional diagrams generated from the gauge link in Figs.~\ref{fig:Feynman}(j), \ref{fig:Feynman}(k) and \ref{fig:Feynman}(l), to obtain the expressions for the vertices we need to calculate $\int\!\dd^4{a}\, F(a)\, e^{ika} I(x+a,x)$.
Using the identity
\bea \label{eq:vertex}
\int\!\dd^4{a}\, F(a)\, e^{ika} I(x+a,x)
&=& \int\!\dd^4{a} \int\!\frac{\dd^4{p}}{(2\pi)^4}\,
\widetilde{F}(p^2)\, e^{ipa}\, e^{ika}\, I(x+a,x)
\nonumber\\
&=&\int\!\dd^4{a} \int\!\frac{\dd^4{p}}{(2\pi)^4}\,
\Big( \widetilde{F}(\partial_a^2)\, e^{ipa} \Big) e^{ika}\, I(x+a,x)
\eea
and partial integration, the crucial point is to calculate $\widetilde{F}(\partial_a^2)\, e^{ika}\, I(x+a,x)$.
Following the derivation in Ref.~\cite{Faessler3}, we can show that
\be
\widetilde{F}(\partial_a^2)\, e^{ika}\, I(x+a,x) 
= e^{ika}\, \widetilde{F}({\cal D}_a^2)\, I(x+a,x),
\ee 
where ${\cal D}_a = \partial_a + ik_a$. 
Using a Taylor expansion, we can write 
\be
\widetilde{F}({\cal D}_a^2)\, I(x+a,x) 
= \sum_0^\infty \frac{\widetilde{F}^{(0)}(0)}{n!}\, {\cal D}_a^{2n}\, I(x+a,x).
\ee
Based on the path-independent definition of the derivative of $I(x+a,x)$ \cite{Terning}, 
    $(\partial/\partial a^\mu) I(x+a,x) = {\mathscr A}_\mu (x+a)$, 
one finds that 
\be \label{eq:Da2I}
{\cal D}_a^2\, I(x+a,x) 
= L({\mathscr A}) - k^2 I(x+a,x),
\ee
where we define 
$L({\mathscr A}) \equiv \partial^a {\mathscr A}_a(x+a) + 2ik^a {\mathscr A}_a(x+a)$.
Iterating Eq.~(\ref{eq:Da2I}), we have
\bea
\big({\cal D}_a^2\big)^2\, I(x+a,x)
&=& ({\cal D}_a^2-k^2) L({\mathscr A})\,
+\, (-k^2)^2\, I(x+a,x)
\nonumber\\
\big({\cal D}_a^2\big)^3\, I(x+a,x)
&=& ({\cal D}_a^4 - {\cal D}_a^2 k^2 + k^4) L({\mathscr A})\,
+\, (-k^2)^3\, I(x+a,x) 
\nonumber\\
&\cdots&
\nonumber\\ 
\big({\cal D}_a^2\big)^n\, I(x+a,x)
&=& \sum_{l=0}^{n-1}\big({\cal D}_a^2\big)^{n-1-l}\, (-k^2)^l\, L({\mathscr A})\, +\, (-k^2)^n\, I(x+a,x)
\nonumber\\
&=& n \int_0^1 \dd{t} \big( {\cal D}_a^2 t - k^2\bar t\, \big)^{n-1} L({\mathscr A})\, +\, (-k^2)^n\, I(x+a,x),
\eea
which then allows us to write
\bea\label{eq:FtildeDa}
\widetilde{F}({\cal D}_a^2) I(x+a,x)
&=& \int_0^1 \dd{t} \widetilde{F}'\big( {\cal D}_a^2\, t - k^2\,\bar t\, \big) L({\mathscr A})\,
 +\, \widetilde{F}(-k^2) I(x+a,x)
\nonumber\\
&=& i\int\!\frac{\dd^4{q}}{(2\pi)^4}\, {\mathscr A}_\mu(q)\, e^{iq(x+a)}\, (2k^\mu+q^\mu)
\int_0^1\dd{t} \widetilde{F}'\big( -(k+q)^2\, t - k^2\, \bar t\, \big)
\nonumber\\
&& +\ \widetilde{F}(-k^2)\, I(x+a,x).
\eea
After integrating over $p$ and $a$ in Eq.~(\ref{eq:vertex}), the last term in Eq.~(\ref{eq:FtildeDa}) vanishes because of the $\delta$-function, $\delta(a)$.
The vertex corresponding to Fig.~\ref{fig:Feynman}(j) can therefore be written as
\be
V_{\rm (j)}^\mu
= -\frac{e_\phi^q\, C_{B\phi}}{f} (\slashed{k}+\slashed{q}) \gamma_5 
\frac{2k^\mu + q^\mu}{2k\cdot q + q^2}
\left[ \widetilde{F}(k+q) - \widetilde{F}(k)\right] \widetilde{F}(q).
\ee
Using the same technique, we can find the corresponding expression for the octet-decuplet transition vertex in Fig.~\ref{fig:Feynman}(k),
\be
V_{\rm (k)}^{\mu\nu}= -\frac{e_\phi^q\, C_{T\phi}}{f} 
\Big[ k^\mu + q^\mu - \big( Z - \tfrac12 \big) \gamma^\mu (\slashed{k}+\slashed{q})
\Big]
\frac{2k^\nu + q^\nu}{2k\cdot q + q^2}
\left[ \widetilde{F}(k+q)-\widetilde{F}(k)\right] \widetilde{F}(q).
\ee
Finally, the vertex with two mesons and one photon from the gauge link in Fig.~\ref{fig:Feynman}(l) is given by 
\bea\label{eq:vertex_l}
V_{\rm (l)}^\mu &=& \frac{e_\phi^q\, C_{\phi\phi^\dag}}{2f^2}
(\slashed{k}_1+\slashed{k}_2)
\bigg\{ 
\frac{2k_1^\mu +q^\mu}{2k_1\cdot q + q^2}
\left[ \widetilde{F}(k_1+q)-\widetilde{F}(k_1) \right]
\widetilde{F}(k_2)\widetilde{F}(q) 
\nonumber\\
&& \hspace*{2.9cm}
+\, \frac{2k_2^\mu +q^\mu}{2k_2\cdot q + q^2}
\left[ \widetilde{F}(k_2+q)-\widetilde{F}(k_2) \right]
\widetilde{F}(k_1) \widetilde{F}(q) 
\bigg\}.
\eea

With the vertices in Eqs.~(\ref{eq:vertex_abc}) -- (\ref{eq:vertex_l}), we have the complete set of Feynman rules which are needed for the calculation of the nucleon properties in the following sections.
Compared with the traditional effective theory, for the nonlocal Lagrangian at each vertex a momentum dependent vertex function appears, generated from the correlation function in the nonlocal Lagrangian.
The correlation function reflects the non-pointlike behavior of the particles and, in principle, is independent of the hadron type.
In our calculations we therefore set the correlation functions to be the same for all hadrons.
We should also note that the nonlocal magnetic interaction has not been included in the above, but can be easily added within the same approach.

\subsection{\it Nonlocal methods at the quark level}

The nonlocal Lagrangian (\ref{eq:nonlocal}) is based on the framework of chiral effective field theory and is expressed in terms of hadronic degrees of freedom.
This nonlocal realization is not unique, however, and nonlocal Lagrangians can also be constructed from quark degrees of freedom. 
In this subsection we will discuss the nonlocal gauge invariant Lagrangian in the approach of Terning \cite{Terning}, as well as in the nonlocal quark-meson coupling model \cite{Faessler, Ivanov}.

The nonlocal action for quarks interacting with photons in Ref.~\cite{Terning} is introduced by the nonlocal mass term written as (omitting the $F_{\mu\nu}F^{\mu\nu}$ term) 
\be \label{eq:nlTerning}
S = \int\!\dd^4{x}\, \bar{q}(x) \gamma^\mu (\partial_\mu - i e_q \mathscr{A}_\mu)\, q(x)\,
+\, \int\!\dd^4{x}\!\int\!\dd^4{y}\, \bar{q}(x)\, \Sigma(x-y)\,
\exp\Big[-ie_q \int_x^y\dd{\omega}^\nu \mathscr{A}_\nu(\omega) \Big] q(y),
\ee
where $\Sigma(x-y)$ is the nonlocal quark mass function.
For the case when $\Sigma(x-y) \to \delta(x-y)$, Eq.~(\ref{eq:nlTerning}) will reduce to the local QED action.
The path integral of the gauge field guarantees the action in (\ref{eq:nlTerning}) is invariant under the U(1) transformation
\begin{subequations}
\label{eq:gaugequark}
\bea
q(x)\ \to\ q'(x)
&=& \exp\big[ie_q\theta(x)\big]\, q(x),
\\
\bar{q}'(x)\ \to\ \bar{q}'(x)
&=& \exp\big[ie_q\theta(x)\big]\, \bar{q}(x),
\\
\mathscr{A}_\mu(x)\, \to \mathscr{A}'_\mu(x)\!
&=& \!\mathscr{A}_\mu + \partial_\mu \theta(x).
\eea
\end{subequations}
The first and second terms in Eq.~(\ref{eq:nlTerning}) refer to the local (denoted by $S_{\rm L}$) and nonlocal ($S_{\rm NL}$) contributions.
For the local case, the vertex is obtained using minimal substitution and is given by
\be
V_{\rm L}^\mu=i\Gamma_{\rm L}^\mu (x,y,z)
= - \frac{\delta^3 S_L}{\delta \psi(x)\delta \bar{\psi}(y)\delta \mathscr{A}_\mu (z)}\bigg|_{{\cal A}_\mu=0}\ .
\ee
Fourier transforming with the exponemt $e^{i(p'y - px - qz)}$ and dropping the coefficient $(2\pi)^4\delta (p'-p-q)$, one has
\be
V_{\rm L}^\mu = -i e_q \gamma_\mu.
\ee
The nonlocal contribution $S_{\rm NL}$ can be rewritten in momentum space as
\be
S_{\rm NL} = \int\!\frac{\dd^4{k}}{(2\pi)^4} \int\!\frac{\dd^4{p}}{(2\pi)^4}\,
\bar{q}(k)\, \widetilde{\Sigma}(p)\, \widetilde{F}(k-p,p),
\ee
where $\widetilde{\Sigma}(p)$ and $\widetilde{F}(k-p,p)$ are the Fourier transformations of $\Sigma(x-y)$ and $F(x,y)$, respectively, where $F(x,y)$ is defined as $F(x,y) \equiv \exp \left[ -ie_q \int_x^y\, \dd{\omega}^\nu\, \mathscr{A}_\nu(\omega) \right] q(y)$. 
After Taylor expanding $\widetilde{\Sigma}(p)$ and Fourier transforming back to position space, we can write the nonlocal action as 
\be
S_{\rm NL} = \int\!\dd^4{x}\!\int\dd^4{y}\,
\delta(x-y)\, \bar q(x)\,
\left[
    \sum_{n=0}^\infty \frac{1}{n!}\,
    \widetilde{\Sigma}^{(n)}(0)(-\partial_y^2)^n
\right]
\exp\Big[ -ie_q \int_x^y \dd{\omega}^\nu \mathscr{A}_\nu(\omega) \Big] q(y).
\ee
The nonlocal vertex corresponding to $S_{\rm NL}$ is then given by 
\bea
V_{\rm NL}^\mu
&=& i\Gamma_{\rm NL}^\mu (x,y,z)
= - \frac{\delta^3 S_{\rm NL}}
{\delta q(x) \delta \bar{q}(y)\delta \mathscr{A}_\mu(z)}\bigg|_{\mathscr{A}_\mu=0} \nonumber\\
&=& i e_q \int\!\dd^4{x'}\, \delta(y-x')
\left[
  \sum_{n=0}^\infty \frac{1}{n!}\widetilde{\Sigma}^{(n)}(0)(-\partial_{x'}^2)^n
\right]
\int_y^{x'}\dd{\omega^\mu}\, \delta(z-\omega)\, \delta(x-x')
\nonumber\\
&\equiv& \sum_{n=0}^\infty \Gamma_{{\rm NL},n}^\mu(x,y,z).
\eea
Using a similar approach as in the last subsection and in Refs.~\cite{Faessler3, Terning}, the nonlocal vertex in momentum space is expressed as
\bea
V_{\rm NL}^\mu(p,p+q,q) 
&=& -e_q (2p+q)^\mu \sum_{n=0}^\infty \frac{1}{n!}
\widetilde{\Sigma}^{(n)}(0)\frac{(p+q)^{2n}-p^{2n}}{2p\cdot q + q^2} 
\nonumber\\
&=& -e_q \frac{(2p+q)^\mu}{2p\cdot q + q^2} 
\left[ \widetilde{\Sigma}(p+q) - \widetilde{\Sigma}(p) \right].
\eea
The total vertex $V^\mu = V^\mu_{\rm L} + V^\mu_{\rm NL}$ satisfies the Ward-Takahashi-Green identity,
\be
q_\mu V^\mu
= e_q \left[ S^{-1} (p+q) - S^{-1}(p)\right]
= -ie_q \slashed{q} 
- e_q\left[\widetilde{\Sigma}(p+q) - \widetilde{\Sigma}(p)\right],
\ee
where $S^{-1}(p) = -i\slashed{p} - \widetilde{\Sigma}(p)$ is the inverse of the quark propagator for the nonlocal Lagrangian.

For the quark-meson interaction, the nonlocal Lagrangian in the approach of Terning is \cite{Terning}
\bea \label{eq:LqmT}
{\cal L}^{\rm T}(x,y)
&=& g_{\rm qm}\, \bar{q}_2(x)\gamma_5 \Sigma(x-y)
\bigg[
M(x) \exp\Big( ie_{q_1} \int_y^x\dd{\omega^\nu} \mathscr{A}_\nu \Big)
\nonumber \\
&& \hspace*{3.6cm} 
+\,  \exp\Big(-ie_{q_2} \int_y^x\dd{\omega^\nu} \mathscr{A}_\nu \Big) M(y)
\bigg] q_1(y),
\eea
where $M(x)$ is the meson field and $g_{\rm qm}$ is the quark-meson coupling constant.
The Lagrangian is locally invariant under the transformation of (\ref{eq:gaugequark}) and
\be
M(x)\ \to\ M'(x) = \exp\left[ ie_M\theta(x) \right] M(x),
\ee
where $e_M$ is the meson charge.
In analogy to the quark-photon vertex, the quark-meson-photon vertex can be written as
\bea \label{eq:vertexT}
V^{\rm T}(p,p+k+q,k,q)
&=& g_{\rm qm}\gamma_5 
\bigg\{ 
e_{q_2}\, \frac{2p^\mu+2k^\mu+q^\mu}{2(p+k)\cdot q + q^2}
\Big[ \widetilde{\Sigma}(p+k+q) - \widetilde{\Sigma}(p+k) \Big]
\nonumber\\
& & \hspace*{1.2cm}
-\, e_{q_1}\, \frac{2p^\mu+q^\mu}{2p\cdot q + q^2}
\Big[ \widetilde{\Sigma}(p+q) - \widetilde{\Sigma}(p) \Big] 
\bigg\},
\eea
where $p$ and $p+k+q$ are the initial and final quark momenta, and $k$ and $q$ are the meson and photon momenta, respectively.

In the above nonlocal realization, the two quark fields are at spacetime points $x$ and $y$, respectively.
The meson field is at the same position as one of them. 
In particular, the nonlocal mass term leads to a modified propagator.
In the nonlocal quark-meson coupling model \cite{Faessler, Ivanov}, the propagator is the same as in the local case because the free Lagrangian is local and only the interacting Lagrangian is nonlocal, written as 
\bea
{\cal L}^{\rm QMC}(x)
&=& g_{\rm qm}\, M(x) 
\int\dd^4{x_1}\!\int\dd^4{x_2}\,
H(x,x_1,x_2)\, \bar{q}_2(x_2)
\exp\Big(-ie_{q_2} \int_{x_2}^x \dd{\omega^\nu} \mathscr{A}_\nu \Big)
\gamma^5 
\nonumber\\
&& \hspace*{4.4cm}
\times\,
\exp\Big( ie_{q_1} \int_{x_1}^x \dd{\omega^\nu} \mathscr{A}_\nu \Big)\, q_1(x_1),
\eea
where $H(x,x_1,x_2)$ is given by 
\be
H(x,x_1,x_2) = \delta(x-\beta_{21}x_1 - \beta_{12}x_2)\, \Phi\left( (x_1 - x_2)^2 \right).
\ee 
Here $\Phi\left( (x_1 - x_2)^2 \right)$ is the correlation function of two constituent quarks with masses $m_1$ and $m_2$, and $\beta_{ij}=m_j/(m_i+m_j)$.
In contrast to Terning's Lagrangian ${\cal L}^{\rm T}(x,y)$ in Eq.~(\ref{eq:LqmT}), the meson field in the nonlocal quark-meson coupling model is at the center of mass of the two-quark system.
The corresponding quark-meson-photon vertex is obtained as
\bea \label{eq:Vqmb}
V^{\rm QMC}(p,p+k+q,k,q)
&=& g_{\rm qm}\, 
\gamma_5
\bigg\{
e_{q_2}\, 
\frac{2\beta_{12}p^\mu+2\beta_{12}\beta_{21}k^\mu+(1-\beta_{21}^2)q^\mu}
     {2\beta_{12}p\cdot q + 2\beta_{12}\beta_{21} k\cdot q + (1-\beta_{21}^2)q^2}
\nonumber\\
& & \hspace*{1.2cm}
\times\,
\Big[ \widetilde{\Phi}\big(p+q+\beta_{21}k\big)
    - \widetilde{\Phi}\big(p+\beta_{21}(k+q)\big)
\Big]
\nonumber\\
& & \hspace*{-2cm}
-\ e_{q_1}
\frac{2\beta_{21}p^\mu+2\beta_{21}^2k^\mu+\beta_{21}^2q^\mu}
     {2\beta_{21}p\cdot q + 2\beta_{21}^2 k\cdot q + \beta_{21}^2 q^2}
\Big[ \widetilde{\Phi}\big(p+\beta_{21}(k+q)\big) 
    - \widetilde{\Phi}\big(p+\beta_{21}k\big)
\Big] 
\bigg\},
\eea
where $\widetilde{\Phi}$ is the Fourier transform of $\Phi$.
Comparing (\ref{eq:Vqmb}) with Eq.~(\ref{eq:vertexT}), in momentum space the vertex in the nonlocal quark-meson coupling model is slightly different from Terning's nonlocal vertex. 
Our nonlocal chiral effective theory is on the hadron degrees of freedom, where baryons are at coordinate $x$, while the meson and photon are at coordinates $x+a$ and $x+b$, respectively.
As a result, the nonlocal vertex in momentum space depends on the momentum of the meson and photon fields, while in both Terning's approach and the nonlocal quark-meson coupling model, the vertex is related to the momentum of the quark fields.

For the baryon-meson-photon interaction, both the regular vertex from the minimal substitution and the additional vertex from the expansion of the gauge link exist in the nonlocal effective theory, whereas only the nonlocal vertex from the gauge link appears in Terning's approach and nonlocal quark-meson coupling model.
Because of the correlation function in the vertex, the loop integrals in both the nonlocal chiral effective theory and the nonlocal quark-meson coupling model are convergent. 
However, the divergence may still exist in the loop integrals with Terning's nonlocal Lagrangian due to the cancellation of the regulators in the propagator and vertex. 

\section{Nucleon form factors}
\label{Sec.4}

\subsection{\it Nucleon electromagnetic form factors}

In this section we will discuss the electromagnetic form factors of the proton and neutron.
The Dirac and Pauli form factors of the nucleon are defined as
\begin{equation}\label{eq:f1f2}
\Gamma^\mu \equiv \langle N(p^\prime)|J^\mu|N(p) \rangle
= \bar{u}(p^\prime) 
  \left[ \gamma^\mu F_1^N(Q^2)
        + \frac{i\sigma^{\mu\nu} q_\nu}{2M_N} F_2^N(Q^2)
  \right] u(p),
\end{equation}
where $q=p^\prime-p$ and $Q^2=-q^2$.
The electric and magnetic form factors are defined as the combinations of the Dirac and Pauli form factors as
\begin{equation}
G_E^N(Q^2) = F_1^N(Q^2) - \frac{Q^2}{4M_N^2} F_2^N(Q^2)~~~~ {\rm and} ~~~~
G_M^N(Q^2) = F_1^N(Q^2) + F_2^N(Q^2).
\end{equation}
In terms of the electromagnetic form factors, the electric (charge) and magnetic radii of the nucleon are, respectively, obtained via
\be
\langle r^2_E \rangle_N
= -6 \frac{d G^N_E(Q^2)}{d Q^2} \bigg|_{Q^2=0},
~~~~		
\langle r^2_M \rangle_N
= \frac{-6}{G^N_M(0)}\frac{d G^N_M(Q^2)}{d Q^2} \bigg|_{Q^2=0}.
\ee

\begin{figure}[tb] 
\begin{center}
\epsfig{file=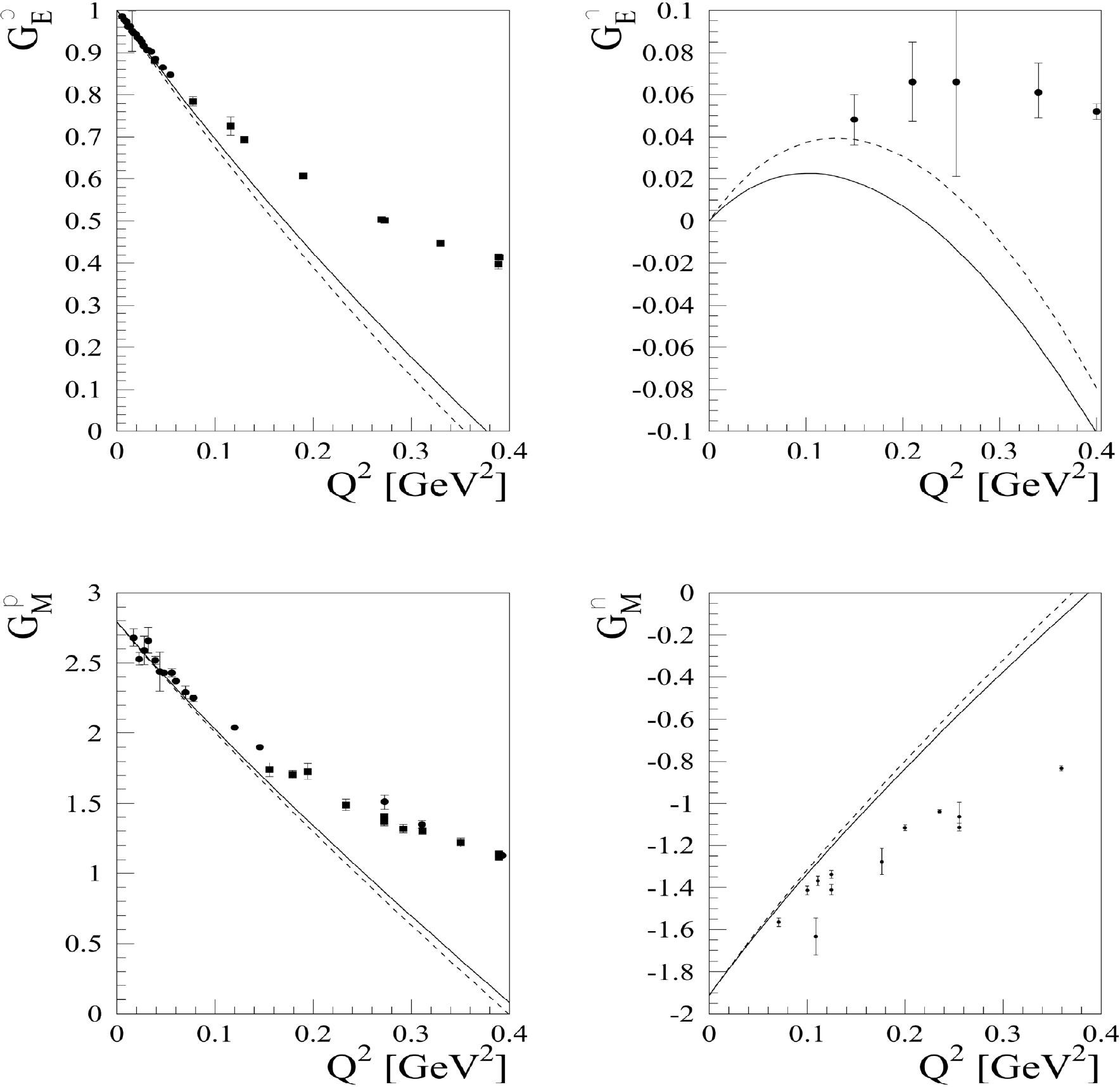,scale=0.70}
\caption{The electric (top row) and magnetic (bottom row) form factors of the proton (left column) and neutron (right column) at ${\cal O}(q^4)$ in the EOMS (solid lines) and infrared (dashed lines) schemes, compared with experimental data (points with error bars). (Figure from Ref.~\cite{Fuchs2}.)}
\label{fig:GEMDR}
\end{center}
\end{figure}

The nucleon electromagnetic form factors have been calculated in chiral perturbation theory up to ${\cal O}(q^4)$.
It was found that the experimental data can only be described well for momenta $Q^2 \lesssim 0.1$~GeV$^2$ in either the infrared or EOMS regularization schemes \cite{Fuchs2}.
Figure \ref{fig:GEMDR} summarises the results for the form factors up to ${\cal O}(q^4)$ \cite{Fuchs2}.
It is clear for all the electric and magnetic form factors of the proton and neutron that the discrepancy between the theoretical calculation and experimental results rapidly increases when $Q^2 > 0.1$~GeV$^2$.
There is not much difference between the results in the infrared and EOMS regularization schemes, as they have the same nonanalytic terms.
The difference caused by the different subtracted terms can be reduced by properly choosing the low energy constants.
Although the results are closer to experiment at $Q^2 \lesssim 0.4$~GeV$^2$ when vector mesons are explicitly included \cite{Kubis2}, it is generally difficult for traditional $\chi$PT to describe the nucleon form factors quantitatively at relatively large momentum transfer, $Q^2 \gtrsim 1$~GeV$^2$.

The nonlocal Lagrangian and electric currents have been discussed in the previous section.
To study nucleon form factors within the nonlocal chiral effective theory, as in local $\chi$PT, the octet, decuplet and octet-decuplet transition operators for magnetic interactions are needed in the one-loop calculations. 
The baryon octet anomalous magnetic Lagrangian is written as \cite{He,Yang2}
\begin{equation}
\label{eq:magB}
\mathcal{L}_{\mathrm{oct}} 
= \frac{e}{4M_B}
\Big(
  c_1\,{\rm Tr}
  \left[\bar{B} \sigma^{\mu\nu}\left\{F^+_{\mu\nu},B\right\}\right] 
+ c_2\,{\rm Tr}
  \left[\bar{B}\sigma^{\mu\nu} \left[F^+_{\mu\nu},B\right]\right] 
+ c_3\,{\rm Tr}
  \left[\bar{B}\sigma^{\mu\nu}B\right]\,{\rm Tr}\left[F^+_{\mu\nu}\right]
\Big),
\end{equation}
where
\begin{equation}
F^+_{\mu\nu} = -\frac12\left(u^\dag F_{\mu\nu}\,Q_c\,u + u
		F_{\mu\nu}\,Q_c\,u^\dag\right).
\end{equation}
Here, $F_{\mu\nu}$ is the external electromagnetic field strength tensor interacting with a quark of flavor $q = u, d$ or $s$.
At the lowest order, the contribution of quark $q$ with unit charge to the octet magnetic moments can be obtained by replacing the charge matrix $Q_c$ with the corresponding diagonal quark matrices 
$\lambda_q = {\rm diag}(\delta_{qu}, \delta_{qd}, \delta_{qs})$.
Taking the nucleon as an example, after expanding Eq.~(\ref{eq:magB}), we find the quark flavor decomposition for the proton and neutron magnetic moments is given by
\begin{subequations}
\bea\label{treemagchiPT}
F_2^{p,u}&\!=\!&c_1+c_2+c_3,
~~~~~~~F_2^{p,d}=c_3,
~~~~~~~~~~~~~~~~~~~~~F_2^{p,s}=c_1-c_2+c_3, 
\\
F_2^{n,u}&\!=\!&c_3,
~~~~~~~~~~~~~~~~~~~F_2^{n,d}=c_1+c_2+c_3,
~~~~~~~~~F_2^{n,s}=c_1-c_2+c_3.
\eea
\end{subequations}
Comparing with the results of the constituent quark model where $F_2^{p,s}=F_2^{n,s}=0$, we obtain the relation $c_3=c_2-c_1$.
On par with the octet anomalous magnetic moment operators, the decuplet anomalous magnetic moment operator is expressed as \cite{He,Yang2}
\begin{equation}\label{eq:magT}
\mathcal{L}_{\mathrm{dec}}
= - \frac{ie F_2^T}{4M_T}\,
	\overline{T}_{\mu}^{abc} \sigma_{\rho\lambda} F^{\rho\lambda} T^{\mu,abc},
\end{equation}
and the transition magnetic operator is written as \cite{He,Yang2}
\begin{equation} \label{eq:magBT}
\mathcal{L}_{\mathrm{trans}}
= i\frac{e}{4M_B} \mu_T F_{\mu\nu}
\Big(
  \epsilon^{ijk} Q_{c,il}\bar{B}_{jm} \gamma^\mu \gamma_5\, T^{\nu,klm} 
+ \epsilon^{ijk} Q_{c,li}\overline{T}^{\mu,klm}\,\gamma^\nu\gamma_5\, B_{mj}
\Big).
\end{equation}
The anomalous magnetic moments of baryons can also be expressed in terms of quark magnetic moments,~$\mu_q$.
For example, for the nucleon one has $\mu_p = (4\mu_u - \mu_d)/3$ and $\mu_n = (4\mu_d - \mu_u)/3$, while for the $\Delta^{++}$ the magnetic moment is given by $\mu_{\Delta^{++}} = 3\mu_u$.
Using SU(3) flavor symmetry, $\mu_u = -2\mu_d = -2\mu_s$, while $\mu_T$ and $F_2^T$, as well as $\mu_q$, can be written in terms of $c_1$ or $c_2$.
For example, one has $\mu_u = 2c_1/3$, $\mu_T= 4 c_1$, and $F_2^{\Delta^{++}} = \mu_{\Delta^{++}}-2 = 2c_1-2$. 
The parameters $c_1$ and $c_2$ can then be determined by the experimental magnetic moments of the proton and neutron.

The nonlocal electromagnetic interaction between a proton and a photon can be written as
\be
{\cal L}_{{\rm em}, p}^{({\rm nonloc})} 
= -e \int\dd^4{a}\, \bar{p}(x)\gamma^\mu p(x){\mathscr A}_\mu(x+a)F_1(a) 
+ \frac{(c_1 + 3c_2)e}{12 M_N} 
  \int\dd^4{a}\, \bar{p}(x)\sigma^{\mu\nu}p(x)\, F_{\mu\nu}(x+a) F_2(a),
\ee
where $F_1(a)$ and $F_2(a)$ are the correlation functions for the nonlocal electric and magnetic interactions.
The form factors at tree level, which are momentum dependent, can be easily obtained using a Fourier transformation. 
The simplest choice is to assume that the correlation function of the nucleon electromagnetic vertex is the same as that of the nucleon-meson vertex, namely, $F_1(a) = F_2(a) = F(a)$.
For this simple choice, however, the Dirac and Pauli form factors will have the same dependence on the momentum transfer at tree level, and the obtained charge form factor of the proton will fall rapidly with $Q^2$, becoming negative at large $Q^2$.
A better choice may be to assume that the charge and magnetic form factors at tree level have the same momentum dependence as the nucleon-meson vertex, $G_M^{\rm tree}(Q^2) = \mu_p^{\rm tree} G_E^{\rm tree}(Q^2) = \mu_p^{\rm tree} \widetilde{F}(Q^2)$, where $\widetilde{F}(Q^2)$ is the Fourier transform of the correlation function $F(a)$. 
Here, $\mu_p^{\rm tree}= 1 + \frac13 c_1 + c_2$ is the tree level magnetic moment of the proton. 
The corresponding functions $\widetilde{F}_1(Q^2)$ and $\widetilde{F}_2(Q^2)$ can then be written as 
\be \label{eq:F1F2regulator}
\widetilde{F}_1(Q^2)
= \frac{4M_N^2 + Q^2\mu_p^{\rm tree}}{4M_N^2+Q^2}\widetilde{F}(Q^2),
~~~~~~~ \widetilde{F}_2(Q^2) = \frac{4M_N^2}{4M_N^2+Q^2}\widetilde{F}(Q^2).
\ee
In the heavy baryon limit, $Q^2/M_N^2 \to 0$, we note that the above two choices are equivalent.

\begin{figure}[tbph] 
\begin{center}
\epsfig{file=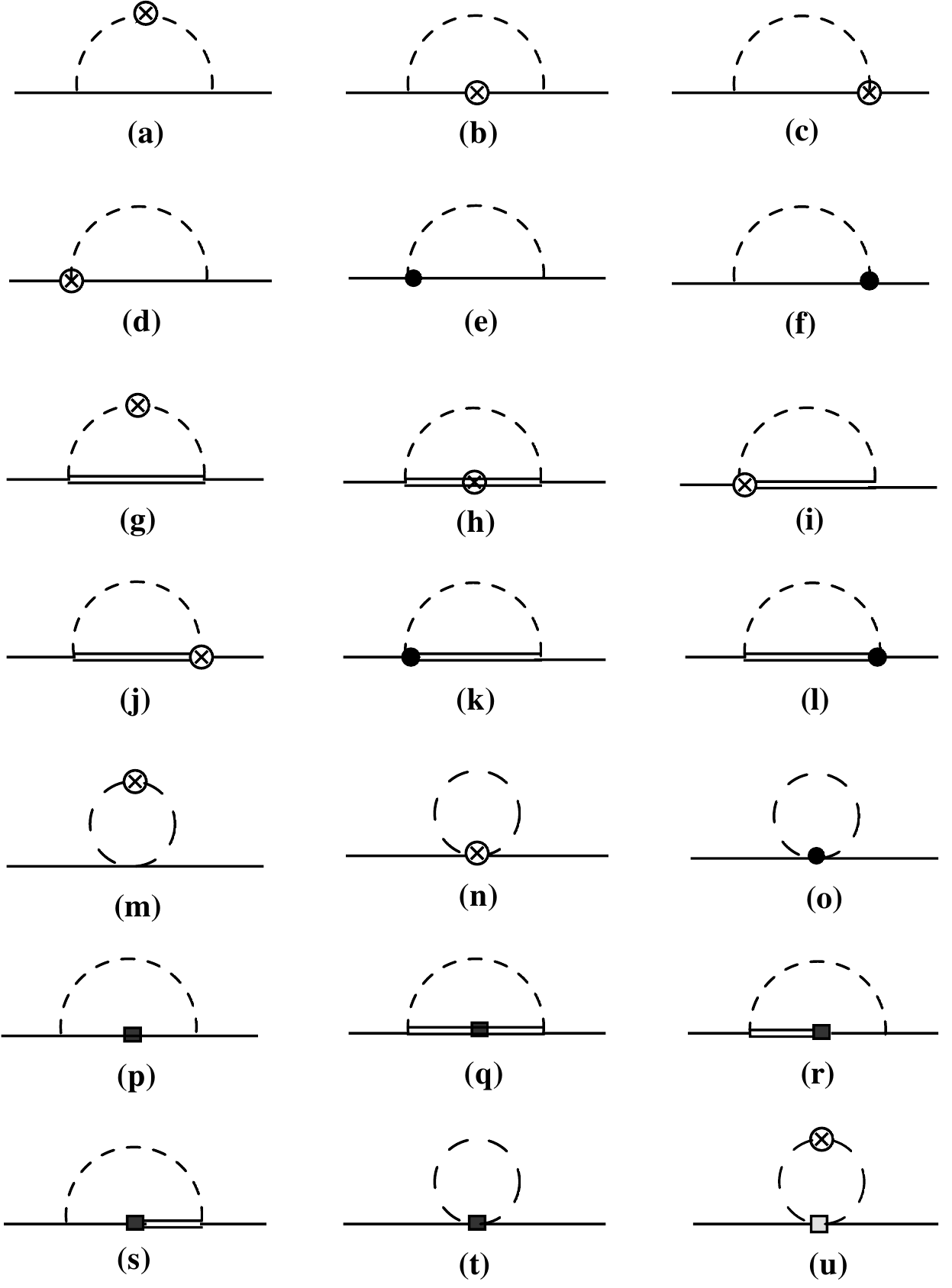,scale=0.8}
\caption{
One-loop Feynman diagrams for the nucleon electromagnetic form factors. The solid, double-solid and dashed lines are for the octet baryons, decuplet baryons and pseudoscalar mesons, respectively.
The circles denote additional gauge link interaction with the external field, while black and gray squares represent the magnetic interactions. (Figure from Ref.~\cite{He5}.)}
\label{fig:FMLOOP}
\end{center}
\end{figure}

According to the nonlocal Lagrangian, the one-loop Feynman diagrams which contribute to the nucleon electromagnetic form factors 
can be summarized as shown in Fig.~\ref{fig:FMLOOP}.
Due to the gauge link, there are additional diagrams where the vertices are represented by black dots. 
These additional diagrams guarantee that the values of the electric form factors of the nucleon at $Q^2=0$ are given by their charges.
To illustrate the gauge invariance in the nonlocal quantum field theory, we consider specifically the pion loop for the external proton in Fig.~\ref{fig:FMLOOP}. 
In particular, we may contrast the neutral pion case versus the charged pion case for the proton form factor computation.

For the neutral pion case, we may take Fig.~\ref{fig:FMLOOP}(b) which provides the expression of the allowed meson loop contributions as
\bea
\Gamma^{\mu,p}_{({\rm b})} 
&=& \frac{(D+F)^2}{4f^2} 
\frac{4M_N^2+\mu_p^{\rm tree}Q^2}{4M_N^2+Q^2}\,
I_{\pi^0 ({\rm b})}^{\mu,p}\
+\frac{(D+F)^2}{2f^2}\frac{\mu_n^{\rm tree}Q^2}{4M_N^2+Q^2}
I_{\pi^+ ({\rm b})}^{\mu,p}\,
\,+\,
\frac{(D+3F)^2}{12f^2} 
\frac{\mu_\Lambda^{\rm tree}Q^2}{4M_\Lambda^2+Q^2}\, 
I_{K ({\rm b})}^{\mu,\Lambda}\
\nonumber\\
&+& \frac{(D-F)^2}{8f^2}\frac{8M_\Sigma^2+(\mu_{\Sigma^0}^{\rm tree}+2\mu_{\Sigma^+}^{\rm tree})Q^2} {4M_\Sigma^2 +Q^2}\,
I_{K ({\rm b})}^{\mu,\Sigma}\
-\ \frac{(D-F)(D+3F)}{4\sqrt{3}f^2}
\frac{\mu_{\Sigma\Lambda}^{\rm tree}Q^2} {4M_\Sigma^2+Q^2}\, I_{K ({\rm b})}^{\mu,\Sigma\Lambda},
\eea
where the tree level magnetic moments $\mu_B^{\rm tree}$ can be obtained from Eq.~(\ref{eq:magB}). 
Here, the generic expression of the loop integral $I_{M ({\rm b})}^{\mu,B}$ is given by
\be
I_{M ({\rm b})}^{\mu,B}
= e\widetilde F(q)\,
\bar{u}(p')
\int\frac{\dd^4{k}}{(2\pi)^4} \slashed{k}\gamma_5 \widetilde{F}(k)
\frac{1}{k^2-m_M^2} \frac{1}{\slashed{p}'-\slashed{k}-M_B}\gamma^\mu \frac{1}{\slashed{p}-\slashed{k}-M_B}\slashed{k}\gamma_5
\widetilde{F}(k)\,
u(p),
\ee
where $m_M$ and $M_B$ are the masses of the intermediate meson and baryon.
In this expression, the $\widetilde{F}(q)$ factor outside the integral is generated from the nonlocal baryon--photon vertex which provides the momentum dependence of the tree level form factors of intermediate baryons. 
The two $\widetilde{F}(k)$ factors inside the integral, however, arise from the two nonlocal baryon--meson vertices, which make the loop integral convergent.
In the local effective theory, both $\widetilde{F}(q)$ and $\widetilde{F}(k)$ are taken as 1. Now, without taking into account the magnetic moments $\mu_B^{\rm tree}$ for simplicity, we can obtain the Ward identity for the neutral pion case as
\be
q_\mu \Gamma_{\pi^0 ({\rm b})}^{\mu,p}
= e\frac{(D+F)^2}{4f^2}\widetilde F(q)\,
\bar{u}(p')
\int\frac{\dd^4{k}}{(2\pi)^4} \slashed{k}\gamma_5 \widetilde{F}(k)
\frac{1}{k^2-m_\pi^2} \frac{1}{\slashed{p}'-\slashed{k}-M_N}\slashed{q}
\frac{1}{\slashed{p} -\slashed{k}-M_N}\slashed{k}\gamma_5
\widetilde{F}(k)\,
u(p),
\ee
where $\slashed{q}$ can be replaced by $\slashed{p}'-\slashed{p}$ to yield
\bea
q_\mu \Gamma_{\pi^0 ({\rm b})}^{\mu,p}
&=& e\frac{(D+F)^2}{4f^2}\widetilde F(q)\,
\bar{u}(p')
\int\frac{\dd^4{k}}{(2\pi)^4} \slashed{k}\gamma_5 \widetilde{F}(k)
\frac{1}{k^2-m_\pi^2}\frac{1}{\slashed{p}-\slashed{k}-M_N}
 \slashed{k}\gamma_5
\widetilde{F}(k)\,
u(p)
\nonumber\\
&-&e\frac{(D+F)^2}{4f^2}\widetilde F(q)\,
\bar{u}(p')
\int\frac{\dd^4{k}}{(2\pi)^4} \slashed{k}\gamma_5 \widetilde{F}(k)
\frac{1}{k^2-m_\pi^2}\frac{1}{\slashed{p}'-\slashed{k}-M_N}
 \slashed{k}\gamma_5
\widetilde{F}(k)\,
u(p)
\nonumber\\
&=&-e\widetilde F(q)\big(\Sigma_{\pi^0}(p')-\Sigma_{\pi^0}(p)\big).
\eea
Here, $\Sigma_{\pi^0}(p)$ represents the neutral pion loop correction to the proton self-energy which corresponds to the wave function renormalization presented explicitly in the Appendix~B of Ref.~\cite{Anatomy} for the local theory derivation.

For the charged pion case, Figs.~\ref{fig:FMLOOP}(a), \ref{fig:FMLOOP}(c), \ref{fig:FMLOOP}(d), \ref{fig:FMLOOP}(e) and \ref{fig:FMLOOP}(f) will provide the contributions to the proton form factors.
Without taking into account $\mu_B^{\rm tree}$ for simplicity again, the contribution of Fig.~\ref{fig:FMLOOP}(a) can be expressed as 
\be
\Gamma^{\mu,p}_{({\rm a})}
= \frac{(D+F)^2}{2f^2} I^{\mu,n}_{\pi^+ ({\rm a})} 
+ \frac{(D+3F)^2}{12f^2} I^{\mu,\Lambda^+}_{K ({\rm a})} 
+ \frac{(D-F)^2}{4f^2} I^{\mu,\Sigma}_{K^+ ({\rm a})},
\ee
where the generic expression of the loop integral $I^{\mu,B}_{M ({\rm a})}$ is given by
\be
I^{\mu,B}_{M ({\rm a})}
= e\widetilde{F}(q)\,
\bar{u}(p')\int\!\frac{\dd^4{k}}{(2\pi)^4}(\slashed{k}+\slashed{q})\gamma_5\,
\widetilde F(q+k)\frac{1}{(k+q)^2-m_\pi^2}(2k+q)^\mu\frac{1}{k^2-m_\pi^2}\frac{1}{\slashed{p}-\slashed{k}-
M_N}
\slashed{k}\gamma_5\widetilde F(k)\,
u(p).
\ee
The contribution from Figs.~\ref{fig:FMLOOP}(c) and \ref{fig:FMLOOP}(d) can be expressed as
\be
\Gamma^{\mu,p}_{({\rm c})+({\rm d})}
= -\frac{(D+F)^2}{2f^2} I^{\mu,n}_{\pi^+ ({\rm c})+({\rm d})} 
- \frac{(D+3F)^2}{12f^2} I^{\mu,\Lambda}_{K^+ ({\rm c})+({\rm d})} 
- \frac{(D-F)^2}{4f^2} I^{\mu,\Sigma}_{K^+ ({\rm c})+({\rm d})},
\ee
where the integral $I^{\mu,B}_{M ({\rm c})+({\rm d})}$ is given by
\bea
I^{\mu,B}_{M ({\rm c})+({\rm d})}
&=& e\widetilde{F}(q)\,
\bar{u}(p')\int\!\frac{\dd^4{k}}{(2\pi)^4}\slashed{k}\gamma_5 \widetilde F(k)
\frac{1}{\slashed{p}'-\slashed{k}-M_B}
\frac{1}{k^2-m_M^2} \gamma^\mu\gamma_5 
\widetilde F(k-q)\,
u(p)     
\nonumber\\
&+& e\widetilde{F}(q)\,
\bar{u}(p')\int\!\frac{\dd^4{k}}{(2\pi)^4} 
\gamma^\mu\gamma_5 \widetilde F(k+q)\frac{1}
{\slashed{p}-\slashed{k}-M_B}
\frac{1}{k^2-m_M^2}\slashed{k}\gamma_5 \widetilde F(k)\,
u(p).
\eea

For the additional diagrams in Figs.~\ref{fig:FMLOOP}(e) and \ref{fig:FMLOOP}(f) from the nonlocal interactions, which do not exist in the local theory, the contribution is given by
\be
\Gamma^{\mu,p}_{({\rm e})+({\rm f})}
= -\frac{(D+F)^2}{2f^2} I^{\mu,n}_{\pi^+ ({\rm e})+({\rm f})} 
- \frac{(D+3F)^2}{12f^2} I^{\mu,\Lambda}_{K ({\rm e})+({\rm f})} 
- \frac{(D-F)^2}{4f^2} I^{\mu,\Sigma}_{K ({\rm e})+({\rm f})},
\ee
where the generic expression of the integral $I^{\mu,B}_{M ({\rm e})+({\rm f})}$ is given by 
\bea
I^{\mu,B}_{M ({\rm e})+({\rm f})}
&=& e\widetilde{F}(q)\,
\bar{u}(p')
\int\frac{\dd^4{k}}{(2\pi)^4} \slashed{k}\gamma_5
\frac{2k^\mu - q^\mu}{2k\cdot q - q^2}
\left[ \widetilde{F}(k)-\widetilde{F}(k-q)\right]
\frac{1}{\slashed{p}'-\slashed{k}-M_B}\frac{1}{k^2-m_M^2}\slashed{k}\gamma_5\widetilde{F}(k)\,
u(p)
\nonumber\\
&+&   
e\widetilde{F}(q)\,
\bar{u}(p')
\int\frac{\dd^4{k}}{(2\pi)^4} \slashed{k}\gamma_5
\frac{2k^\mu + q^\mu}{2k\cdot q + q^2}
\left[ \widetilde{F}(k+q)-\widetilde{F}(k)\right]
\frac{1}{\slashed{p}-\slashed{k}-M_B}\frac{1}{k^2-m_M^2}\slashed{k}\gamma_5
\widetilde{F}(k)\,
u(p).
\nonumber\\
\eea
Note here that the $I^{\mu,N}_{\pi^0({\rm e})+({\rm f})}$ term is absent as the nonlocal baryon--meson vertex is proportional to the meson charge.
For the charged pion case, we thus have the following contributions to the Ward identity,
\bea
&& q_\mu
\left( 
  \Gamma_{\pi^+({\rm a})}^{\mu,n} 
+ \Gamma_{\pi^+({\rm c})+({\rm d})}^{\mu,n}
+ \Gamma_{\pi^+({\rm e})+({\rm f})}^{\mu,n}
\right)
\nonumber\\
&&=e\frac{(D+F)^2}{2f^2}\widetilde{F}(q)\,
\bar{u}(p')\int\!\frac{\dd^4{k}}{(2\pi)^4}(\slashed{k}+\slashed{q})\gamma_5\,
\widetilde F(q+k)\frac{1}{k^2-m_\pi^2}
\frac{1}{\slashed{p}-\slashed{k}-M_N}
\slashed{k}\gamma_5\widetilde F(k)\,
u(p)
\nonumber\\
&&-e\frac{(D+F)^2}{2f^2}\widetilde{F}(q)\,
\bar{u}(p')\int\!\frac{\dd^4{k}}{(2\pi)^4}(\slashed{k}+\slashed{q})\gamma_5\,
\widetilde F(q+k)\frac{1}{(k+q)^2-m_\pi^2}
\frac{1}{\slashed{p}-\slashed{k}-M_N}
\slashed{k}\gamma_5\widetilde F(k)\,
u(p)
\nonumber\\
&&-e\frac{(D+F)^2}{2f^2}\widetilde{F}(q)\,
\bar{u}(p')\int\!\frac{\dd^4{k}}{(2\pi)^4}\slashed{k}\gamma_5 \widetilde F(k)
\frac{1}{\slashed{p}'-\slashed{k}-M_N}
\frac{1}{k^2-m_\pi^2} \slashed{q}\gamma_5 
\widetilde F(k-q)\,
u(p)    
\nonumber\\
&&-e\frac{(D+F)^2}{2f^2}\widetilde{F}(q)\,
\bar{u}(p')\int\!\frac{\dd^4{k}}{(2\pi)^4} 
\slashed{q}\gamma_5 \widetilde F(k+q)\frac{1}
{\slashed{p}-\slashed{k}-M_N}
\frac{1}{k^2-m_\pi^2}\slashed{k}\gamma_5 \widetilde F(k)\,
u(p)
\nonumber\\
&&-e\frac{(D+F)^2}{2f^2}\widetilde{F}(q)\,
\bar{u}(p')
\int\frac{\dd^4{k}}{(2\pi)^4} \slashed{k}\gamma_5
\left[ \widetilde{F}(k)-\widetilde{F}(k-q)\right]
\frac{1}{\slashed{p}'-\slashed{k}-M_N}\frac{1}{k^2-m_\pi^2}\slashed{k}\gamma_5\widetilde{F}(k)\,
u(p)
\nonumber\\
&&-e\frac{(D+F)^2}{2f^2}\widetilde{F}(q)\,
\bar{u}(p')
\int\frac{\dd^4{k}}{(2\pi)^4} \slashed{k}\gamma_5
\left[ \widetilde{F}(k+q)-\widetilde{F}(k)\right]
\frac{1}{\slashed{p}-\slashed{k}-M_N}\frac{1}{k^2-m_\pi^2}\slashed{k}\gamma_5
\widetilde{F}(k)\,
u(p),~~~~
\eea
where the first term can be cancelled with the fourth term and the first part of the last term.
After shifting the momentum $k \to k-q$ in the second term, we can see its cancellation with the third term and the second part of the fifth term. 
The remaining terms can then be expressed as
\bea
&& q_\mu
\left(
  \Gamma_{\pi^+({\rm a})}^{\mu,n} 
+ \Gamma_{\pi^+({\rm c})+({\rm d})}^{\mu,n}
+ \Gamma_{\pi^+({\rm e})+({\rm f})}^{\mu,n}
\right)
\nonumber\\
&&=e\frac{(D+F)^2}{2f^2}\widetilde{F}(q)\,
\bar{u}(p')
\int\frac{\dd^4{k}}{(2\pi)^4} \slashed{k}\gamma_5 \widetilde{F}(k)
\frac{1}{\slashed{p}-\slashed{k}-M_N}\frac{1}{k^2-m_\pi^2}\slashed{k}\gamma_5\widetilde{F}(k)\,
u(p)
\nonumber\\
&&-e\frac{(D+F)^2}{2f^2}\widetilde{F}(q)\,
\bar{u}(p')
\int\frac{\dd^4{k}}{(2\pi)^4} \slashed{k}\gamma_5 \widetilde{F}(k)
\frac{1}{\slashed{p}'-\slashed{k}-M_N}\frac{1}{k^2-m_\pi^2}\slashed{k}\gamma_5\widetilde{F}(k)\,
u(p)
\nonumber\\
&&=-e\widetilde F(q)\big(\Sigma_{\pi^+}(p')-\Sigma_{\pi^+}(p)\big),
\eea
where $\Sigma_{\pi^+}(p)$ represents the charged pion loop contribution to the proton self-energy corresponding to the wave function renormalization presented again explicitly in Appendix~B of Ref.~\cite{Anatomy} for the local theory derivation. 
We note here the cancellation of the terms among $q_\mu I_{M ({\rm a})}^{\mu,B}$, $q_\mu I_{M ({\rm c})+({\rm d})}^{\mu,B}$ and $q_\mu I^{\mu,B}_{M ({\rm e})+({\rm f})}$, demonstrating the gauge invariance without taking the local limit, $\widetilde{F}(k)=1$.
This assures that the Ward identity is satisfied in the nonlocal theory and the values of electric form factors of the nucleon at $Q^2=0$ must be their charges with the additional diagrams generated in the nonlocal Lagrangian.

Having now established the correct Ward identity both for the neutral pion case and the charged pion case, we briefly describe the numerical computation in the nonlocal theory. 
In practice, we take a dipole form for $\widetilde{F}(k)$,
\be \label{DipoleFF}
\widetilde{F}(k) = \left(\frac{\Lambda^2}{k^2-m_j^2- \Lambda^2}\right)^2,
\ee
where $\Lambda$ is a free parameter, while $m_j=m_M$ and $m_j=0$ for the baryon--meson and baryon--photon vertices, respectively.
The results are found to be very close to the experimental nucleon form factors when $\Lambda \approx 0.9$~GeV.
Further details of the numerical results for the nucleon form factors can be found in Ref.~\cite{He}.

\begin{figure}[]
\begin{center}
\hspace{-0.1in}\includegraphics[scale=0.4]{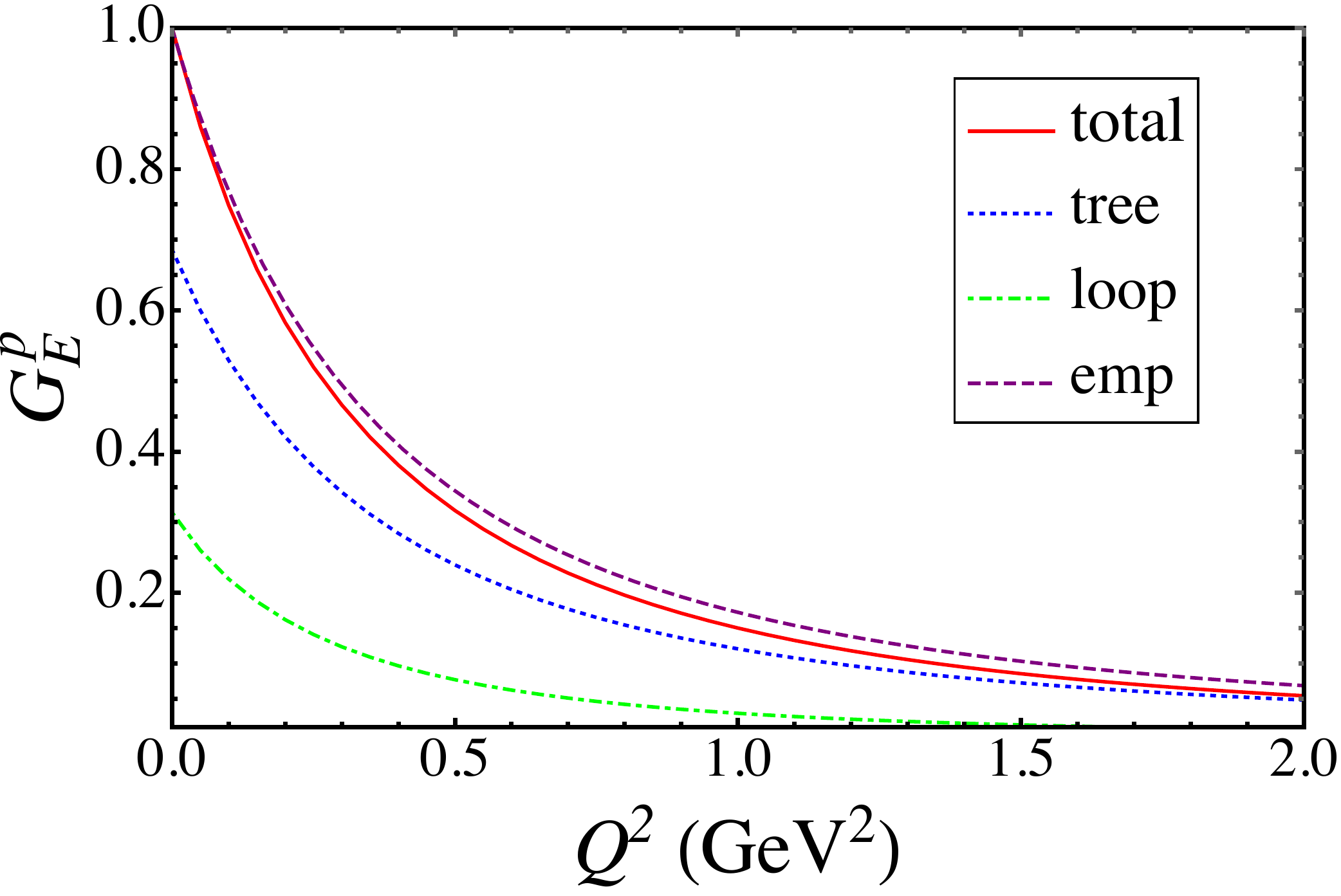}
\hspace{0.3in}\includegraphics[scale=0.4]{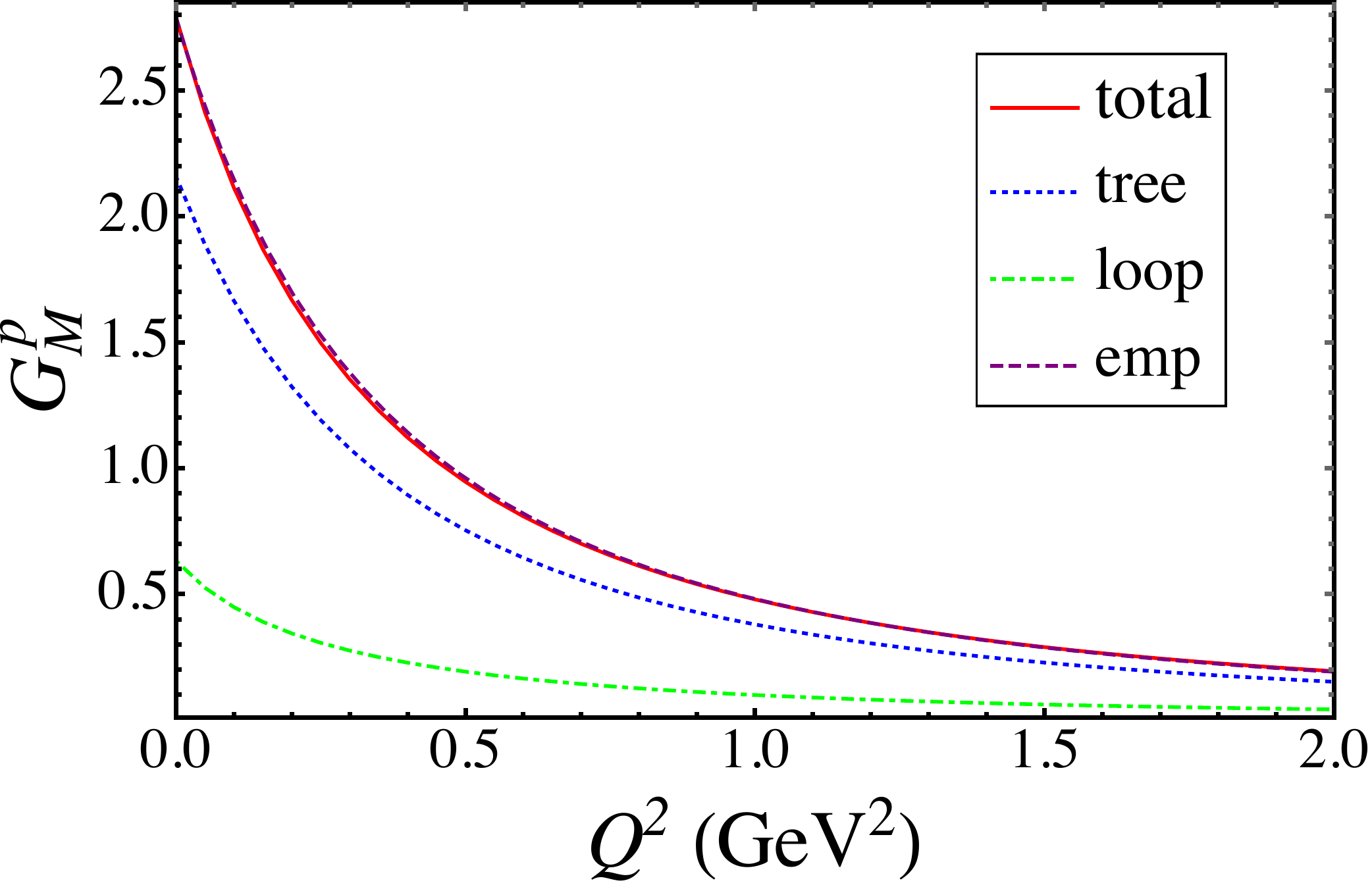}
\caption{The proton electric (left) and magnetic (right) form factors versus momentum transfer $Q^2$ with $\Lambda= 0.9$~GeV, showing the tree (blue dotted lines), loop (green dot-dashed lines) and total (red solid lines) contributions, compared with the empirical result (magenta dashed lines). (Figure from Ref.~\cite{He}.)}
\label{fig:GEMP}
\end{center}
\end{figure}

The calculated proton electric and magnetic form factors $G^p_E$ and $G_M^p$ are shown in Fig.~\ref{fig:GEMP} as a function of $Q^2$. 
The calculation is found to describe the empirical result,
    $G_{M,\rm emp}^p = 2.79\, G_{E,\rm emp}^p$, where
    $G_{E,\rm emp}^p = 1/(1 + Q^2/0.71\,{\rm GeV}^2)^2$,
relatively well, with the loop contributing around 30\% of the total.
As noted above, the nonlocal Lagrangian generates the covariant regulator which makes the loop integral convergent on the one hand, while also generating the $Q^2$ dependent contribution at tree level on the other. 
In contrast to the conventional $\chi$PT, the tree level contribution is not expanded in powers of momentum transfer.
As a result, both the tree and loop contributions decrease smoothly with increasing $Q^2$. 
The net results are close to the empirical values up to $Q^2 = 2$~GeV$^2$.
For the electric form factor, at $Q^2 = 0$ the sum of the tree and loop contributions to the proton charge is unity, as required.
The proton magnetic radius is 0.848~fm in our calculation, which is close to the experimental value. 
Additional diagrams generated from the expansion of the gauge link are crucial to attain the renormalized proton charge 1, as explained earlier. Compared to the magnetic form factor, the charge form factor decreases slightly faster.
As a result, the obtained charge radius 0.857~fm is a little larger than the magnetic radius.

\begin{figure}[]
\begin{center}
\hspace{-0.1in}\includegraphics[scale=0.4]{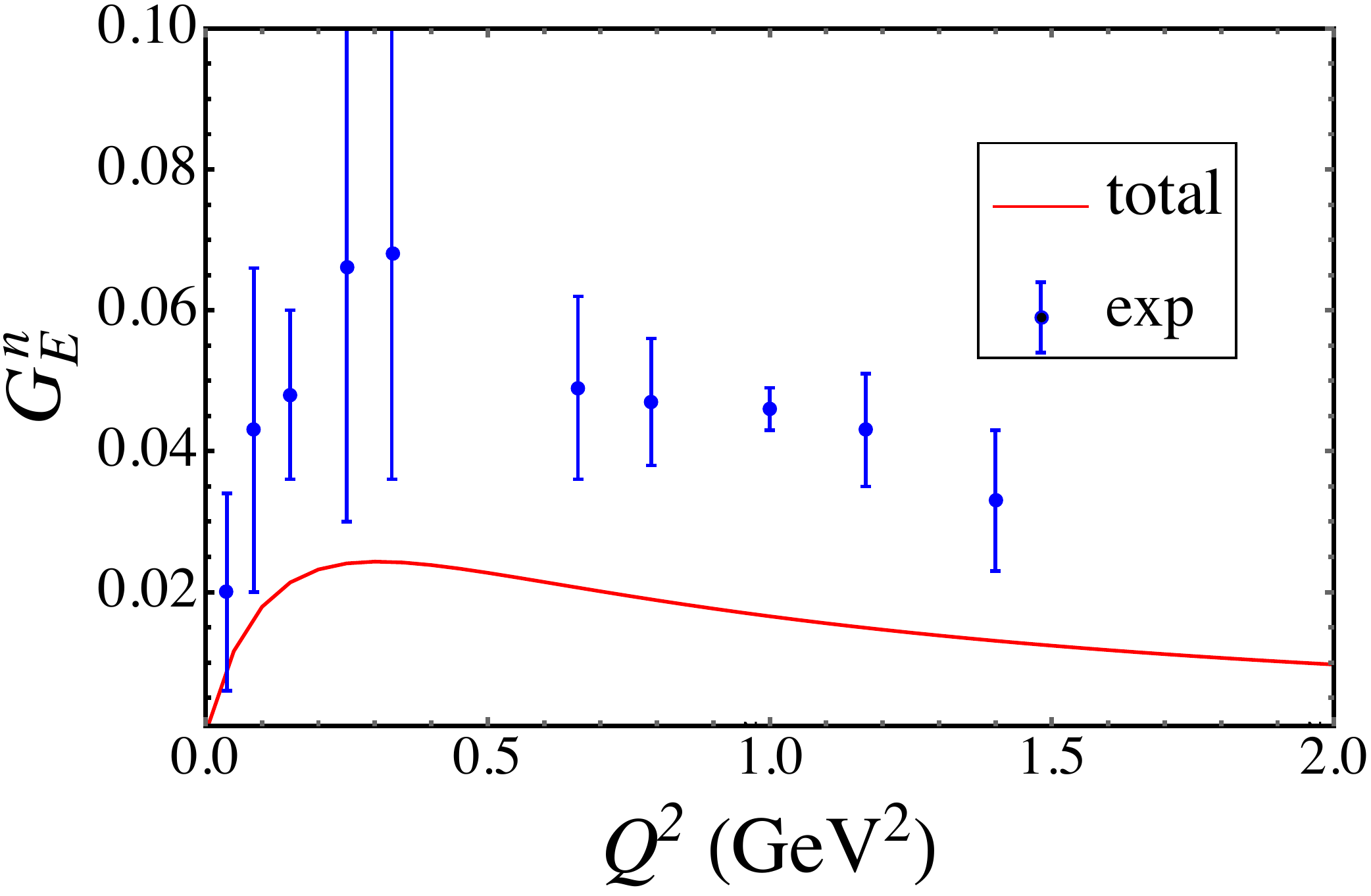}
\hspace{0.3in}\includegraphics[scale=0.4]{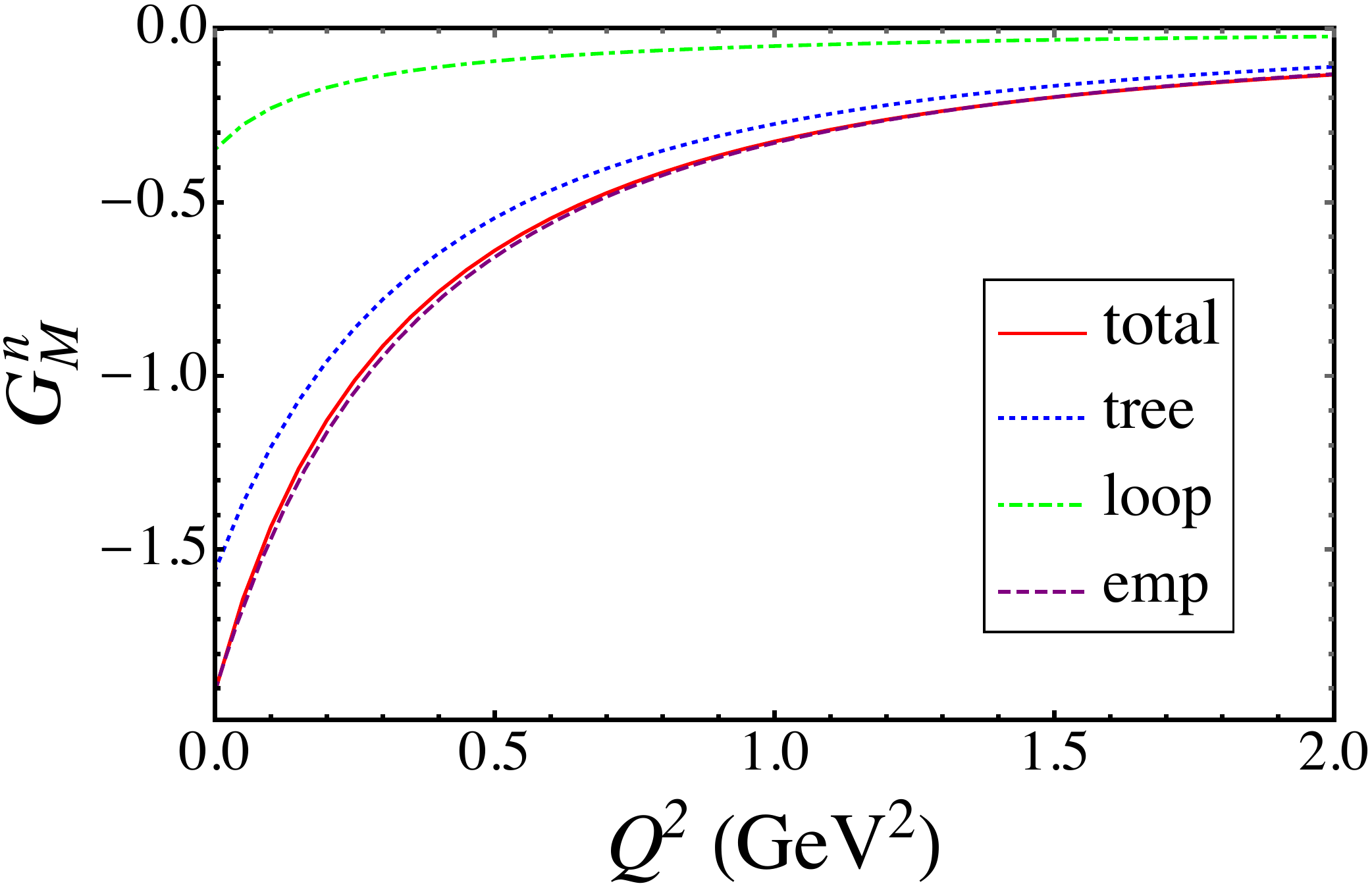}
\caption{The neutron electric (left) and magnetic (right) form factors versus momentum transfer $Q^2$ with $\Lambda= 0.9$~GeV for the tree (blue dotted lines), loop (green dot-dashed lines) and total (red solid lines) contributions, and compared with the empirical result (magenta dashed lines).
The tree contribution to $G_E^n$ is zero. (Figure from Ref.~\cite{He}.)}
\label{fig:GEMN}
\end{center}
\end{figure}

The calculated neutron electric and magnetic form factors $G^n_E$ and $G_M^n$ are shown in Fig.~\ref{fig:GEMN} versus $Q^2$.
For the neutron electric form factor, all the contributions are entirely from the loop diagrams. 
The neutron charge radius $\langle (r^n_M)^2 \rangle = -0.077$~fm$^2$, which is a little smaller in magnitude than the experimental value $-0.11$~fm$^2$.
The calculated electric form factor of neutron in Fig.~\ref{fig:GEMN} is lower than the experimental results, but the overall behavior is qualitatively similar.
Due to the introduction of the gauge link, $G_E^n(0)$ is automatically zero.
For $G_M^n$, both the tree and loop contributions are important, and the total result is very close to the empirical parametrization 
    $G_{M,\rm emp}^n = -1.91/(1 + Q^2/0.71\,{\rm GeV}^2)^2$ 
up to $Q^2 = 2$~GeV$^2$. 
The magnetic radius of neutron is estimated to be 0.867~fm. 
From Figs.~\ref{fig:GEMP} and \ref{fig:GEMN}, we can see that the loop diagrams contribute about $25\%-30\%$ to the proton electromagnetic form factors as well as the neutron magnetic form factor, while $70\%-75\%$ of the form factors are provided by the tree level contribution.

To summarize this section, we have shown the advantage of the nonlocal chiral effective theory in studying the nucleon form factors. 
As in the local $\chi$PT, the parameters $c_1$ and $c_2$ are determined by the nucleon magnetic moments, and the only free parameter, $\Lambda$, is chosen to obtain reasonable agreement with the $Q^2$ dependence of the nucleon form factors.
Compared with the results of Fig.~\ref{fig:GEMDR} in local $\chi$PT, where there are four other free parameters (besides $c_1$ and $c_2$) that are fixed by the experimental charge and magnetic radii of proton and neutron, the nonlocal Lagrangian clearly has fewer free parameters. 
The nucleon radii in the nonlocal case are calculated instead of fitted.

We also note that since in the nonlocal Lagrangian the baryon field is at the coordinate $x$, while the meson or photon field is at $x+a$, a Taylor series expansion of the meson field $\phi(x+a)$ at $x$ could lead to an expansion of the nonlocal Lagrangian as an infinite set of higher order local ones. 
We did not perform such expansion, but the nonlocal effect may be understood as a form of resummation of the higher order contributions. 
Effectively, the nonlocal gauge invariant Lagrangian makes it possible to study hadron properties at the relatively larger momentum transfer compared with ordinary local $\chi$PT.
Not only are the ultraviolet divergences absent, but also the numerical results appear reasonable for the nonlocal theory. 
In the following, we will apply the nonlocal chiral effective theory to study more specific nucleon and other baryon observables.

\subsection{\it Strange form factors}

It is well known that a complete characterization of the nucleon substructure requires to go beyond the three valence quarks. 
One of the great challenges of modern hadron physics is to unravel the precise role of hidden flavors in the structure of the nucleon.
In particular, the strange quark contribution to the nucleon form factors has attracted a lot of interest because it comes purely from the sea quark sector.
Experimental measurements on the strange form factors are very challenging since the strange quark contributions to both electric and magnetic form factors are very small.
In 2006, a combined analysis of HAPPEX, SAMPLE and A4 experiments around $Q^2 = 0.1$~GeV$^2$ showed that $G_E^s$ prefers a somewhat negative value and $G_M^s$ prefers a somewhat positive value, although $G_E^s = G_M^s = 0$ was still compatible with the data at the 95\% C.L. \cite{Aniol}.
In 2009, however, measurements of the parity-violating asymmetry in elastic electron scattering on hydrogen at forward and backward angles for $Q^2 = 0.22$~GeV$^2$ indicated that $G_E^s$ was positive and $G_M^s$ negative \cite{Baunack}, opposite to the earlier experimental analysis.

\begin{figure}[tb] 
\begin{center}
\epsfig{file=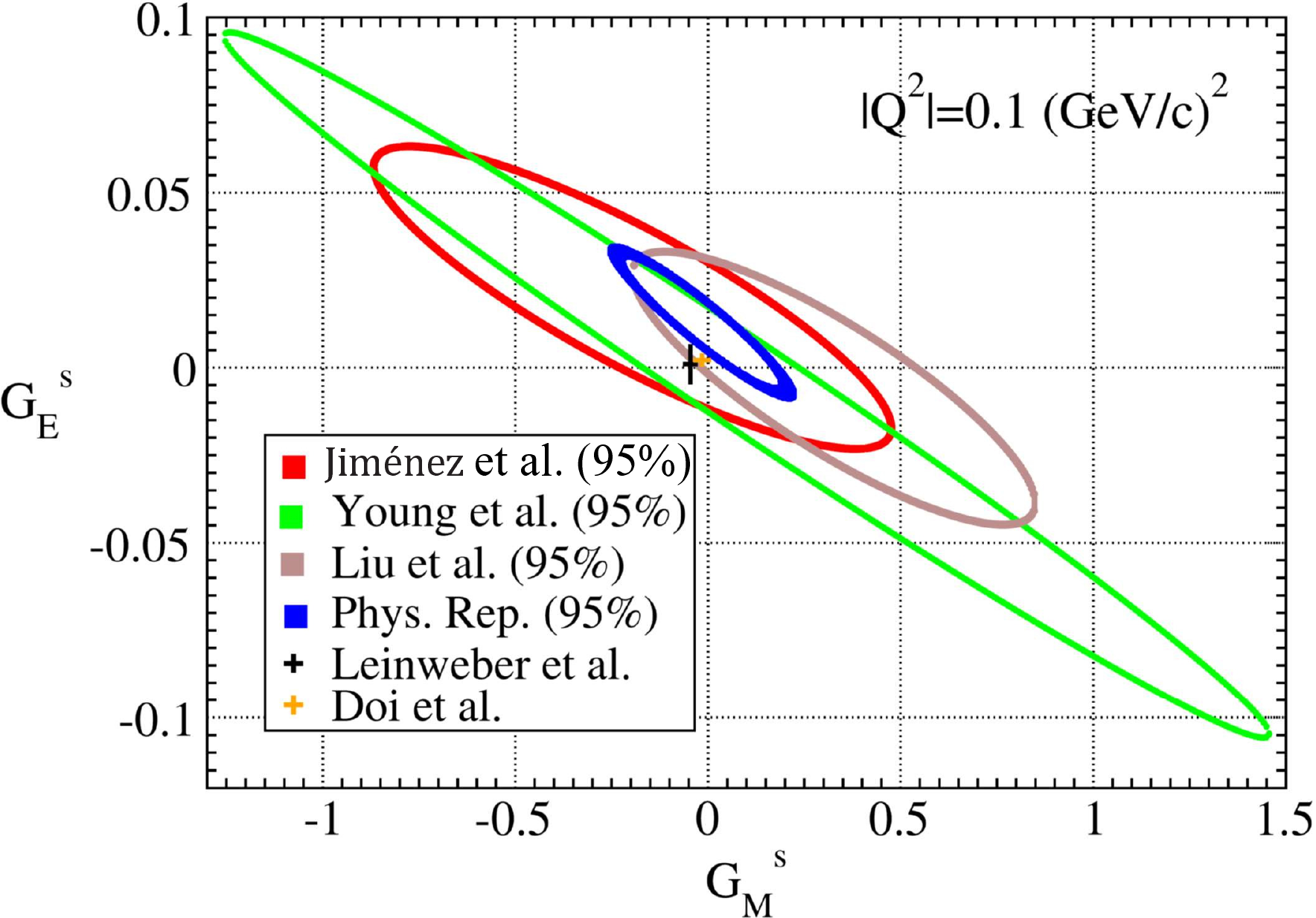,scale=0.6}
\caption{95\% confidence level constraint ellipses in the plane $G^s_E-G^s_M$ at $Q^2=0.1$~GeV$^2$ for various analyses: red, green, brown and blue ellipses from Refs.~\cite{Jimenez}, \cite{Young}, \cite{Liu} and \cite{Jimenez2}, respectively. Theoretical predictions are indicated by the black and orange crosses. (Figure from Ref.~\cite{Jimenez}.)}
\label{fig:GSGLOBAL}
\end{center}
\end{figure}

In 2014, a statistical analysis of the full set of parity-violating asymmetry data for elastic electron scattering including the high precision measurement from the $Q$-weak experiment was performed \cite{Jimenez}.
These estimates favored nonzero vector strangeness, specifically, positive (negative) values for the electric (magnetic) strange form factors. Figure~\ref{fig:GSGLOBAL} shows the constraint on $G_E^s$ and $G_M^s$ from  Ref.~\cite{Jimenez}.
The more recent result from Gonzalez-Jimenez {\it et al.} \cite{Jimenez} (red ellipse) is much larger than their earlier result \cite{Jimenez2} (blue ellipse) due to the presence of more free parameters in the new fit. 
Compared with the analysis by Liu {\it et al.} \cite{Liu} (brown ellipse) and Young {\it et al.} \cite{Young} (green ellipse), where a total of 10 data points in the range of $0.091 \leq Q^2 \leq 0.126$~GeV$^2$ and 19 data points in the range of $0.038 \leq Q^2 \leq 0.299$~GeV$^2$ were used, respectively, more experimental data and free parameters were included in the global analysis of Gonzalez-Jimenez {\it et al.} \cite{Jimenez}.

The strange form factors were also investigated in chiral perturbation theory \cite{Hemmert}. 
However, there are unknown low energy constants appearing in the chiral Lagrangian, which has limited the ability to compute the strange magnetic form factor or the charge radius.
The quantity that one wishes to predict -- the strangeness vector current matrix element -- is the same quantity that one needs to know in order to make a prediction \cite{Musolf, Kubis3}.
For example, from Eq.~(\ref{eq:magB}), one can see that the last term
has no contribution to the magnetic moments of octet baryons because the trace of the charge matrix $Q_c = {\rm diag}(2/3,-1/3,-1/3)$ is zero. 
The baryon magnetic moments are only related to $c_1$ and $c_2$, for instance, $F_2^p = (1/3) c_1 + c_2$, $F_2^n = -(2/3) c_1$.
In the local chiral perturbation theory, the loop integrals for the form factors are divergent and the sum of the loop and tree level contributions leads to the renormalized $c_1^R$ and $c_2^R$, which can be determined by fitting the magnetic moments of the nucleon or octet baryons.
However, the electromagnetic behavior of baryons cannot determine the parameter $c_3^R$.
This can only be determined by experiments measuring directly the strange magnetic moment of the nucleon.
As a result, the calculated strange electric and magnetic form factors can be both positive and negative with large error bars reflecting the large uncertainties in the experimental data \cite{Hemmert}.

\begin{figure}[]
\begin{center}
\hspace{-0.1in}\includegraphics[scale=0.4]{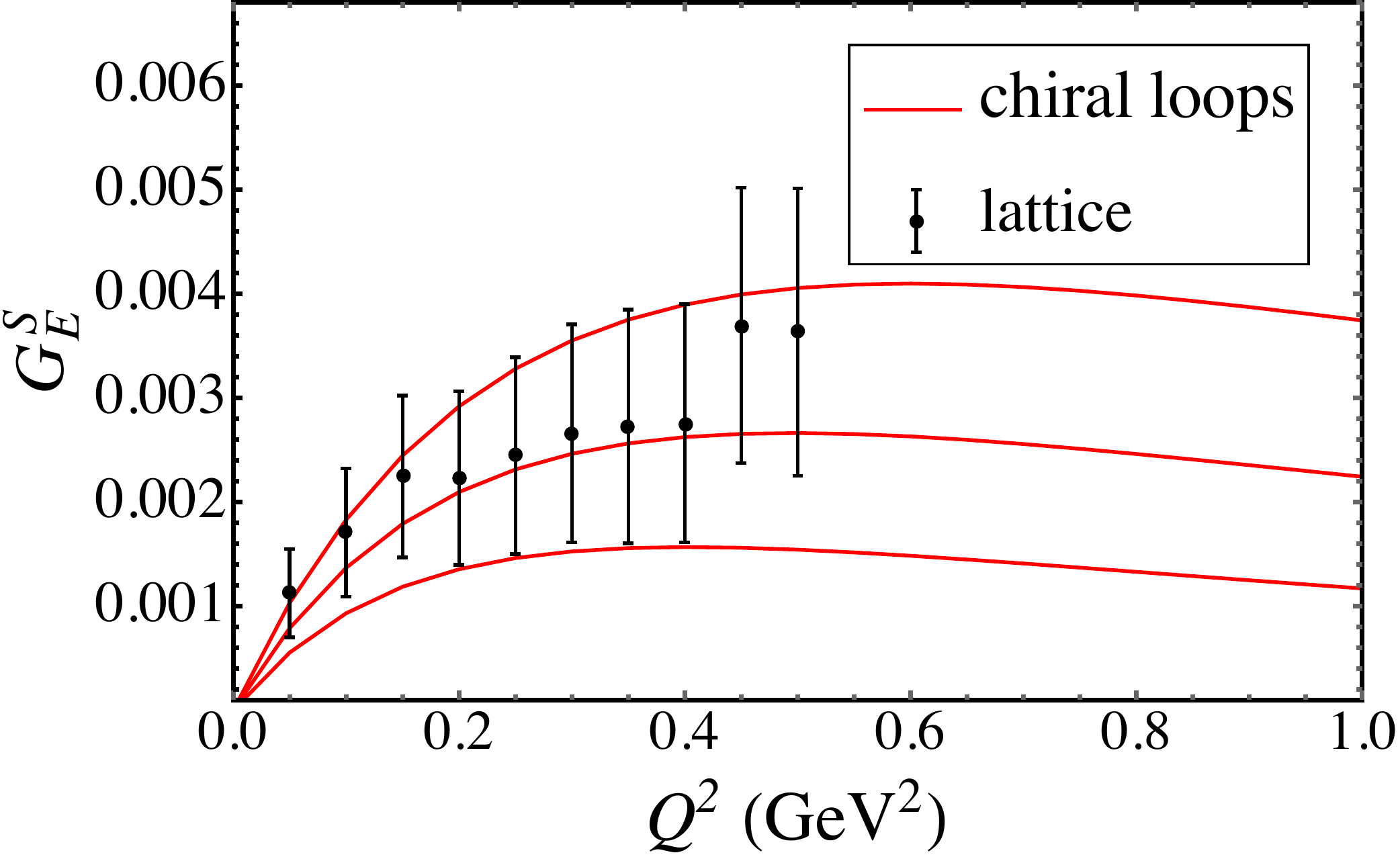}
\hspace{0.3in}\includegraphics[scale=0.4]{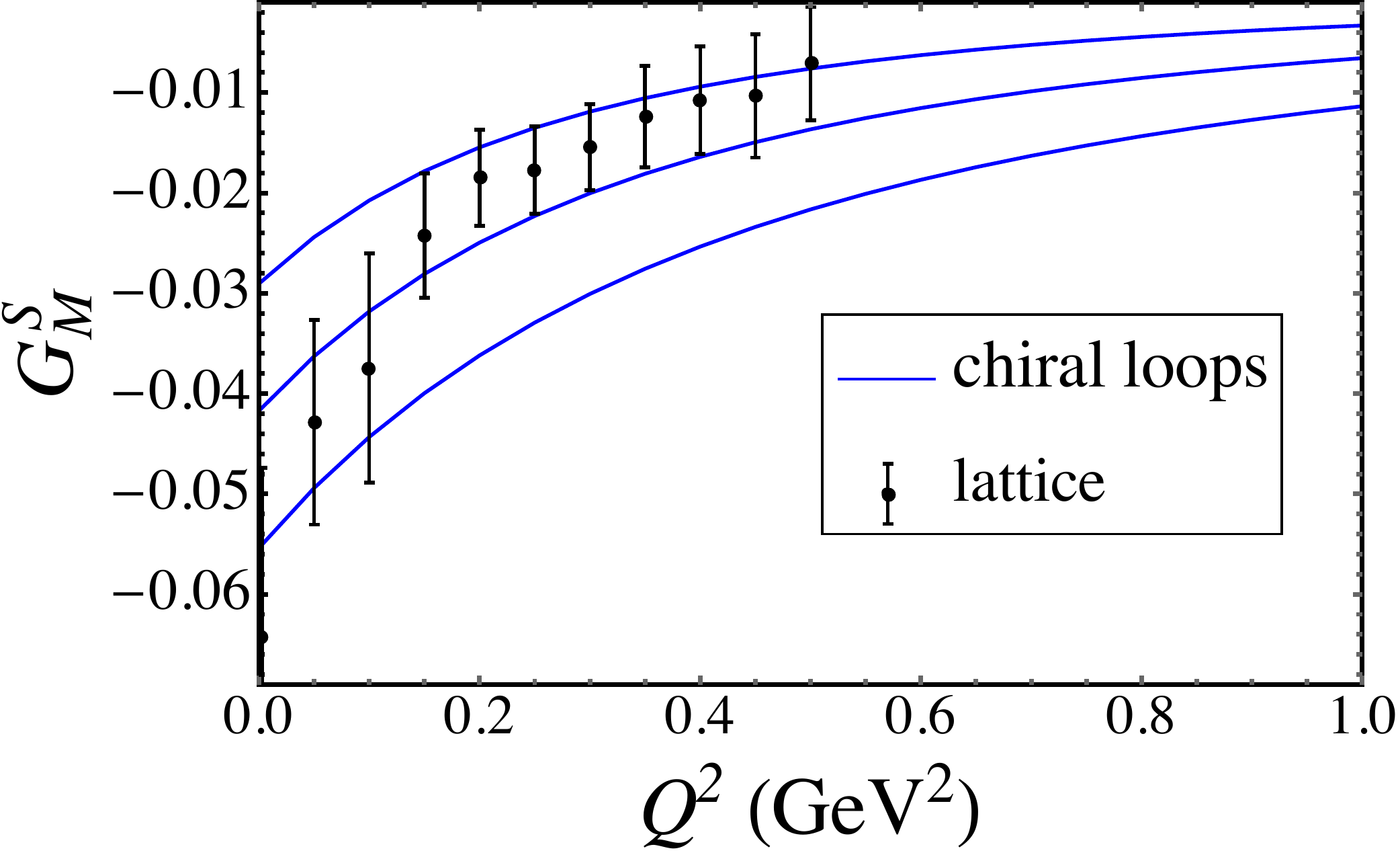}
\caption{The strange electric $G_E^s$ and magnetic $G_M^s$ form factors of the nucleon versus momentum transfer squared $Q^2$ for different $\Lambda$ values. The three solid lines from top to bottom on the left figure (from bottom to top on the right figure) correspond to $\Lambda=1, 0.9$ and 0.8~GeV, respectively. The data with error bars are from the lattice simulation of Ref.~\cite{Sufian}. (Figure from Ref.~\cite{He2}.)}
\label{fig:GSNL}
\end{center}
\end{figure}

In the nonlocal chiral effective theory, both the loop and tree level contributions are finite.
Therefore, in practice, $c_3$ can be chosen to be $c_2-c_1$, such that the tree level contribution is zero for the bare proton and neutron.
This is impossible in the local theory, where the tree level term must be present to cancel with the infinity from the loop integral.
This is also an advantage of the nonlocal effective theory, which can actually allow one to make predictions for the strange form factors. 
With the magnetic moments of valence strange quark in bare hyperons determined by the electromagnetic moments of baryons, one can obtain the strange form factors of the nucleon \cite{He2}.

The strange electric and magnetic form factors $G^s_E(Q^2)$ and $G^s_M(Q^2)$ are plotted in Fig.~\ref{fig:GSNL}.
Three solid lines from top to bottom on the left panel (from bottom to top on the right panel) represent results with $\Lambda = 1, 0.9$ and 0.8~GeV, respectively, and the data with error bars are from the lattice simulation at the physical pion mass from Ref.~\cite{Sufian}.
At $Q^2 = 0$, the strange electric form factor is normalized to $G^s_E(0)=0$. This is verified only when the additional diagrams generated from the expansion of the gauge link are included. 
The strange charge form factor first increases and then decreases with increasing $Q^2$.
At finite $Q^2$, $G^s_E(Q^2)$ is always a small positive number.
The strange magnetic form factor is negative and decreases in magnitude with increasing $Q^2$.
At zero momentum transfer, for $\Lambda = 0.9(1)$ GeV the strange magnetic moment is $G^s_M(0) = -0.041^{(12)}_{(14)}$. 
From Fig.~\ref{fig:GSNL}, one can see that the calculated results for both the electric and magnetic form factors are in good agreement with the lattice data. 
From the strange form factors, the strange radii can be obtained as
\be
\langle r_E^2 \rangle_s
= -6 \frac{d G^s_E(Q^2)}{d Q^2} \bigg|_{Q^2=0},
~~~~~~~~
\langle r_M^2 \rangle_s
= -6 \frac{d G^s_M(Q^2)}{d Q^2} \bigg|_{Q^2=0}.
\ee
For $\Lambda = 0.9(1)$ GeV, one finds 
    $\langle r_E^2 \rangle_s = -0.004(1)$~fm$^2$ and 
    $\langle r_M^2 \rangle_s = -0.028(3)$~fm$^2$.

\begin{figure}[]
\begin{center}
\hspace{-0.1in}\includegraphics[scale=0.4]{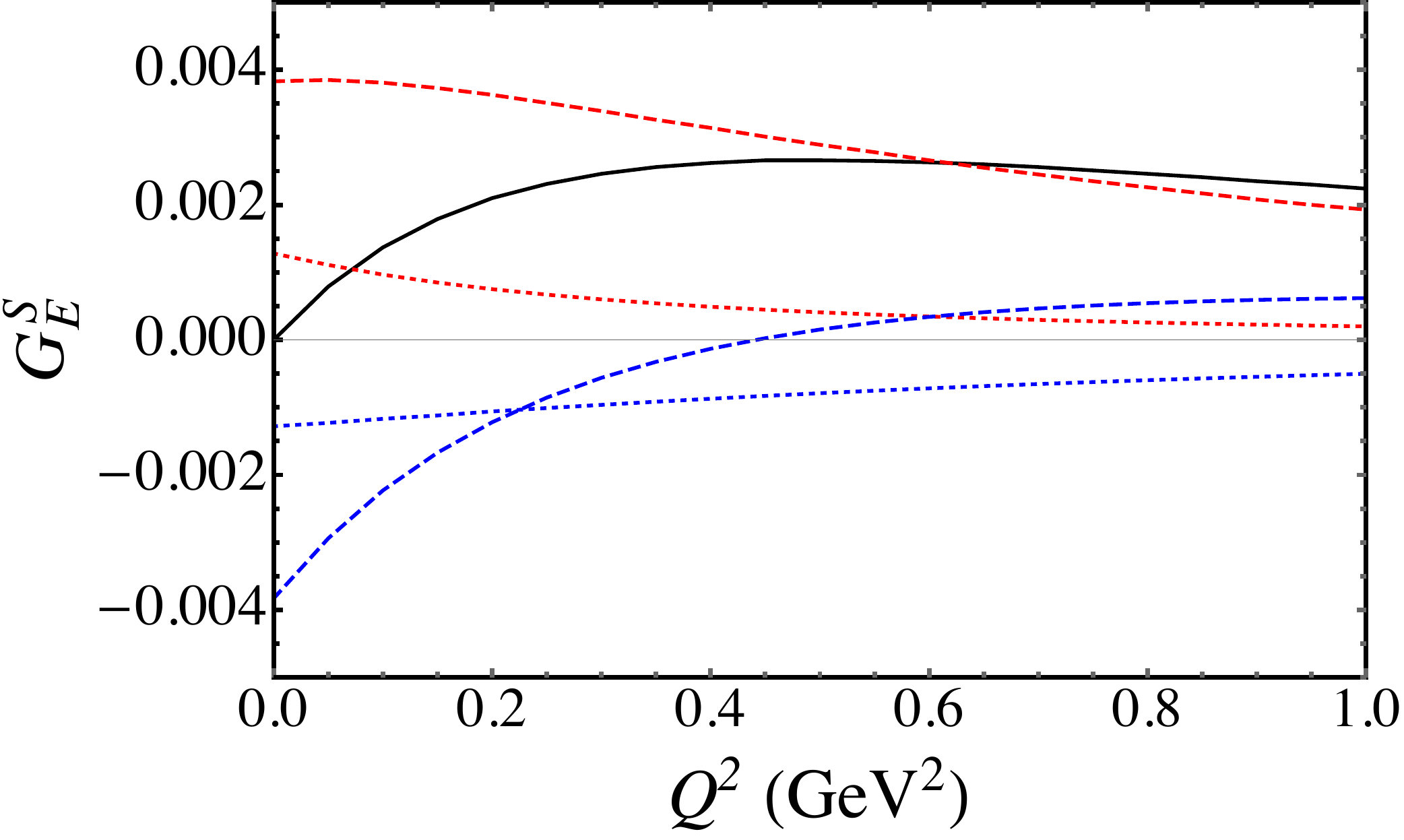}
\hspace{0.3in}\includegraphics[scale=0.4]{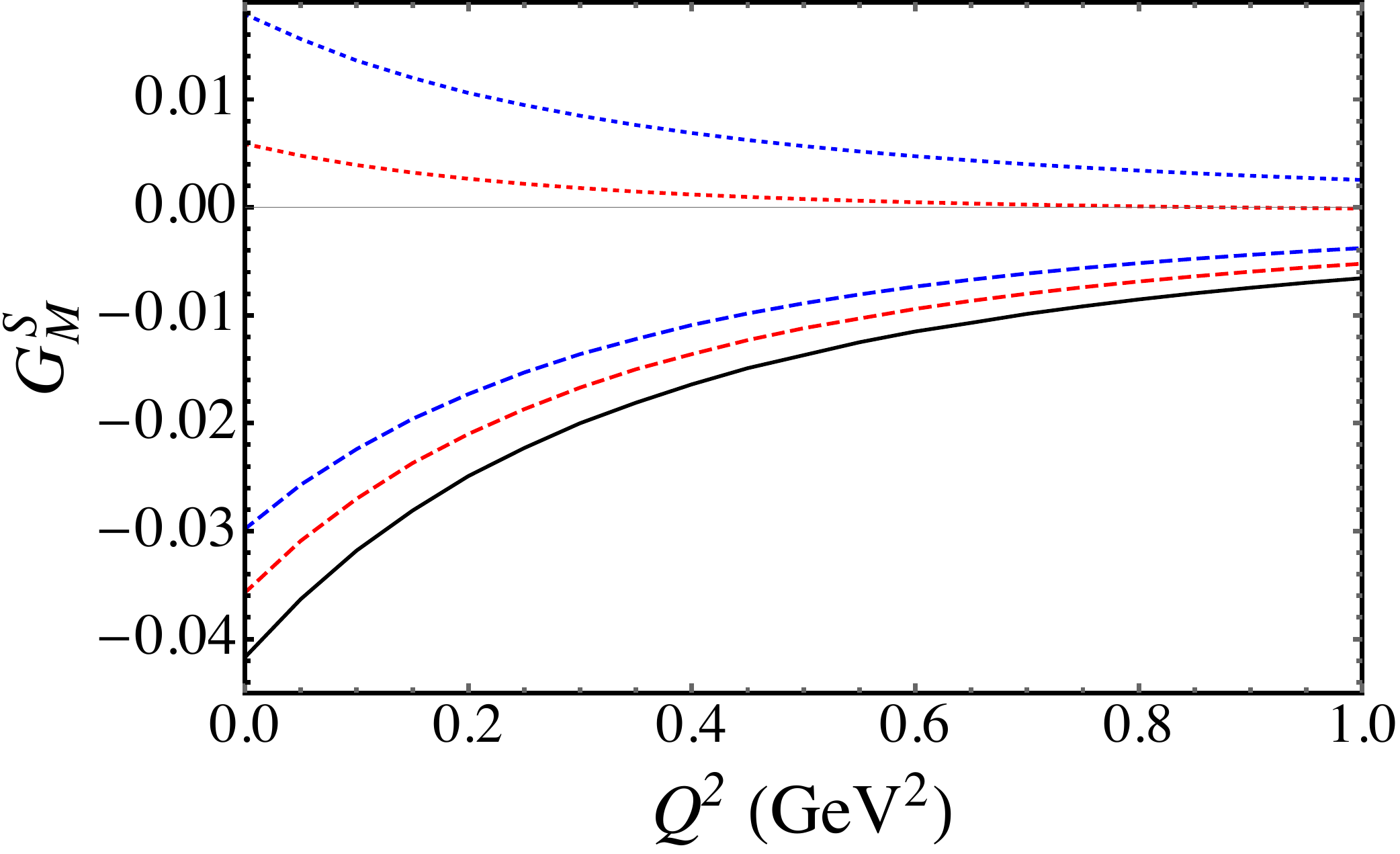}
\caption{Individual contributions to the strange form factors of the nucleon with $\Lambda = 0.9$~GeV, including the baryon octet (dashed lines), decuplet (dotted lines) and total (solid lines) contributions. The contributions from the regular diagrams (red lines) and additional diagrams from the gauge link (blue lines) are indicated. (Figure from Ref.~\cite{He2}.)}
\label{fig:GSPART}
\end{center}
\end{figure}

The individual contributions to the strange form factors from the baryon octet and decuplet intermediate states, as well as from the regular and additional diagrams, are shown in Fig.~\ref{fig:GSPART} for $\Lambda = 0.9$~GeV.
For the strange electric form factor at zero momentum transfer, $G^s_E(0)$, the contributions from the regular and additional gauge link diagrams cancel, resulting in a net zero strangeness at $Q^2=0$.
This is guaranteed by the U(1) gauge symmetry of the strange quark.
The octet contribution is dominant especially at small momentum transfers, and both the regular and additional diagrams are important numerically.
For the strange magnetic form factor, $G_M^s(Q^2)$, the contributions from the octet and decuplet intermediate states have different signs.
As for the $G_E^s(Q^2)$ case, the octet contribution is larger than the decuplet, while the regular and additional gauge link diagrams provide comparable contributions.

\newpage
\subsection{\it Light sea quark form factors}

\begin{figure}[t] 
\begin{center}
\epsfig{file=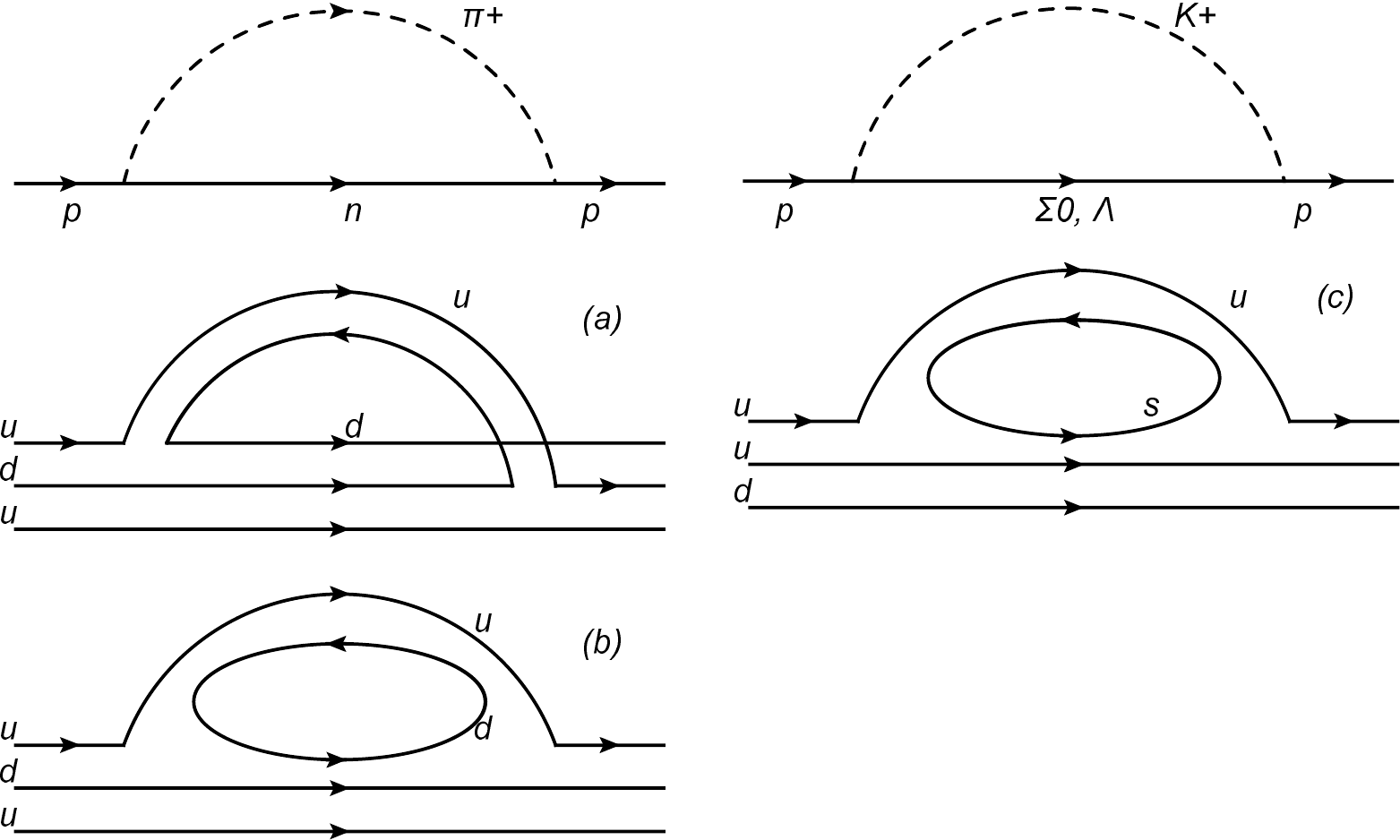,scale=0.6}
\caption{Quark flow diagrams for the $\pi^+$ (left) and $K^+$ (right) mesons: (a) is the connected diagram; (b) and (c) are the disconnected sea diagrams  for $\pi^+$ and $K^+$, respectively. (Figure from Ref.~\cite{Yang}.)}
\label{fig:quarkflow}
\end{center}
\end{figure}

In the last two subsections we have shown that nonlocal chiral effective theory can provide good descriptions of nucleon electromagnetic and strange form factors up to relatively large momentum transfers.
In this subsection we will focus on the light sea quark contributions to the nucleon form factors.
Although it is challenging experimentally to separate the contributions of sea quarks, especially the light sea quarks, in lattice calculations it is possible to simulate these quantities directly.
Early lattice calculations of the electromagnetic form factors were performed in the quenched approximation with large pion (or quark) masses.
With improvements in computing capability and algorithms, current simulations on the lattice can also compute the disconnected contributions of light and strange quarks at the physical pion mass \cite{Sufian, Alexandrou4}, making comparisons of results from EFT and from lattice more relevant.

To calculate the light sea quark contributions requires the coupling constants for the disconnected diagrams. 
The coefficients for the connected and disconnected diagrams can be obtained as shown in Ref.~\cite{Leinweber2} using the quark flows illustrated in Fig.~\ref{fig:quarkflow}.
The results are the same as those extracted within the graded symmetry formalism in quenched chiral perturbation theory \cite{Bernard4}. 
In Fig.~\ref{fig:quarkflow} (from Ref.~\cite{Yang}) we display the rainbow diagram using the quark flows to exemplify the method of separating the quenched and sea quark contributions.
The coefficient from the $\pi^+$ loop diagram in full QCD effective theory is $(D+F)^2$, and corresponds to the sum of the coefficients from Figs.~\ref{fig:quarkflow}(a) and \ref{fig:quarkflow}(b).
The coefficient of Fig.~\ref{fig:quarkflow}(b) for the sea quark contribution is the same as that of Fig.~\ref{fig:quarkflow}(c) for the $K^+$ loop from SU(3) symmetry, which can be verified directly from the chiral Lagrangian.
The coefficient of the quenched sector can thus be obtained by subtracting the coefficient of the sea diagram from the total coefficient.

\begin{table}[t]
\begin{center}
\caption{The coefficients for the $\pi$ and $K$ meson loops for the full, quenched, and sea diagrams.}
\label{tab:quench}
\begin{tabular}{c|ccc}
\hline \\[-2mm]
meson & full coefficient  &  quenched diagram   &  sea diagram    \\[2mm]
\hline \\[-2mm]
$\pi^0$ & $\frac{1}{2} (D+F)^2$ & $-\frac{1}{3} D^2 + 2 D F - F^2$ & \begin{tabular}[c]{@{}c@{}}$\frac13(D^2+3F^2)$~~ $[u]$ \\[1mm] $\frac{1}{2} (D-F)^2$~~~~~ $[d]$ \end{tabular} \\[6mm]
$\pi^+$ & $(D+F)^2$ & $\frac{1}{3} \left(D^2+6 D F-3 F^2\right)$ & $\frac{2}{3} \left(D^2+3 F^2\right)$ \\[2mm]
$\pi^-$ & $0$ & $-(D-F)^2$ & $(D-F)^2$ \\[2mm]
$K^0$ & $(D-F)^2$ & $0$ & $(D-F)^2$ \\[2mm]
$K^+$ & $\frac{2}{3} \left(D^2+3 F^2\right)$ & $0$ & $\frac{2}{3} \left(D^2+3 F^2\right)$ \\[2mm] \hline
\end{tabular}
\end{center}
\end{table}

The coefficients for the $\pi$ and $K$ meson loops for both quenched and sea quark flow diagrams are listed in Table~\ref{tab:quench}. 
From here one can see that the coefficients for the disconnected $u$ and $d$ quark diagrams of the $\pi^0$ loop are only half of the corresponding coefficients of the $K$ loop.
This is due to the fact that the $\eta$ meson is not included for this estimation: if the $\pi$, $K$ and $\eta$ mesons were all degenerate, then the loop contribution of the $d$-sea quark in the proton would be the same as the $s$ quark. 
For the $u$-sea quark, there is no octet $uuu$ state, meaning that the mass of octet $uuu$ state is infinite. 
If we set arbitrarily the octet $uuu$ mass to be the same as the nucleon, then the contributions of the $u$-sea and $d$-sea quarks would be equivalent.

\begin{figure}[t]
\begin{center}
\hspace{-0.1in}\includegraphics[scale=0.94]{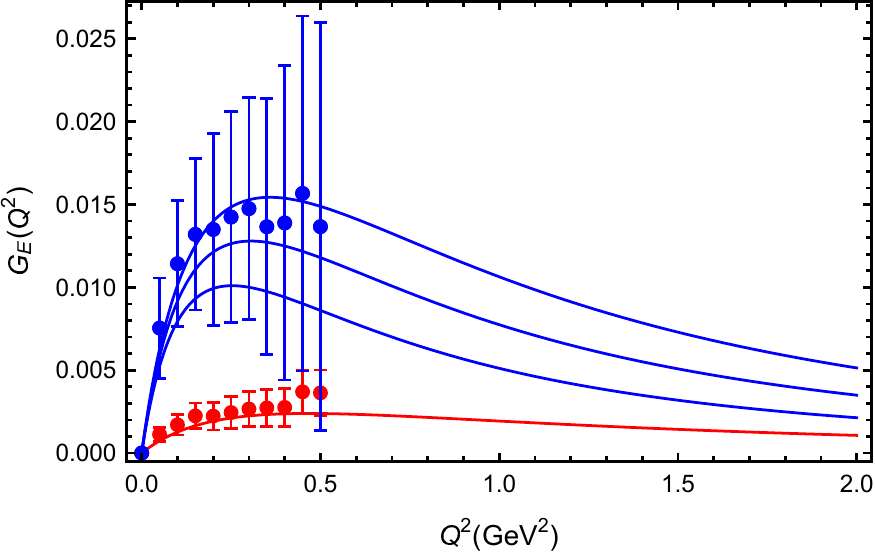}
\hspace{0.3in}\includegraphics[scale=0.94]{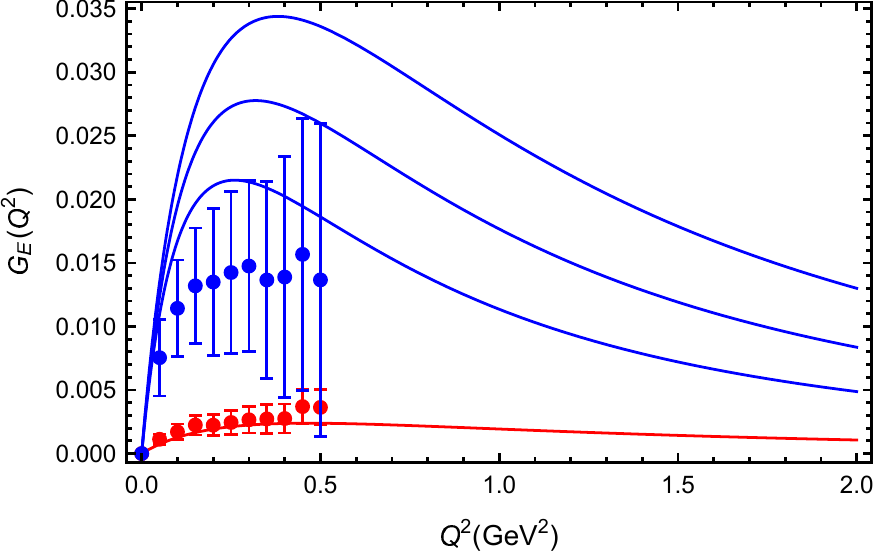}
\caption{Light sea quark contributions to the proton electric form factor versus $Q^2$ for the $u$ quark (left) and $d$ quark (right), for $\Lambda=1.0$~GeV (top blue lines), 0.9~GeV (middle blue lines) and 0.8~GeV (bottom blue lines).  The strange quark contribution (red lines) corresponds to $\Lambda=0.9$~GeV. The points with error bars are from lattice simulations \cite{Sufian}. (Figure from Ref.~\cite{Yang}.)}
\label{fig:udE}
\end{center}
\end{figure}

\begin{figure}[h] 
\begin{center}
\hspace{-0.1in}\includegraphics[scale=0.94]{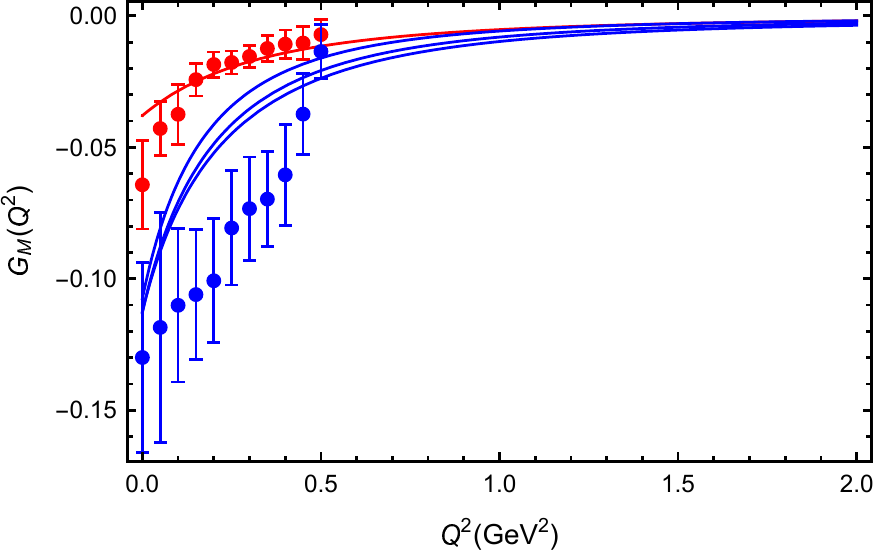}
\hspace{0.3in}\includegraphics[scale=0.93]{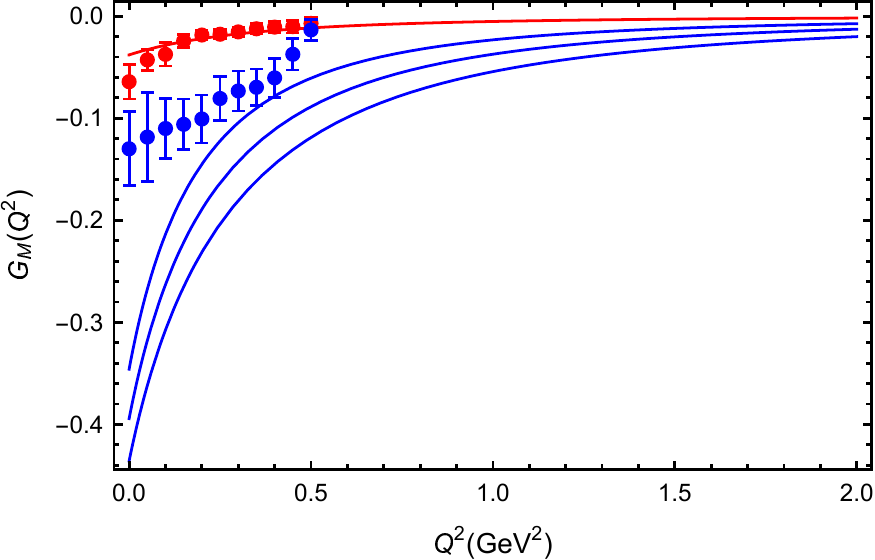}
\caption{Light sea quark contributions to the proton magnetic form factor versus $Q^2$ for the $u$ quark (left) and $d$ quark (right), for $\Lambda=1.0$~GeV (bottom blue lines), 0.9~GeV (middle blue lines) and 0.8~GeV (top blue lines). The strange quark contribution (red lines) corresponds to $\Lambda=0.9$~GeV. The points with error bars are from lattice simulations \cite{Sufian}. (Figure from Ref.~\cite{Yang}.)}
\label{fig:udM}
\end{center}
\end{figure}

The sea quark form factors calculated in the nonlocal chiral effective theory can be found in Ref.~\cite{Yang}.
In Fig.~\ref{fig:udE} we show the contributions of the light sea quarks with unit charge to the proton electric form factor.
The contributions from $u$ and $d$ sea quarks are shown for $\Lambda=1.0$, 0.9, and 0.8~GeV, and compared with the strange quark contribution for $\Lambda=0.9$~GeV, as well as with recent lattice QCD data \cite{Sufian}.
Since the valence contributions of the $u$ and $d$ quarks are not included, the electric form factor of sea quark is zero at $Q^2=0$.
From the figure, one can see that the strange quark form factor can be described very well.
The calculated $u$ quark result is also compatible with the lattice data, while the $d$ quark contribution is larger than the lattice results.
The larger sea contribution of the $d$ quark than that of the $u$ quark is due to the fact that there is no intermediate octet contribution for the $u$-sea quark.
The $u$-sea contribution comes only from the decuplet intermediate states except the $\pi^0$ loop, where the contribution is identical for the $u$ and $d$ quarks. 
Similar results were found for the $\bar{d}-\bar{u}$ asymmetry in the proton, which will be discussed in Sec.~\ref{Sec.5} below.
Compared with the light sea quark form factors, the strange quark contribution to the electric form factor is about $5-10$ times smaller due to the suppression of the $K$ meson loop.

The sea quark contributions to the proton magnetic form factor of the $u$ and $d$ quarks with the unit charge are plotted in Fig.~\ref{fig:udM}.
As for the electric case, the calculated magnetic form factor for strange quarks is in a good agreement with the lattice data, while the contributions from the light sea quarks show some discrepancy with the data.
The magnitude of the (negative) magnetic form factors of all the sea quarks decrease monotonously with increasing $Q^2$.
The absolute values of the $u$-sea quark contribution are smaller than the lattice results for the light sea quarks, while the absolute values of the $d$-sea quark contribution are larger than the lattice data.
The magnitude of magnetic moments of both $u$-sea and $d$-sea quarks are larger than the strange quark, especially at small $Q^2$. 
For $Q^2=0$, the magnetic moments of the $u$-sea and $d$-sea quarks are $-0.11$ and $-0.39$, respectively, while the strange magnetic moment is about $-0.04$.

We should note, however, that the lattice results for the light sea quarks cannot distinguish between the contributions from $u$-sea quark and $d$-sea quarks.
Therefore, the lattice data on the light sea quarks can be regarded as an average of the $u$-sea and $d$-sea quark contributions, and better agreement between the nonlocal EFT calculation and lattice data would be obtained for the isoscalar flavor combination.
In addition, the obvious difference between the form factors of the $u$-sea and $d$-sea quarks cannot be attributed to the mass difference between the $u$ and $d$ quarks; any mass difference would be associated with charge symmetry violation, leading to differences between $G^{p,u}_{E,M}$ and $G^{n,d}_{E,M}$, which are expected to be very small.
In contrast, the large difference between $G^{p,u}_{E,M}$ and $G^{p,d}_{E,M}$ is associated with the effects of the nonperturbative valence quark environment.

The results calculated with the nonlocal EFT are also consistent with the lattice data from Ref.~\cite{Alexandrou4}, where the electric and magnetic form factors of light and strange quarks have opposite signs because the charges of light and strange quarks are included in their currents.
In the nonlocal calculation of the form factors of sea quarks with unit charge, the signs of the form factors are the same for all three flavors.
Since the contributions from the strange quark are much smaller, it is challenging both for experiments and lattice simulations to obtained accurate values of the strange form factors.
Even an unambiguous determination of the sign of the strange quark form factors is an important step in the quest to understand the structure of the nucleon.

In this respect, the quantity $G^u_{\Sigma^-}$, denoting the $u$ quark contribution to the electromagnetic form factors of $\Sigma^-$ hyperon (or similarly $G^d_{\Sigma^+}$ denoting the $d$ quark contribution to the electromagnetic form factors of the $\Sigma^+$) could be more useful physical observables, especially for lattice simulations \cite{Wang5}.
These are similar to the strange form factors in the sense that both of the quantities arise purely from disconnected sea quark contributions.
However, in the EFT framework, $G^u_{\Sigma^-}$ and $G^d_{\Sigma^+}$ are generated by the $\pi$ meson loop, whose contributions should be much larger than the strange form factors generated from the $K$ meson loop.
They can therefore serve as ideal quantities for future lattice simulations, and help to shed light on the sign of the strange quark form factors.

\begin{figure}[t] 
\begin{center}
\epsfig{file=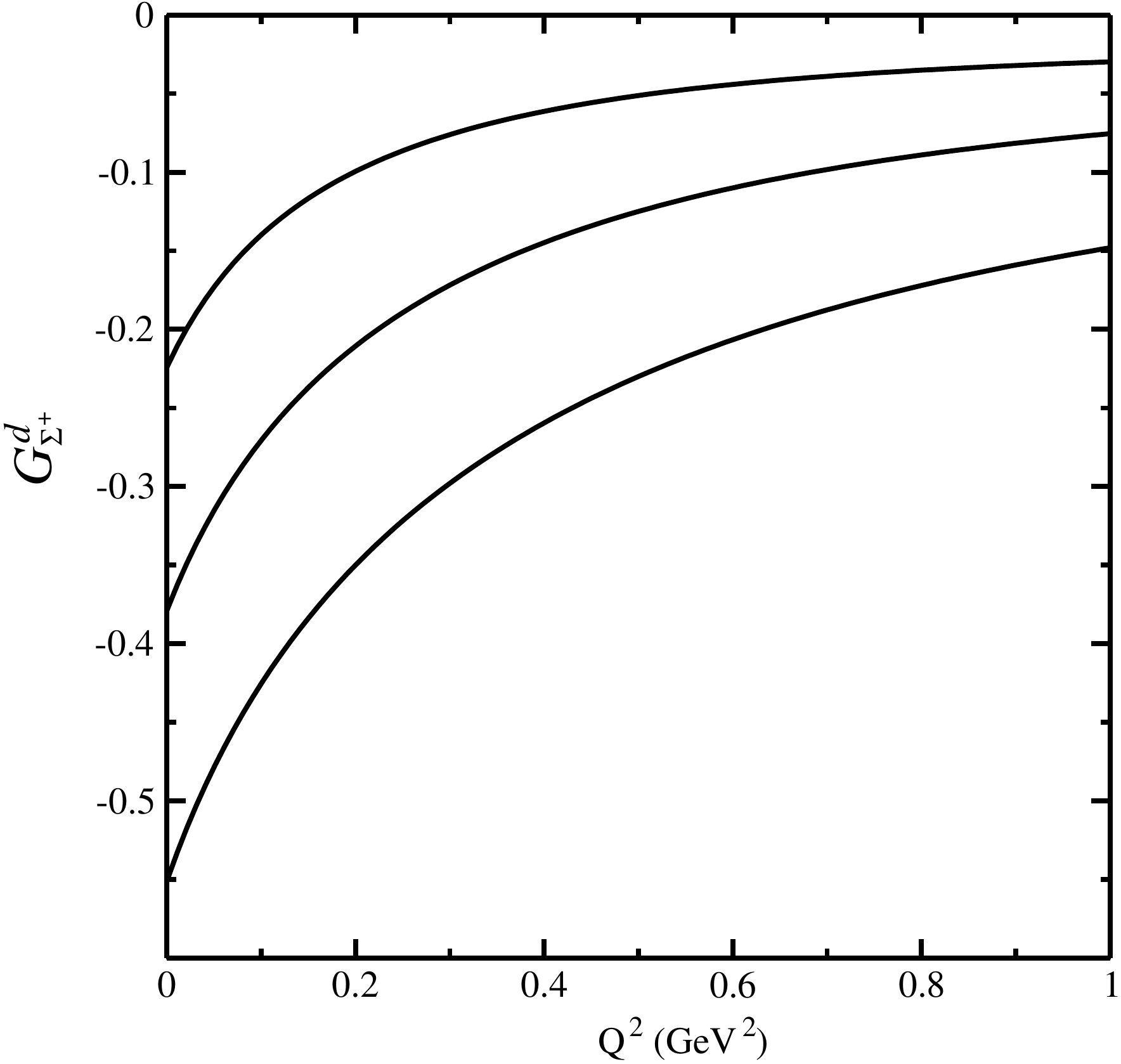,scale=0.42}
\caption{The $Q^2$ dependence of the $d$ quark contribution to the magnetic form factor of the $\Sigma^+$ hyperon for $\Lambda = 0.6$~GeV (upper line), 0.8~GeV (middle line) and 1.0~GeV (lower line). (Figure from Ref.~\cite{Wang5}.)} 
\label{fig:sigma}
\end{center}
\end{figure}

In Fig.~\ref{fig:sigma} the magnetic form factor $G^d_{\Sigma^+}(Q^2)$ calculated with the chiral effective theory in FRR is shown for $\Lambda = 0.6$, 0.8, and 1~GeV \cite{Wang5}.
It is apparent that the magnetic form factor does not change sign with increasing $Q^2$ for any choice of $\Lambda$.
This is similar to the case of the strange magnetic form factor of the nucleon. However, the absolute value of $G^d_{\Sigma^+}$ is about one order of magnitude larger than the strange magnetic form factor of the nucleon.
Since its absolute value decreases with increasing $Q^2$, it would be preferable to measure the magnetic form factor at low $Q^2$ values.
At $Q^2 = 0$, the magnetic moment is calculated to be
$\mu^d_{\Sigma^+} = G^d_{\Sigma^+}(0) = -0.38$ for $\Lambda=0.8$~GeV.
Varying $\Lambda$ from 0.6~GeV to 1~GeV, the magnetic moment $\mu^d_{\Sigma^+}$ changes from $-0.22$ to $-0.55$.
This sensitivity suggests that it may be feasible to study this form factor in lattice QCD simulations.

\subsection{\it Form factors of octet baryons}

In this subsection we extend the discussion of nucleon form factors to form factors of all the octet baryons.
Since the strong baryon--meson interaction and the electromagnetic interaction are included in the chiral SU(3) Lagrangian, it is straightforward to generalize the nucleon one-loop calculations to the other octet baryons, without introducing any additional parameters. 
The only difference is that the two parameters $c_1$ and $c_2$ are now determined by fitting the magnetic moments of the octet baryons, while in the nucleon case they were constrained by the nucleon magnetic moments.
The predictions for the electromagnetic form factors and radii of octet baryons will not only shed light on their structure, but also verify the applicability of the nonlocal chiral effective theory itself.
It was found in Ref.~\cite{Yang2} that numerical results for the octet baryons were in close agreement with the experimental data and comparable with the data from lattice simulations.
They were also consistent with the results from the local chiral effective field theory (at low $Q^2$). 
The main results of Ref.~\cite{Yang2} are summarized in the following.

\begin{table}[h]
\caption{Magnetic moments $\mu_B$ of octet baryons (in units of the nucleon magneton $\mu_N$). Listed are the results from the nonlocal theory, lattice simulations, $\chi$PT with the IR and EOMS schemes, the NJL and PCQM models, as well as experimental data.}
\resizebox{\textwidth}{30mm}{
	\begin{tabular}{l|l|l|l|l|l|l|l|l}
		\hline 
 & Nonlocal \cite{Yang2} & Latt. \cite{Lin5} & Latt. \cite{Shanahan2}& IR \cite{Kubis} & EOMS \cite{Blin} & NJL \cite{Serrano} & PCQM \cite{Liu2} & Exp.~\cite{PDG}  \\ \hline 
$\mu_p$	  & ~~\,$2.644(159)$ & ~~\,$2.4(2)$ & ~~\,$2.3(3)$ & ~~\,2.61 & ~~\,2.79 & ~~\,$2.78$ & ~~\,$2.735(121)$ & ~~\,2.793 
\\[2mm] 
$\mu_n$	  & $-1.984(216)$ & $-1.59(17)$ & $-1.45(17)$ & $-1.69$ & $-1.913$ & $-1.81$ & $-1.956(103)$ & $-1.913$ \\ [2mm]
$\mu_{\Sigma^+}$ & ~~\,$2.421 (147)$ & ~~\,$2.27(16)$ & ~~\,$2.12(18)$ & ~~\,$2.53$ & ~~\,2.1(4) & ~~\,$2.62$ & ~~\,$2.537(201)$ & ~~\,$2.458(10)$ \\ [2mm]
$\mu_{\Sigma^0}$ & ~~\,$0.584 (77)$ & ~~~~\,$-$ & ~~~~\,$-$ & ~~\,0.76 & ~~\,0.5(2) & ~~~~\,$-$ & ~~\,$0.838(91)$ & ~~~~\,$-$ \\ [2mm]
$\mu_{\Sigma^-}$ & $-1.253(8)$ & $-0.88(8)$ & $-0.85(10)$ & $-1.00$ & $-1.1(1) $ & $-1.62$ & $-0.861(40)$ & $-1.160(25)$ \\ [2mm]
$\mu_\Lambda$	 & $-0.594(57)$ & ~~~~\,$-$ & ~~~~\,$-$  & $-0.76$ & $-0.5(2)$ & ~~~~\,$-$ & $-0.867(74)$ & $-0.613(4)$ \\ [2mm]
$\mu_{\Xi^0}$ & $-1.380(169)$ & $-1.32(4)$ & $-1.07(7) $ & $-1.51$ & $-1.0(4)$ & $-1.14$ & $-1.690(142)$ & $-1.250(14)$ \\ [2mm]
$\mu_{\Xi^-}$ & $-0.725(77)$ & $-0.71(3)$ & $-0.57(5)$ & $-0.93$ & $-0.7(1)$ & $-0.67$ & $-0.840(87)$ & $-0.651(80)$ \\ [2mm]
\hline
	\end{tabular}}
	\label{Mag}
\end{table}

In Table~\ref{Mag} the octet baryon magnetic moments obtained from the nonlocal chiral effective theory are listed, with uncertainties estimated by varying $\Lambda$ from 0.8 to 1.0~GeV.
The results are compared with two lattice simulations from Refs.~\cite{Lin5, Shanahan2}, $\chi$PT calculations with IR \cite{Kubis} and EOMS \cite{Blin} regularization, the NJL and PCQM models \cite{Serrano, Liu2}, and with the experimental data \cite{PDG}.
From the table, one can see that all the magnetic moments of the octet baryons are reasonably well reproduced. 
The largest deviation from the experimental values is for the $\Xi$ hyperons, where the calculated central values of the magnetic moments $\mu_{\Xi^0}$ and $\mu_{\Xi^-}$ are about $10\%$ larger than the empirical ones. 
For the other baryons, the deviations from the experimental values are less than $\approx 5\%$.
Taking into account the error estimation, the calculated magnetic moments of octet baryons are in quite good agreement with the experimental results.
The results from the lattice simulations are somewhat smaller, which may be partially attributed to the effect from the large pion mass as well as from the quenched approximation.

\begin{figure}[tb]
\begin{center}
\hspace{-0.1in}\includegraphics[scale=0.56]{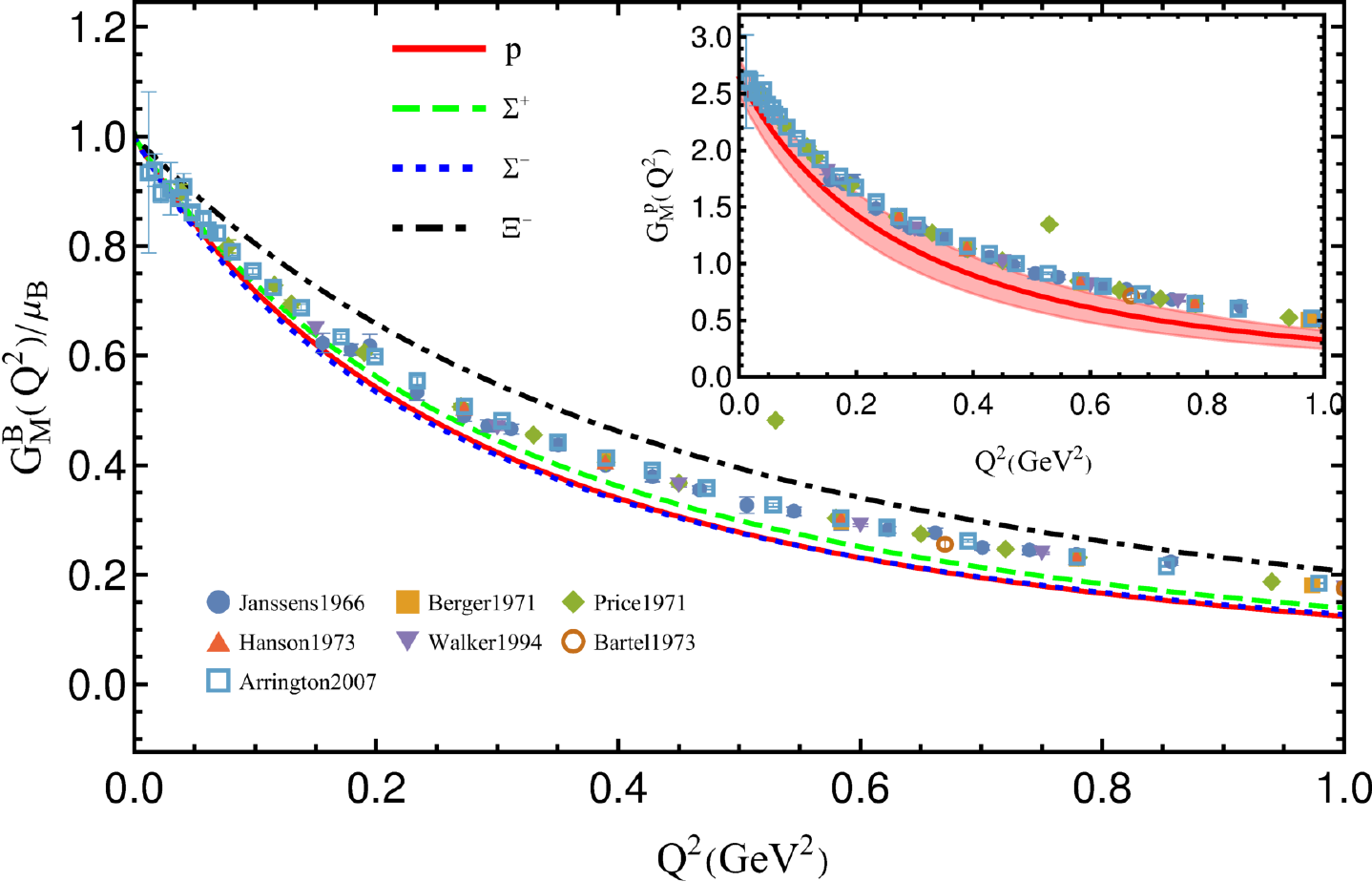}
\hspace{0.3in}\includegraphics[scale=0.56]{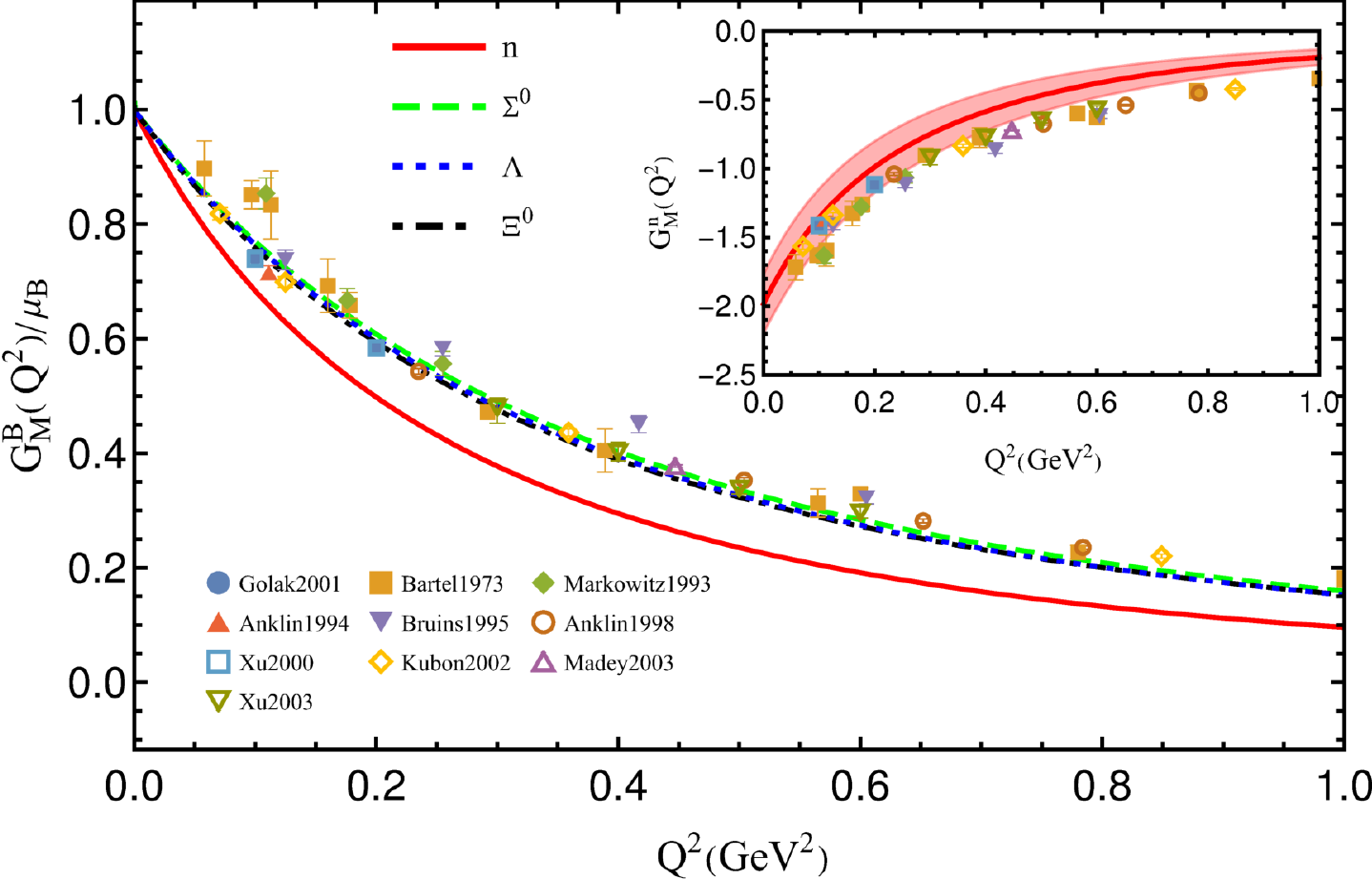}
\caption{Normalized magnetic form factors of octet baryons $G_M^B/\mu_B$ versus $Q^2$ for charged (left panel) and neutral (right panel) baryons. The magnetic form factors of the proton and neutron, with $\Lambda$ varying from 0.8 to 1~GeV, are shown in the insets. (Figure from Ref.~\cite{Yang2}.)}
\label{fig:octetM}
\end{center}
\end{figure}

The normalized magnetic form factors of the charged and neutral octet baryons are plotted in Fig.~\ref{fig:octetM} as a function of the momentum transfer squared, $Q^2$.
The proton and neutron magnetic form factors, with the $\Lambda$ parameter varying from 0.8 to 1~GeV, are also plotted in the insets of Fig.~\ref{fig:octetM}.
Considering the uncertainty on the $\Lambda$ parameter, the agreement between the calculated proton magnetic form factor and experiment is reasonable. 
The other form factors of the charged baryons have a similar momentum dependence as the proton's magnetic form factor.
Among them, the magnetic form factor of the $\Xi^-$ hyperon decreases more slowly with increasing $Q^2$.

For the normalized magnetic form factors of the neutral octet baryons in Fig.~\ref{fig:octetM}, the shapes of the $\Sigma^0$, $\Lambda$ and $\Xi^0$ magnetic from factors are similar to each other.
The neutron magnetic form factor appears to lie below the experimental data, dropping faster than the other neutral baryon form factors.
Taking the uncertainty of the $\Lambda$ parameter into account, the calculated neutron magnetic form factor, shown in the inset for $\Lambda$ varying from 0.8 to 1~GeV) is again in reasonable agreement with the experimental data.
The smaller normalized neutron magnetic form factor relative to the data is partially due to the larger value of the magnetic moment $\mu_n$ from the theory compared with the experimental value. 
All of the form factors of the octet baryons have a dipole-like momentum dependence.
Since the method of the calculation is the same for nucleons and other octet baryons, we expect the nonlocal effective theory to provide reasonable descriptions for all the octet baryons.

\begin{table}[btph]
\caption{Octet baryon magnetic radii $\langle r_M^2 \rangle_B$ (in units of fm$^2$). Listed are results from the nonlocal theory, lattice simulations, $\chi$PT with IR and EOMS schemes, NJL and PCQM models, and experimental data.}\label{radiiM}
\resizebox{\textwidth}{30mm}{
	\begin{tabular}{l|l|l|l|l|l|l|l|l}
\hline
 & Nonlocal \cite{Yang2} & Latt. \cite{Boinepalli} & Latt. \cite{Shanahan2} & IR \cite{Kubis} & EOMS \cite{Blin}& NJL \cite{Serrano} & PCQM \cite{Liu2} & Exp. \cite{PDG}  \\ \hline  
$\langle r_M^2 \rangle_p$ & ~~\,$0.785(132)$ & ~~\,$0.470(48)$ & ~~\,$0.71(8)$ & ~~\,0.699 & ~~\,0.9(2) & ~~\,$0.76$ & ~~\,$0.909(84)$ & ~~\,$0.72(4)$ \\ [2mm]
$\langle r_M^2 \rangle_n$ & ~~\,$0.845(148)$ & ~~\,$0.478(50)$ & ~~\,$0.86(9)$ & ~~\,0.790 & ~~\,0.8(2) & ~~\,$0.83$ & ~~\,$0.922(79)$ & ~~\,$0.75(2)$ \\ [2mm]
$\langle r_M^2 \rangle_{\Sigma^+}$ & ~~\,$0.765(131)$ & ~~\,$0.466(42)$ & ~~\,$0.66(5)$ & ~~\,$0.80(5)$ & ~~\,1.2(2) & ~~\,$0.77$ & ~~\,$0.885(94)$ & ~~~~\,$-$ \\ [2mm]
$\langle r_M^2 \rangle_{\Sigma^0}$ & ~~\,$0.618(124)$ & ~~\,$0.432(38)$ & ~~~~\,$-$ & ~~\,$0.45(8) $ & ~~\,1.1(2) & ~~~~\,$-$ & ~~\,$0.851(102)$ & ~~~~\,$-$ \\ [2mm]
$\langle r_M^2 \rangle_{\Sigma^-}$ & ~~\,$0.901(119)$ & ~~\,$0.483(49)$ & ~~\,$1.05(9)$ & ~~\,$1.20(13) $ & ~~\,1.2(2) & ~~\,$0.92$ & ~~\,$0.951(83)$ & ~~~~\,$-$ \\ [2mm]
$\langle r_M^2 \rangle_\Lambda$	 & ~~\,$0.620(126)$ & ~~\,$0.347(24)$ & ~~~~\,$-$ & ~~\,$0.48(9)$ & ~~\,0.6(2) & ~~~~\,$-$ & ~~\,$0.852(103)$ & ~~~~\,$-$ \\ [2mm]
$\langle r_M^2 \rangle_{\Xi^0}$  & ~~\,$0.657(128)$ & ~~\,$0.384(22)$ & ~~\,$0.53(5)$ & ~~\,$0.61(12)$ & ~~\,0.7(3) & ~~\,$0.44$ & ~~\,$0.871(99)$ & ~~~~\,$-$ \\ [2mm]
$\langle r_M^2 \rangle_{\Xi^-}$ & ~~\,$0.534(135)$ & ~~\,$0.336(18)$ & ~~\,$0.44(5)$ & ~~\,$0.50(16)$ & ~~\,0.8(1) & ~~\,$0.26$ & ~~\,$0.840(109)$ & ~~~~\,$-$ \\ [2mm]
\hline
	\end{tabular}}
\end{table}

Table~\ref{radiiM} provides a summary of the magnetic radii determined from the slopes of the calculated magnetic form factors in Fig.~\ref{fig:octetM} at zero momentum transfer, compared with results from lattice simulations, chiral perturbation theory, and phenomenological quark models.
Although the central values of the nonlocal effective theory predictions for the proton and neutron slightly larger than the experimental values, the results are reasonable within the quoted uncertainties.
The magnetic radii of the octet baryons vary from 
$\langle r_M^2 \rangle_B \approx 0.5$~fm$^2$ to 0.9~fm$^2$, but do not exhibit any simple dependence on the baryon or quark mass. 
However, it is quite remarkable that the order of the values from largest to smallest appears almost equivalent for each of the different methods, even though the values themselves from the different calculations are quite different.
For instance, the $\Sigma^-$ and $\Xi^-$ hyperons have the largest and smallest magnetic radii, respectively, regardless of the theoretical methods.
The neutron magnetic radius appears as the second largest among the octet baryons.

\begin{figure}[tb]
\begin{center}
\hspace{-0.1in}\includegraphics[scale=0.56]{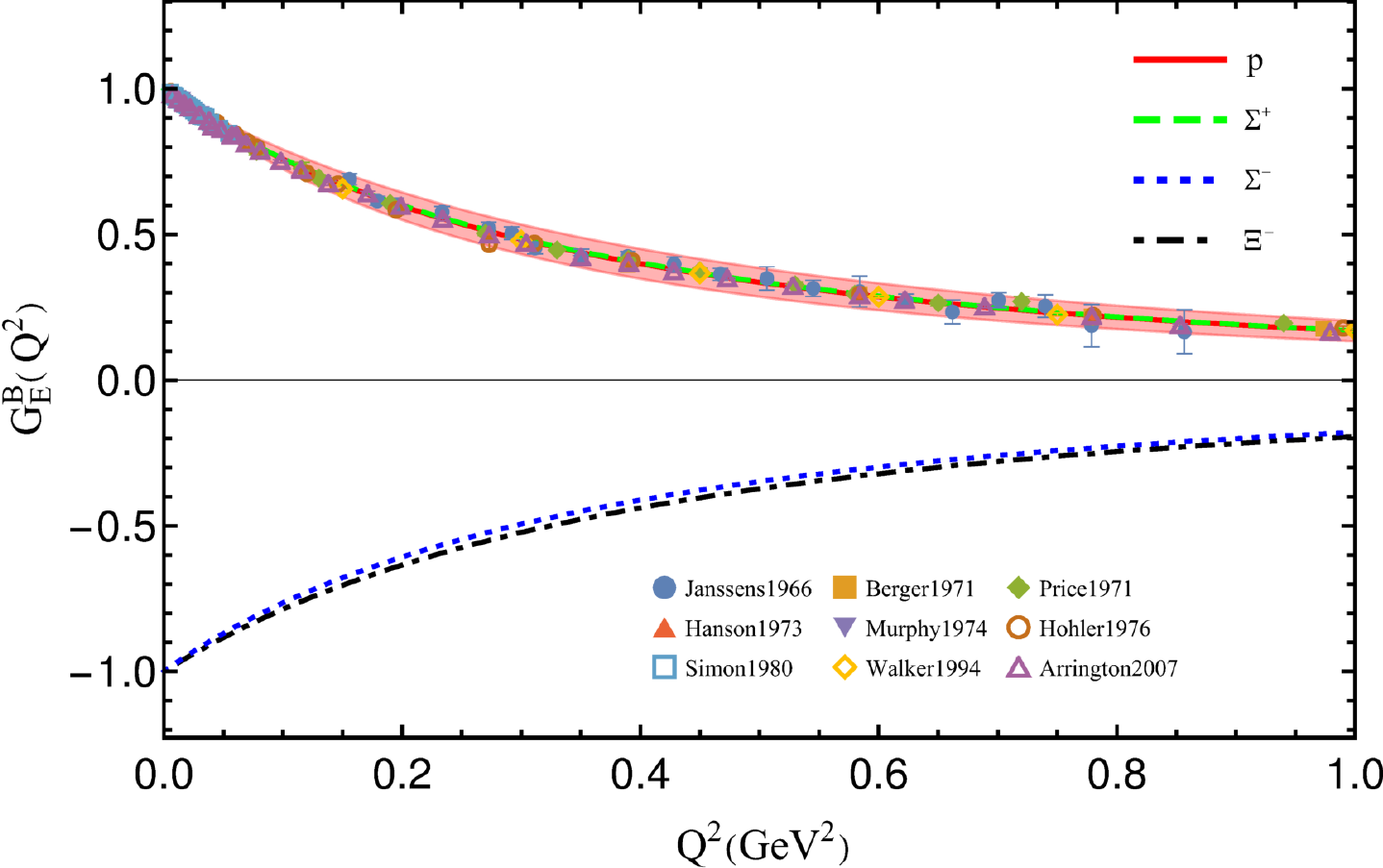}
\hspace{0.3in}\includegraphics[scale=0.56]{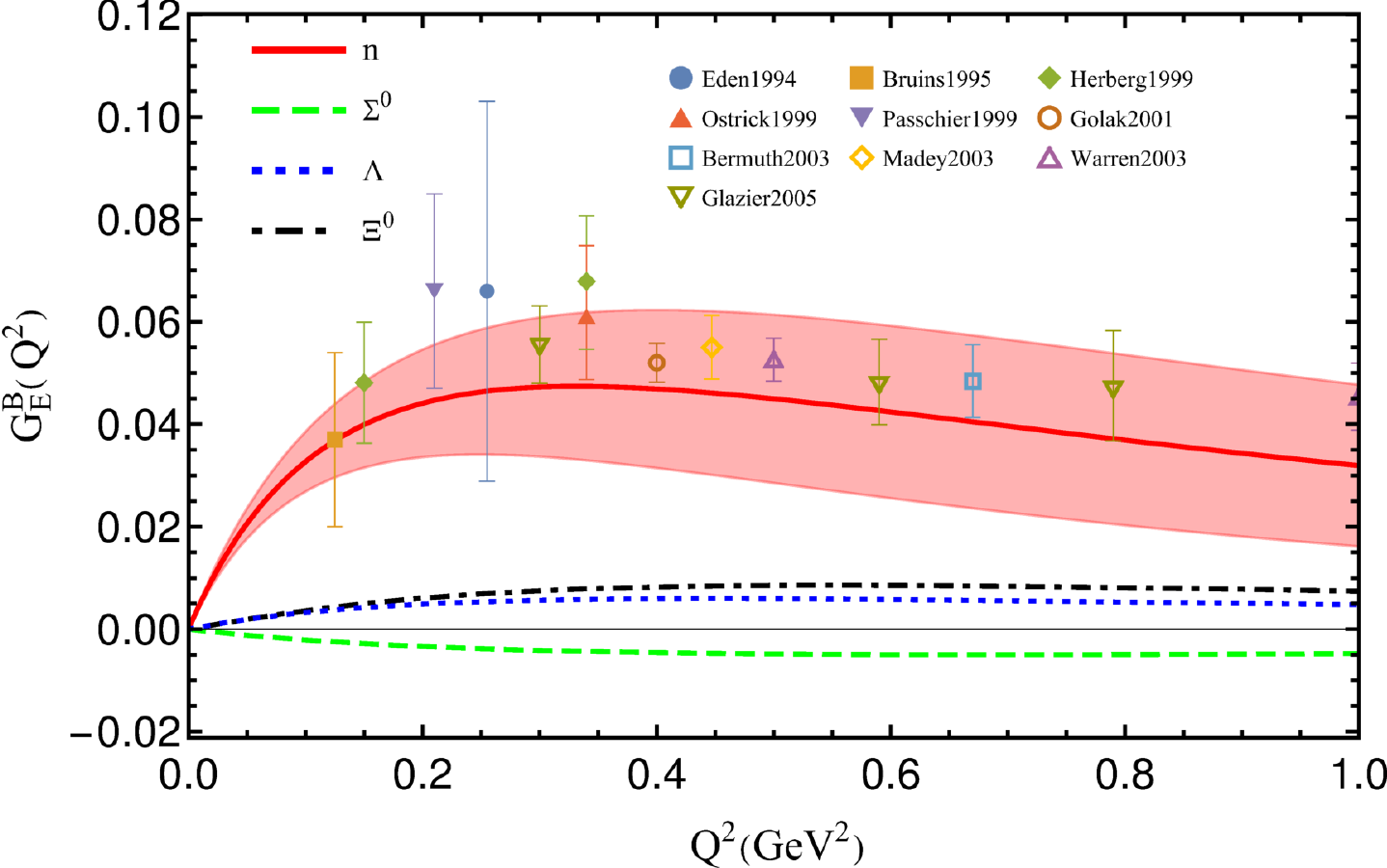}
\caption{Electric form factors of octet baryons $G_E^B$ versus $Q^2$ for charged baryons (left panel) and neutral baryons (right panel). The bands on the proton and neutron electric form factors correspond to $\Lambda$ varying between 0.8 and 1~GeV. (Figure from Ref.~\cite{Yang2}.)}
\label{fig:octetE}
\end{center}
\end{figure}

The electric form factors of the charged and neutral octet baryons are shown in Fig.~\ref{fig:octetE}.
As for the magnetic form factors, the uncertainty bands on the proton and neutron electric form factors are shown for $0.8 \leq \Lambda \leq 1$~GeV.
Because of the additional interaction which makes the nonlocal Lagrangian locally gauge invariant, the electric form factors are correctly normalized to their respective charges at $Q^2=0$.
The proton charge form factor is in very good agreement with the experimental data.
The momentum dependence of the electric form factors of the other charged baryons is very similar.
For the neutral baryon electric form factors, again due to charge conservation, the form factors vanish at zero momentum transfer.
The calculated neutron electric form factor is consistent with the experimental data, albeit with larger relative uncertainties.
The form factors of the other neutral baryons are quite small in magnitude.
One should note that there are no tree level contributions to the electric form factors of any of the neutral baryons, and the entire contributions arise from the loop diagrams.
Among them, the neutron has the largest contribution from the pion loop diagrams, with the corresponding pion loop contributions for the other neutral baryons relatively small because of their small coupling constants.

Finally, Table~\ref{radiiE} summarizes the charge radii of the octet baryons.
The results from the nonlocal theory are comparable with the experimental data for the nucleon and $\Sigma^-$ baryon. 
A small proton charge radius, $\langle r_E \rangle_p = 0.831(7)(12)$, or equivalently, $\langle r_E^2 \rangle_p = 0.691(32)$, reported recently in Ref.~\cite{Xiong}, is also close to the value $\langle r_E^2 \rangle_p = 0.729(112)$ from the nonlocal theory.
The charge radii vary around 0.6 and 0.7~fm$^2$ for the charged baryons, in contrast to the magnetic radii, which have a rather large variation for different charged baryons. 
From the table, one can see that the predictions for the neutral baryons (both sign and magnitude) are quite different from each other, while the calculated charge radii of charged baryons from different models are more or less comparable to each other. 
Certainly, more scrutinized investigations of the form factors as well as the radii of the octet baryons are necessary, along with further experiments and lattice simulations.

\begin{table}[htbp]
\caption{Octet charge radii $\langle r_E^2 \rangle_B$ (in units of fm$^2$). Listed are results from the nonlocal theory, lattice simulations, $\chi$PT with IR and EOMS schemes, NJL and PCQM models, and experimental data.}
\resizebox{\textwidth}{30mm}{
	\begin{tabular}{l|l|l|l|l|l|l|l|l}
	\hline 
 & Nonlocal \cite{Yang2} & Latt.~\cite{Wangradii} & Latt.~\cite{Shanahan} & IR \cite{Kubis} & EOMS \cite{Blin} & NJL \cite{Serrano} & PCQM \cite{Liu2} & Exp.~\cite{PDG} \\ \hline 
$\langle r_E^2\rangle_p$	 & ~~\,$0.729(112)$ & ~~\,$0.685(66)$ & ~~\,$0.76(10)$ & ~~\,0.717 & ~~\,0.878 & ~~\,$0.76$ & ~~\,$0.767(113)$ & ~~\,$0.707(1)$ \\ [2mm]
$\langle r_E^2 \rangle_n$ & $-0.146(18)$ & $-0.158(33)$ & ~~~~\,$-$ & $-0.113$ & ~~\,$0.03(7)$ & $-0.14$ & $-0.014(1)$ & $-0.116(2)$ \\ [2mm]
$\langle r_E^2 \rangle_{\Sigma^+}$ & ~~\,$0.719(116)$ & ~~\,$0.749(72)$ & ~~\,$0.61(8)$ & ~~\,$0.60(2)$ & ~~\,0.99(3) & ~~\,$0.92$ & ~~\,$0.781(108)$ & ~~~~\,$-$ \\[2mm] 
$\langle r_E^2 \rangle_{\Sigma^0}$ & ~~\,$0.010(4)$ & ~~~~\,$-$ & ~~~~\,$-$ & $-0.03(1)$ & ~~\,0.10(2) & ~~~~\,$-$ & ~~\,$0$ & ~~~~\,$-$ \\ [2mm]
$\langle r_E^2 \rangle_{\Sigma^-}$	& ~~\,$0.700(124)$ & ~~\,$0.657(58)$ & ~~\,$0.45(3)$ & ~~\,$0.67(3)$ & ~~\,0.780 & ~~\,$0.74$ & ~~\,$0.781(63)$ & ~~\,$0.61(16)$ \\ [2mm]
$\langle r_E^2 \rangle_\Lambda$	 & $-0.015(4)$ & ~~\,$0.010(9)$ & ~~~~\,$-$ & ~~\,$0.11(2)$ & ~~\,0.18(1) & ~~~~\,$-$ & ~~\,$0$ & ~~~~\,$-$ \\ [2mm]
$\langle r_E^2 \rangle_{\Xi^0}$ & $-0.015(7)$ & ~~\,$0.082(29)$ & ~~~~\,$-$ & ~~\,$0.13(3)$ & ~~\,0.36(2) & ~~\,$0.24$ & ~~\,$0.014(8)$ & ~~~~\,$-$ \\[2mm] 
$\langle r_E^2 \rangle_{\Xi^-}$ & ~~\,$0.601(127)$ & ~~\,$0.502(47)$ & ~~\,$0.37(2)$ & ~~\,$0.49(5)$ & ~~\,0.61(1) & ~~\,$0.58$ & ~~\,$0.767(113)$ & ~~~~\,$-$ \\[2mm] 
\hline
	\end{tabular}}
	\label{radiiE}
\end{table}

\newpage
\section{Parton distribution functions}
\label{Sec.5}

Based on the same nonlocal Lagrangian as in Sec.~\ref{Sec.3}, we will extend in this section our study of meson loop contributions to parton distribution functions in the proton.
We will focus on the sea quark sector, including the unpolarized $\bar{u}$, $\bar{d}$, $s$ and $\bar{s}$ PDFs, and the polarized $s$ quark PDF.
In particular, we will consider differences between different PDFs, or asymmetries, which allow one to isolate nonperturbative QCD effects from the more standard perturbative gluon radiation.
The framework for the calculation of the sea quark PDFs is the convolution formalism, which combines the proton $\to$ baryon $+$ meson splitting functions with the valence quark and antiquark PDFs in the intermediate state mesons and baryons.
This is similar to the framework used for form factors in the previous section, where the sea quark form factors were obtained from the one-loop calculation using the form factors of valence quarks in the intermediate states.

\subsection{\it Convolution formulas} 

The quark PDFs are defined by the nucleon matrix elements of bilocal field operators as 
\begin{eqnarray}\label{eq:PDF}
 \int\frac{\dd{\lambda}}{2\pi}\, e^{-i\lambda x p_+}\,
\langle N(p)|\, \bar q(\tfrac12 \lambda n)\, \slashed{n}\, q(-\tfrac12\lambda n)\, |N(p)\rangle 
= \bar u(p)\, \slashed n\, q(x)\, u(p),
\end{eqnarray}
where $n_\mu$ is the light-cone vector projection of the ``plus" component of momenta, and $x=k^+/p^+$ is the nucleon light-cone momentum fraction carried by the interacting quark.
Using the crossing symmetry properties of the spin-averaged PDFs, $q(-x) = -\bar q(x)$, the $n$-th Mellin moment ($n \geq 1$) of the distribution for a given flavor $q$ ($q = u, d, s, \ldots$) is given by
\begin{eqnarray}
Q^{(n-1)}
&=& \int_0^1 \dd{x}\, x^{n-1}
    \big[ q(x) + (-1)^n\, \bar{q}(x) \big].
\label{eq:mom_def}
\end{eqnarray}
In the operator product expansion, the moments $Q^{(n-1)}$ are related to matrix elements of local twist-two, spin-$n$ operators $\mathcal{O}_q^{\mu_1 \cdots \mu_n}$ between nucleon states with momentum~$p$,
\begin{equation}
\langle N(p) |\, \mathcal{O}_q^{\mu_1 \cdots \mu_n}\, | N(p) \rangle
= 2\, Q^{(n-1)}\, p^{\mu_1} \cdots p^{\mu_n},
\end{equation}
where the operators are given by
\begin{equation}
\label{eq:Oq}
\mathcal{O}^{\mu_1 \cdots \mu_n}_q
= i^{n-1}\, \bar{q} \gamma^{ \{ \mu_1 }\overleftrightarrow{D}^{\mu_2}
  \cdots \overleftrightarrow{D}^{ \mu_n \} } q\, ,
\end{equation}
with
$\overleftrightarrow{D}
 = \frac{1}{2} \big( \overrightarrow{D} - \overleftarrow{D} \big)$,
and the braces $\{\, \cdots \}$ denote symmetrization of Lorentz indices.
In the effective theory, the quark operators ${\cal O}_q$ are matched to hadronic operators ${\cal O}_j$ with the same quantum numbers~\cite{Salamu2, XGWang, Chen3},
\begin{equation}
\mathcal{O}^{\mu_1 \cdots \mu_n}_q
= \sum_{j} Q^{(n-1)}_{j}\ \mathcal{O}^{\mu_1 \cdots \mu_n}_j,
\label{eq:PDFmatch}
\end{equation}
where the coefficients $Q^{(n-1)}_{j}$ are the $n$-th moments of the PDF $q_j(x)$ in the hadronic configuration $j$,
\begin{eqnarray}
\label{eq:cnqj}
Q^{(n-1)}_j &=& \int_{-1}^1 \dd{x}\, x^{n-1}\, q_j(x).
\end{eqnarray}
The nucleon matrix elements of the hadronic operators ${\cal O}_j^{\mu_1 \cdots \mu_n}$ are given in terms of moments of the splitting functions $f_j(y)$,
\begin{equation}
\langle N(p) |\, \mathcal{O}_j^{\mu_1 \cdots \mu_n}\, | N(p) \rangle
= 2\, f_j^{(n)}\,
  p^{ \{ \mu_1 } \cdots p^{ \mu_n\} },
\label{eq:fjn_def}
\end{equation}
where
\begin{eqnarray}
f_j^{(n)} &=& \int_{-1}^{1} \dd{y}\, y^{n-1} f_j(y),
\label{eq:fjn}
\end{eqnarray}
with $y$ the light-cone momentum fraction of the nucleon carried by the hadronic state $j$.
Assuming Eq.~(\ref{eq:fjn_def}) holds also for the off-shell nucleon states, the operator relation in Eq.~(\ref{eq:PDFmatch}) then gives rise to the convolution formula for the PDFs
\begin{equation}
q(x)\ =\ \sum_j \big[ f_j \otimes q_j^v \big](x)\
\equiv\ \sum_j
        \int_0^1 \dd{y} \int_0^1 \dd{z}\, \delta(x-yz)\, f_j(y)\, q_j^v(z),
\label{eq:conv}
\end{equation}
where $q_j^v \equiv q_j - \bar{q}_j$ is the valence distribution
for the quark flavor $q$ in the hadron $j$.

The valence distributions in the intermediate state hadronic configurations are needed as input to the calculation of the sea quark distributions in the proton.
Considering intermediate octet baryon states as an example, the most general expression for the quark vector operator $\mathcal{O}^{\mu_1 \cdots \mu_n}_q$ in terms of hadronic operators is then written as
\begin{equation}
\begin{split}
\label{eq:vector-op}
\mathcal{O}^{\mu_1 \cdots \mu_n}_q
&= \Big[
  \alpha^{(n)}
  (\overline{B} \gamma^{\mu_1} {B} \lambda^q_+)
+ \beta^{(n)}
  (\overline{B} \gamma^{\mu_1} \lambda^q_+ {B})
+ \sigma^{(n)}
  (\overline{B} \gamma^{\mu_1} {B})\,
  \mathrm{Tr}[\lambda^q_+]
\Big]\, p^{\mu_2} \cdots p^{\mu_n}			\\
&+\,
\Big[
  \bar\alpha^{(n)}
  (\overline{B} \gamma^{\mu_1} \gamma_5 {B} \lambda^q_-)
+ \bar\beta^{(n)}
  (\overline{B} \gamma^{\mu_1} \gamma_5 \lambda^q_- {B})
+ \bar\sigma^{(n)}
  (\overline{B} \gamma^{\mu_1} \gamma_5 {B})\,
  \mathrm{Tr}[\lambda^q_-]
\Big]\, p^{\mu_2} \cdots p^{\mu_n}			\\
&+\,
\mathrm{permutations} - \mathrm{Tr},
\end{split}
\end{equation}
where ``Tr'' denotes traces over Lorentz indices.
The flavor operators $\lambda_\pm^q$ are defined by
\begin{equation}
\lambda_\pm^q
= \frac{1}{2}
  \big( u \lambda^q u^\dag \pm u^\dag \lambda^q u \big) \, ,
\end{equation}
where $\lambda^q = {\rm diag}(\delta_{qu}, \delta_{qd}, \delta_{qs})$ are diagonal $3 \times 3$ quark flavor matrices.
The coefficients $\{ \alpha^{(n)}, \beta^{(n)}, \sigma^{(n)} \}$ and $\{ \bar\alpha^{(n)}, \bar\beta^{(n)}, \bar\sigma^{(n)} \}$ are related to moments of the spin-averaged and spin-dependent PDFs in octet baryons, respectively.
After expanding the flavor matrices, the operator $\mathcal{O}^{\mu_1 \cdots \mu_n}_q$ can be rearranged in the form
\begin{equation}\label{eq:vector-operator}
\mathcal{O}^{\mu_1 \cdots \mu_n}_q
= Q^{(n-1)}_B
    \mathcal{O}^{\mu_1 \cdots \mu_n}_{B}
+ Q^{(n-1)}_{B\phi\phi}
    \mathcal{O}^{\mu_1 \cdots \mu_n}_{B\phi\phi}
+ Q_{B\phi}^{(n-1)}\,
    \mathcal{O}^{\mu_1 \cdots \mu_n}_{B\phi},
\end{equation}
where the individual vector hadronic operators are given by
\begin{subequations}
\begin{eqnarray}\label{eq:had_ops}
\mathcal{O}^{\mu_1 \cdots \mu_n}_{B}
&=& \big( \overline{B} \gamma^{\mu_1} B \big)
    p^{\mu_2} \cdots p^{\mu_n},
\\
\mathcal{O}^{\mu_1 \cdots \mu_n}_{B \phi \phi}
&=& \frac{1}{f^2_\phi}
    \big( \overline{B} \gamma^{\mu_1} B \bar{\phi}\, \phi \big)\,
    p^{\mu_2} \cdots p^{\mu_n},
\\
\mathcal{O}^{\mu_1 \cdots \mu_n}_{B \phi}
&=& \frac{i}{f_\phi}
    \big( \overline{B}' \gamma^{\mu_1}\gamma_5 B \phi
        - \overline{B}  \gamma^{\mu_1}\gamma_5 B' \bar{\phi}
    \big)\,
    p^{\mu_2} \cdots p^{\mu_n}.
\end{eqnarray}
\end{subequations}
For proton PDFs, as we have in our case, $B'$ corresponds to a proton.
The coefficients $Q_j^{(n-1)}$ of each of the operators are defined in terms of Mellin moments of the corresponding parton distributions in the intermediate mesons and baryons, as in Eq.~(\ref{eq:cnqj}),
\begin{subequations}
\begin{eqnarray}\label{eq:Qn-1}
Q^{(n-1)}_B
&=& \int_{-1}^1 \dd x\, x^{n-1}\, q_B(x),
\\
Q^{(n-1)}_{B\phi\phi}
&=& \int_{-1}^1 \dd x\, x^{n-1}\, q_{\phi}^{\rm tad}(x), 
\\
Q_{B\phi}^{(n-1)}
&=& \int_{-1}^1 \dd x\, x^{n-1}\, q_B^{(\rm KR)}(x),
\end{eqnarray}
\end{subequations}
where the PDFs are the valence distributions in the intermediate configurations.

\begin{table}[thb]
\begin{center}
\caption{Moments $Q^{(n-1)}_{B}$ of the $u$, $d$ and $s$ quark distributions in octet baryons.}
\begin{tabular}{c|c|c|c} \hline 
 &  &  &  \\
$B$ & $U^{(n-1)}_{B}$ & $D^{(n-1)}_{B}$ & $S^{(n-1)}_{B}$ \\ 
 &  &  &  \\
\hline 
 &  &  &  \\
$p$ & $\alpha^{(n)}+\beta^{(n)}+\sigma^{(n)}$ & $\sigma^{(n)}$ & $\alpha^{(n)}-\beta^{(n)}+\sigma^{(n)}$ \\ [2mm]
$n$ & $\sigma^{(n)}$ & $\alpha^{(n)}+\beta^{(n)}+\sigma^{(n)}$ & $\alpha^{(n)}-\beta^{(n)}+\sigma^{(n)}$ \\ [2mm]
$\Sigma^+$ & $\alpha^{(n)}+\beta^{(n)}+\sigma^{(n)}$ & $\alpha^{(n)}-\beta^{(n)}+\sigma^{(n)}$ & $\sigma^{(n)}$ \\ [2mm]
$\Sigma^0$ & $\alpha^{(n)}+\sigma^{(n)}$ & $\alpha^{(n)}+\sigma^{(n)}$ & $\sigma^{(n)}$ \\ [2mm]
$\Sigma^-$ & $\alpha^{(n)}-\beta^{(n)}+\sigma^{(n)}$ & $\alpha^{(n)}+\beta^{(n)}+\sigma^{(n)}$ & $\sigma^{(n)}$ \\ [2mm]
$\Lambda$ & $\frac13\alpha^{(n)}+\sigma^{(n)}$ & $\frac13\alpha^{(n)}+\sigma^{(n)}$ & $\frac43\alpha^{(n)}+\sigma^{(n)}$ \\ [2mm]
$\Lambda\Sigma^0$ & $\frac{1}{\sqrt{3}}\alpha^{(n)}$ & $-\frac{1}{\sqrt{3}}\alpha^{(n)}$ & $0$ \\
 &  &  &  \\ \hline
\end{tabular}
\label{tab:PDFoct}
\end{center}
\end{table}

\begin{table}[thpb]
\begin{center}
\caption{Moments $Q^{(n-1)}_{B\phi\phi}$ of the $u$, $d$ and $s$ quark distributions arising from the  $BB\phi\phi$ tadpole vertex.}
\footnotesize {
\begin{tabular}{c|cc|cc|cc} \hline 
 &  &  &  \\
$B$ & \multicolumn{2}{|c}{$U^{(n-1)}_{B\phi\phi}$} & \multicolumn{2}{|c} {$D^{(n-1)}_{B\phi\phi}$} & \multicolumn{2}{|c} {$S^{(n-1)}_{B\phi\phi}$} \\ 
& $\pi^+\pi^-$ & $K^+K^-$ & $\pi^+ \pi^-$ & $K^0 \overline{K}^0$ & $K^0 \overline{K}^0$ & $K^+K^-$ \\
\hline 
 &  &  &  &  &  &  \\
$p$ & $\frac{1}{2}(\alpha^{(n)}+\beta^{(n)})$ & $\beta^{(n)}$ & $-\frac{1}{2}(\alpha^{(n)}+\beta^{(n)})$ & $-\frac{1}{2}(\alpha^{(n)}-\beta^{(n)})$ & $\frac{1}{2}(\alpha^{(n)}-\beta^{(n)})$ & $-\beta^{(n)}$ \\ [2mm]
$n$ & $-\frac{1}{2}(\alpha^{(n)}+\beta^{(n)})$ & $-\frac{1}{2}(\alpha^{(n)}-\beta^{(n)})$ & $\frac{1}{2}(\alpha^{(n)}+\beta^{(n)})$ & $\beta^{(n)}$ & $-\beta^{(n)}$ & $\frac{1}{2}(\alpha^{(n)}-\beta^{(n)})$\\ [2mm]
$\Sigma^+$ & $\beta^{(n)}$ & $\frac{1}{2}(\alpha^{(n)}+\beta^{(n)})$ & $-\beta^{(n)}$ & $\frac{1}{2}(\alpha^{(n)}-\beta^{(n)})$ & $-\frac{1}{2}(\alpha^{(n)}-\beta^{(n)})$ & $-\frac{1}{2}(\alpha^{(n)}+\beta^{(n)})$ \\ [2mm]
$\Sigma^0$ & $0$ & $\frac{1}{2}\alpha^{(n)}$ & $0$  & $\frac{1}{2}\alpha^{(n)}$ & $-\frac{1}{2}\alpha^{(n)}$ & $-\frac{1}{2}\alpha^{(n)}$ \\ [2mm]
$\Sigma^-$ & $-\beta^{(n)}$ & $\frac{1}{2}(\alpha^{(n)}-\beta^{(n)})$ & $\beta^{(n)}$ & $\frac{1}{2}(\alpha^{(n)}+\beta^{(n)})$ & $-\frac{1}{2}(\alpha^{(n)}+\beta^{(n)})$ & $-\frac{1}{2}(\alpha^{(n)}-\beta^{(n)})$ \\ [2mm]
$\Lambda$ & $0$ & $-\frac{1}{2}\alpha^{(n)}$ & $0$ & $-\frac{1}{2}\alpha^{(n)}$ & $\frac{1}{2}\alpha^{(n)}$ & $\frac{1}{2}\alpha^{(n)}$ \\ [2mm]
$\Lambda\Sigma^0$ & $\frac{1}{\sqrt{3}}\alpha^{(n)}$ & $\frac{1}{2\sqrt{3}}\alpha^{(n)}$ & $-\frac{1}{\sqrt{3}}\alpha^{(n)}$ & $-\frac{1}{2\sqrt{3}}\alpha^{(n)}$ & $\frac{1}{2\sqrt{3}}\alpha^{(n)}$ & $-\frac{1}{2\sqrt{3}}\alpha^{(n)}$ \\ 
 &  &  &  &  &  &  \\ \hline 
\end{tabular}
\label{tab:PDFtad}
}
\end{center}
\end{table}

Each of the moments $Q^{(n-1)}_j$ can be expressed in terms of the coefficients $\{ \alpha^{(n)}, \beta^{(n)}, \sigma^{(n)}\}$ and $\{ \bar{\alpha}^{(n)}, \bar{\beta}^{(n)}, \bar{\sigma}^{(n)}\}$ from Eq.~(\ref{eq:vector-op}).
For example, the moments $Q^{(n-1)}_B$ for $u$, $d$ and $s$ quark are listed in Table \ref{tab:PDFoct}.
Solving for the coefficients, one can write these as linear combinations of the individual $u$, $d$ and $s$ quark moments in the proton,
\begin{subequations} \label{eq:momentB}
\begin{eqnarray} 
\alpha^{(n)} &=& \frac{1}{2}(U_{p}^{(n-1)}+S_{p}^{(n-1)}) - D_{p}^{(n-1)},
\\
\beta^{(n)} &=& \frac{1}{2}(U_{p}^{(n-1)}-S_{p}^{(n-1)}),
\\
\sigma^{(n)} &=& D_{p}^{(n-1)}.
\end{eqnarray}
\end{subequations}

The moments of $Q^{(n-1)}_{B\phi\phi}$ of the tadpole diagram for $u$, $d$ and $s$ quark are also expressed in terms of $\alpha^{(n)}$, $\beta^{(n)}$ and $\sigma^{(n)}$, and are listed in Table \ref{tab:PDFtad}. 
Here one can see that there is no contribution to the $u$-quark moments from $K^0 \overline {K}^0$, as well as no contribution to the $d$-quark moments from $K^+K^-$, or contribution to the $s$-quark moments from $\pi^+\pi^-$.
Furthermore, the moments $Q^{(n-1)}_{B\phi\phi}$ from the tadpole diagram can also be expressed in terms of the quark moments in the proton.

The moments $Q^{(n-1)}_{B\phi}$ of the Kroll-Ruderman terms are expressed in terms of $\bar{\alpha}^{(n)}$, $\bar{\beta}^{(n)}$ and $\bar{\sigma}^{(n)}$, which are related to the moments of the spin-dependent PDFs in the proton.
The above relationships are only related to the flavor structure of the operators and do not depend on the structure of the $\gamma$ matrices.
The coefficients $\bar{\alpha}^{(n)}$, $\bar{\beta}^{(n)}$ and $\bar{\sigma}^{(n)}$ therefore have the relationships similar to those in Eqs.~(\ref{eq:momentB}),
\begin{subequations}
\label{eq;momentBpo}
\begin{eqnarray} 
\bar{\alpha}^{(n)}
&=& \frac12 
(\Delta U_{p}^{(n-1)} + \Delta S_{p}^{(n-1)}) - \Delta D_{p}^{(n-1)},
\\
\bar{\beta}^{(n)} 
&=& \frac12
(\Delta U_{p}^{(n-1)}-\Delta S_{p}^{(n-1)}),
\\
\bar{\sigma}^{(n)} 
&=& \Delta D_{p}^{(n-1)}.
\end{eqnarray}
\end{subequations}
where $\Delta Q_p^{(n-1)}$ is defined in terms of the spin-dependent PDFs in analogy with Eq.~(\ref{eq:mom_def}).

Because the relations for the moments are valid for any order of $n$, corresponding relations must also be true for the actual PDFs in terms of which they are defined.
With these relationships, the PDFs of quarks in different hadronic configurations can be expressed in terms of spin-averaged and soin-dependent quark distributions in the proton $q_p(x)$ and $\Delta q_p(x)$, for which we use parameterizations from Refs.~\cite{Diehl, Martin, Leader4}.
Using the convolution form and the valence quark distributions in proton as input, we can then calculate the sea quark distributions with the splitting functions derived from the operators in terms of hadronic degrees of freedom.
In the following subsections we will discuss the unpolarized and polarized PDFs of sea quarks in the proton.

\subsection{\it Unpolarized distributions}

\subsubsection{\it $\bar{d} - \bar{u}$ asymmetry}

The observation of the $\bar{d}-\bar{u}$ flavor asymmetry in the light quark sea of the proton has been one of the seminal results in hadronic physics over the past three decades, leading to reevaluations of our understanding of the quark structure of the nucleon.
The convolution form for the $\bar{d}-\bar{u}$ asymmetry can be written explicitly as
\be \label{eq:d-ucon}
\bar{d}(x) - \bar{u}(x) 
= \left[
\big( f_{\pi^+ n}^{(\rm rbw)} + f_{\pi^+ \Delta^0}^{(\rm rbw)} - f_{\pi^- \Delta^{++}}^{(\rm rbw)}+ f_\phi^{(\rm bub)} 
\big) \otimes q_\pi^v 
\right](x).
\ee
The rainbow and bubble Feynman diagrams for the splitting functions are the same as Figs.~\ref{fig:FMLOOP}(a),  \ref{fig:FMLOOP}(g) and \ref{fig:FMLOOP}(m). 
The splitting function corresponding to Fig.~\ref{fig:FMLOOP}(a) for the intermediate state neutron is given by a sum of nucleon on-shell and $\delta$-function contributions,
\begin{eqnarray}
f^{\rm (rbw)}_{\pi^+ n}(y)
&=& \frac{2 (D+F)^2 M_N^2}{(4\pi f)^2}
    \Big[ f^{\rm (on)}_N(y)
        + f^{(\delta)}_\pi(y)
        - \delta f^{(\delta)}_\pi(y)
    \Big],
\label{eq:piN-rbw}
\end{eqnarray}
where the explicit forms for the basis functions $f^{\rm (on)}_N$, $f^{(\delta)}_\pi$, and $\delta f^{(\delta)}_\pi$ are given in Ref.~\cite{Salamu} for the dipole regulator $\widetilde{F}(k)$ with the nonlocal Lagrangian. 
The on-shell function $f^{\rm (on)}_N$ is nonzero for $y>0$, while the local $f^{(\delta)}_\pi$ and nonlocal $\delta f^{(\delta)}_\pi$ functions are proportional to $\delta(y)$ and hence have contributions to $\bar d-\bar u$ only at $x=0$~\cite{Burkardt, Salamu3}. 
In the point-like limit, where the cutoff in the regulator $\Lambda \to \infty$, the nonlocal function $\delta f^{(\delta)}_\pi$ vanishes.
However, at finite $\Lambda$ values it remains nonzero.

For the $\pi \Delta$ intermediate state contributions to the asymmetry in Eq.~(\ref{eq:d-ucon}), the splitting function for the rainbow diagram in Fig.~\ref{fig:FMLOOP}(g) includes several regular and $\delta$-function terms,
\begin{eqnarray}
f^{\rm (rbw)}_{\pi^- \Delta^{++}}(y)\
=\ 3 f^{\rm (rbw)}_{\pi^+ \Delta^0}(y)\
&=& \frac{{\cal C}^2 \overline{M}^2}{2 (4\pi f)^2}
\Big[ f^{\rm (on)}_\Delta(y)
     + f^{\rm (on\, end)}_\Delta(y)
     - \frac{1}{18} f^{(\delta)}_\Delta(y)           \notag\\
& & \hspace*{1.5cm}
   +\, \frac{2 M_N^2 \big( \overline{M}^2 - m_\pi^2 \big)}
            {3 M_\Delta^2\, \overline{M}^2}
       \big( f^{(\delta)}_\pi(y) - \delta f^{(\delta)}_\pi(y) \big)
\Big],
\label{eq:piD-rbw}
\end{eqnarray}
where $\overline{M} = M_N + M_\Delta$.
As for the $\pi N$ case, the on-shell function for the $\Delta$ intermediate state, $f^{\rm (on)}_\Delta$, is nonzero for $y > 0$, with a shape that is qualitatively similar to $f^{\rm (on)}_N$. 
The on-shell end-point function $f^{\rm (on\, end)}_\Delta$ also has a similar shape for finite $\Lambda$, but in the $\Lambda \to \infty$ limit is associated with an end-point singularity that gives a $\delta$-function at $y=1$.
The functions $f^{(\delta)}_\pi$ and $\delta f^{(\delta)}_\pi$ are equivalent to those in Eq.~(\ref{eq:piN-rbw}), while $f^{(\delta)}_\Delta$ is a new function that appears only for the decuplet intermediate state.

The contribution to the $\bar d-\bar u$ asymmetry from the bubble diagram in Fig.~\ref{fig:FMLOOP}(m) is given by the same combination of basis $\delta$-function contributions as for the rainbow diagrams,
\begin{eqnarray}
f^{\rm (bub)}_\pi(y)
&=&-\frac{2 M_N^2}{(4\pi f)^2}
    \Big[ f^{(\delta)}_\pi(y) - \delta f^{(\delta)}_\pi(y)
    \Big].
\label{eq:f-bub}
\end{eqnarray}
Although the $\delta$-function term gives a nonzero PDF only at $x = 0$, it contributes to the integral of $\bar d-\bar u$ over $x$, and will therefore indirectly affect the normalization of the spitting function for $x > 0$.
Because experimental cross sections are in practice available only for $x > 0$, the $\delta$-function pieces are difficult to constrain directly.
The advantage of the nonlocal approach employed here is that by consistently introducing a vertex correlator in coordinate space in the nonlocal Lagrangian~\cite{Salamu}, the same regulator function appears in all splitting functions derived from the fundamental interaction, which in our case is parameterized through the single parameter $\Lambda$.

Because of the presence of the regulator, the end-point term $\delta(1-y)$ in the local theory turns into a smooth function of $y$ in the nonlocal theory, making the results more realistic.
However, since the regulator is a function of the meson momentum $k$, the $\delta(y)$ term still exists in the nonlocal approach.
To eliminate both $\delta(y)$ and $\delta(1-y)$ terms in the splitting functions, more complicated correlation functions are necessary.
On the other hand, the leading nonanalytic behavior of the splitting functions is the same in both the local and nonlocal versions of the $\chi$PT.

\begin{figure}[t] 
\begin{center}
\epsfig{file=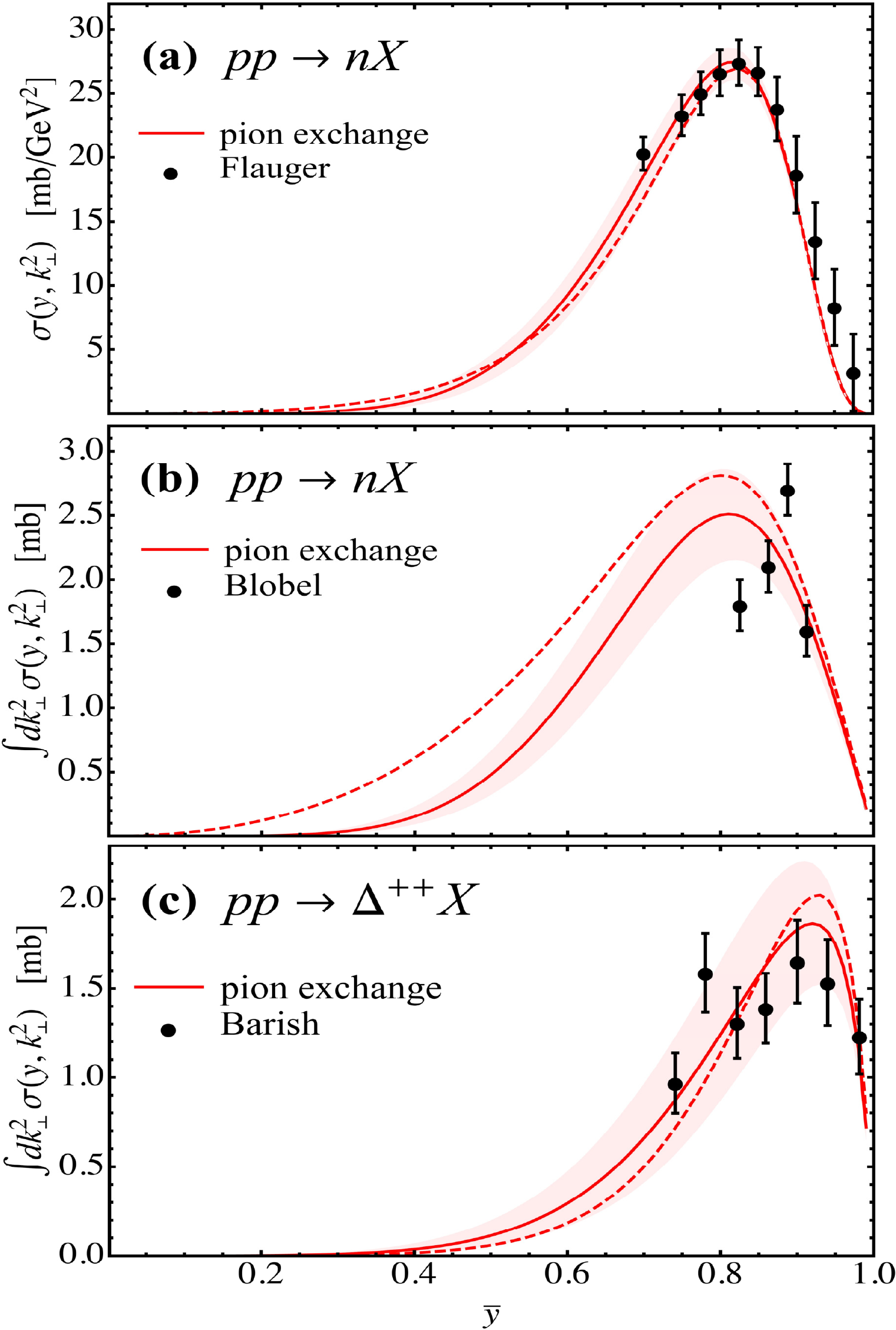,scale=0.6}
\caption{Differential inclusive hadron production cross section
	$\sigma(y,k_\bot^2)$ versus $\bar{y}$ for
   {\bf (a)} $pp \to n X$ at $k_\bot^2 = 0$
	\cite{Flauger};	
   {\bf (b)} $pp \to n X$ integrated over $k_\bot^2$
	\cite{Blobel};		
   {\bf (c)} $pp \to \Delta^{++} X$ integrated over $k_\bot^2$
	\cite{Barish2},	
	compared with the fitted nonlocal pion exchange contributions for
	$\Lambda_{\pi N}=1.0(1)$~GeV and
	$\Lambda_{\pi\Delta}=0.9(1)$~GeV
	(solid red lines and pink $1\sigma$ uncertainty bands) and with Pauli-Villars regularization (dashed red lines). 
	(Figure from Ref.~\cite{Salamu2}.)}
\label{fig:crosspi}
\end{center}
\end{figure}

In the numerical form factor calculations in the previous section, the parameter $\Lambda$ was chosen to be $\sim 1$~GeV, and taken to be the same for all the regulators of the baryon-meson interaction. 
Alternatively, $\Lambda$ can also be chosen to be different for octet baryon-meson and octet-decuplet-meson interactions, constrained by the 
cross sections for the inclusive baryon production processes $pp \to n X$ and $pp \to \Delta^{++} X$ \cite{Salamu2, XGWang, XGWang2, Holtmann}.
The results for the differential neutron production cross section are shown in Fig.~\ref{fig:crosspi} versus $\bar{y}\ (\equiv 1-y)$ \cite{Salamu2}.
In Fig.~\ref{fig:crosspi}(a) the results are compared with the neutron production data from the ISR at CERN at energies $\sqrt{s}$ between $\approx 31$ and 63~GeV for $0^\circ$ neutron production angles, or \mbox{$k_\bot^2=0$}~\cite{Flauger}. 
Data from the hydrogen bubble chamber experiment at the CERN proton synchrotron at $\sqrt{s}$ around 5 and 7 GeV \cite{Blobel} are shown in Fig.~\ref{fig:crosspi}(b) for the $k_\bot$-integrated neutron cross section. 
Because the pion-exchange processes is dominant only at large $\bar{y}$ \cite{McKenney}, only data in the region $\bar{y} > 0.7$ are included. 
A good description of the single and double differential neutron data can be achieved with the parameter $\Lambda_{\pi N} = 1.0(1)$~GeV.
For the inclusive production of decuplet baryons, the $k_\bot^2$-integrated $\Delta^{++}$ cross section is shown in Fig.~\ref{fig:crosspi}(c) compared with hydrogen bubble chamber data taken at Fermilab for $\sqrt{s} \approx 20$~GeV~\cite{Barish2}. 
A good fit to the data is obtained with a value of the decuplet $\Lambda_{\pi\Delta} = 0.9(1)$~GeV, which is slightly smaller than that for the neutron production cross sections.
For comparison, the hadron production cross sections were also computed in Refs.~\cite{XGWang, XGWang2} using Pauli-Villars regularization for the local effective theory.

\begin{figure}[t] 
\begin{center}
\epsfig{file=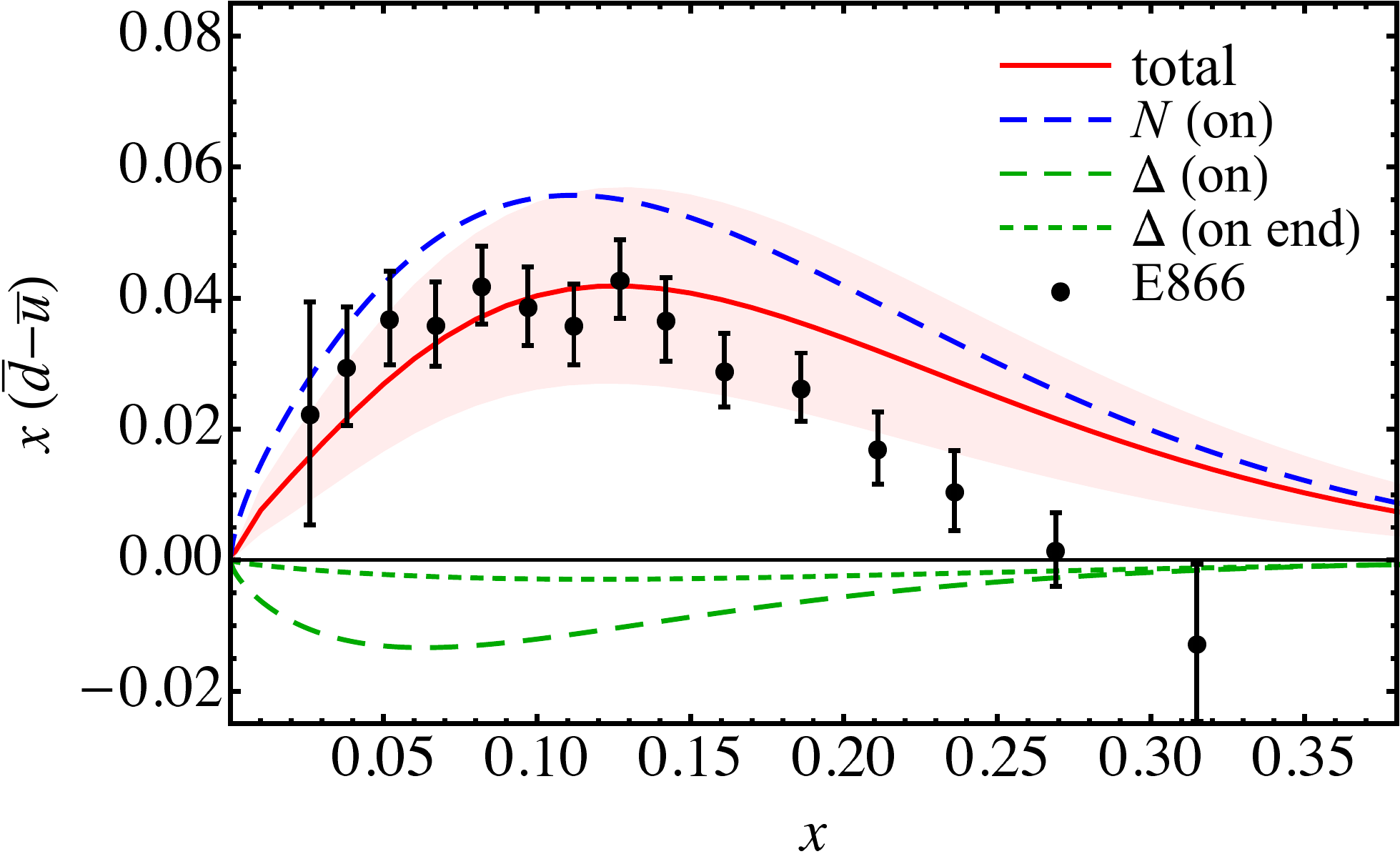,scale=0.6}
\caption{The flavor asymmetry of the proton $x(\bar d-\bar u)$ versus $x$ with the parameters $\Lambda_{\pi N} = 1.0(1)$~GeV and $\Lambda_{\pi \Delta} = 0.9(1)$~GeV, including nucleon on-shell (dashed blue), $\Delta$ on-shell (dashed green), and $\Delta$ end-point (dotted green) contributions, and compared with the asymmetry extracted from the Fermilab E615 Drell-Yan experiment~\cite{Towell}. (Figure from Ref.~\cite{Salamu2}.)} 
\label{fig:dbub}
\end{center}
\end{figure}

Using the values of $\Lambda_{\pi N}$ and $\Lambda_{\pi \Delta}$ for the nonlocal calculation constrained by the inclusive $pp$ cross sections, the flavor asymmetry $\bar d-\bar u$ can be evaluated from the convolution of the splitting functions and the pion PDF in Eq.~(\ref{eq:d-ucon})~\cite{Salamu2}. 
The results for $x(\bar d-\bar u)$ are shown in Fig.~\ref{fig:dbub}, compared with the asymmetry extracted from the E866 Drell-Yan lepton-pair production experiment from Fermilab~\cite{Towell, Salamu2}.
At nonzero $x$ values only the on-shell nucleon and $\Delta$ and end-point $\Delta$ terms contribute to the asymmetry, each of which is indicated in Fig.~\ref{fig:dbub}.
The positive nucleon on-shell term makes the largest contribution, which is partially cancelled by the negative $\Delta$ contributions.
The end-point term is relatively small compared with the on-shell $\Delta$ component.

Although the $\delta$-function contributions to the flavor asymmetry are not directly visible in Fig.~\ref{fig:dbub}, their effect can be seen in the lowest moment of the asymmetry,
\begin{equation}
\langle \bar d - \bar u \rangle
\equiv \int_0^1 \dd{x} \big( \bar d(x)-\bar u(x) \big).
\label{eq:dbubmom}
\end{equation}
For the best fit values $\Lambda_{\pi N}=1.0(1)$~GeV and $\Lambda_{\pi \Delta}=0.9(1)$~GeV, the contributions from the individual terms in Eqs.~(\ref{eq:d-ucon})--(\ref{eq:f-bub}) are listed in Table~\ref{tab:du}, along with the combined contributions from the $x>0$ and $x=0$ terms, and the local and nonlocal terms, to the total integrated result.
The nucleon on-shell term is the most important component, with a contribution that is within $\approx 20\%$ of the total integrated value $\langle \bar d - \bar u \rangle = 0.128^{(44)}_{(42)}$, where the errors here reflect the uncertainties on the cutoff parameters. 
The on-shell and end-point $\pi \Delta$ terms yield overall negative contributions, with magnitude $\approx 30\%$ of the on-shell $\pi N$.
The various $\delta$-function terms from all three diagrams cancel to a considerable degree, with the $x=0$ contribution making up $\approx 20\%$ of the total. 
Furthermore, the breakdown into the local and nonlocal pieces shows that the latter is negative with magnitude $\approx 20\%$ of the local.
Note that the more recent SeaQuest experiment at Fermilab~\cite{SeaQuest} found evidence for a larger, more positive $\bar d-\bar u$ asymmetry at the higher-$x$ values, bringing the data and theory in Fig.~\ref{fig:dbub} closer to agreement.

\begin{table}[hbt]
\begin{center}
\caption{Contributions to the integral
	$\langle \bar d-\bar u \rangle \equiv \int_0^1 \dd{x} (\bar d-\bar u)$
	from the $\pi N$ rainbow, $\pi \Delta$ rainbow and$\pi$ bubble diagrams,
	for the best fit parameters $\Lambda_{\pi N}=1.0(1)$~GeV and $\Lambda_{\pi \Delta}=0.9(1)$~GeV. The contributions from the various terms in Eqs.~(\ref{eq:d-ucon})--(\ref{eq:f-bub}) are listed	individually, as are the combined contributions from	$x>0$ and $x=0$, and the local and nonlocal terms,	to the total. Note that some numbers do not sum to the
	totals because of rounding.}
\begin{tabular}{lll}				\hline \\[-2mm]
     ~diagram~
   &
   & ~~$\langle \bar d-\bar u \rangle$
\\ [2mm] \hline \\ [-2mm]
     ~$\pi N$ (rbw)~
   & ~$f_N^{\rm (on)}$~
   & ~~\,$0.152^{(32)}_{(30)}$~		\\[2mm]
   & ~$f_\pi^{(\delta)}$~
   & $-0.079^{(20)}_{(18)}$~~~		\\[2mm]
   & ~$\delta f_\pi^{(\delta)}$~
   & ~~\,$0.044^{(10)}_{(9)}$~		\\[2mm]
     ~total~$\pi N$~
   & 
   & ~~\,$0.116(22)$~		\\[2mm] \hline \\[-2mm]
     ~$\pi \Delta$ (rbw)~
   & ~$f_\Delta^{\rm (on)}$~
   & $-0.042(12)$~~~		\\[2mm]
   & ~$f_\Delta^{\rm (on\,end)}$~
   & $-0.008^{(4)}_{(3)}$~~~		\\[2mm]
   & ~$f_\Delta^{(\delta)}$~
   & ~~\,$0.002(1)$~		\\[2mm]
   & ~$f_\pi^{(\delta)}$~
   & ~~\,$0.038(10)$~		\\[2mm]
   & ~$\delta f_\pi^{(\delta)}$~
   & $-0.021(5)$~~~		\\[2mm]
     ~total~$\pi \Delta$~
   & 
   & $-0.032(10)$~~~		\\[2mm] \hline \\[-2mm]
     ~$\pi$~(bub)
   & ~$f_\pi^{(\delta)}$~
   & ~~\,$0.099^{(25)}_{(22)}$~		\\[2mm]
   & ~$\delta f_\pi^{(\delta)}$~
   & $-0.054^{(13)}_{(12)}$~~~		\\[2mm]
     ~total~$\pi$~bubble~~~
   &
   & ~~\,$0.044^{(12)}_{(10)}$~		\\[2mm] \hline \\[-2mm]
     ~{\bf total}~
   & 
   & ~~$\bm{0.128^{(44)}_{(42)}$}~		\\[2mm] \hline
\end{tabular}
\label{tab:du}
\end{center}
\end{table}

\newpage
\subsubsection{\it $s - \bar{s}$ asymmetry}

Within a similar approach one can consider the contribution of meson loops to the $s-\bar{s}$ asymmetry in the proton.
The contribution to the antistrange PDF in the proton from kaon loops in Fig.~\ref{fig:FMLOOP} can be written as
\begin{equation}
\begin{split}
\bar{s}(x)
= \Big[
  \Big(   \sum_{\phi B} f^{\rm (rbw)}_{\phi B}
	+ \sum_{\phi T} f^{\rm (rbw)}_{\phi T}
	+ \sum_{\phi} f^{\rm (bub)}_\phi
  \Big) \otimes \bar{s}_K
  \Big](x),
\end{split}
\label{eq:antistrange}
\end{equation}
where the sums are over the states
	$\phi B = \{ K^+ \Lambda, K^+ \Sigma^0, K^0 \Sigma^+ \}$
for the kaon-octet baryon rainbow diagram [Fig.~\ref{fig:FMLOOP}(a)],
	$\phi T = \{ K^+ \Sigma^{*0}, K^0 \Sigma^{*+} \}$
for the kaon-decuplet baryon rainbow diagram [Fig.~\ref{fig:FMLOOP}(g)], and for the $\phi = K^+ (K^-)$ and $K^0 (\overline{K}^0)$ loop in the bubble diagram [Fig.~\ref{fig:FMLOOP}(m)].
In analogy with the light sea quark PDFs, the splitting functions for the antistrange PDF also include the on-shell and $\delta$-function terms. 
The splitting functions are the same as those in the last subsection, except for the coefficients~\cite{Salamu2}.

For the strange quark PDF, the diagrams associated with the rainbow, Kroll-Ruderman, tadpole and additional diagrams from the gauge link shown in Fig.~\ref{fig:FMLOOP} have contributions, except the diagrams with magnetic interactions.
Assuming that all nonperturbatively generated strangeness resides in the intermediate state hyperons, the strange quark PDF in the proton can be written as
\begin{equation}
\begin{split}
s(x)
&= \sum_{B\phi}
   \Big\{
   \Big[ \bar{f}^{\rm (rbw)}_{B\phi} \otimes s_B \Big](x)
 + \Big[ \bar{f}^{\rm (KR)}_B \otimes s^{\rm (KR)}_B \Big](x)
 + \Big[ \delta \bar{f}^{\rm (KR)}_B \otimes s^{(\delta)}_B \Big](x)
   \Big\}		\\
&+\, \sum_{T\phi}
   \Big\{
   \Big[ \bar{f}^{\rm (rbw)}_{T\phi} \otimes s_T \Big](x)
 + \Big[ \bar{f}^{\rm (KR)}_T \otimes s^{\rm (KR)}_T \Big](x)
 + \Big[ \delta \bar{f}^{\rm (KR)}_T \otimes s^{(\delta)}_T \Big](x)
   \Big\}		\\
&+\, \sum_\phi
   \Big\{
   \Big[ \bar{f}^{\rm (tad)}_\phi \otimes s^{\rm (tad)}_\phi \Big](x)
 + \Big[ \delta \bar{f}^{\rm (tad)}_\phi \otimes s^{(\delta)}_\phi \Big](x)
   \Big\},
\end{split}
\label{eq:strange}
\end{equation}
where the sums are over the octet baryon--meson states
	$B \phi = \{ \Lambda K^+, \Sigma^0 K^+, \Sigma^+ K^0 \}$,
decuplet baryon--meson states
	$T \phi = \{ \Sigma^{*0} K^+, \Sigma^{*+} K^0 \}$,
and mesons $\phi = K^+ (K^-)$ and $K^0 (\overline{K}^0)$ for the tadpole contributions.
The splitting functions for all the one-loop diagrams in Eq.~(\ref{eq:strange}) use the shorthand notation ${\bar f}_j(y) \equiv f_j(1-y)$.

For the octet hyperon rainbow diagram, Fig.~\ref{fig:FMLOOP}(b), the splitting function for the intermediate $\Sigma^+$ can be written in terms of the on-shell, off-shell and $\delta$-function basis functions as
\be \label{eq:f-rbw-Bphi}
f^{\rm (rbw)}_{\Sigma^+ K^0}(y)
= \frac{(D\, -\, F)^2 (M_N+M_\Sigma)^2}{2(4\pi f)^2}\,
    \Big[ f^{\rm (on)}_\Sigma(y)
	+ f^{\rm (off)}_\Sigma(y)
	+ 4\, \delta f^{\rm (off)}_\Sigma(y)
	- f^{(\delta)}_K(y)
    \Big],
\ee
where the functions $f^{\rm (on)}_{\Sigma}$, $f^{\rm (off)}_{\Sigma}$, $\delta f^{\rm (off)}_{\Sigma}$ and $f^{(\delta)}_K$ are given in Ref.~\cite{Salamu}.
For the octet Kroll-Ruderman diagrams in Figs.~\ref{fig:FMLOOP}(c)--(f), 
the local and nonlocal splitting functions $f^{\rm (KR)}_{\Sigma}$ and $\delta f^{\rm (KR)}_{\Sigma}$ are given by
\be \label{eq:f-KR-Bphi}
f^{\rm (KR)}_{\Sigma^+}(y)
= \frac{(D\,-F\,)^2 (M_N+M_\Sigma)^2}{2(4\pi f)^2}\,
    \Big[- f^{\rm (off)}_\Sigma(y)
	 + 2 f^{(\delta)}_K(y)
    \Big],
\ee
and
\be \label{eq:f-link-Bphi}
\delta f^{\rm (KR)}_{\Sigma^+}(y)
= \frac{(D\,-F\,)^2 (M_N+M_\Sigma)^2}{2(4\pi f)^2}\,
    \left[-4\, \delta f^{\rm (off)}_\Sigma(y)\,
	  -\,  \delta f^{(\delta)}_K(y)
    \right],
\ee
respectively.

For the decuplet hyperon contributions, the respective splitting functions for the intermediate $\Sigma^{*+}$ are given by
\bea
f^{\rm (rbw)}_{\Sigma^{*+} K^0}(y)
&=& \frac{{\cal C}^2 (M_N+M_{\Sigma^*})^2}{6 (4\pi f)^2}	
\Big[
  f^{\rm (on)}_{\Sigma^*}(y)
+ f^{\rm (on\, end)}_{\Sigma^*}(y)
- 2 f^{\rm (off)}_{\Sigma^*}(y)
- 2 f^{\rm (off\, end)}_{\Sigma^*}(y) \notag \\
&& \hspace*{-1.5cm}
+ 4\, \delta\!f^{\rm (off)}_{\Sigma^*}(y)			
+ \frac{1}{18} f^{(\delta)}_{\Sigma^*}(y)
- \frac{1}{6}\, \delta\!f^{(\delta)}_{\Sigma^*}(y)
- \frac{(M_N+M_\Sigma)^2 [(M_N+M_{\Sigma^*})^2 + 3 m_K^2]}
       {6 M_{\Sigma^*}^2\, (M_N+M_{\Sigma^*})^2}\, f^{(\delta)}_K(y)
\Big]						
\label{eq:dsigma}
\eea
for the decuplet rainbow diagram in Fig.~\ref{fig:FMLOOP}(h),
\bea
\hspace*{-0.3cm}
f^{(\rm KR)}_{\Sigma^{*+}}(y)
&=&\frac{{\cal C}^2 (M_N+M_{\Sigma^*})^2}{6 (4\pi f)^2}
\Big[
  2 f^{\rm (off)}_{\Sigma^*}(y)
+ 2 f^{\rm (off\, end)}_{\Sigma^*}(y)			\nonumber \\ 
&-& \frac19
    \big( f^{(\delta)}_{\Sigma^*}(y)
        - \delta\!f^{(\delta)}_{\Sigma^*}(y)
    \big)
 +  \frac{(M_N+M_\Sigma)^2 [(M_N+M_{\Sigma^*})^2 + m_K^2]}
         {3 M_{\Sigma^*}^2\, (M_N+M_{\Sigma^*})^2}\, f^{(\delta)}_K(y)
\Big]					
\label{eq:fTKR}
\eea
for the Kroll-Ruderman diagram in Figs.~\ref{fig:FMLOOP}(i) and \ref{fig:FMLOOP}(j), and
\begin{eqnarray}
\hspace*{-0.5cm}
\delta\!f^{(\rm KR)}_{\Sigma^{*+}}(y)
&=& \frac{{\cal C}^2 (M_N+M_{\Sigma^*})^2}{6 (4\pi f)^2}
\Big[
  - 4\, \delta\!f^{\rm (off)}_{\Sigma^*}(y)
  + \frac{1}{18}\, \delta\!f^{(\delta)}_{\Sigma^*}(y)	\notag\\
& & \hspace*{3cm}
  -\, \frac{(M_N+M_{\Sigma})^2 [(M_N+M_{\Sigma^*})^2 - m_K^2]}
           {6 M_{\Sigma^*}^2\, (M_N+M_{\Sigma^*})^2}\,
    \delta\!f^{(\delta)}_K(y)
\Big]
\label{eq:fdfTKR}
\end{eqnarray}
for the additional Kroll-Ruderman diagram in Figs.~\ref{fig:FMLOOP}(k) and \ref{fig:FMLOOP}(l).
The expressions for the decuplet basis functions
  $f^{\rm (on)}_{\Sigma^*}$,
  $f^{\rm (on\, end)}_{\Sigma^*}$,
  $f^{\rm (off)}_{\Sigma^*}$,
  $f^{\rm (off\, end)}_{\Sigma^*}$ and
  $f^{\rm (\delta)}_{\Sigma^*}$,
as well as the nonlocal functions
  $\delta f^{\rm (off)}_{\Sigma^*}$ and
  $\delta f^{\rm (\delta)}_{\Sigma^*}$,
are also given in Ref.~\cite{Salamu}.

Finally, for the local and nonlocal tadpole contributions to the strange quark PDF from Figs.~\ref{fig:FMLOOP}(n) and \ref{fig:FMLOOP}(o), the splitting functions are given by
\begin{eqnarray}
\label{eq:f-tad}
f^{\rm (tad)}_{K^+}(y)
&=& 2\, f^{\rm (tad)}_{K^0}(y)\
 =\ -\frac{ (M_N+M_\Sigma)^2}{(4\pi f)^2}\,
    f^{(\delta)}_K(y),				\\
\delta\!f^{\rm (tad)}_{K^+}(y)
&=& 2\, \delta\!f^{\rm (tad)}_{K^0}(y)\
 =\ \frac{(M_N +M_\Sigma)^2}{(4\pi f)^2}\,
    \delta\!f^{(\delta)}_K(y),
\label{eq:ad45}
\end{eqnarray}
in terms of the local and nonlocal basis functions $f^{(\delta)}_K$
and $\delta\!f^{(\delta)}_K$.

\begin{figure}[h] 
\begin{center}
\epsfig{file=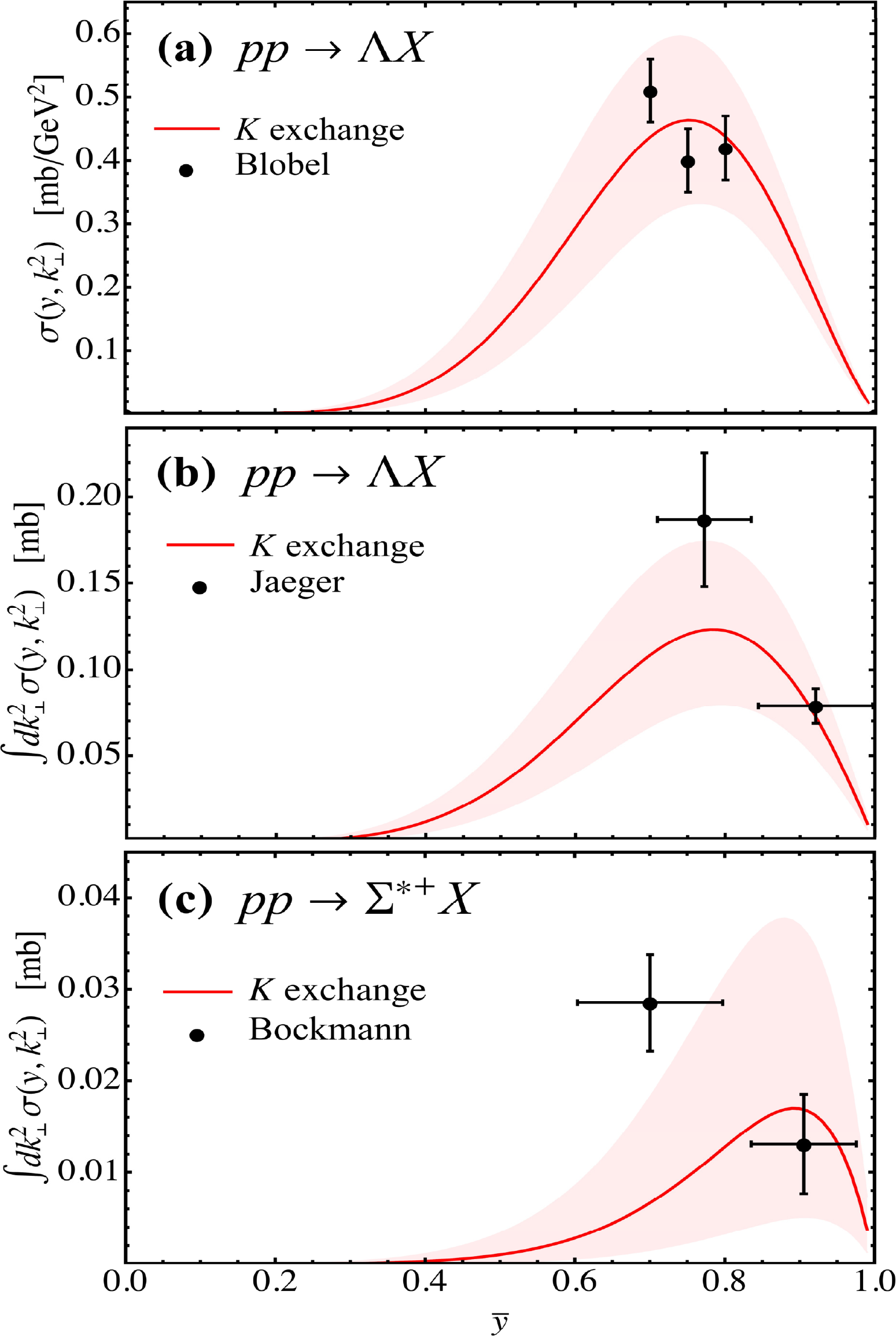,scale=0.55}
\caption{Differential hadron production cross section $\sigma(y,k_\bot^2)$ versus $\bar{y}$ for {\bf (a)} $pp \to \Lambda X$ at $k_\bot = 0.075$~GeV \cite{Blobel};	
{\bf (b)} $pp \to \Lambda X$ integrated over $k_\bot^2$
\cite{Jaeger};	{\bf (c)} $pp \to \Sigma^{*+} X$ integrated over $k_\bot^2$	\cite{Bockmann}, compared with the fitted nonlocal $K$ exchange contributions for regulator parameters $\Lambda_{K\Lambda}=1.1(1)$~GeV and $\Lambda_{K\Sigma^*}=0.8(1)$~GeV (solid red lines and pink $1\sigma$ uncertainty bands). (Figure from Ref.~\cite{Salamu2}.)}
\label{fig:crossK}
\end{center}
\end{figure}

To determine the cutoff parameter in the regulator for the kaon-hyperon-nucleon vertices, inclusive hyperon production cross sections in $pp$ collisions are considered, in analogy with the neutron and $\Delta$ production above \cite{Salamu2}. 
Data on inclusive $\Lambda$ production are from Refs.~\cite{Blobel, Jaeger, Bockmann}.
In Fig.~\ref{fig:crossK}, the inclusive $pp \to \Lambda X$ and $\Sigma^{*+} X$ cross sections for $\bar{y} > 0.7$ are fitted with the calculated splitting functions.
The best fit to the CERN bubble chamber $\Lambda$ production data from Ref.~\cite{Blobel} at $k_\bot = 0.075$~GeV [Fig.~\ref{fig:crossK}(a)] and the $k_\perp$-integrated data from Ref.~\cite{Jaeger} [Fig.~\ref{fig:crossK}(b)] yields a dipole regulator mass $\Lambda_{K\Lambda} = 1.1(1)$~GeV.
Comparison of the singly differential decuplet $\Sigma^{*+}$ production data at large~$\bar{y}$ [Fig.~\ref{fig:crossK}(c)] with the kaon exchange cross section gives a best fit for the decuplet regulator mass of $\Lambda_{K\Sigma^*} = 0.8(1)$~GeV.

\begin{figure}[tb]
\centering
\begin{tabular}{ccc}
\hspace{0.0cm}{\epsfxsize=3.0in\epsfbox{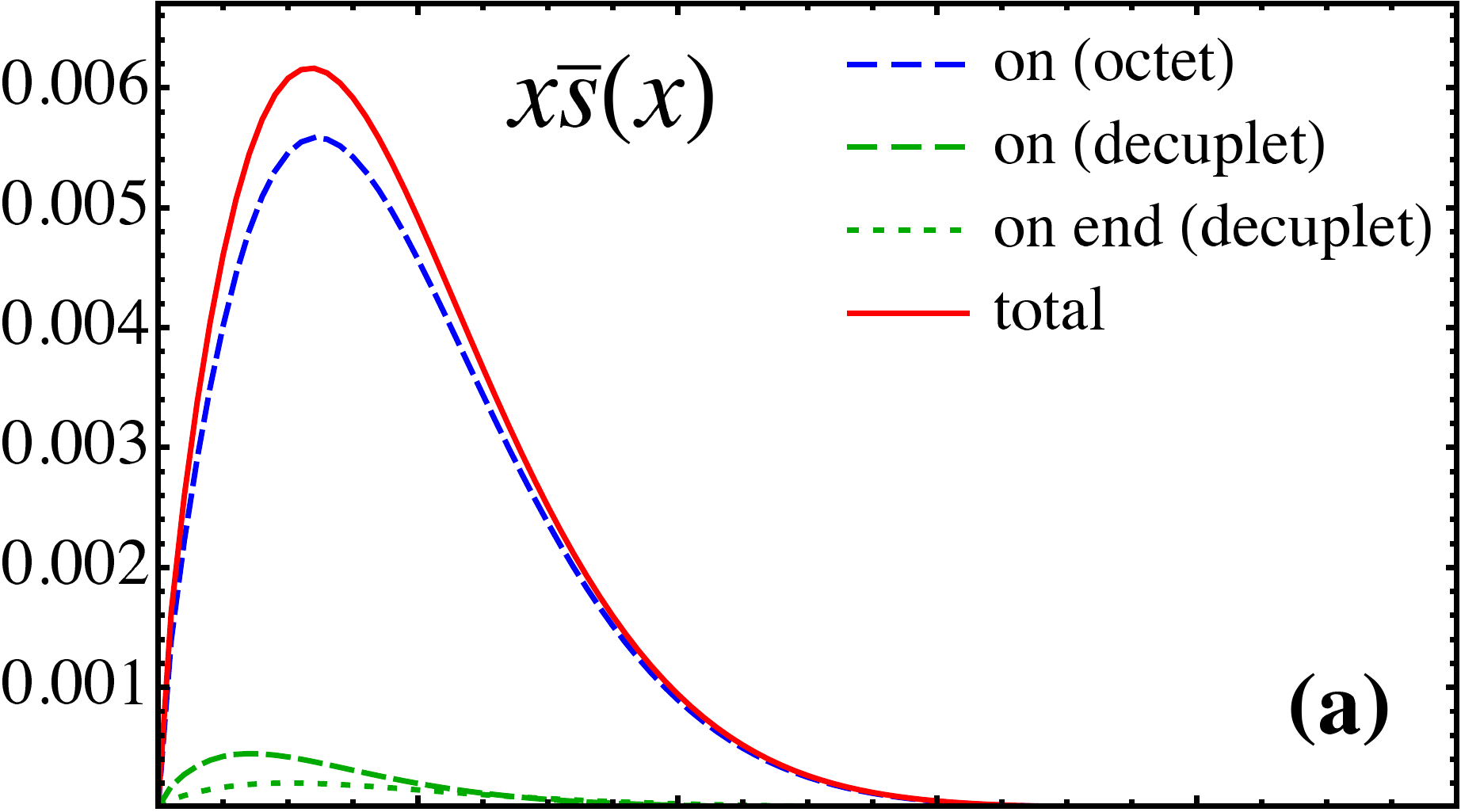}\vspace*{-0.0cm}}&
\hspace{0.1cm}{\epsfxsize=3.1in\epsfbox{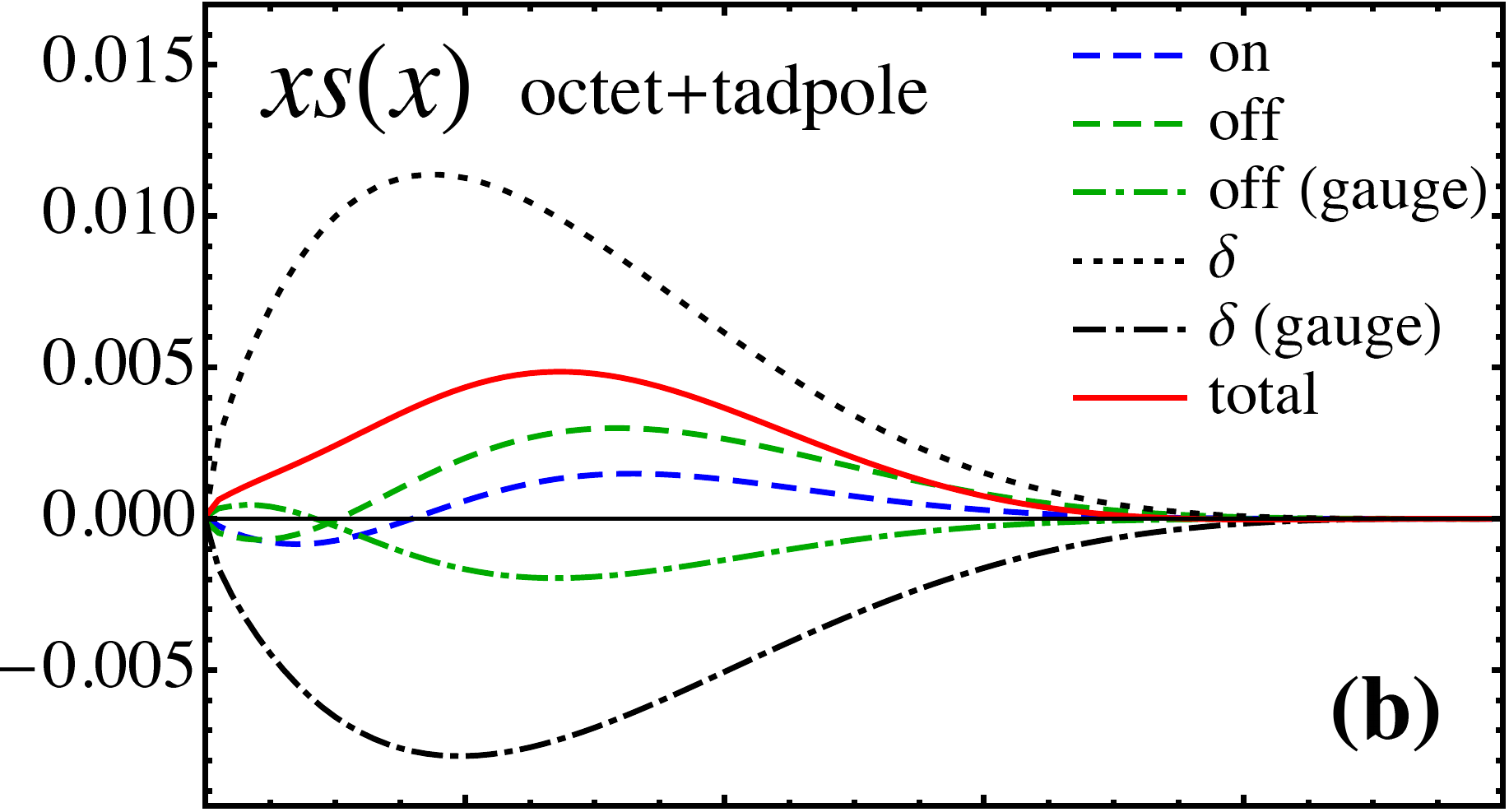}}&\\
\hspace{-0.1cm}{\epsfxsize=3.2in\epsfbox{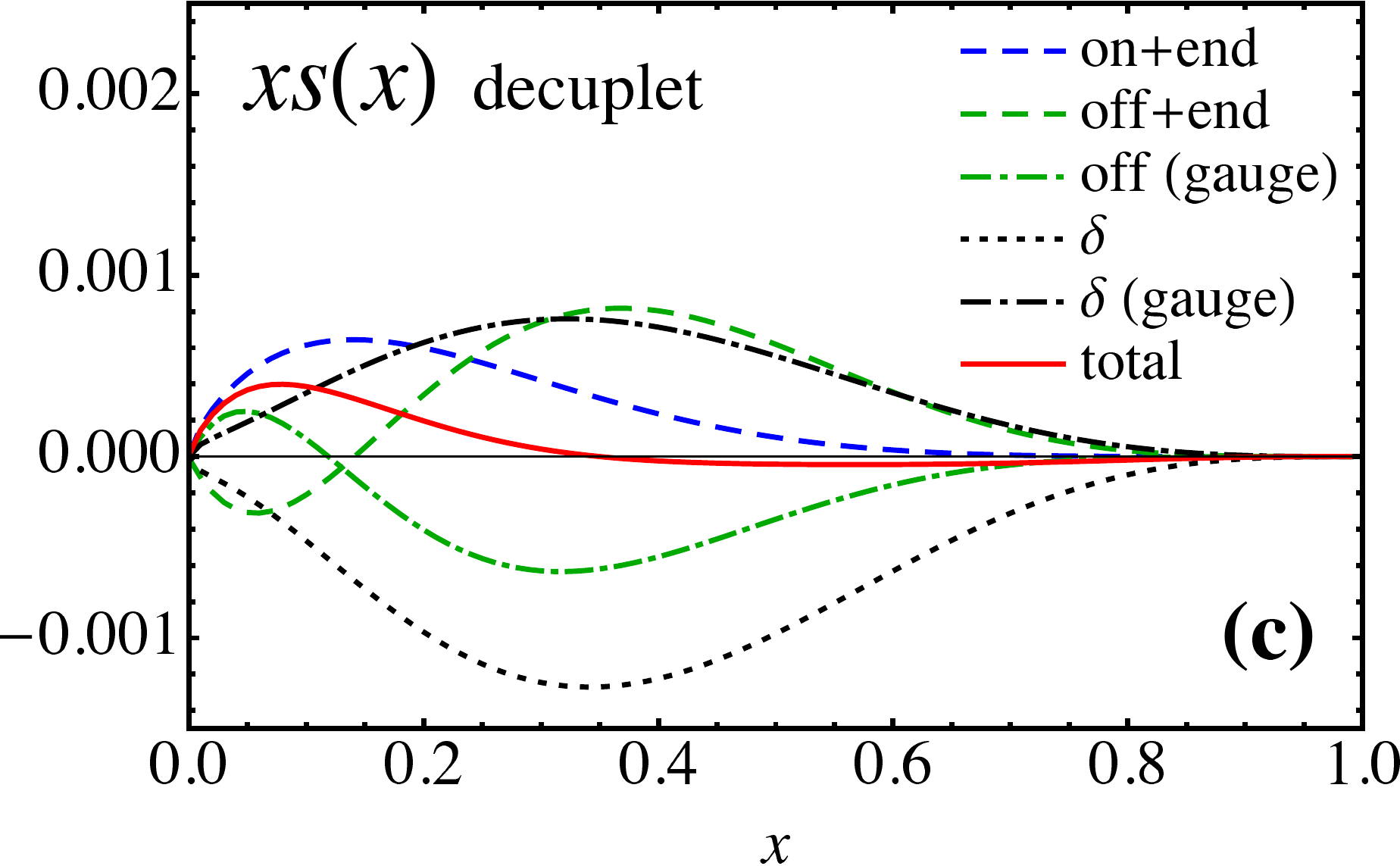}}&
\hspace{0.5cm}{\epsfxsize=3.1in\epsfbox{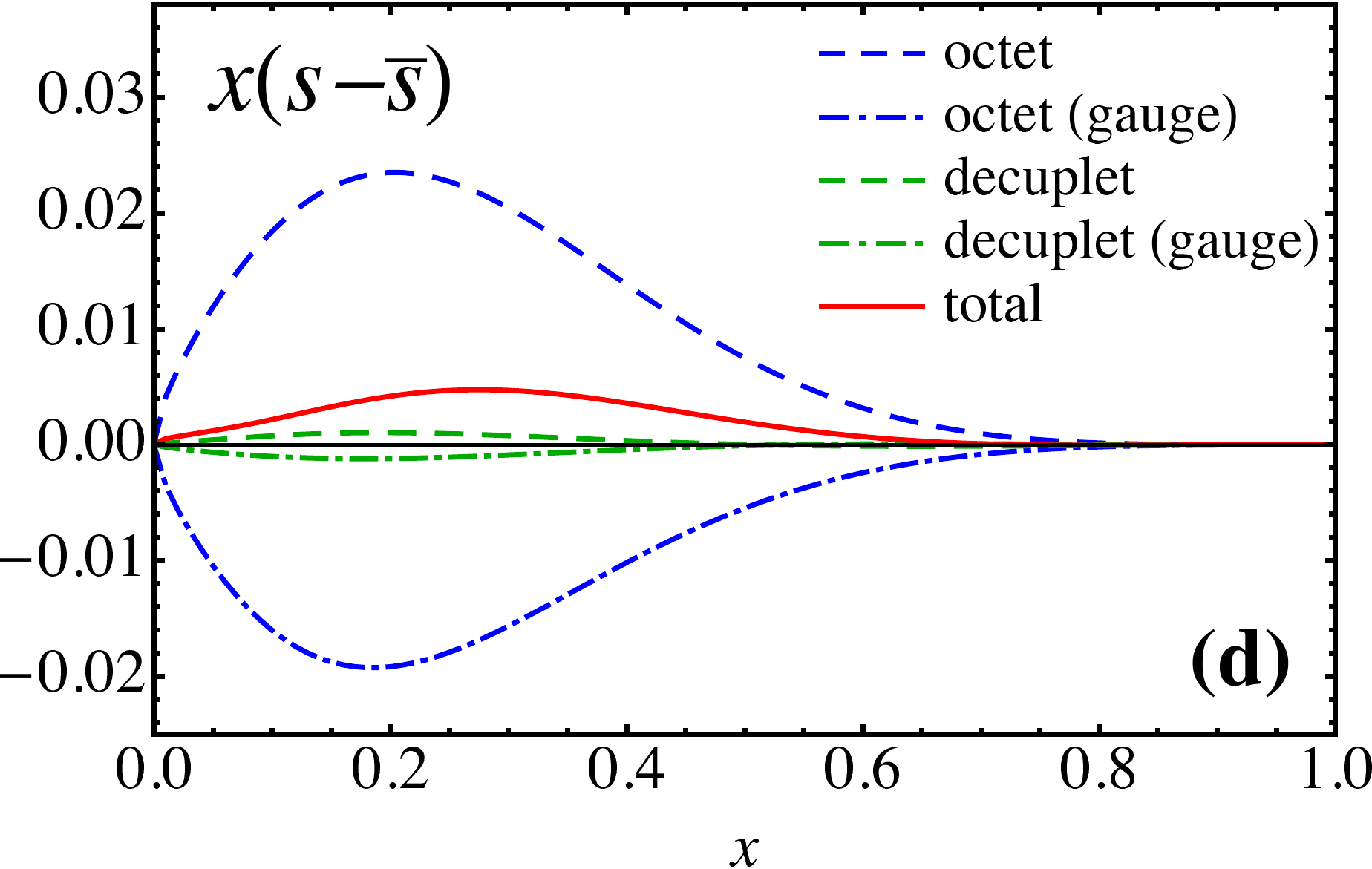}}&
\end{tabular}
\caption{Kaon loop contributions to {\bf (a)} antistrange PDF $x \bar s$ from the octet and decuplet rainbow diagrams [Fig.~\ref{fig:FMLOOP}(a) and (g)]; {\bf (b)} strange quark PDF $xs$ from the octet rainbow [Fig.~\ref{fig:FMLOOP}(b)], Kroll-Ruderman [Fig.~\ref{fig:FMLOOP}(c)-(f)], and tadpole [Fig.~\ref{fig:FMLOOP}(n)-(o)] diagrams; {\bf (c)} strange PDF $xs$ from the decuplet rainbow [Fig.~\ref{fig:FMLOOP}(h)] and Kroll-Ruderman [Fig.~\ref{fig:FMLOOP}(i)-(l)] diagrams; {\bf (d)} strange asymmetry $x(s-\bar{s})$, showing the local and nonlocal (gauge) octet and decuplet contributions, along with the total asymmetry. The PDFs are computed with the best fit regulator parameters $\Lambda_{K\Lambda} = \Lambda_{K\Sigma} = 1.1$~GeV and $\Lambda_{K\Sigma^*} = 0.8$~GeV.
(Figure from Ref.~\cite{Salamu2}.)}
\label{fig:xstrange}
\end{figure}

The kaon loop contributions to the strange and antistrange distributions in the proton can be calculated with the determined $\Lambda$ \cite{Salamu2}.
In Fig.~\ref{fig:xstrange} the various octet and decuplet contributions to $xs$ and $x\bar s$ are shown for the best fit parameters $\Lambda_{K\Lambda} = \Lambda_{K\Sigma} = 1.1$~GeV and $\Lambda_{K\Sigma^*} = 0.8$~GeV.
For the $x \bar s$ PDF in Fig.~\ref{fig:xstrange}(a), the octet on-shell contribution from the rainbow diagram [Fig.~\ref{fig:FMLOOP}(a)] dominates over the decuplet on-shell and end-point terms from the decuplet rainbow diagram [Fig.~\ref{fig:FMLOOP}(g)].
The resulting $x\bar s$ distribution peaks at $x \approx 0.1$ and essentially vanishes beyond $x \approx 0.6$. 
The $\delta$-function terms from the rainbow diagrams as well as from the kaon bubble diagram [Fig.~\ref{fig:FMLOOP}(m)] do not appear in Fig.~\ref{fig:xstrange}(a), as they have contributions to $\bar s$ only at $x=0$.

For the strange quark distribution, from the convolution in Eq.~(\ref{eq:strange}) one finds that all terms from each of the rainbow, Kroll-Ruderman and tadpole diagrams have nonzero contributions at $x > 0$. 
Since there are many individual terms, we display ones involving octet+tadpole and decuplet baryons separately in Figs.~\ref{fig:xstrange}(b) and \ref{fig:xstrange}(c), respectively. 
In contrast to the antistrange case, where the on-shell term is dominant, there are sizeable contributions to the strange distribution from many of the terms, with nontrivial cancellations between them.
A qualitatively similar scenario is evident in Fig.~\ref{fig:xstrange}(c) for the decuplet intermediate state contributions to $xs$, where the individual on-shell, off-shell, $\delta$-function and gauge link terms are shown.

Finally, the resulting asymmetry $x(s-\bar s)$ in Fig.~\ref{fig:xstrange}(d) reflects the interplay between the $\bar s$ PDF, which is dominant at low $x$, and the $s$-quark PDF, which extends to larger values of $x$. 
A key feature of this result is the strong cancellation between positive local and negative nonlocal, gauge-link dependent contributions, in both the octet and decuplet channels.
The net effect is then a small positive $x(s-\bar s)$ asymmetry, peaking at $x \approx 0.2-0.3$, and about an order of magnitude smaller than the asymmetry between the $\bar d$ and $\bar u$ PDFs resulting from pion loops.

In addition to the shape, it is instructive also to examine the contributions of the various terms to the lowest moments of the $s$ and $\bar s$ PDFs, and, in particular, the average number of strange and antistrange quarks,
\begin{equation}
\langle s \rangle = \int_0^1 \dd{x}\, s(x), \ \ \ \
\langle \bar s \rangle = \int_0^1 \dd{x}\, \bar s(x),
\label{eq:S0}
\end{equation}
and the average momentum carried by these,
\begin{equation}
\langle x s \rangle = \int_0^1 \dd{x}\, x s(x), \ \ \ \
\langle x \bar s \rangle = \int_0^1 \dd{x}\, x \bar s(x).
\label{eq:S1}
\end{equation}
Although the shapes of the $s$ and $\bar s$ distributions themselves are obviously rather different, the numbers of $s$ and $\bar s$ quarks in the nucleon are the same, $\langle s \rangle = \langle \bar s \rangle$, which is guaranteed by the local gauge invariance.
The zero net strangeness can be verified by explicitly summing the contributions to $\langle s \rangle$ and $\langle \bar s \rangle$ from the various diagrams in Fig.~\ref{fig:FMLOOP}, as Table~\ref{tab:ssbar} indicates.

\begin{table}[tbp]
\begin{center}
\caption{Contributions from octet $Y=\Lambda, \Sigma^0, \Sigma^+$
	and decuplet $Y^*=\Sigma^{*0}, \Sigma^{*+}$ hyperons to the
	average number
	(in units of $10^{-2}$)
	and momentum carried
	(in units of $10^{-3}$)
	by $s$ and $\bar s$ quarks in the nucleon,
	for dipole regulator mass parameters
	$\Lambda_{KY} = 1.1(1)$~GeV and
        $\Lambda_{KY^*} = 0.8(1)$~GeV.}
{\footnotesize
\begin{tabular}{llll|llll}	\hline 
   &
   & ~~~~$\langle \bar s \rangle$
   & ~~~~$\langle x \bar s \rangle$
   &
   &
   & ~~~~$\langle s \rangle$
   & ~~~~$\langle x s \rangle$	\\
   &
   & ~$(\times 10^{-2})$~
   & ~$(\times 10^{-3})$~
   &
   &
   & ~$(\times 10^{-2})$~
   & ~$(\times 10^{-3})$~	\\ \hline
    ~$KY{\rm (rbw)}$~
  & ~$f_Y^{\rm (on)}$~
  & ~~\,$1.39^{(69)}_{(54)}$
  & ~~\,$1.33^{(74)}_{(56)}$
  & ~$YK{\rm (rbw)}$~
  & ~$f_Y^{\rm (on)}$~
  & ~~\,$1.39^{(69)}_{(54)}$
  & ~~\,$1.67^{(78)}_{(63)}$			\\ [1mm]
  & ~$f_K^{(\delta)}$~
  & $-1.66^{(79)}_{(63)}$~~
  & ~~\,0
  &
  & ~$f_Y^{\rm (off)}$~
  & $-4.01^{(168)}_{(142)}$~~
  & $-5.35^{(212)}_{(183)}$~~			\\[1mm]
  & ~$\delta f_K^{(\delta)}$~
  & ~~\,$1.12^{(50)}_{(41)}$
  & ~~\,0
  &
  & ~$\delta f_Y^{\rm (off)}$~
  & ~~\,$2.70^{(107)}_{(92)}$
  & ~~\,$3.12^{(113)}_{(102)}$			\\ [1mm]
  &
  &
  &
  &
  & ~$ f_K^{(\delta)}$~
  & ~~\,$1.66^{(79)}_{(63)}$
  & ~~\,$2.82^{(135)}_{(107)}$			\\[1mm]
  &
  &
  &
  & ~$YK{\rm (KR)}$~
  & ~$f_Y^{\rm (off)}$~
  & ~~\,$4.01^{(168)}_{(142)}$
  & ~~\,$6.29^{(250)}_{(215)}$			\\[1mm]
  &
  &
  &
  &
  & ~$f_K^{(\delta)}$~
  & $-3.31^{(158)}_{(126)}$~~
  & $-6.66^{(318}_{(253)}$~~			\\[1mm]
  &
  &
  &
  & ~$YK{\rm (\delta KR)}$~
  & ~$\delta f_Y^{\rm (off)}$~
  & $-2.70^{(107)}_{(92)}$~~
  & $-3.68^{(133)}_{(120)}$~~			\\[1mm]
  &
  &
  &
  &
  & ~$\delta f_K^{(\delta)}$~
  & ~~\,$1.12^{(50)}_{(41)}$
  & ~~\,$2.24^{(101)}_{(82)}$			\\ [1mm] 
   ~total~octet
  &
  & ~~\,$0.85^{(40)}_{(32)}$
  & ~~\,$1.33^{(74)}_{(56)}$
  & ~total~octet
  &
  & ~~\,$0.85^{(40)}_{(32)}$
  & ~~\,$0.46(14)$			\\ [1mm] \hline
    ~$K {\rm (bub)}$~
  & ~$f_K^{(\delta)}$~
  & ~~\,$4.85^{(232)}_{(184)}$
  & ~~\,0
  & ~$K {\rm (tad)}$~
  & ~$f_K^{(\delta)}$~
  & ~~\,$4.85^{(232)}_{(184)}$
  & ~~\,$7.87^{(376)}_{(298)}$               	\\[1mm]
  & ~$\delta f_K^{(\delta)}$~
  & $-3.27^{(147)}_{(120)}$~~
  & ~~\,0
  & ~$K {\rm (\delta tad)}$~
  & ~$\delta f_K^{(\delta)}$~
  & $-3.27^{(147)}_{(120)}$~~
  & $-5.30^{(238)}_{(194)}$~~			\\ [1mm] 
   ~total~bubble
  &
  & ~~\,$1.59^{(85)}_{(64)}$
  & ~~\,0
  & ~total~tadpole
  &
  & ~~\,$1.59^{(85)}_{(64)}$
  & ~~\,$2.57^{(138}_{(104)}$			\\[1mm] \hline
    ~$KY^*{\rm (rbw)}$~
  & ~$f_{Y^*}^{\rm (on)}$~
  & ~~\,$0.09^{(13)}_{(7)}$
  & ~~\,$0.06^{(9)}_{(4)}$
  & ~$Y^*K{\rm (rbw)}$~
  & ~$f_{Y^*}^{\rm (on)}$~
  & ~~\,$0.09^{(13)}_{(7)}$
  & ~~\,$0.10^{(14)}_{(8)}$			\\[1mm]
  & ~$f_{Y^*}^{\rm (on\,end)}$~
  & ~~\,$0.04^{(7)}_{(3)}$
  & ~~\,$0.03^{(6)}_{(3)}$
  &
  & ~$f_{Y^*}^{\rm (on\,end)}$~
  & ~~\,$0.04^{(7)}_{(3)}$
  & ~~\,$0.04^{(7)}_{(3)}$			\\[1mm]
  & ~$f_{Y^*}^{(\delta)}$~
  & $-0.01(1)$~~
  & ~~\,0
  &
  & ~$f_{Y^*}^{\rm (off)}$~
  & $-0.59^{(72)}_{(42)}$~~
  & $-0.75^{(89)}_{(52)}$~~			\\[1mm]
  & ~$f_K^{(\delta)}$~
  & $-0.15^{(20)}_{(11)}$~~
  & ~~\,0
  &
  & ~$f_{Y^*}^{\rm (off\,end)}$~
  & ~~\,$0.17^{(23)}_{(12)}$
  & ~~\,$0.21^{(29)}_{(15)}$			\\[1mm]
  & ~$\delta f_K^{(\delta)}$~
  & ~~\,$0.11^{(14)}_{(8)}$
  & ~~\,0
  &
  & ~$\delta f_{Y^*}^{\rm (off)}$~
  & ~~\,$0.34^{(45)}_{(24)}$
  & ~~\,$0.38^{(47)}_{(27)}$			\\[1mm]
  &
  &
  &
  &
  & ~$f_K^{(\delta)}$~
  & ~~\,$0.18^{(24)}_{(13)}$
  & ~~\,$0.26^{(34)}_{(19)}$			\\[1mm]
  &
  &
  &
  &
  & ~$f_{Y^*}^{(\delta)}$~
  & ~~\,$0.01(1)$
  & ~~\,$0.01^{(2)}_{(1)}$			\\[1mm]
  &
  &
  &
  &
  & ~$\delta f_{Y^*}^{(\delta)}$~
  & $-0.07^{(11)}_{(5)}$~~
  & $-0.10^{(16)}_{(7)}$~~			\\[1mm]
  &
  &
  &
  & ~$Y^* K{\rm (KR)}$~
  & ~$f_{Y^*}^{\rm (off)}$~
  & ~~\,$0.59^{(72)}_{(42)}$
  & ~~\,$1.02^{(121)}_{(71)}$			\\[1mm]
  &
  &
  &
  &
  & ~$f_{Y^*}^{\rm (off\,end)}$~
  & $-0.17^{(23)}_{(12)}$~~
  & $-0.29^{(39)}_{(21)}$~~			\\[1mm]
  &
  &
  &
  &
  & ~$f_K^{(\delta)}$~
  & $-0.34^{(44)}_{(24)}$~~
  & $-0.65^{(85)}_{(47)}$~~			\\[1mm]
  &
  &
  &
  &
  & ~$f_{Y^*}^{(\delta)}$~
  & $-0.02^{(3)}_{(1)}$~~
  & $-0.03^{(5)}_{(2)}$~~			\\[1mm]
  &
  &
  &
  &
  & ~$\delta f_{Y^*}^{(\delta)}$~
  & ~~\,$0.05^{(8)}_{(3)}$
  & ~~\,$0.09^{(15)}_{(7)}$			\\[1mm]
  &
  &
  &
  & ~$Y^* K{\rm (\delta KR)}$~
  & ~$\delta f_{Y^*}^{\rm (off)}$~
  & $-0.34^{(45)}_{(24)}$~~
  & $-0.51^{(63)}_{(36)}$~~			\\[1mm]
  &
  &
  &
  &
  & ~$\delta f_K^{(\delta)}$~
  & ~~\,$0.11^{(14)}_{(8)}$
  & ~~\,$0.22^{(27)}_{(15)}$			\\[1mm]
  &
  &
  &
  &
  & ~$\delta f_{Y^*}^{(\delta)}$~
  & ~~\,$0.02^{(4)}_{(2)}$
  & ~~\,$0.05^{(7)}_{(3)}$			\\ [1mm] 
   ~total~decuplet
  &
  & ~~\,$0.08^{(12)}_{(6)}$
  & ~~\,$0.09^{(15)}_{(7)}$
  & ~total~decuplet
  &
  & ~~\,$0.08^{(12)}_{(6)}$
  & ~~\,$0.04^{(4)}_{(3)}$			\\ [1mm] \hline
   ~{\bf total}
  &
  & ~~\bm{$2.51^{(136)}_{(102)}$}
  & ~~\bm{$1.42^{(89)}_{(62)}$}
  &~{\bf total}
  &
  & ~~\bm{$2.51^{(136)}_{(102)}$}
  & ~~\bm{$3.08^{(155)}_{(120)}$}		
\\ [1mm] \hline 
\end{tabular}
}
\label{tab:ssbar}
\end{center}
\end{table}

While the lowest moments of the $s$ and $\bar s$ are constrained to be equal, there is no such requirement for higher moments, including the $x$-weighted moment corresponding to the momentum
carried by $s$ and $\bar s$ quarks. 
Since the total $s-\bar s$ asymmetry is found to be mostly positive over the range of $x$ relevant in this analysis, not surprisingly the total $\langle x (s-\bar s) \rangle$ moment is also positive. Including the uncertainties on the kaon-nucleon-hyperon vertex regulator parameters, the combined asymmetry is \cite{Salamu2}
\begin{equation}
\langle x (s - \bar s) \rangle
= 1.66^{(81)}_{(74)}\, \times 10^{-3}.
\label{eq:xs-sbar}
\end{equation}
An asymmetry of this magnitude will be challenging to determine experimentally.

\subsection{\it Polarized distributions}

For the contributions from meson loops to polarized quark distributions in the nucleon, in this section we focus on the strange quark polarization, $\Delta s$.
Among the $u$, $d$ and $s$ flavors, the contribution to the proton spin from the strange quark is the least well determined, and phenomenological studies often rely on assumptions such as SU(3) flavor symmetry and equivalence of the strange and antistrange polarizations, $\Delta s = \Delta \bar{s}$, to simplify the analyses.
In many of the studies which have made these assumptions, the integrated strange quark polarization has typically been found to be in the vicinity of $\Delta s^+ \equiv \Delta s + \Delta \bar s \approx -0.1$.
Recent direct lattice simulations of disconnected loop contributions have yielded slightly smaller magnitudes for the strange quark polarization, 
$\Delta s^+_{\rm latt} = -0.046(8)$~\cite{Alexandrou:2020sml},
while an analysis of the spin problem taking into account the angular momentum carried by the meson cloud~\cite{Thomas:2008ga, Schreiber:1988uw, Myhrer:2007cf}, suggests a value of order $-0.01$~\cite{Bass:2009ed, Yamaguchi:1989sx}.
The recent JAM global QCD analysis, which used inclusive and semi-inclusive DIS data in order to relax the SU(3) symmetry constraint, also supports a smaller magnitude for the strange polarization,
    $\Delta s^+_{\rm JAM} = -0.03(10)$~\cite{Ethier}
at a scale of $Q^2=1$~GeV$^2$, but with a larger uncertainty.
A review of the status and results from the global QCD analysis and lattice QCD communities can be found in Ref.~\cite{Lin:2018}.

To calculate the polarized parton distributions, we need the axial-vector current (operator) which couples to the external axial-vector field.
Compared with the unpolarized PDF, where the vector current for the nonlocal Lagrangian was obtained from the locally gauge invariant Lagrangian, the current for the axial-vector current is a little simpler, since in this case no path integral of the gauge field is needed.
From the Lagrangian Eqs.~(\ref{eq:LOCT}) and (\ref{eq:LDEC}) one can obtain the axial-vector current that couples to the external axial-vector field
$a_\mu^a$ as
\begin{eqnarray}
J_A^{\mu,a}
&=& \frac{1}{2}{\rm Tr}
   \big[
   \bar B \gamma^\mu
   \left[ u \lambda^a u^\dagger - u^\dagger \lambda^a u, B
   \right]	
      + \frac{D}{2}{\rm Tr}
   \big[
   \bar B \gamma^\mu \gamma_5
   \left\{ u \lambda^a u^\dagger + u^\dagger \lambda^a u, B
   \right\}
   \big]						\notag\\
&+&
   \frac{F}{2}{\rm Tr}
   \big[
   \bar B \gamma^\mu \gamma_5
   \left[ u \lambda^a u^\dagger + u^\dagger \lambda^a u, B
   \right]
   \big]				
+ \frac{F-D}{2}{\rm Tr}
   \big[ \bar B \gamma^\mu \gamma_5 B \big ]
   {\rm Tr} \big[ u \lambda^a u^\dagger + u^\dagger \lambda^a u  \big]	\notag\\
&+& \frac{\cal C}{2}
   \left(
   \overline{T}_\nu \Theta^{\nu\mu}
   (u \lambda^a u^\dagger + u^\dagger \lambda^a u) B
   + {\rm h.c.}
   \right)						
+
   \frac{{\cal H}}{2}\,
   \overline{T}_\nu \gamma^{\nu\alpha\mu}
   \left( u \lambda^a u^\dagger + u^\dagger \lambda^a u, T_\alpha
   \right).
\label{eq:ch1}
\end{eqnarray}
The currents for given quark flavors are then expressed as combinations of the above currents with $a=0, 3$ and 8 as
\begin{subequations} \label{eq:ch3}
\begin{eqnarray} 
J_A^{\mu,u} &=&\frac{1}{3} J_A^{\mu,0}
	 + \frac{1}{2} J_A^{\mu,3}
	 + \frac{1}{2\sqrt3} J_A^{\mu,8},
\\
J_A^{\mu,d} &=&\frac{1}{3} J_A^{\mu,0}
	 - \frac{1}{2} J_A^{\mu,3}
	 + \frac{1}{2\sqrt3} J_A^{\mu,8},
\\
J_A^{\mu,s} &=&\frac{1}{3} J_A^{\mu,0}
	 - \frac{1}{\sqrt3} J_A^{\mu,8} .
\end{eqnarray}
\end{subequations}
Using Eqs.~(\ref{eq:ch1}) and (\ref{eq:ch3}), the strange quark current $J_A^{\mu,s}$ can be written explicitly as
\begin{eqnarray}
\label{eq:js}
J_A^{\mu,s}
&=& (F-D)\, \overline \Sigma^+ \gamma^\mu \gamma^5 \Sigma^+ 
 +  (F-D)\, \overline \Sigma^- \gamma^\mu \gamma^5 \Sigma^- 
 +  (F-D)\, \overline \Sigma^0 \gamma^\mu \gamma^5 \Sigma^0
+ \Big( F + \frac{D}{3}\Big) \bar\Lambda \gamma^\mu \gamma^5 \Lambda
\notag \\
&+&  \frac{1}{2f^2}
  \Big( 2F \bar p \gamma^\mu \gamma^5 p\, K^+ K^-
     + (F-D)\, \bar p \gamma^\mu \gamma^5 p\, K^0 K^0
  \Big)	
\notag \\
&+& \frac{{\cal H}}{3}\,
\Big(
   \overline{\Sigma}_\alpha^{*+} g^{\alpha\beta}\gamma^\mu \gamma^5 \Sigma_\beta^{*+}
 + \overline{\Sigma}_\alpha^{*0} g^{\alpha\beta}\gamma^\mu \gamma^5 \Sigma_\beta^{*0}
 + \overline{\Sigma}_\alpha^{*-} g^{\alpha\beta}\gamma^\mu \gamma^5 \Sigma_\beta^{*-}
\Big)
\notag\\
&-& \frac{i}{\sqrt2 f}
\Big(
  \bar p \gamma^\mu \Sigma^+\, K^0
+ \frac{1}{\sqrt2} \bar p \gamma^\mu \Sigma^0\, K^+
+ \frac{\sqrt{3}}{\sqrt2} \bar p \gamma^\mu \Lambda\, K^+ + {\rm H.c.} 
\Big)
\notag\\
&+&
  \frac{\cal C}{\sqrt{3}}
  \left(- \overline \Sigma^{+}\, \Theta^{\mu\nu} \Sigma_\nu^{*+}
  + \overline \Sigma^0\, \Theta^{\mu\nu} \Sigma_\nu^{*0}
  + \overline \Sigma^-\, \Theta^{\mu\nu} \Sigma_\nu^{*-}
  + {\rm H.c.}
  \right),
\end{eqnarray}
where the terms involving the doubly-strange baryons $\Xi^{0,-}$ and $\Xi^{*0,-}$ and the triply-strange $\Omega^-$ are omitted as they do not couple to the nucleon initial states.
The current obtained from the Lagrangian is equivalent to the operators ${\cal O}_{\Delta q}^{\mu_1}$ which is expressed in terms of $\bar{\alpha}^{(1)}, \bar{\beta}^{(1)}, \bar{\sigma}^{(1)}$, $\alpha^{(1)},\beta^{(1)}, \sigma^{(1)}$ and $\bar{\gamma}^{(1)},\bar{\omega}^{(1)}$ \cite{XGWang3}, with
\bea
\bar{\alpha}^{(1)} &=& 2F +\tfrac23 D, ~~~ 
\bar{\beta}^{(1)}\  =\ F - \tfrac53 D, ~~~
\bar{\sigma}^{(1)}\ =\ 0,
\nonumber \\
\alpha^{(1)} &=& 2, \hspace*{1.85cm}
\beta^{(1)}\  =\ 1, \hspace*{1.65cm}
\sigma^{(1)}\ =\ 0,
\\
\bar{\gamma}^{(1)} &=& {\cal H}, \hspace*{1.7cm}
\bar{\omega}^{(1)}\ =\ -{\cal C}.
\nonumber 
\eea

In analogy with the unpolarized case, the convolution relation for the spin-dependent PDFs in the nucleon can be written as
\begin{equation}\label{eq:convolution-1}
\Delta q(x)
= \sum_j \big[\Delta f_j \otimes \Delta q_j^+\big](x)
\equiv \sum_j \int_0^1 \dd{y}\, \int_0^1 \dd{z}\, 
  \delta(x-yz)\, \Delta f_j(y)\, \Delta q_j^+(z),
\end{equation}
where $\Delta q_j^+ = \Delta q_j + \Delta \bar{q}_j$ is the spin-dependent valence quark distribution for quark flavor $q$ in the hadronic configuration $j$.
For the strange quark, the splitting functions $\Delta f_j(y)$ can be calculated from the matrix elements of the axial-vector current $J_A^{\mu,s}$.
The Feynman diagrams including the octet rainbow, tadpole, Kroll-Ruderman, decuplet rainbow, and octet-decuplet transition diagrams are included in Fig.~\ref{fig:FMLOOP}, without gange link diagrams. 
The convolution form then gives the strange quark PDF in terms of the explicit hadronic configurations as
\begin{eqnarray}
\label{eq:convolution-2}
\Delta s(x)
&=& \sum_{B\phi}
    \Big( \Delta \bar{f}_{B\phi}^{(\rm rbw)} \otimes \Delta s_B
        + \Delta \bar{f}_{B\phi}^{(\rm KR)}  \otimes \Delta s_B^{(\rm KR)}
    \Big)
 +\ \sum_\phi \Delta \bar{f}_\phi^{(\rm tad)} \otimes \Delta s_\phi^{(\rm tad)}\nonumber\\
&+& \sum_{T \phi}
    \Delta \bar{f}_{T \phi}^{(\rm rbw)} \otimes \Delta s_{T}
    +\sum_{T B \phi} \Delta \bar{f}_{TB \phi} \otimes \Delta s_{T B},
\end{eqnarray}
where for notational convenience we also define the splitting functions $\Delta \bar{f}(y)\equiv \Delta f(\bar{y})$ with $\bar{y}\equiv 1-y$ the baryon momentum fraction when the meson carries momentum fraction $y$, as for unpolarized PDFs.

For the octet hyperon rainbow diagrams, Fig.~\ref{fig:FMLOOP}(b), the splitting function for the intermediate state $\Sigma^+$ can be written in terms of the on-shell, off-shell and $\delta$-function basis functions as
\be \label{eq:DeltasB}
\Delta f^{\rm (rbw)}_{\Sigma^+ K^0}(y)
= \frac{C_{B\phi}^2 (M_N+M_\Sigma)^2}{(4\pi f)^2}\,
    \Big[
      \Delta f^{\rm (on)}_\Sigma(y)
	+ \Delta f^{\rm (off)}_\Sigma(y)
	+ \Delta f^{(\delta)}_\Sigma(y)
    \Big],				
\ee
where the functions $\Delta f^{\rm (on)}_{\Sigma}$, $\Delta f^{\rm (off)}_{\Sigma}$ and $\Delta f^{(\delta)}_\Sigma$ are given in Refs.~\cite{XGWang3, He6}.
The tadpole contributions to the splitting functions from Fig.~\ref{fig:FMLOOP}(n) are given by
\be
\label{eq:Deltastad}
\Delta f^{\rm (tad)}_{K^+}(y)
= 2\, f^{\rm (tad)}_{K^0}(y)\
 =\ -\frac{ (M_N+M_\Sigma)^2}{(4\pi f)^2}\,
    f^{(\delta)}_K(y),
\ee
where the tadpole function $\Delta f_\phi^{(\delta)}$ is related to the $\delta$-function term in the rainbow diagram in Eq.~(\ref{eq:DeltasB}), 
\begin{equation}
\Delta f_\phi^{(\delta)}(y) = - \Delta f_B^{(\delta)}(y).
\end{equation}
For the octet Kroll-Ruderman diagrams in Figs.~\ref{fig:FMLOOP}(c) and \ref{fig:FMLOOP}(d), the splitting function 
is given by
\be \label{eq:DeltasKR}
\Delta f^{\rm (KR)}_{\Sigma^+}(y)
= \frac{C_{B\phi}(M_N+M_\Sigma)^2}{(4\pi f)^2}\,
    \Big[\Delta f^{\rm (off)}_\Sigma(y)
	 + 2 \Delta f^{(\delta)}_\Sigma(y)
    \Big].
\ee
For the decuplet hyperon contributions, the splitting functions for the intermediate 
$\Sigma^{*+}$ are given by
\bea
\Delta f^{\rm (rbw)}_{\Sigma^{*+} K^0}(y)
&=& \frac{C_{T\phi}^2 (M_N+M_\Sigma^*)^2}{(4\pi f)^2}	
\Big[
\Delta  f^{\rm (on)}_{\Sigma^*}(y)
+ \Delta f^{\rm (off)}_{\Sigma^*}(y)
+ \Delta f^{(\delta)}_{\Sigma^*}(y)
\Big]						
\label{eq:DeltasT}
\eea
for the decuplet rainbow diagram in Figs.~\ref{fig:FMLOOP}(h), and
\bea
\hspace*{-0.3cm}
\Delta f^{\rm (rbw)}_{\Sigma^{*+}\Sigma^+ K^0}(y)
&=&\frac{C_{B\phi}C_{T\phi} (M_N+M_\Sigma^*)(M_\Sigma+M_\Sigma^*)}{(4\pi f)^2}
\Big[
\Delta  f^{\rm (on)}_{\Sigma^*\Sigma}(y)
+ \Delta f^{\rm (off)}_{\Sigma^*\Sigma}(y)
+ \Delta f^{(\delta)}_{\Sigma^*\Sigma}(y)
\Big]					
\label{eq:DeltasBT}
\eea
for the octet-decuplet transition diagram in Figs.~\ref{fig:FMLOOP}(r) and \ref{fig:FMLOOP}(s), where the external field is an axial-vector.
The expressions for the decuplet basis functions
  $\Delta f^{\rm (on)}_{\Sigma^*}$,
  $\Delta f^{\rm (off)}_{\Sigma^*}$,
  $\Delta f^{\rm (\delta)}_{\Sigma^*}$,
  $\Delta f^{\rm (on)}_{\Sigma^*\Sigma}$,
  $\Delta f^{\rm (off)}_{\Sigma^*\Sigma}$,
and $\Delta f^{\rm (\delta)}_{\Sigma^*\Sigma}$ are also given in Refs.~\cite{XGWang3, He6}.

\begin{figure}[t]
\begin{minipage}[b]{.45\linewidth}
\hspace*{-0.3cm}\includegraphics[width=1.1\textwidth, height=5.5cm]{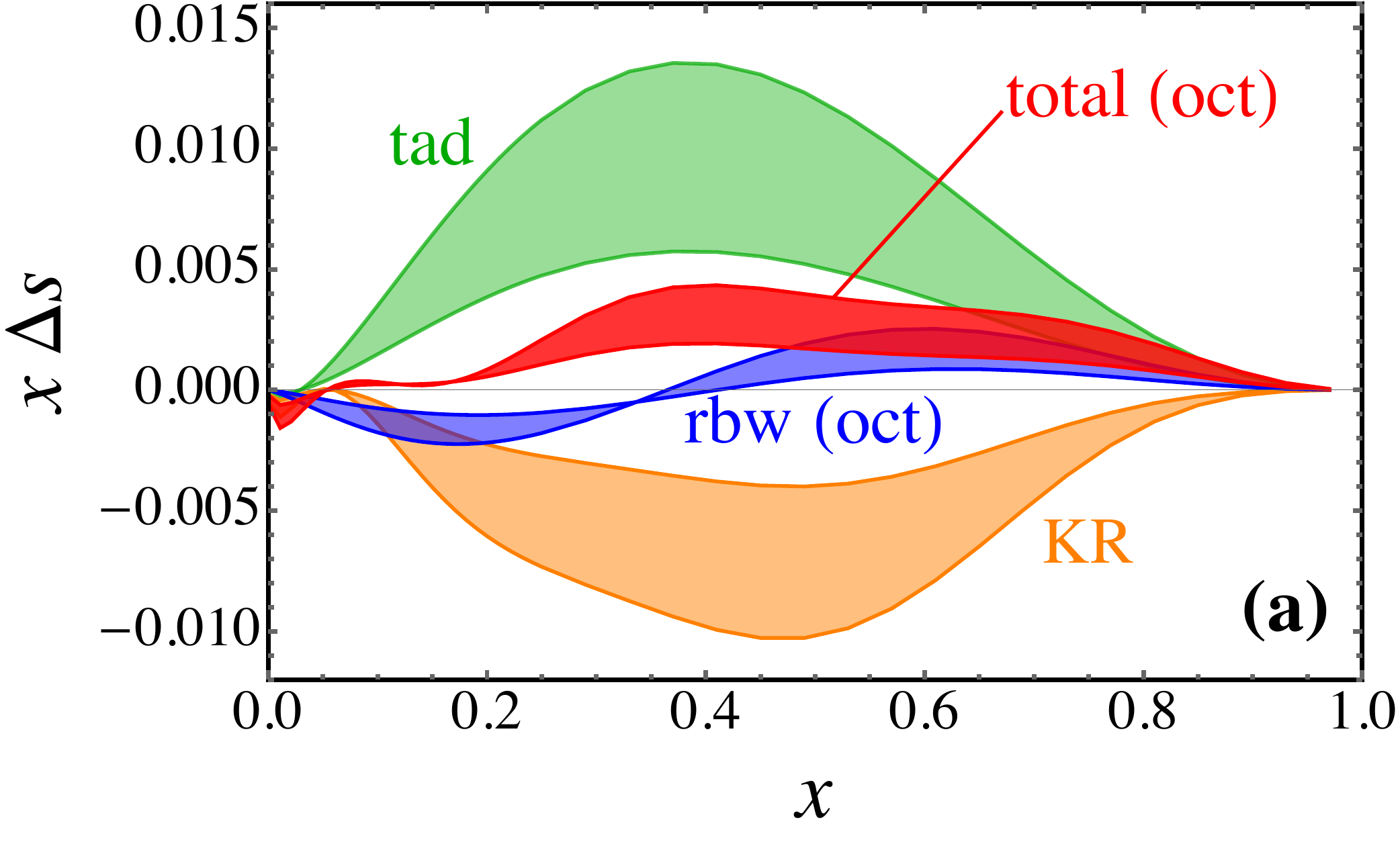}
\end{minipage}
\hfill
\begin{minipage}[b]{.45\linewidth}   
\hspace*{-0.95cm} \includegraphics[width=1.1\textwidth, height=5.5cm]{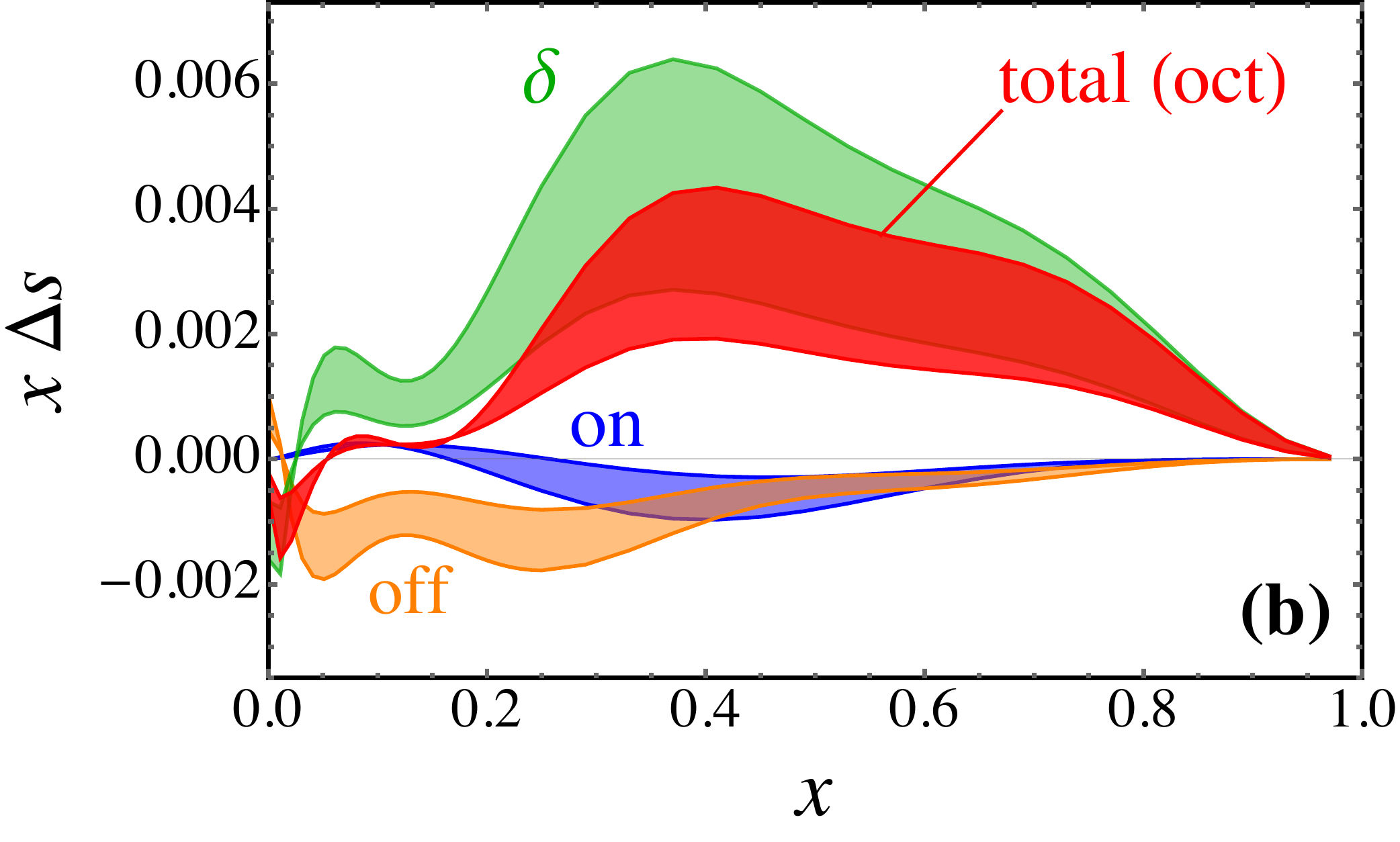} 
\end{minipage}  
\\[0.3cm]
\begin{minipage}[t]{.45\linewidth}
\hspace*{-0.5cm}\includegraphics[width=1.12\textwidth, height=5.5cm]{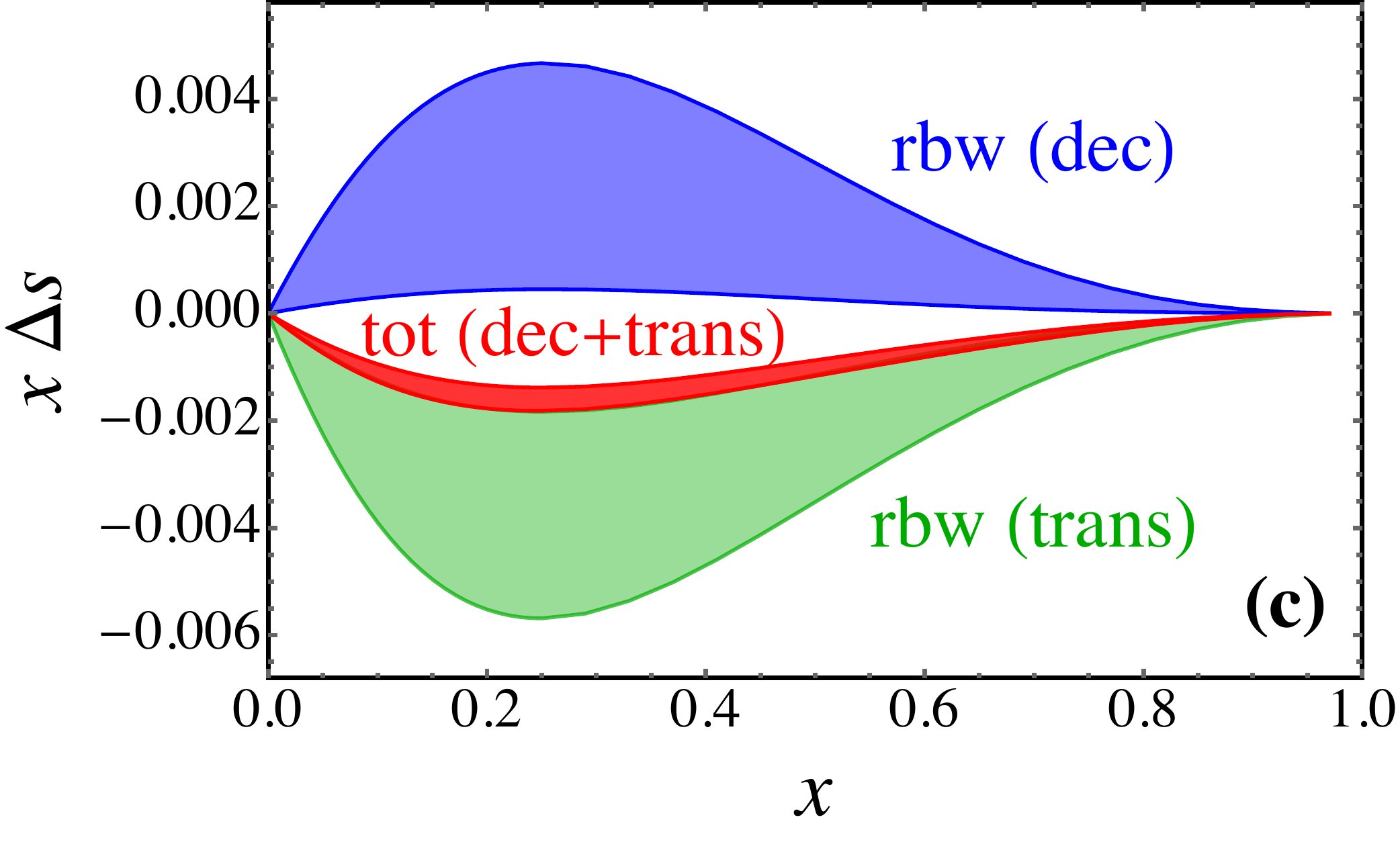}
\end{minipage}
\hfill
\begin{minipage}[t]{.45\linewidth}   
\hspace*{-0.85cm} \includegraphics[width=1.1\textwidth, height=5.5cm]{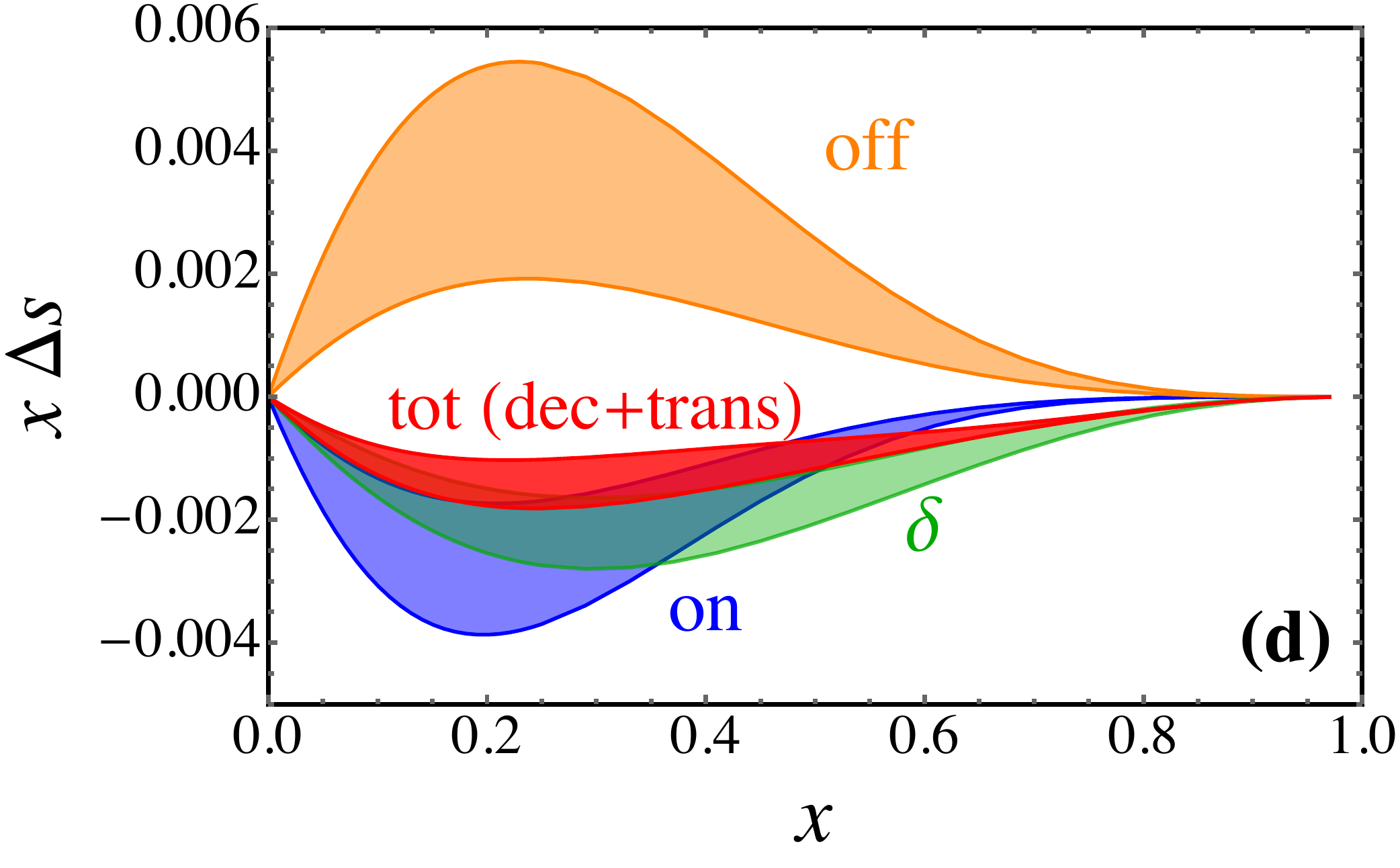}  
   \vspace{-12pt}
\end{minipage} 
\caption{Contributions to the $x\Delta s$ distribution in the proton at $Q^2=1$~GeV$^2$ from various meson loop diagrams with octet intermediate states [{\bf (a), (b)}] and decuplet (and decuplet-octet interference) states [{\bf (c), (d)}]. The bands for the octet and decuplet contributions correspond to the range of $\Lambda=1.0$~GeV to 1.2~GeV and 0.7~GeV to 0.9~GeV, respectively. (Figure from Ref.~\cite{He6}.)}
\label{fig:xDeltas}
\end{figure}

With the same parameters $\Lambda$ as determined for the unpolarized strange quark PDF, the contributions to the polarized $x\Delta s$ PDF from the various terms in Eq.~(\ref{eq:convolution-2}) are shown in Fig.~\ref{fig:xDeltas}, for decompositions in terms of types of diagrams and types of functions \cite{He6}. 
For the intermediate octet states, there are large cancellations between the positive tadpole and negative KR diagrams, while the rainbow diagram makes a relatively small contribution.
The total octet contribution is negative for very small $x$ and positive for $x > 0.1$. 
The total on-shell, off-shell, and $\delta$-function contributions for the octet intermediate-state and tadpole diagrams are seen to change sign with $x$.
The $\delta$-function contribution is negative when $x$ is small, but is positive and dominant when $x>0.2$. 
The on-shell contribution changes smoothly with $x$, and is positive at small $x$ and negative at large $x$. 
The behavior of the total contribution is mainly determined by the $\delta$-function contribution. 
For the diagrams involving intermediate states with decuplet baryons, there are cancellations between positive decuplet rainbow and negative octet-decuplet transition contributions, resulting in a total negative contribution.
In contrast to the octet case, the off-shell contributions are positive, but cancelled somewhat by the negative on-shell and $\delta$-function terms.
The net result is a total negative effect, with magnitude comparable to that from the octet. 

\begin{figure}[t] 
\begin{minipage}[b]{.45\linewidth}
\hspace*{-0.55cm}\includegraphics[width=1.125\textwidth, height=5.8cm]{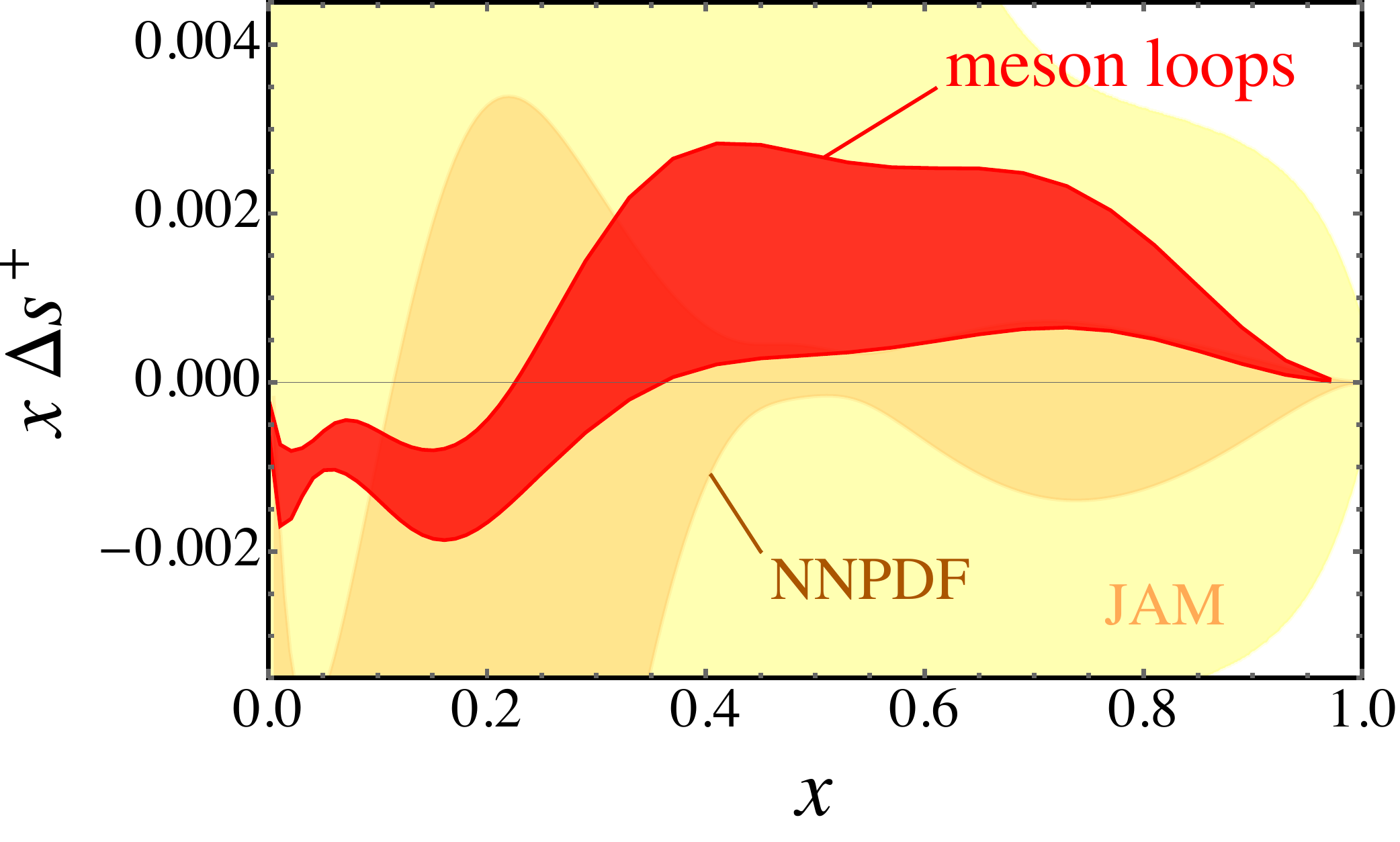} 
 \vspace{-6pt}
\end{minipage}
\hfill
\begin{minipage}[b]{.45\linewidth}   
\hspace*{-1.1cm} \includegraphics[width=1.1\textwidth, height=5.5cm]{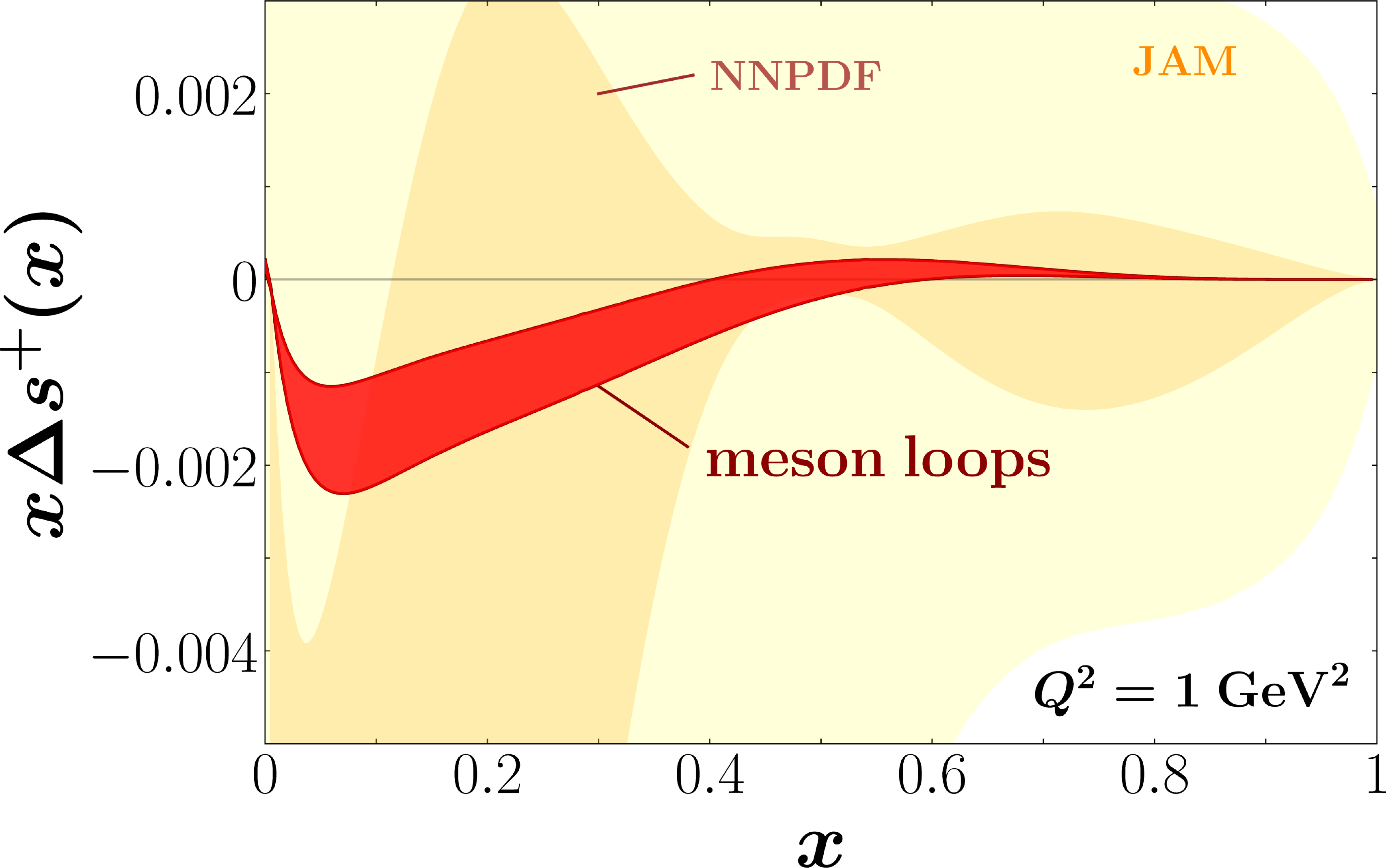} 
   \vspace{15pt}
\end{minipage}  
\caption{Comparison of the calculated total meson loop contributions to the polarized strange quark PDF in the nonlocal effective theory (left) and in the local theory with PV regularization (right) with $x \Delta s^+$ from the NNPDF~\cite{NNPDF:2014, Hartland:2013} and JAM~\cite{Ethier} global QCD analyses at $Q^2 = 1$~GeV$^2$.
The red bands corresponds to the ranges $\Lambda=1.0-1.2$~GeV for octet and $0.7-0.9$~GeV for decuplet baryons for the nonlocal theory, and the ranges $\{\mu_1,\mu_2\} = \{545,600\}$\,MeV to $\{526,894\}$\,MeV for octet baryons and $\mu=762$\,MeV for decuplet baryons for PV regularization.
(Figures from Refs.~\cite{XGWang3, He6}.)}
\label{fig:xDeltast}
\end{figure}

The total contribution to $x\Delta(s)$ is shown in Fig.~\ref{fig:xDeltast} \cite{He6}, along with phenomenological PDFs from the NNPDF~\cite{NNPDF:2014, Hartland:2013} and JAM~\cite{Ethier} analyses at $Q^2=1$~GeV$^2$, and compared with results from a calculation with Pauli-Villars regularization \cite{XGWang3}. 
The magnitude of the total result in the nonlocal effective theory is comparable to that with PV regularization, although there are quantitative differences especially at high $x$.
The first moment of $\Delta s(x)$, or the total spin carried by strange quarks in the proton, is in the range $[-0.0051, -0.0026]$ in the nonlocal theory \cite{He6}.
The small magnitude of the calculated strange polarization, in both approaches, compared with the uncertainty bands of the global parameterizations reflects the relatively weak constraints on $\Delta s$ that exist from current experiments. 
The JAM study~\cite{Ethier}, in particular, performed a dedicated analysis of the strange quark PDF using data from inclusive and semi-inclusive DIS, without imposing the commonly used assumption about SU(3) flavor symmetry for the axial charges, leading to a significantly larger uncertainty on $\Delta s$.

\section{GPDs and TMDs}
\label{Sec.6}

\subsection{\it Convolution formulas}

In this section we will extend our discussion from ordinary PDFs  to generalized parton distributions and transverse momentum dependent distributions, focusing on the sea quark GPDs and Sivers functions. 
Let's first discuss the convolution form for the generalized parton distributions. 
The GPDs are defined by the matrix elements of bi-local field operators as 
\begin{eqnarray}\label{eq:GPD}
\int\frac{\dd{\lambda}}{2\pi}\, e^{-i \lambda x P_+}
\langle N(p')|\, \bar q(\tfrac12 \lambda n)\, \slashed{n}\, q(-\tfrac12 \lambda n)\, |N(p)\rangle
= \bar u(p') 
    \bigg[
    \slashed n H^q(x,\xi,t) + \frac{i\sigma^{\mu\nu}n_\mu q_\nu}{2M_N} E^q(x,\xi,t)
\bigg] u(p),
\end{eqnarray}
where $x=k^+/P^+$ is the quark light-cone momentum fraction, $P=(p+p')/2$ is the average momentum of the initial and final state nucleons, and $\xi=-q^+/2P^+$ is the skewness parameter.
Because of Lorentz invariance, the GPDs $H^q$ and $E^q$ can only depend on the kinematical variables $x$, $\xi$ and $t \equiv \Delta^2 = (p'-p)^2$.
(Note that here we use $\Delta^2$ for the momentum transfer squared between the hadrons rather than $Q^2$, which is reserved for the virtuality of the incoming photon.)
After integrating over $x$, the electromagnetic form factors can be obtained as
\begin{eqnarray}
F_1^q(t)=\int_{-1}^1\,\dd{x} H^q(x,\xi,t),~~~~~~~
F_2^q(t)=\int_{-1}^1\,\dd{x} E^q(x,\xi,t).
\end{eqnarray}
The combinations of the above form factors can generate the electric and magnetic form factors. 
In the following, we restrict ourselves to the case $\xi=0$, and consider the GPDs in the $\Delta^+ = 0$ frame \cite{Brodsky:1997de}, defined by
\begin{equation}
p^\mu = \big(P^+,P^-,-\tfrac12 {\bm \Delta_\perp}\big),
\qquad
p'^\mu = \big(P^+,P^-,\tfrac12 {\bm \Delta_\perp}\big),
\qquad
\Delta^\mu = \big(0,\,0,\,\bm \Delta_\perp\big).
\end{equation}

We define the $n$-th Mellin moments of the GPDs $H^q(x,\xi,t)$ and $E^q(x,\xi,t)$ by
\begin{subequations}\label{eq:HEqn}
\begin{eqnarray}
H_q^n(\xi,t)\ \equiv\ \int_{-1}^1 \dd{x} x^{n-1} H^q(x,\xi,t)
&=& \sum_{i=0, {\rm even}}^{n-1} (-2\xi)^i A^{ni}_q(t) 
 + (-2\xi)^n C^n_q(t)\Big|_{n\ {\rm even}},
\\
E_q^n(\xi,t)\ \equiv\ \int_{-1}^1 \dd{x} x^{n-1} E^q(x,\xi,t)
&=& \sum_{i=0, {\rm even}}^{n-1} (-2\xi)^i B^{ni}_q(t) 
 - (-2\xi)^n C^n_q(t)\Big|_{n\ {\rm even}},
\end{eqnarray}
\end{subequations}
where $A_q^{ni}(t)$, $B_q^{ni}(t)$ and $C_q^{n}(t)$ are the generalized form factors. 
These form factors can be related to the matrix elements of local twist-2 operators ${\cal O}_{q}^{\mu\mu_1 \cdots \mu_n}$ between nucleon states \cite{Ji1998},
\begin{eqnarray}
\langle N(p') | \mathcal{O}_{q}^{\mu \mu_1 \cdots \mu_{n-1}} | N(p) \rangle
&=& \bar{u}(p')\sum_{i=0,\rm even}^{n-1} 
A_q^{ni}(t)\, \gamma^{ \{ \mu } \Delta^{\mu_1} \cdots \Delta^{\mu_i} P^{\mu_{i+1}} \cdots P^{\mu_{n-1} \} } u(p) 
\nonumber \\
&-& \frac{i}{2M_N}\bar{u}(p')\sum_{i=0,\rm even}^{n-1} 
B_q^{ni}(t)\, \Delta_\nu \sigma^{ \nu \{ \mu} \Delta^{\mu_1} \cdots \Delta^{\mu_i} P^{\mu_{i+1}} \cdots P^{\mu_{n-1} \} } u(p) 
\nonumber \\
&+& \frac{1}{M_N}\bar{u}(p')\,
C_q^n(t)\Big|_{n\ {\rm even}}\, \Delta^{\{ \mu}\cdots\Delta^{\mu_{n-1} \}} u(p),
\end{eqnarray}
where the symmetric and traceless operators are defined, in analogy with Eq.~(\ref{eq:Oq}), as
\begin{equation}\label{eq:O-q}
\mathcal{O}^{\mu\mu_1 \cdots \mu_{n-1}}_q\
=\ i^{n-1} \bar{q} \gamma^{ \{\mu }
  \overleftrightarrow{D}^{\mu_1} \cdots
  \overleftrightarrow{D}^{\mu_{n-1}\} } q,
\end{equation}
with the braces $\{ \cdots \}$ representing symmetrization over the indices $\mu_i$ and subtraction of traces.
As for the PDF case, in the effective theory the quark operators are matched to hadronic operators with the same quantum numbers \cite{Chen3, He5},
\begin{equation}\label{eq:operator-match}
\mathcal{O}^{\mu\mu_1\cdots \mu_{n-1}}_q\
=\ \sum_j Q^{(n-1)}_j \mathcal{O}^{\mu\mu_1 \cdots \mu_{n-1}}_j,
\end{equation}
where the subscript $j$ labels different types of hadronic operators.
The coefficients $Q^{(n-1)}_j$ can be defined through the $n$-th moments of the GPDs in the hadronic configuration~$j$. 
Matrix elements of the hadronic operators $\mathcal{O}^{\mu\mu_1 \cdots \mu_{n-1}}_j$ are related to the moments of the hadronic splitting functions $f_j$ and $g_j$, which are defined by
\begin{equation} \label{eq:splitting}
\bar{u}(p')
\bigg[ \gamma^+ f_j(y,t) + \frac{i\sigma^{+\nu}\Delta_\nu}{2M_N} g_j(y,t)
\bigg] u(p)
= \int \dd^4{k}\, \widetilde{\Gamma}^+_j(k)\, \delta\Big(y-\frac{k^+}{P^+}\Big)
\equiv \Gamma_j^+, 
\end{equation}
where $\Gamma^+_j$ is the contribution to the vector vertex in Eq.~(\ref{eq:f1f2}) from hadronic configuration~$j$, $k$ is the internal meson momentum, and $y$ is the light-cone momentum fraction of the nucleon carried by the hadronic state $j$.
The moments of the splitting functions are then given by
\begin{eqnarray}
f^{(n)}_j = \int^1_{-1} \dd{y} y^{n-1} f_j(y,t),
\qquad
g^{(n)}_j = \int^1_{-1} \dd{y} y^{n-1} g_j(y,t).
\end{eqnarray}
For $n=1$, the $y$ integral of the splitting functions leads to the corresponding form factors.
Inserting Eq.~(\ref{eq:operator-match}) into the nucleon states and contracting both sides with $n_\mu n_{\mu_1}\cdots n_{\mu_{n-1}}$, we have
\begin{equation}
\label{eq:fj}
H_q^n\, \bar{u}(p') \slashed {n} u(p) 
+ \frac{i}{2M_N} E_q^n\, \bar{u}(p') n_{\mu}\sigma^{\mu\nu}\Delta_\nu u(p)
= \sum_j Q^{(n-1)}_j \left[ f^{(n)}_j \bar{u}(p') \slashed {n} u(p) 
+ \frac{i}{2M_N} g^{(n)}_j \bar{u}(p') n_{\mu}\sigma^{\mu\nu}\Delta_\nu u(p) \right],
\end{equation}
where the moments $H_q^n$ and $E_q^n$ are defined in Eqs.~(\ref{eq:HEqn}).
Since Eq.~(\ref{eq:fj}) is valid for any order of $n$, the operator relation in Eq.~(\ref{eq:operator-match}) then gives rise to the convolution formula for the GPDs at zero skewness,
\begin{subequations}
\begin{eqnarray} \label{eq:convolution}
\hspace*{-1.5cm}
&& H^q(x,t) \equiv H^q(x,\xi=0,t)
= \sum_j \big[f_j \otimes q_j^v\big](x,t)
\equiv \sum_j \int_0^1 \dd{y} \int_0^1 \dd{z}\, 
  \delta(x-yz)\, f_j(y,t)\, q_j^v(z,t),
\\
\hspace*{-1.5cm}
&& E^q(x,t) \equiv E^q(x,\xi=0,t)\,
= \sum_j \big[g_j \otimes q_j^v\big](x,t)\,
\equiv \sum_j \int_0^1 \dd{y} \int_0^1 \dd{z}\, 
  \delta(x-yz)\, g_j(y,t)\, q_j^v(z,t),
\end{eqnarray} 
\end{subequations}
where $q^v_j(x,t) = [q_j-\bar{q}_j](x,t) \equiv q^v_j(x,\xi\!=\!0,t)$ is the GPDs of valence quark $q$ in the hadron configuration~$j$ at zero skewness.
In our calculation, the cross symmetry $q(-x,t) = -\bar{q}(x,t)$ has been applied so that the above integrals are from 0 to 1.

\subsection{\it Valence GPDs in bare baryons}
\label{ssec:valenceGPDs}

In this subsection, we derive the relationships for the input GPDs of different quark flavors.
The twist-two operators for the PDFs and GPDs were discussed in detail in Refs.~\cite{Salamu2,XGWang,He5}. 
Here, we take the intermediate octet baryons as the hadronic configuration as an example.
The most general vector operator for an octet baryon intermediate state is written as 
\begin{eqnarray}
\label{eq:GPD-op}
{\cal O}_{q,B}^{\mu\mu_1\cdots\mu_{n-1}}
&=& \sum_{i=0,\rm even}^{n-1}
\Big( 
  \alpha_A^{(ni)}\, 
  {\rm Tr}\left[ \bar{B} \gamma^\mu \left\{\lambda_q^+, B\right\} \right]
+ \beta_A^{(ni)}\,
  {\rm Tr}\left[ \bar{B} \gamma^\mu \left[\lambda_q^+,B\right] \right]   \nonumber \\
& & \hspace*{1.3cm}
+\, \sigma_A^{(ni)}\,
  {\rm Tr}\left[ \bar{B} \gamma^\mu B \right] 
  {\rm Tr}\left[ \lambda_q^+ \right]
\Big)\,
\Delta^{\mu_1} \cdots \Delta^{\mu_i} P^{\mu_i+1}\cdots P^{\mu_{n-1}}
\nonumber \\
&+& \sum_{i=0,\rm even}^{n-1}\frac{i}{2M_B}
\Big(
  \alpha_B^{(ni)}\,
  {\rm Tr}\left[ \bar{B} \sigma^{\mu\nu} \left\{\lambda_q^+,B\right\} \right]
+ \beta_B^{(ni)}\,
  {\rm Tr}\left[ \bar{B} \sigma^{\mu\nu} \left[ \lambda_q^+,B \right] \right]
\nonumber \\
& & \hspace*{2.2cm}
+\, \sigma_B^{(ni)}\,
  {\rm Tr}\left[ \bar{B} \sigma^{\mu\nu}B \right]
  {\rm Tr}\left[ \lambda_q^+ \right]
\Big)\,
\Delta^\nu \Delta^{\mu_1} \cdots \Delta^{\mu_i} P^{\mu_i+1}\cdots P^{\mu_{n-1}}     
\nonumber \\
&+& \frac{1}{M_B} 
\Big(
  \alpha_C^{(n)}\Big|_{n\, \rm even}\,
  {\rm Tr}\left[ \bar{B} \left\{\lambda_q^+,B\right\} \right]
+ \beta_C^{(n)}\Big|_{n\, \rm even}\,
  {\rm Tr}\left[ \bar{B} \left[\lambda_q^+,B \right] \right] 
\nonumber \\
& & \hspace*{0.8cm}
+\, \sigma_C^{(n)}\Big|_{n\, \rm even}\,
  {\rm Tr}\left[ \bar{B} B \right]
  {\rm Tr}\left[ \lambda_q^+ \right]
\Big)\,
\Delta^\mu \Delta^{\mu_1} \cdots \Delta^{\mu_n}.
\end{eqnarray}
Contracting both sides with $n_\mu n_{\mu_1}\cdots n_{\mu_{n-1}}$, we have
\begin{eqnarray} \label{eq:operator}
n_{\mu} n_{\mu_1}\cdots n_{\mu_{n-1}}{\cal O}_{q,B}^{\mu\mu_1\cdots\mu_{n-1}} 
&=& \alpha^{(n)} 
{\rm Tr}\left[\bar{B} \slashed {n}
\left\{\lambda_q^+,B\right\}\right]
+\beta^{(n)} {\rm Tr}\left[\bar{B}
\slashed{n} \left[\lambda_q^+,B \right]\right] 
+ \sigma^{(n)}{\rm Tr}\left[\bar{B}
\slashed{n}B \right]{\rm Tr}\left[\lambda_q^+\right] 
\nonumber \\
&&\hspace*{-5cm}
+\, \frac{i\alpha_{\rm mag}^{(n)}}{2M_B}{\rm Tr}\left[\bar{B} n_\mu\sigma^{\mu\nu}\Delta_\nu
\left\{\lambda_q^+,B\right\}\right] 
+ \frac{i\beta_{\rm mag}^{(n)}}{2M_B}
{\rm Tr}\left[\bar{B}
n_\mu\sigma^{\mu\nu}\Delta_\nu \left[\lambda_q^+,B \right]\right]  
+ \frac{i\sigma_{\rm mag}^{(n)}}{2M_B}
{\rm Tr}\left[\bar{B}
n_\mu\sigma^{\mu\nu}\Delta_\nu B \right]{\rm Tr}\left[\lambda_q^+\right], \nonumber \\
\end{eqnarray}
where $X^{(n)}$ and $X_{\rm mag}^{(n)}$ ($X=\alpha,\beta,\sigma$) are expressed as
\begin{subequations}
\begin{eqnarray}
X^{(n)} 
&=& \sum_{i=0,{\rm even}}^{n-1} (-2\xi)^i X_A^{(ni)} 
+(-2\xi)^n X_C^{(n)}\Big|_{n\, {\rm even}},
\\
X_{\rm mag}^{(n)} 
&=& \sum_{i=0,{\rm even}}^{n-1} (-2\xi)^i X_B^{(ni)} 
-(-2\xi)^n X_C^{(n)}\Big|_{n\, {\rm even}}.
\end{eqnarray}
\end{subequations}
The coefficients $\{ \alpha^{(n)}, \beta^{(n)}, \sigma^{(n)} \}$
are related to the moments of the spin-averaged GPD $H^q(x,\xi,t)$, while $\{ \alpha_{\rm mag}^{(n)}, \beta_{\rm mag}^{(n)}, \sigma_{\rm mag}^{(n)} \}$
are related to moments of the spin-flip GPD $E^q(x,\xi,t)$.
With the simplification of the flavor matrices, the operator (\ref{eq:operator}) can be rewritten as
\begin{equation}
n_{\mu}n_{\mu_1}\cdots n_{\mu_{n-1}}{\cal O}_{q,B}^{\mu\mu_1\cdots\mu_{n-1}}
= Q_{B}^{(n-1)} {\cal O}_B
+ Q_{B, \rm mag}^{(n-1)} {\cal O}_{B, \rm mag}
+ Q_{B\phi\phi^\dag}^{(n-1)} {\cal O}_{B\phi\phi^\dag}
+ Q_{B\phi\phi^\dag, \rm mag}^{(n-1)} {\cal O}_{B\phi\phi^\dag, \rm mag},
\end{equation}
where the hadronic operators are given by
\begin{subequations}
\begin{eqnarray}
{\cal O}_{B} 
= \bar{B} \slashed{n} B,~~~ 
&&
{\cal O}_{B, \rm mag} 
= \frac{i}{2M_B}\bar{B} n_\mu\sigma^{\mu\nu} \Delta_\nu B,
\\
{\cal O}_{B\phi\phi^\dag} 
= \frac{1}{f^2} \bar{B} \slashed{n} B\phi\phi^\dag ,~~~
&&
{\cal O}_{B\phi\phi^\dag, \rm mag} 
= -\frac{i}{2M_Bf^2}\bar{B} n_\mu\sigma^{\mu\nu} \Delta_\nu B\phi\phi^\dag .
\end{eqnarray}
\end{subequations}
The coefficients $Q^{(n-1)}_{j}$ in front of each of the operators are defined in terms of Mellin moments of the corresponding PDFs in the intermediate hadron states,
\begin{subequations}
\begin{eqnarray}
\int _{-1}^1 \dd{x} x^{n-1} H^q_B(x,\xi,t) 
&=& Q^{(n-1)}_{B},
\\
\int _{-1}^1 \dd{x} x^{n-1} E^q_B(x,\xi,t) 
&=& Q^{(n-1)}_{B,\rm mag},
\\
\int _{-1}^1 \dd{x} x^{n-1} H^{q,\rm tad}_{\phi\phi^\dag}(x,\xi,t)
&=& Q^{(n-1)}_{B\phi\phi^\dag},
\\
\int _{-1}^1 \dd{x} x^{n-1} E^{q,\rm tad}_{\phi\phi^\dag}(x,\xi,t) 
&=& Q^{(n-1)}_{B\phi\phi^\dag,\rm mag},
\end{eqnarray}
\end{subequations}
where each of the moments $Q^{(n-1)}_j$ can be expressed in terms of the coefficients
   $\{ \alpha^{(n)}, \beta^{(n)}, \sigma^{(n)}\}$ and $\{\alpha_{\rm mag}^{(n)}, \beta_{\rm mag}^{(n)}, \sigma_{\rm mag}^{(n)} \}$. 
Since the relationships between the quark moments in the different configurations $Q^{(n-1)}_{B}$ and $Q^{(n-1)}_{B\phi\phi^\dag}$ do not depend on the momentum transfer $t$, they are the same for GPDs case.
For the moments $Q_{B, \rm mag}^{(n-1)}$, they have the same relationships as $Q_B^{(n-1)}$ because the corresponding operators have the same flavor structure,
\begin{subequations}
\begin{eqnarray} 
\label{eq:momentmagB}
\alpha_{\rm mag}^{(n)}
&=& \frac12 \Big( U_{p,{\rm mag}}^{(n-1)} + S_{p,{\rm mag}}^{(n-1)} \Big) - D_{p,{\rm mag}}^{(n-1)},
\\
\beta_{\rm mag}^{(n)} 
&=& \frac12 \Big( U_{p,{\rm mag}}^{(n-1)} - S_{p,{\rm mag}}^{(n-1)} \Big), 
\\
\sigma_{\rm mag}^{(n)}
&=& D_{p,{\rm mag}}^{(n-1)}.
\end{eqnarray}
\end{subequations}
Assuming the strangeness in the bare nucleon state to be zero, we have
\begin{equation}
\sigma^{(n)} = \beta^{(n)} - \alpha^{(n)}, ~~~~~~~
\sigma_{\rm mag}^{(n)} = \beta_{\rm mag}^{(n)} - \alpha_{\rm mag}^{(n)}.
\end{equation}
In particular, for $n=1$ the above expressions are consistent with the Lagrangian for the magnetic interaction, where $c_3=c_2-c_1$.
Similarly, $Q^{(n-1)}_{B\phi\phi^\dag}$ and $Q^{(n-1)}_{B\phi\phi^\dag,\rm mag}$ also have the same relationships.

As in the PDFs case, since the relations for the moments are valid for any order of $n$, those relations must also be true for the GPDs themselves.
With the relationships, the GPDs of quarks in different hadronic configurations can be expressed in terms of quark distributions in proton $H^q_p(x,\xi,t)$, $\widetilde{H}^q_p(x,\xi,t)$ and $E^q_p(x,\xi,t)$, which are parameterized in Refs.~\cite{Diehl, Martin, Leader4}.
With the similar approach as for PDFs, in the next section we discuss the dependence of the parton distributions on the momentum transfer $t$ and transverse momentum $k_T$, as well as the momentum fraction $x$.

\subsection{\it Generalized parton distributions}

\begin{table}[t]
\begin{center}
\caption{Coefficients $C_B^{\rm mag}$, $C_T^{\rm mag}$, $C_{BT}^{\rm mag}$, $C_{\phi\phi^\dag}^{\rm mag}$ and $C_{\phi\phi^\dag}'$ for octet and decuplet baryons in the effective Lagrangian.}
\begin{tabular}{c|ccccccc} \hline \\ [-2mm]  
\hspace*{0.3cm}\bm{$B$} \hspace*{0.3cm}
& \hspace*{0.3cm}\bm{$p$} \hspace*{0.3cm}
& \hspace*{0.3cm}\bm{$n$} \hspace*{0.3cm}
& \hspace*{0.3cm}\bm{$\Sigma^+$}\hspace*{0.3cm}
& \hspace*{0.3cm}\bm{$\Sigma^0$}\hspace*{0.3cm}
& \hspace*{0.3cm}\bm{$\Sigma^-$}\hspace*{0.3cm}
& \hspace*{0.3cm}\bm{$\Lambda$}\hspace*{0.3cm}	
& \hspace*{0.3cm}\bm{$\Lambda\Sigma^0$}\hspace*{0.3cm} 
\\ [2mm]
\hspace*{0.3cm}$C_{B}^{\text {mag}}$\hspace*{0.3cm}
& \hspace*{0.3cm}$\frac13c_1 +c_2$\hspace*{0.3cm}
& \hspace*{0.3cm}$-\frac23 c_1$\hspace*{0.3cm}
& \hspace*{0.3cm}$\frac13c_1 +c_2$\hspace*{0.3cm}
& \hspace*{0.3cm}$\frac13 c_1$\hspace*{0.3cm}
& \hspace*{0.3cm}$\frac13 c_1-c_2$\hspace*{0.3cm}
& \hspace*{0.3cm}$-\frac13 c_1$\hspace*{0.3cm}	
& \hspace*{0.3cm}$\frac{1}{\sqrt{3}} c_1$\hspace*{0.3cm}	 
\\[2mm] \hline \\[-2mm]
\hspace*{0.3cm}\bm{$T$}\hspace*{0.3cm}
& \hspace*{0.3cm}\bm{$\Delta^{++}$}\hspace*{0.3cm}
& \hspace*{0.3cm}\bm{$\Delta^+$}\hspace*{0.3cm}
& \hspace*{0.3cm}\bm{$\Delta^0$}\hspace*{0.3cm}
& \hspace*{0.3cm}\bm{$\Delta^-$} \hspace*{0.3cm}
& \hspace*{0.3cm}\bm{$\Sigma^{*+}}$\hspace*{0.3cm}
& \hspace*{0.3cm}\bm{$\Sigma^{*0}}$\hspace*{0.3cm}
& \hspace*{0.3cm}\bm{$\Sigma^{*-}}$\hspace*{0.3cm}	\\[2mm]
\hspace*{0.3cm}$C_{T}^{\text {mag}}$\hspace*{0.3cm}
& \hspace*{0.3cm}$\frac23F_2^T$\hspace*{0.3cm}
& \hspace*{0.3cm}$\frac13F_2^T$\hspace*{0.3cm}
& \hspace*{0.3cm}$0$\hspace*{0.3cm}
& \hspace*{0.3cm}$-\frac13F_2^T$\hspace*{0.3cm}
& \hspace*{0.3cm}$\frac13F_2^T$\hspace*{0.3cm}
& \hspace*{0.3cm}$0$\hspace*{0.3cm}
& \hspace*{0.3cm}$-\frac13F_2^T$\hspace*{0.3cm}	
\\[2mm] \hline \\[-2mm]
\hspace*{0.3cm}\bm{$BT$}\hspace*{0.3cm}
& \hspace*{0.3cm}\bm{$p\Delta^+$}\hspace*{0.3cm}
& \hspace*{0.3cm}\bm{$n\Delta^0$}\hspace*{0.3cm}
& \hspace*{0.3cm}\bm{$\Sigma^+\Sigma^{*+}$} \hspace*{0.3cm}
& \hspace*{0.3cm}\bm{$\Sigma^0\Sigma^{*0}$}\hspace*{0.3cm}
& \hspace*{0.3cm}\bm{$\Lambda\Sigma^{*0}$}\hspace*{0.3cm}
& \hspace*{0.3cm}\bm{$\Sigma^-\Sigma^{*-}$}\hspace*{0.3cm}	\\[2mm]
\hspace*{0.3cm}$C_{BT}^{\text {mag}}$\hspace*{0.3cm}
& \hspace*{0.3cm}$-\frac{1}{\sqrt{3}}c_4$\hspace*{0.3cm}
& \hspace*{0.3cm}$-\frac{1}{\sqrt{3}}c_4$\hspace*{0.3cm}
& \hspace*{0.3cm}$\frac{1}{\sqrt{3}}c_4$\hspace*{0.3cm}
& \hspace*{0.3cm}$\frac{1}{2\sqrt{3}}c_4$\hspace*{0.3cm}
& \hspace*{0.3cm}$\frac{1}{2}c_4$\hspace*{0.3cm}
& \hspace*{0.3cm}$0$\hspace*{0.3cm}	
\\ [2mm] \hline \\[-2mm]
\bm{$\phi\phi^\dag$}
& \bm{$\pi^+\pi^-$}
& \bm{$K^0 \overline{K}^0$}
& \bm{$K^+K^-$}
&
&						\\ [2mm]
$C_{\phi\phi^\dag}^{\text{mag}}$
& $-\frac{1}{2}(c_1+c_2)$
& $0$
& $-c_2$
&
&						\\ [2mm]
$C_{\phi\phi^\dag}'$
& $4(b_{10}+b_{11})$
& $4(b_{11}-b_{10})$
& $8b_{11}+2b_{9}$ 
&
\\[2mm] \hline
\end{tabular}
\label{tab:Cmag}
\end{center}
\end{table}

The sea quark contributions to the unpolarized and polarized parton distributions in the proton were discussed in Sec.~\ref{Sec.5}.
In the present section we extend that discussion to the case of GPDs of sea quarks in the off-forward scattering direction.
GPDs contain a wealth of information on the partonic structure of the nucleon, which is one of the central themes in hadronic physics.
They have close relationships with electromagnetic form factors, and by integrating GPDs with different powers of the momentum fraction $x$, GPDs can be transformed into Mellin moments.
Since form factors and PDFs are special cases of GPDs, in general GPDs provide more information about the internal structure of the nucleon than either of these limiting quantities.

For unpolarized GPDs at finite $t$, there are two twist-two distributions, $H^q(x,\xi,t)$ and $E^q(x,\xi,t)$, for each quark flavor $q$.
Along with the electric interaction for the PDFs at zero momentum transfer, the Lagrangian for magnetic interactions, Eqs.~(\ref{eq:magB})$-$(\ref{eq:magBT}), is needed in the loop calculations.
Although they do not contribute to the PDF $q(x)$ in the forward limit, $t=0$, magnetic interactions contribute to both $H^q(x,\xi,t)$ and $E^q(x,\xi,t)$ at finite values of $t$.
For the baryon-meson interaction, in addition to the leading order Lagrangian (\ref{eq:Llocal}), there is a higher order contribution which is of the same order in the power counting as the magnetic Lagrangian \cite{Kubis}, 
\begin{eqnarray}\label{eq:adm}
{\cal L'} 
= 2i\, b_9\, 
\sigma^{\mu\nu}{\rm Tr}\left[\bar{B}u_\mu\right]
{\rm Tr}\left[u_\nu B\right]
+ 2i\, b_{10}\,
\sigma^{\mu\nu}{\rm Tr}\left[\bar{B}\{[u_\mu,u_\nu],B\}\right]
+ 2i\, b_{11}\,
\sigma^{\mu\nu}{\rm Tr}\left[\bar{B}[[u_\mu,u_\nu],B]\right],
\end{eqnarray}
where the coefficients $b_9$, $b_{10}$ and $b_{11}$ are determined phenomenologically to be $b_9=1.36$~GeV$^{-1}$, $b_{10}=1.24$~GeV$^{-1}$ and $b_{11}=0.46$~GeV$^{-1}$ \cite{Kubis}.
This interaction contributes to the $E^q(x,\xi,t)$ GPDs of antiquarks in the one-loop bubble diagram, Fig.~\ref{fig:FMLOOP}(u).
After expanding the flavor matrices, the total nonlocal Lagrangian relevant for the calculation of the GPDs can be written as \cite{He5}
\begin{eqnarray}
&&{\cal L}^{\rm (nonloc)}(x)
= \bar B(x) (i\gamma^\mu \mathscr{D}_{\mu} - M_B) B(x)
+ \overline{T}_\mu(x)
  (i\gamma^{\mu\nu\alpha} \mathscr{D}_{\alpha} - M_T \gamma^{\mu\nu})
  T_\nu(x)						\notag\\
&&+\,
  \bar{p}(x)
  \left[
    \frac{C_{B\phi}}{f} \gamma^\mu \gamma^5 B(x)\,
  + \frac{C_{T\phi}}{f} \Theta^{\mu\nu} T_\nu(x)
  \right]\mathscr{D}_{\mu} \left(
  \int \dd^4{a}\, {\cal G}_\phi^q(x,x+a) F(a)\
   \phi(x+a)\right) + {\rm H.c.}		\notag\\
&&+\, 
  \frac{i C_{\phi\phi^\dag}}{2 f^2}\,
  \bar{p}(x) \gamma^\mu p(x)                                  
 \int \dd^4{b}\, {\cal G}_\phi^q(x,x+a) F(a)\phi(x+a) \mathscr{D}_{\mu} \left(
  \int \dd^4{b}\, {\cal G}_\phi^q(x+b,x) F(b) \phi^\dag(x+b)\right)
       + {\rm H.c.}						\notag\\
&&+\, 
  \frac{i C_{\phi\phi^\dag}'}{2 f^2}\,
  \bar{p}(x) \sigma^{\mu\nu} p(x)				
\mathscr{D}_{\mu}\left(\int \dd^4{a}\, {\cal G}_\phi^q(x,x+a) F(a) \phi(x+a)\right) 
\mathscr{D}_{\nu}\left(\int \dd^4{b}\, {\cal G}_\phi^q(x+b,x) F(b)\phi^\dag(x+b) \right)
\notag\\
&&+\, \frac{C_{B}^{\text {mag}}}{4M_B}\,
   \bar B(x) \sigma^{\mu\nu} B(x) F_{\mu\nu}(x) - \frac{C_T^{\rm mag}}{4M_T}
   \overline T_\alpha(x) \sigma^{\mu\nu} T^\alpha(x) F_{\mu\nu}(x)	
+	\frac{iC_{BT}^{\rm mag}}{4M_B}
   \bar B(x) \gamma^\mu\gamma_5 T^\nu (x) F_{\mu\nu}(x)	
\notag\\
&&+\,
  \frac{C_{\phi\phi^\dag}^{\text{mag}}}{4 M_Bf^2}\,
  \bar{p}(x) \sigma^{\mu\nu} p(x) \int \dd^4{a}\!\int \dd^4{b}\, F_{\mu\nu}(x)		
{\cal G}_\phi^q(x+b,x+a) F(a) F(b)\phi(x+a) \phi^\dag(x+b)
\notag\\
&&+\,
  \mathscr{D}_\mu \phi(x) (\mathscr{D}_\mu \phi)^\dag(x)
 + \cdots,
\label{eq:j4}
\end{eqnarray}
where the derivative $\mathscr{D}_{\mu}$ and gauge link ${\cal G}_\phi^q(x,y)$ are defined in Eqs.~(\ref{eq:partial}) and (\ref{eq:link}), respectively.
The coefficients $C_{B\phi}$, $C_{T\phi}$ and $C_{\phi\phi^\dag}$ are listed in Table~\ref{tab:C} and $C_B^{\rm mag}$, $C_T^{\rm mag}$, $C_{BT}^{\rm mag}$, $C_{\phi\phi^\dag}^{\rm mag}$ and $C_{\phi\phi^\dag}'$ in Table~\ref{tab:Cmag}. 
From the Lagrangian (\ref{eq:j4}) one can obtain the vector current interacting with the external field $\mathscr{A}_\mu$.

The convolution form for the unpolarized GPDs of antiquarks in the proton can be expressed as 
\begin{subequations}\label{eq:HEqbarconv}
\begin{eqnarray}
H^{\bar{q}}(x,t)
&=&\sum_{\phi B T} 
\Big[ 
\big(f_{\phi B}^{\rm rbw}+f_{\phi T}^{\rm rbw}+f^{\rm bub}_\phi \big)
\otimes H^{\bar{q}}_\phi
\Big](x,t),
\\
E^{\bar{q}}(x,t)
&=&\sum_{\phi B T} 
\Big[
\big(g_{\phi B}^{\rm rbw}+g_{\phi T}^{\rm rbw} + g^{\rm bub'}_\phi \big) \otimes H^{\bar{q}}_\phi
\Big](x,t),
\end{eqnarray}
\end{subequations}
while for quark GPDs one has a more complicated structure,
\begin{subequations}\label{eq:HEqconv}
\begin{eqnarray}\label{sf}
H^q(x,t) 
&=& Z_2 H^q_0(x,t)
+ \sum_{\phi B T}
\bigg \{ 
\Big[
\big( f_{\phi B}^{\rm rbw}+f_{\phi T}^{\rm rbw}+f^{\rm bub}_\phi \big) \otimes H^q_\phi
\Big](x,t) 
\nonumber \\
&+& 
  \Big[ \bar{f}_{B\phi}^{\rm rbw} \otimes H^q_B \Big](x,t)
+ \Big[ \bar{f}^{\rm KR}_{B\phi} \otimes H^{q,{\rm KR}}_B \Big](x,t)
+ \Big[ \delta \bar{f}^{\rm KR}_{B} \otimes H^{s,{\rm KR}}_B \Big](x,t)  
\nonumber\\
&+&
  \Big[ \bar{f}_{T\phi}^{\rm rbw} \otimes H^q_T \Big](x,t)
+ \Big[ \bar{f}^{\rm KR}_{T\phi} \otimes H^{q,{\rm KR}}_T \Big](x,t)
+ \Big[ \delta\bar{f}^{\rm KR}_{T\phi} \otimes H^{q,{\rm KR}}_T \Big](x,t)
\nonumber\\
&+&
  \Big[ \bar{f}_{B\phi}^{\rm rbw,mag} \otimes E^q_B \Big](x,t)
+ \Big[ \bar{f}_{T\phi}^{\rm rbw,mag} \otimes E^q_T \Big](x,t)
+ \Big[ \bar{f}_{BT}^{\rm rbw,mag} \otimes E^q_{BT} \Big](x,t)
\nonumber  \\
&+&
  \Big[ \bar{f}^{\rm tad}_\phi \otimes H^{q,{\rm tad}}_{\phi\phi^\dag} \Big](x,t)
+ \Big[ \delta\bar{f}^{\rm tad}_\phi \otimes H^{q,{\rm tad}}_{\phi\phi^\dag} \Big](x,t)
\bigg \},
\\
&& \nonumber\\
E^q(x,t) 
&=& Z_2 E^q_0(x,t)
+ \sum_{\phi B T}
\bigg \{ 
\Big[ 
\big( g_{\phi B}^{\rm rbw}+g_{\phi T}^{\rm rbw}+g^{\rm bub'}_\phi \big) \otimes H^q_\phi
\Big](x,t) 
\nonumber \\
&+& 
  \Big[ \bar{g}_{B\phi}^{\rm rbw} \otimes H^q_B \Big](x,t)
+ \Big[ \bar{g}^{\rm KR}_{B\phi} \otimes H^{q,{\rm KR}}_B \Big](x,t)
+ \Big[ \delta \bar{g}^{\rm KR}_{B} \otimes H^{q,{\rm KR}}_B \Big](x,t) 
\nonumber\\
&+&
  \Big[ \bar{g}_{T\phi}^{\rm rbw} \otimes H^q_T \Big](x,t)
+ \Big[ \bar{g}^{\rm KR}_{T\phi} \otimes H^{q,{\rm KR}}_T \Big](x,t)
+ \Big[ \delta\bar{g}^{\rm KR}_{T\phi} \otimes H^{q,{\rm KR}}_T \Big](x,t)
\nonumber\\
&+&
  \Big[ \bar{g}_{B\phi}^{\rm rbw,mag} \otimes E^q_B \Big](x,t)
+ \Big[ \bar{g}_{T\phi}^{\rm rbw,mag} \otimes E^q_T \Big](x,t)
+ \Big[ \bar{g}_{BT}^{\rm rbw,mag} \otimes E^q_{BT} \Big](x,t)
\nonumber\\
&+&
  \Big[ \bar{g}^{\rm tad, mag}_\phi \otimes E^{q,{\rm tad}}_{\phi\phi^\dag} \Big](x,t) 
\bigg \},
\end{eqnarray}
\end{subequations}
where the splitting functions are calculated with the vector current and the corresponding Feynman diagrams include the rainbow, Kroll-Ruderman, tadpole and bubble diagrams, as in Fig.~\ref{fig:FMLOOP}. 
The vertices with the external field include electric, magnetic, and additional interactions arising from the gauge link. 
For the $u$ and $d$ quarks the intermediate states baryons and mesons are the nucleon and $\Delta$ baryons and $\pi$ mesons.
For the strange quark, the intermediate baryons and mesons are the octet and decuplet hyperons and $K$ mesons.
It should be noted that both the electric and magnetic operators make contributions to $H^q(x,t)$ and $E^q(x,t)$ at finite momentum transfer.
At zero momentum transfer, although the second (Pauli) term in the bracket of Eq.~(\ref{eq:GPD}) does not contribute to the matrix element, the function $E^q(x,0)$ itself is nonzero.
In the numerical calculations presented here, the skewness parameter $\xi$ will be set to be zero.
With the relations discussed in Sec.~\ref{ssec:valenceGPDs}, all the input GPDs will be expressed in terms of GPDs of valence quarks in the proton, $H^q_p(x,t)$, $E^q_p(x,t)$ and $\widetilde{H}^q_p(x,t)$ \cite{XGWang, Salamu2, He5}. 
The parameterization expressions for these GPDs can be found in Refs.~\cite{Diehl, Martin, Leader4} at the input scale $\mu_0=1$~GeV.

\begin{figure}[t]
\begin{center}
    \begin{minipage}{0.45\linewidth}
        \centering
        \centerline{
        \includegraphics[width=1\textwidth]{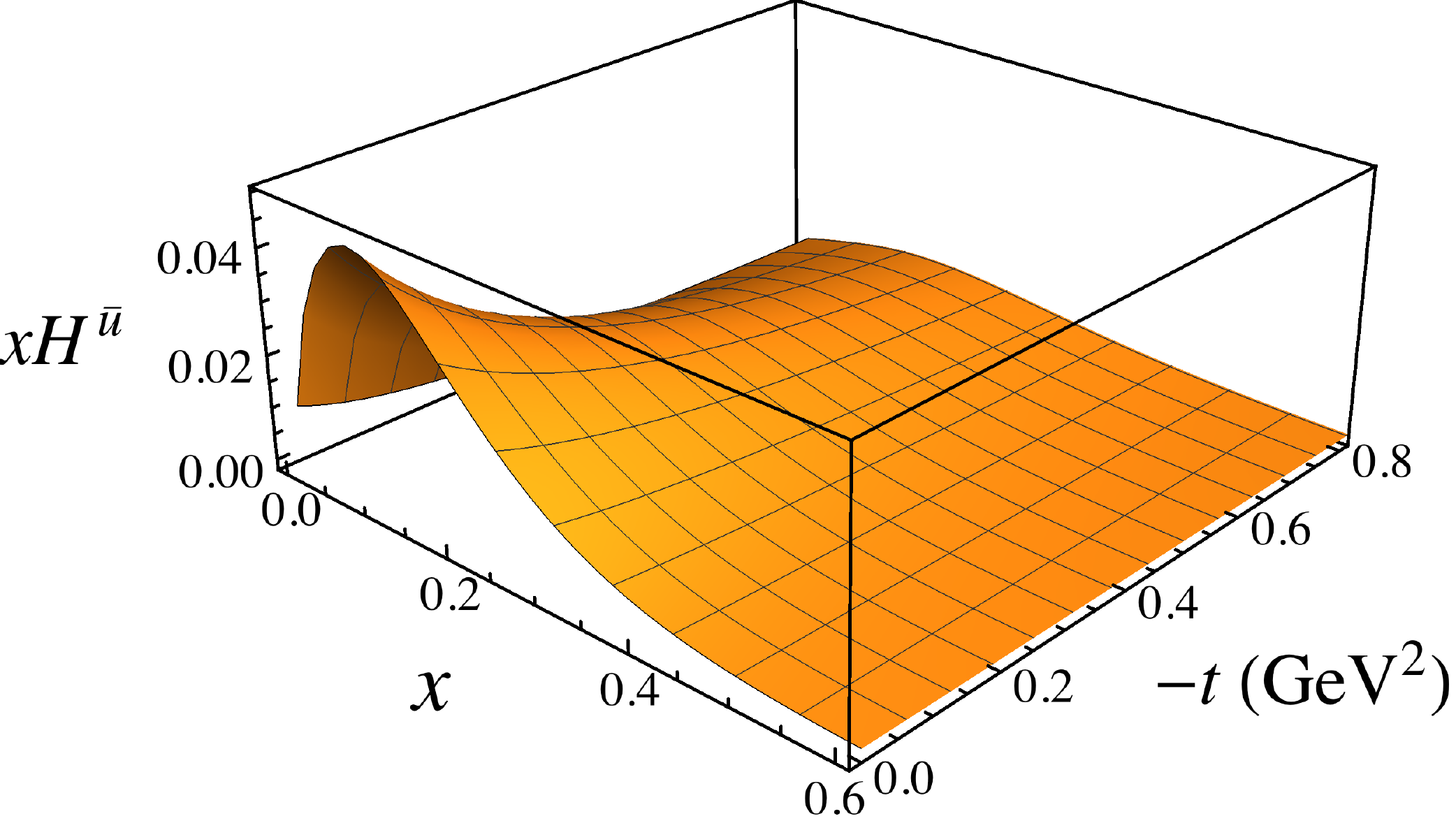}
        }
    \end{minipage}
    \begin{minipage}{0.45\linewidth}
        \centering
        \centerline{
        \includegraphics[width=1\textwidth]{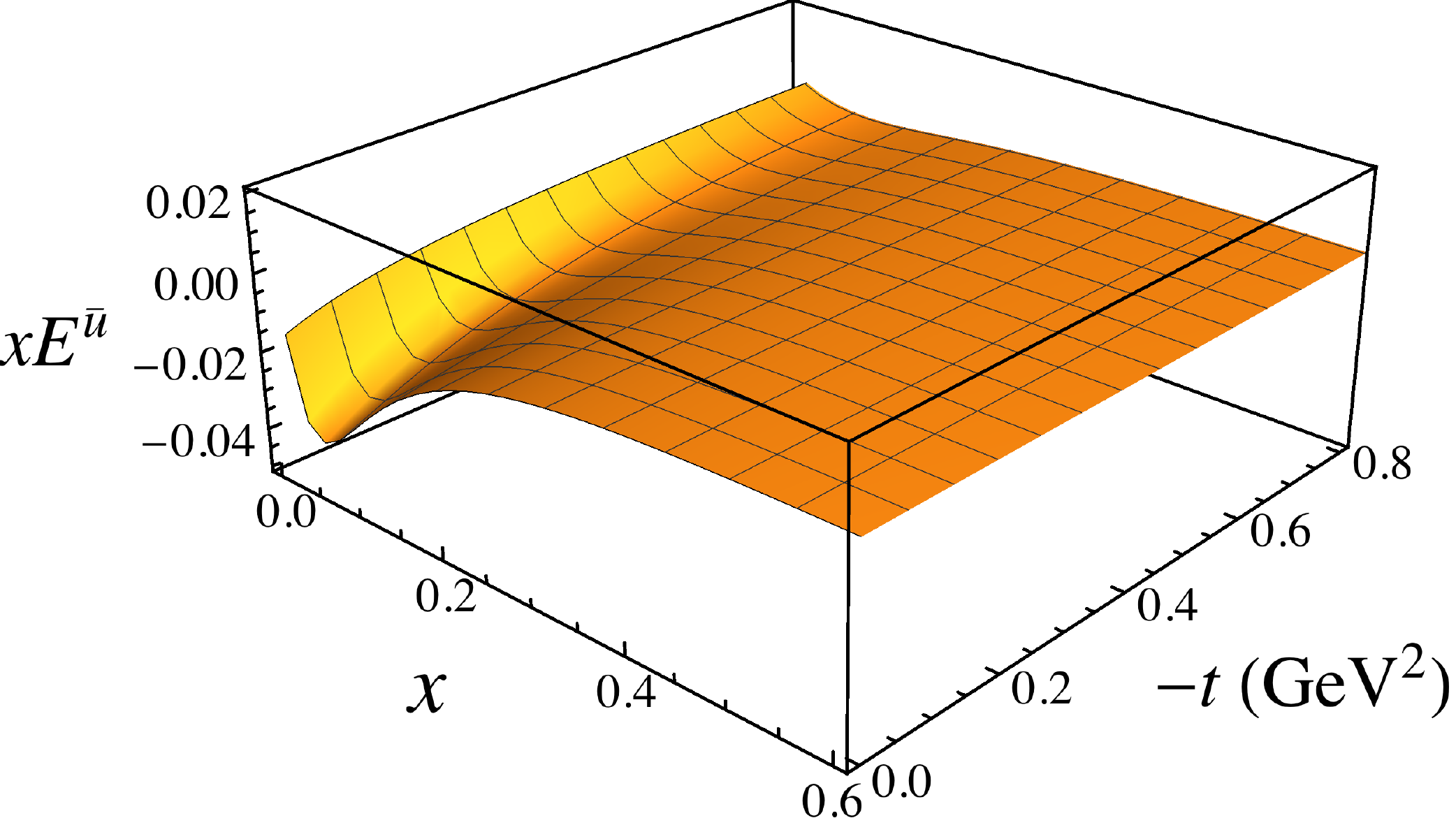}
        }
    \end{minipage}
\\[0.4cm]
    \begin{minipage}{0.45\linewidth}
        \centering
        \centerline{
        \includegraphics[width=1\textwidth]{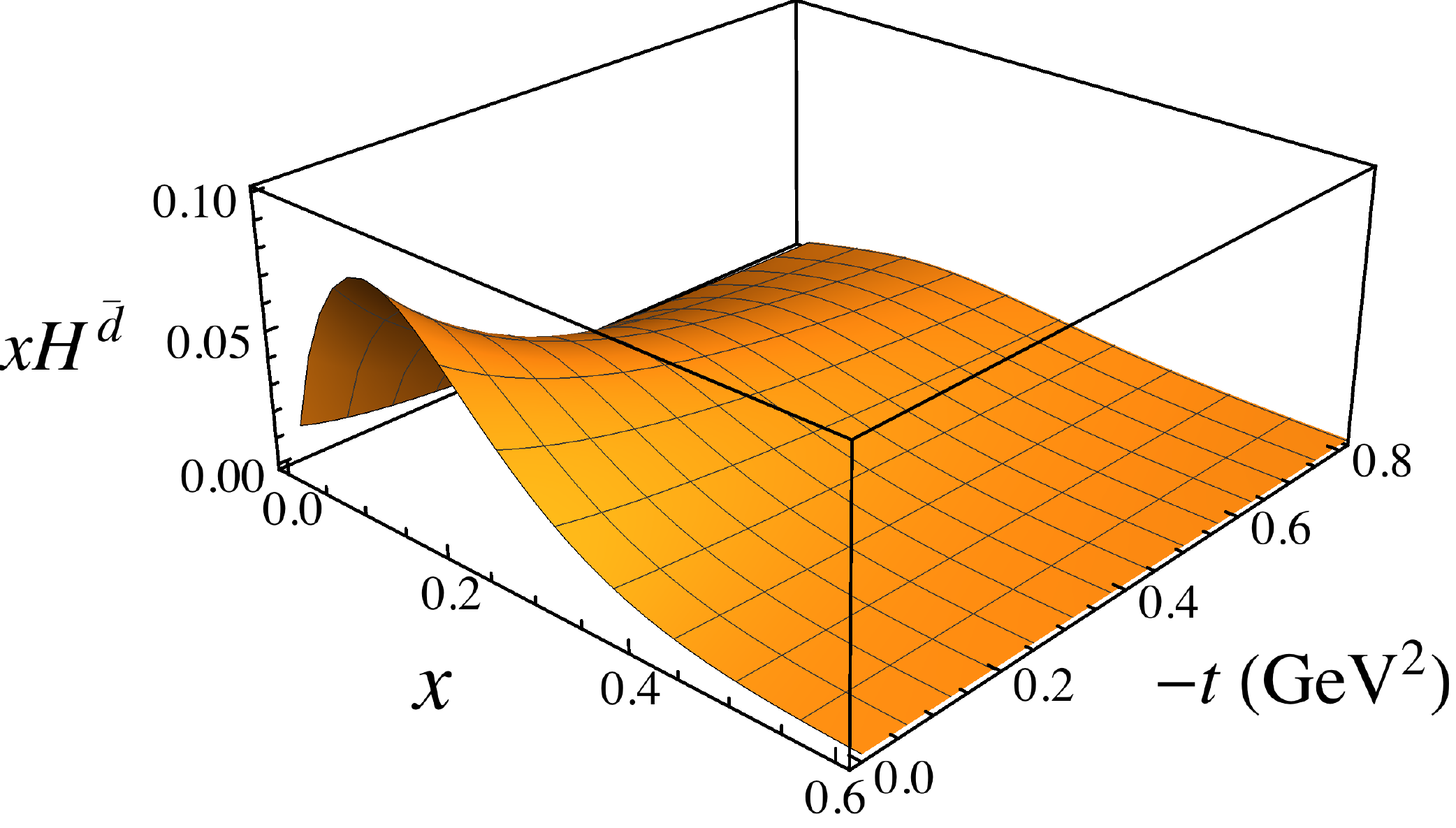}
        }
    \end{minipage}
    \begin{minipage}{0.45\linewidth}
        \centering
        \centerline{
        \includegraphics[width=1\textwidth]{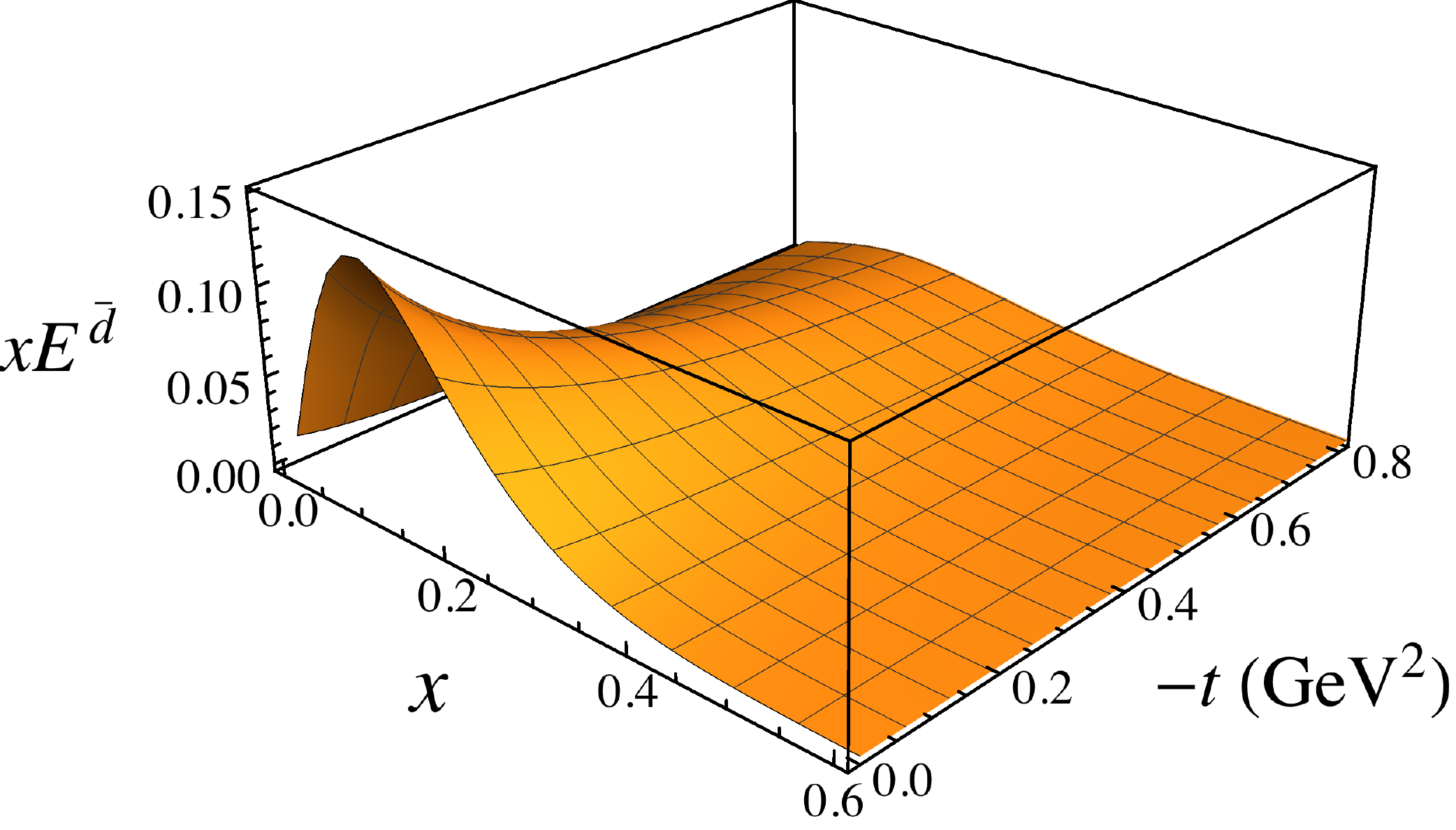}
        }
    \end{minipage}
\caption{The 3D antiquark GPDs $xH^{\bar{q}}$ and $xE^{\bar{q}}$ for $\bar{q}=\bar{u}$ and $\bar d$ versus momentum fraction $x$ and momentum squared transfer $-t$, with $\Lambda=1$~GeV. (Figure from Ref.~\cite{He5}.)}
\label{3dubar}
\end{center}
\end{figure}

With the calculated splitting functions and the valence quark distributions in the bare hadrons as input, we can compute the GPDs of sea quarks in the proton using the convolution forms (\ref{eq:HEqbarconv}) and (\ref{eq:HEqconv}), as described in Ref.~\cite{He5}.
In Fig.~\ref{3dubar} the 3-dimensional plots of the antiquark GPDs $xH^{\bar{q}}$ and $xE^{\bar{q}}$ versus the momentum fraction $x$ and momentum transfer $-t$ are shown for $\bar{q}=\bar{u}$ and $\bar{d}$.
For the $\bar{u}$ distribution, $xH^{\bar{u}}(x,t)$ is positive and peaks at $x \sim 0.1$ for a given $t$, and for fixed $x$ it decreases with increasing $-t$. 
The magnetic GPD $xE^{\bar{u}}(x,t)$ is negative, and peaks at slightly smaller $x$ values compared with $xH^{\bar{u}}$.
For the $\bar d$ flavor, the shape of the $xH^{\bar{d}}$ GPD is similar to that of $xH^{\bar{u}}$, but with a larger magnitude at fixed $x$ and $t$, reflective of the light antiquark flavor asymmetry.
Interestingly, the $xE^{\bar{d}}$ GPD is positive, with absolute value much larger than $xE^{\bar{u}}$.
Note that the $\delta$-function term in the splitting functions gives no contribution to the $xH^{\bar{q}}$ and $xE^{\bar{q}}$ GPDs at nonzero $x$, but contributes to the $x$ integrals of $H^{\bar{q}}$ and $E^{\bar{q}}$.

\begin{figure}[t]
\begin{center}
    \begin{minipage}{0.45\linewidth}
        \centering
        \centerline{
        \includegraphics[width=1\textwidth]{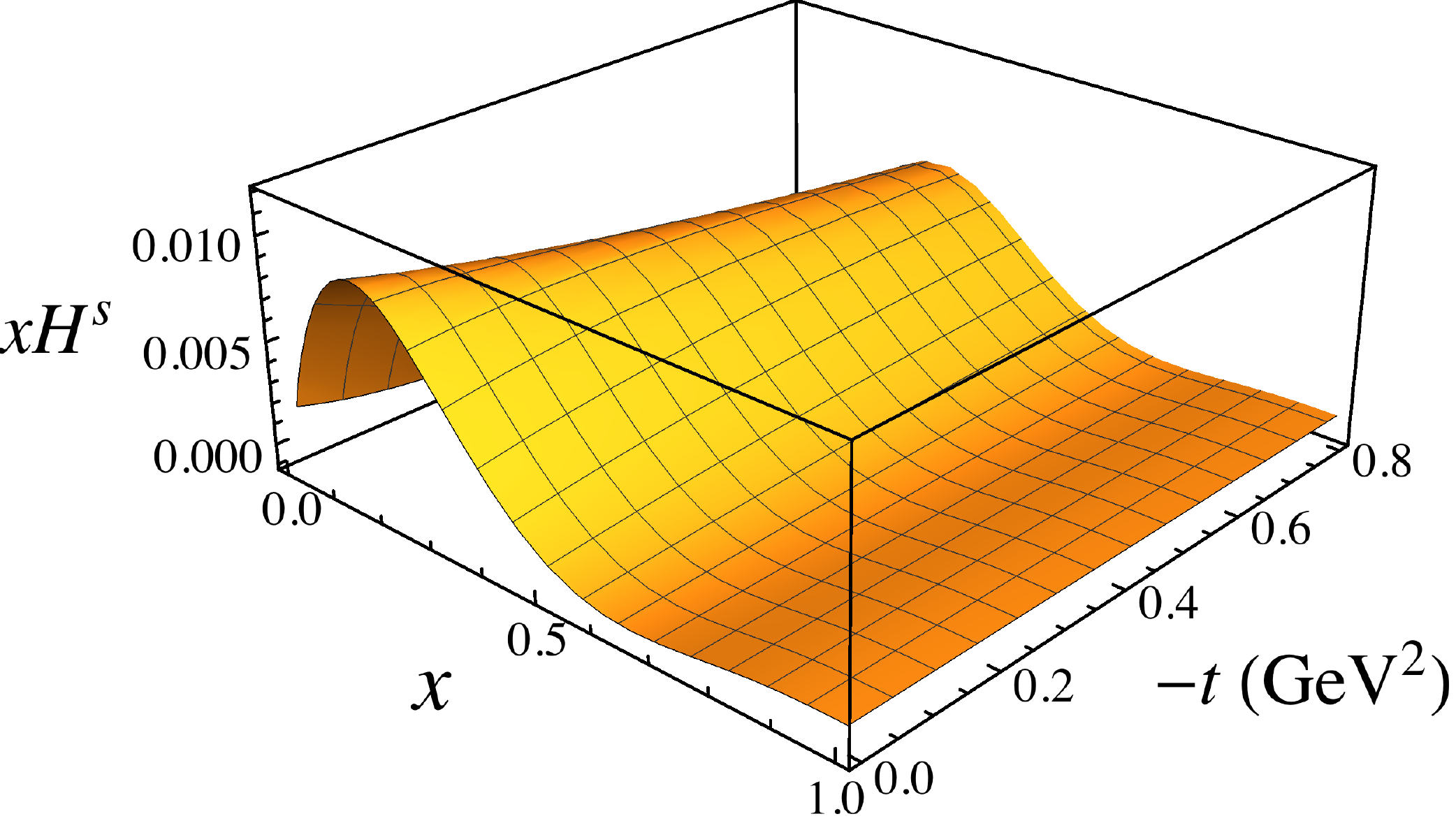}
        }
    \end{minipage}
    \begin{minipage}{0.45\linewidth}
        \centering
        \centerline{
        \includegraphics[width=1\textwidth]{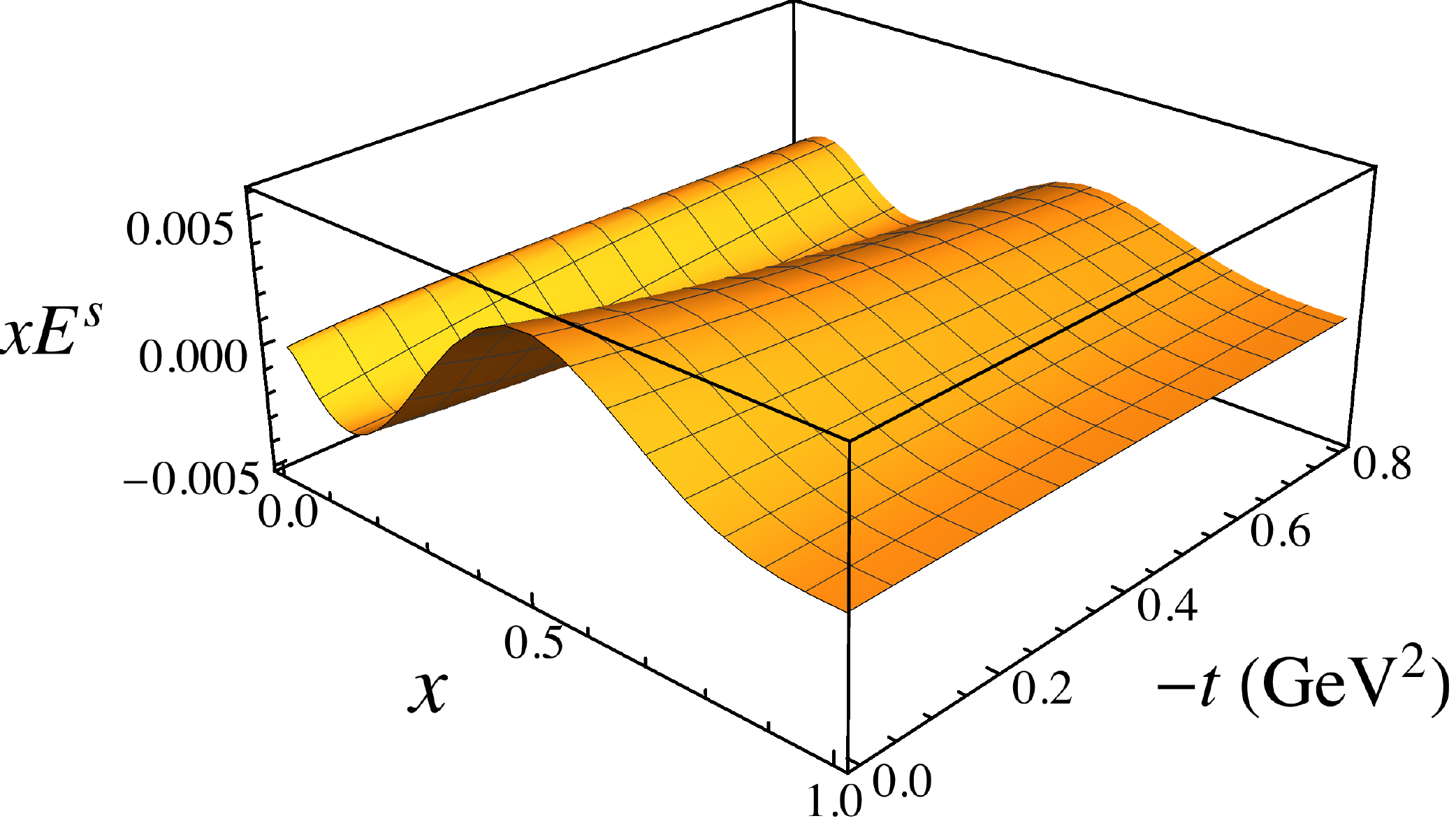}
        }
    \end{minipage}    
\\[0.4cm]
    \begin{minipage}{0.45\linewidth}
        \centering
        \centerline{
        \includegraphics[width=1\textwidth]{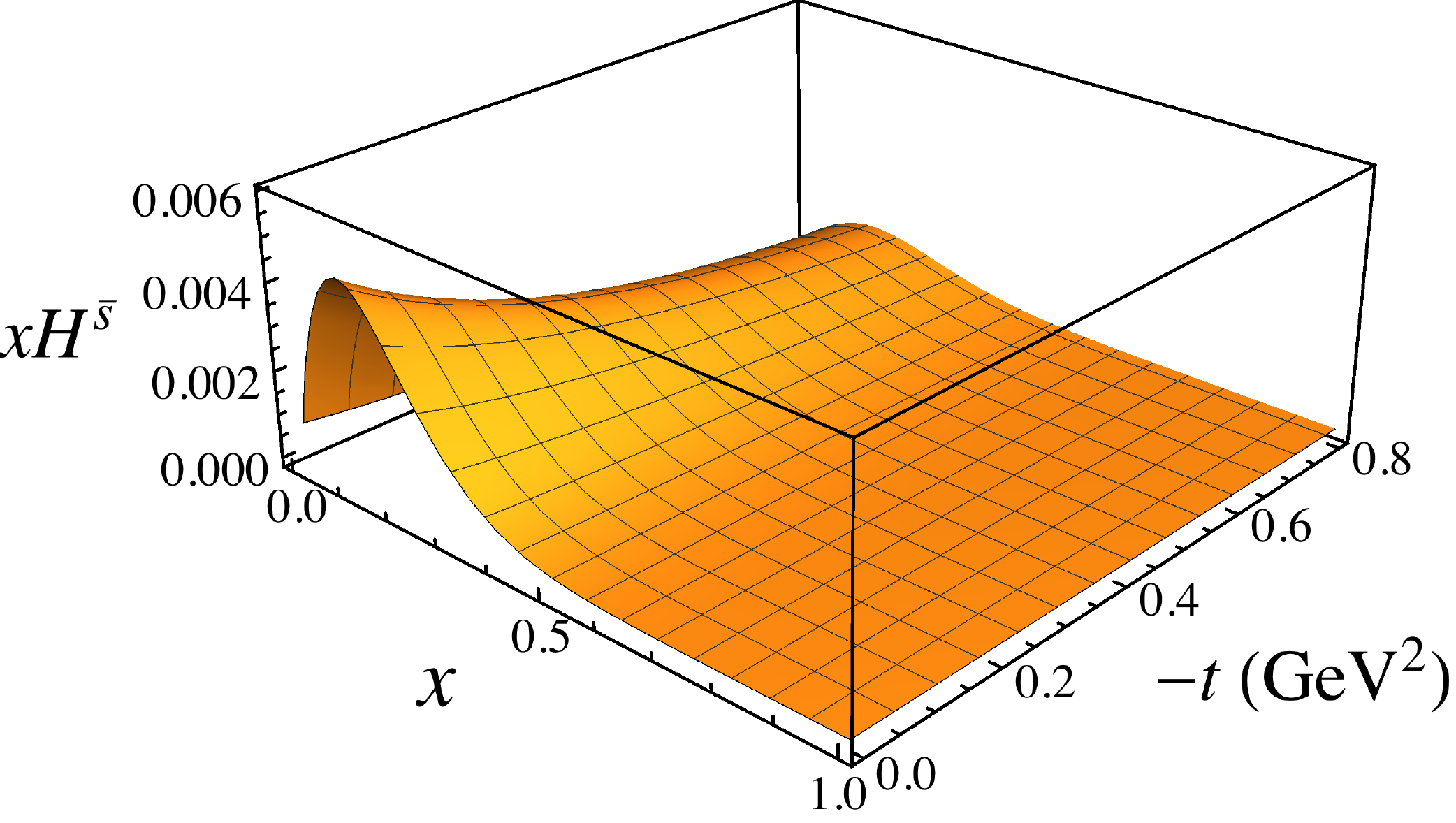}
        }
    \end{minipage}
        \begin{minipage}{0.45\linewidth}
        \centering
        \centerline{
        \includegraphics[width=1\textwidth]{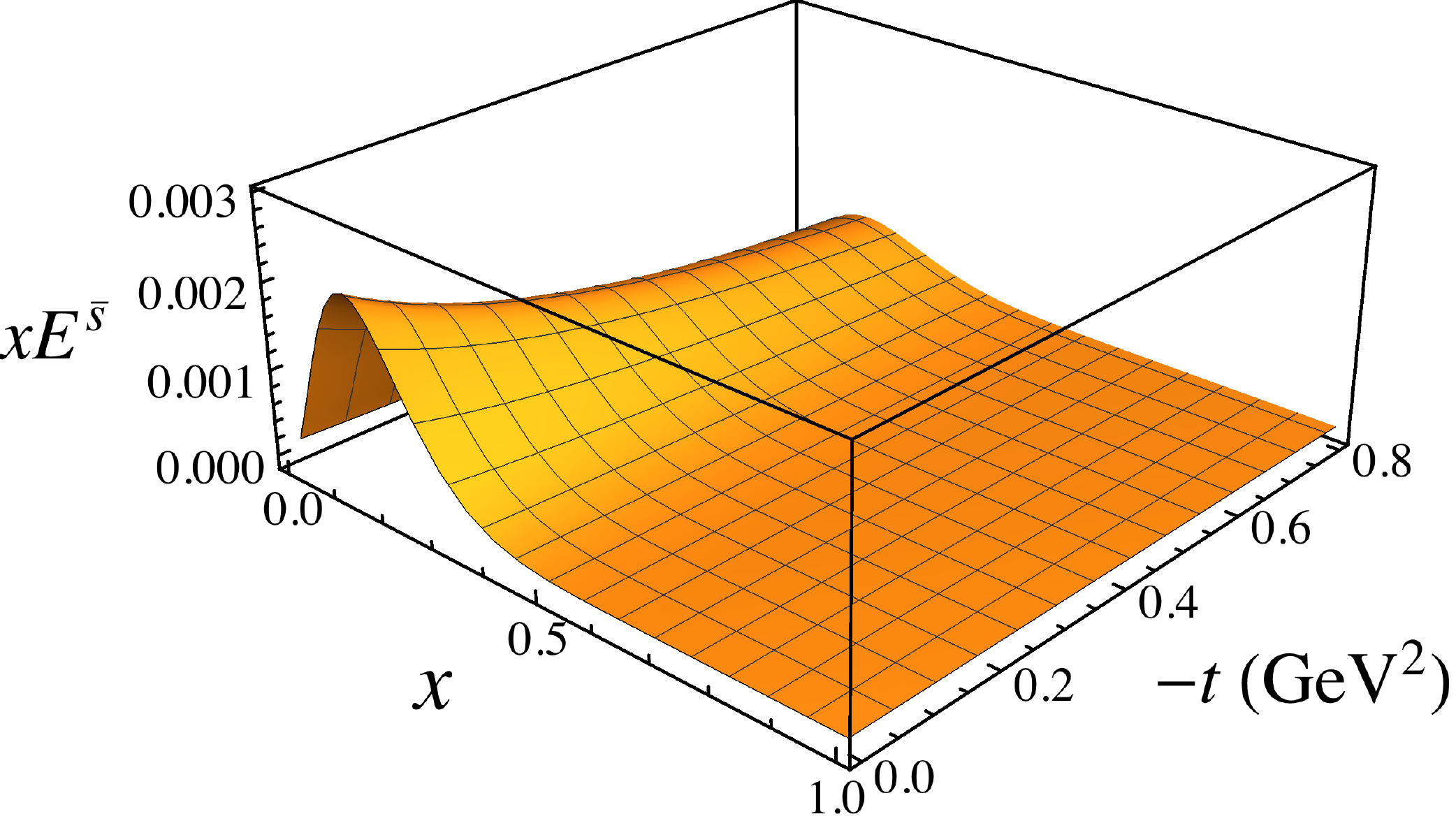}
        }
    \end{minipage}
\caption{The 3D strange and antistrange GPDs $xH^{s,\bar s}$ and $xE^{s,\bar{s}}$ versus momentum fraction $x$ and momentum transfer squared $-t$, with $\Lambda=1$~GeV. (Figure from Ref.~\cite{He5}.)}
\label{3dssbar}
\end{center}
\end{figure}

In Fig.~\ref{3dssbar}, the strange and antistrange GPDs are shown versus $x$ and $t$.
Compared to the light sea quark GPDs, the strange GPDs have significantly smaller magnitude. 
Both the $xH^s$ and $xH^{\bar{s}}$ GPDs are positive, with $xH^s$ larger than $xH^{\bar{s}}$ at small $x$.
When $x$ is large, $x \gtrsim 0.5$, $xH^s$ and $xH^{\bar{s}}$ are close to each other.
However, the $x$ integrals of $H^s$ and $H^{\bar{s}}$ at $t=0$ are the same with the inclusion of the $\delta$-function term, which is consistent with the requirement that the nucleon has zero net strangeness.
In contrast, the behaviors of the $xE^s$ and $xE^{\bar{s}}$ GPDs are quite different.
The sign of $xE^{\bar{s}}$ is the same as that of $xE^{\bar{d}}$, while $xE^s$ is the only one among these sea quark distributions which changes sign with $x$.

\begin{figure}[t] 
\begin{minipage}[b]{.44\linewidth}
\hspace*{0.2cm}\includegraphics[width=1.09\textwidth, height=5.5cm]{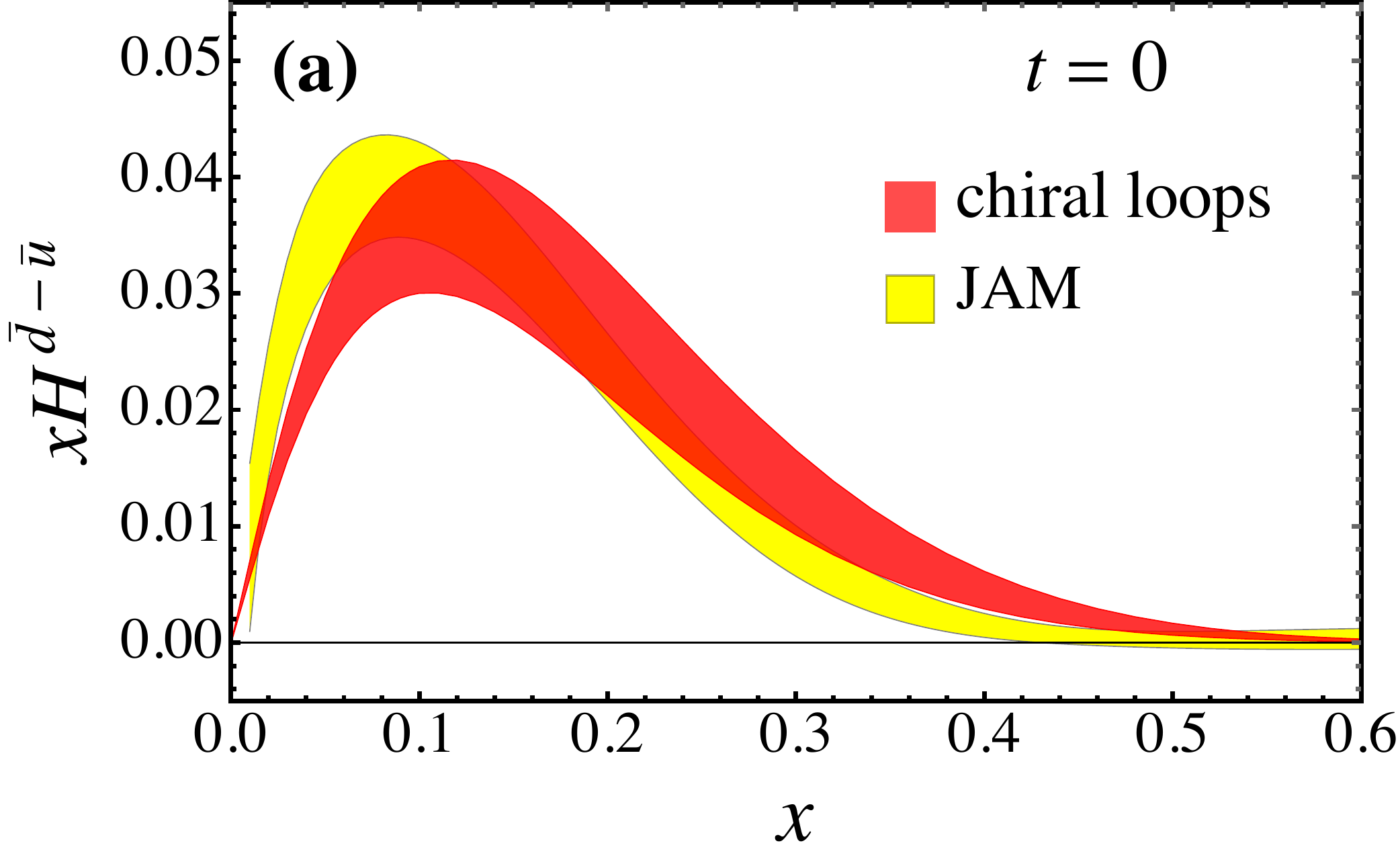}   
 \vspace{0pt}
\end{minipage}
\hfill
\begin{minipage}[b]{.44\linewidth}   
\hspace*{-1.0cm} \includegraphics[width=1.1\textwidth, height=5.5cm]{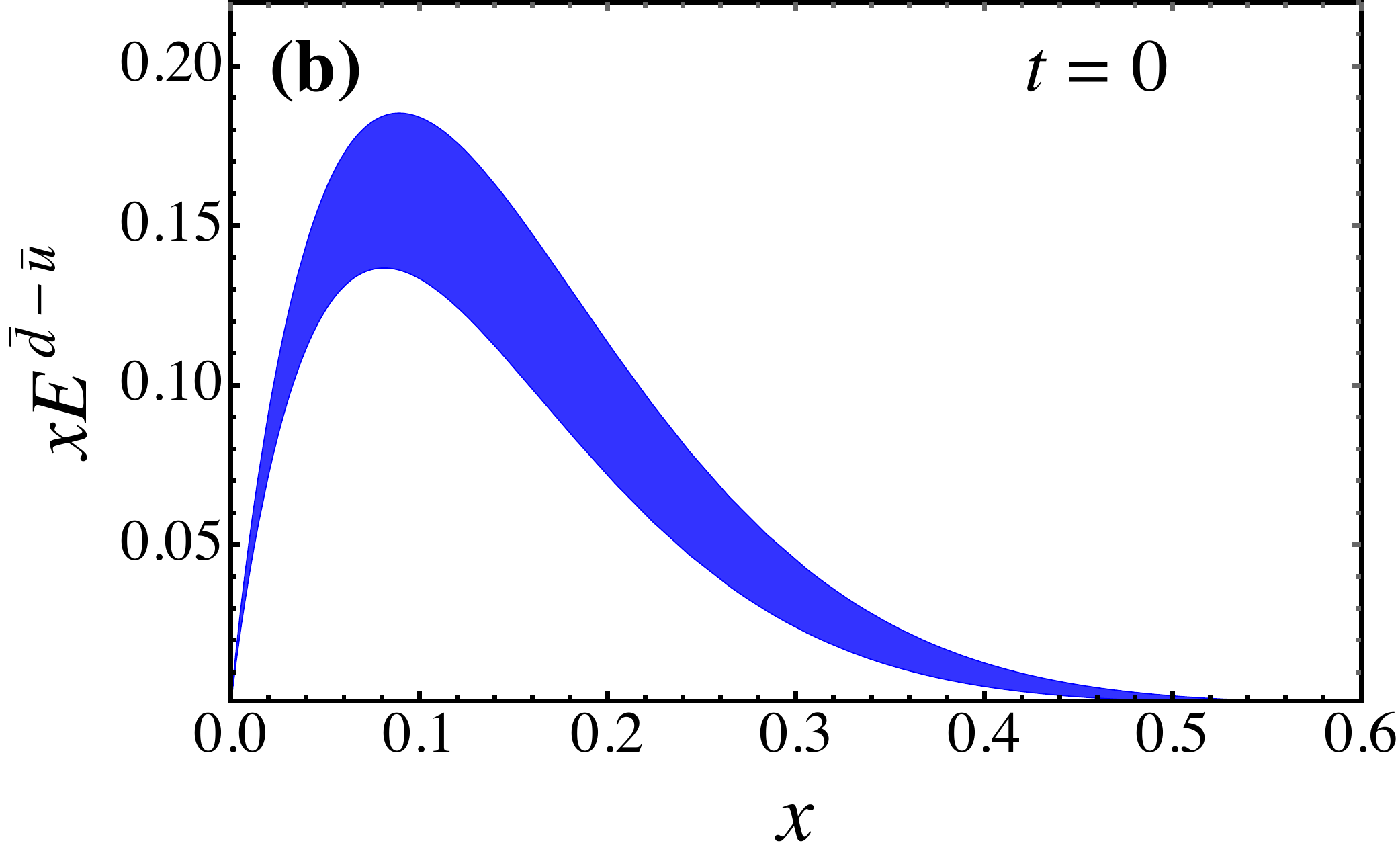} 
   \vspace{0pt}
\end{minipage}  
\\[-0.4cm]
\begin{minipage}[t]{.44\linewidth}
\hspace*{0.1cm}\includegraphics[width=1.11\textwidth, height=5.5cm]{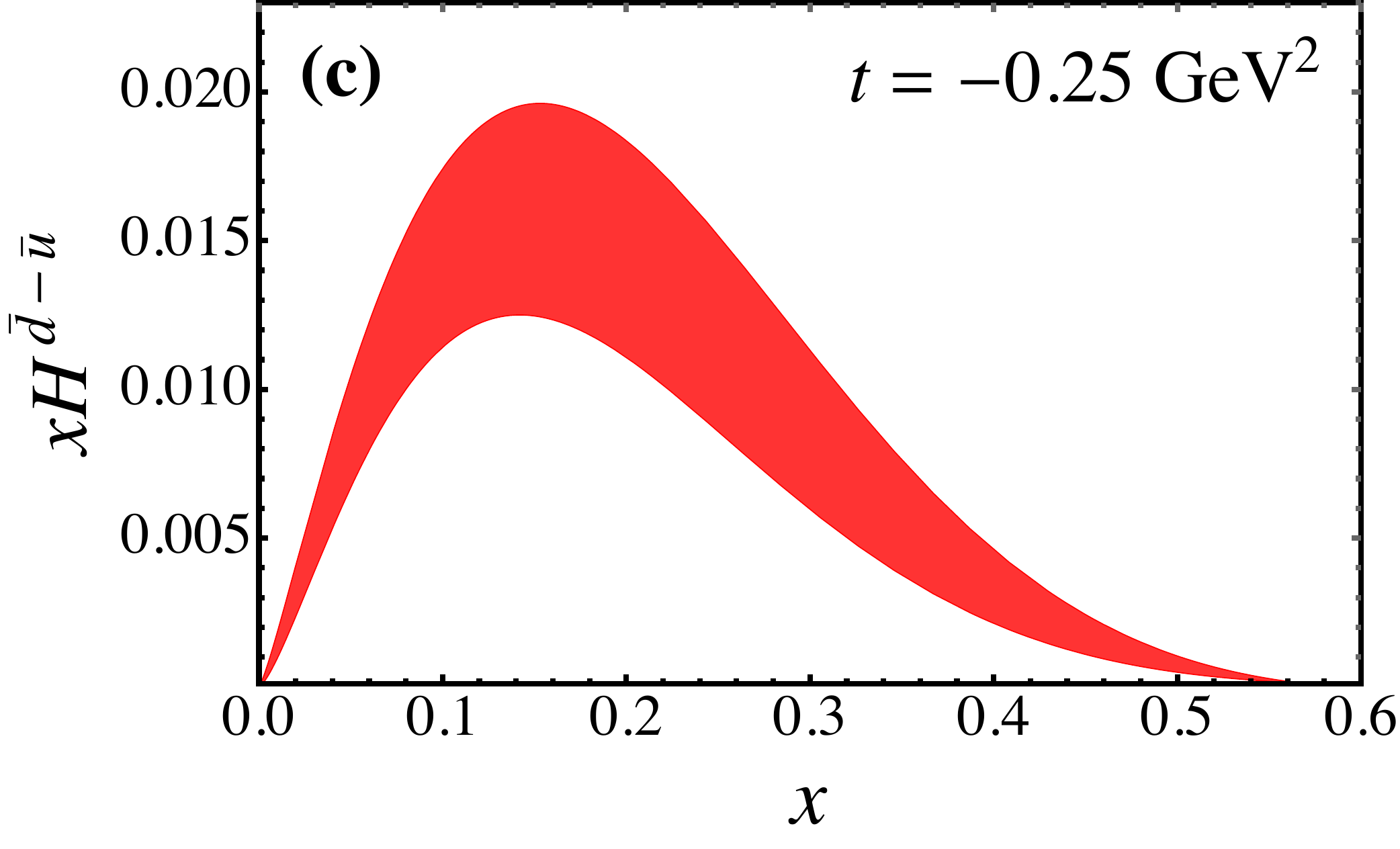}   
 \vspace{0pt}
\end{minipage}
\hfill
\begin{minipage}[t]{.44\linewidth}   
\hspace*{-0.9cm} \includegraphics[width=1.1\textwidth, height=5.5cm]{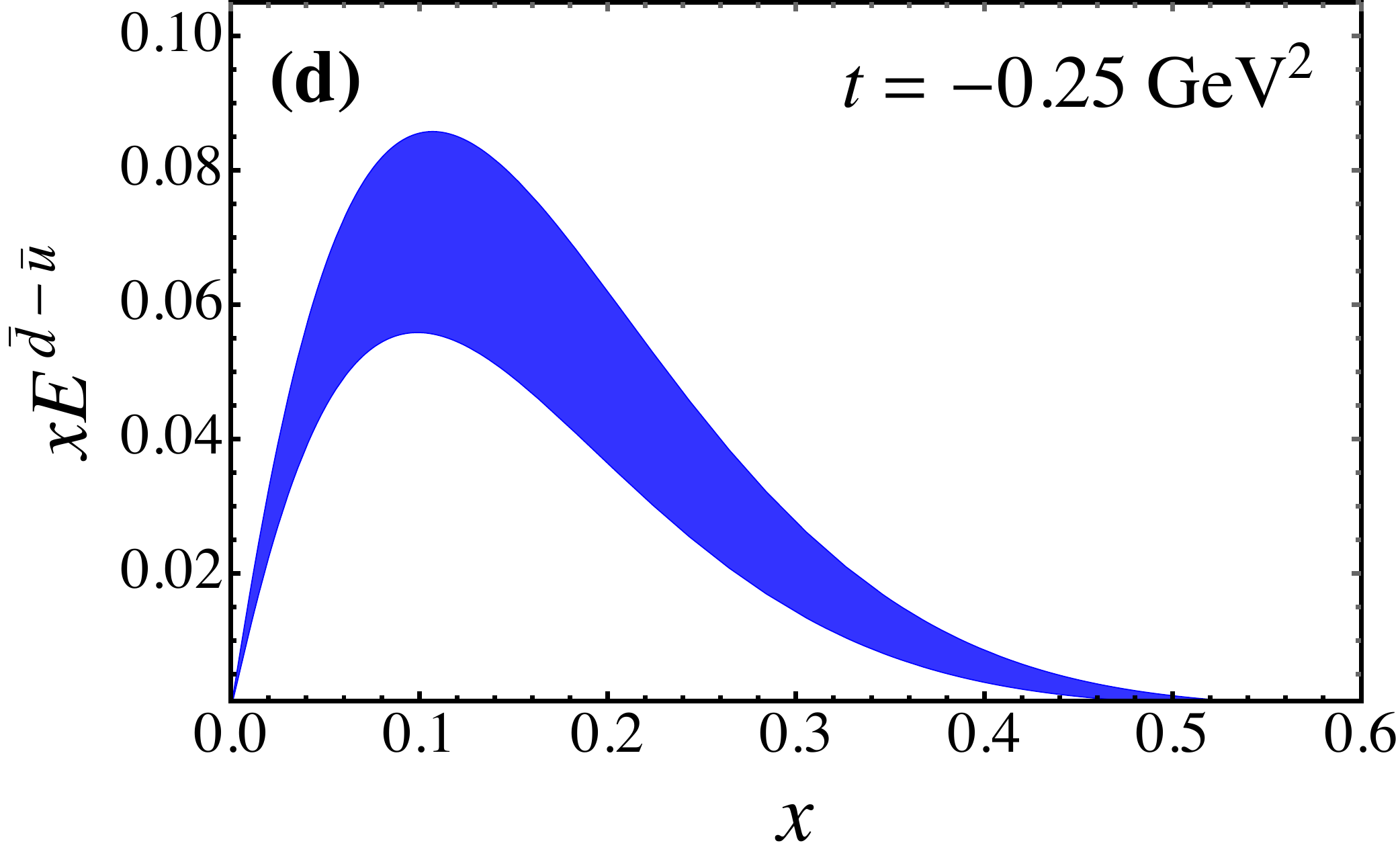}  
   \vspace{0pt}
\end{minipage} 
\vspace*{-0.5cm}
\caption{Light antiquark asymmetries for the electric $xH^{\bar{u}-\bar{d}}$ (red bands) and magnetic $xE^{\bar{u}-\bar{d}}$ (blue bands) GPDs versus  $x$ at $t=0$ [{\bf (a)}, {\bf (b)}] and $t=-0.25$~GeV$^2$ [{\bf (c)}, {\bf (d)}], for cutoff parameter $\Lambda = 1.0(1)$~GeV. The asymmetries are shown at the input scale $Q_0=1$~GeV, except for the electric asymmetry at $t=0$, which is compared with the $x(\bar d-\bar u)$ PDF asymmetry from the JAM global QCD analysis~\cite{Cocuzza} (yellow band) evolved to the scale $Q_0=m_c$. (Figure from Ref.~\cite{He5}.)}
\label{2dd-ut0}
\end{figure}

To more clearly visualize the $\bar{d}-\bar{u}$ asymmetry, in Fig.~\ref{2dd-ut0}(a) and (b) we plot $xH^{\bar{d}-\bar{u}}$ and $xE^{\bar{d}-\bar{u}}$ at $t=0$ for $\Lambda=1.0(1)$~GeV.
The result for $xH^{\bar{d}-\bar{u}}$ is compared with that from the JAM global QCD analysis \cite{Cocuzza}, showing relatively good agreement with the phenomenological values.
For the $x$-integrated asymmetry we find 
    $\int_0^1 \dd{x} H^{\bar{d}-\bar{u}}(x,0) = 0.11(2)$ 
for $\Lambda=1.0(1)$~GeV.
For the $xE^{\bar{d}-\bar{u}}$ asymmetry, although the shape is similar, the magnitude is about 4 times larger than for $xH^{\bar{d}-\bar{u}}$, which suggests it may be a good physical quantity to explore experimentally.
The $xH^{\bar{d}-\bar{u}}$ and $xE^{\bar{d}-\bar{u}}$ asymmetries at $-t=0.25$~GeV$^2$ are shown in Fig.~\ref{2dd-ut0}(c) and (d).
Both of these have smaller magnitude than at zero momentum transfer, as expected, with the peaks slightly shifted to higher $x$.

\begin{figure}[tb]
\begin{minipage}[b]{.445\linewidth}
\hspace*{-0.4cm}\includegraphics[width=1.1\textwidth, height=5.5cm]{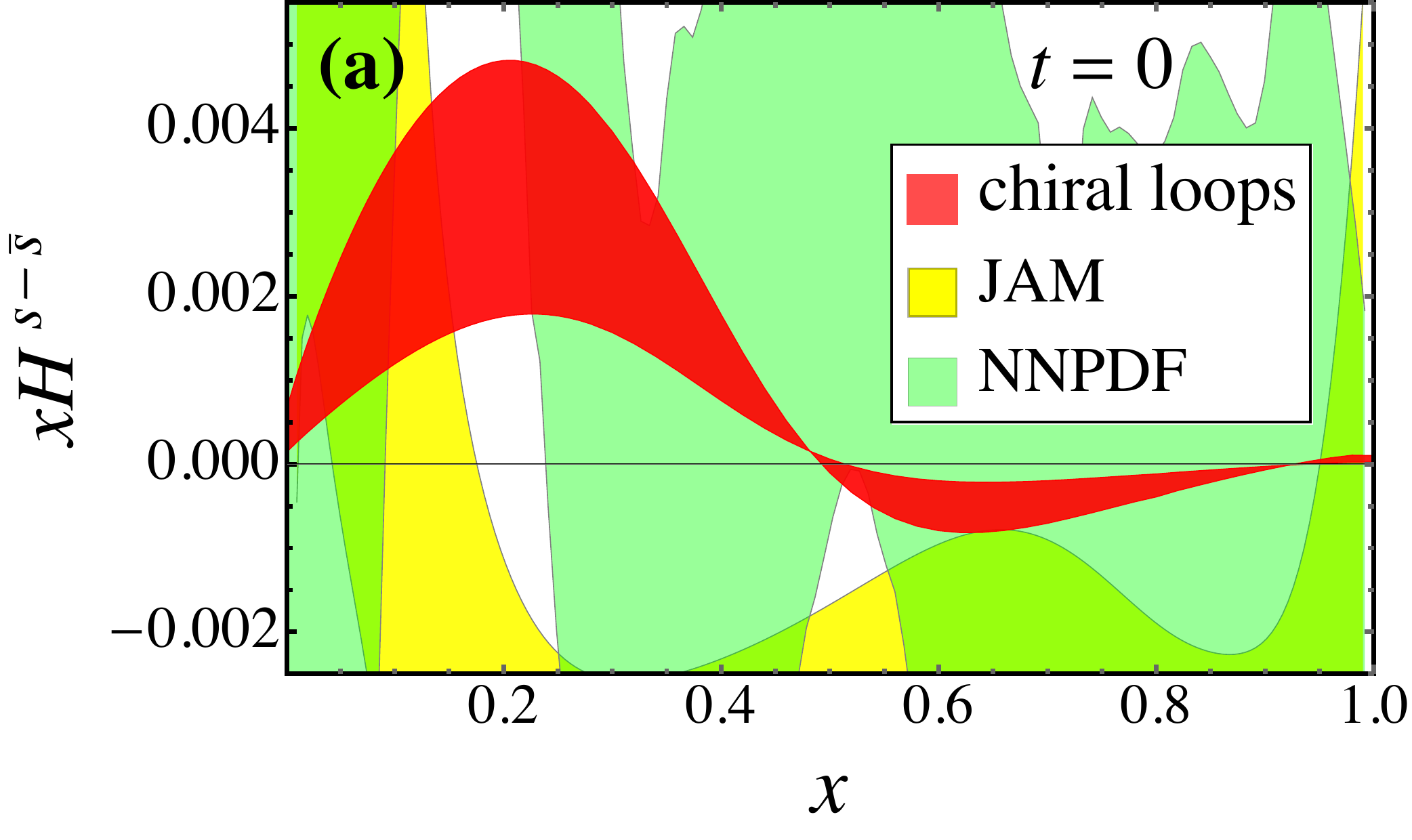}   
 \vspace{0pt}
\end{minipage}
\hfill
\begin{minipage}[b]{.443\linewidth}   
\hspace*{-1.1cm} \includegraphics[width=1.1\textwidth, height=5.5cm]{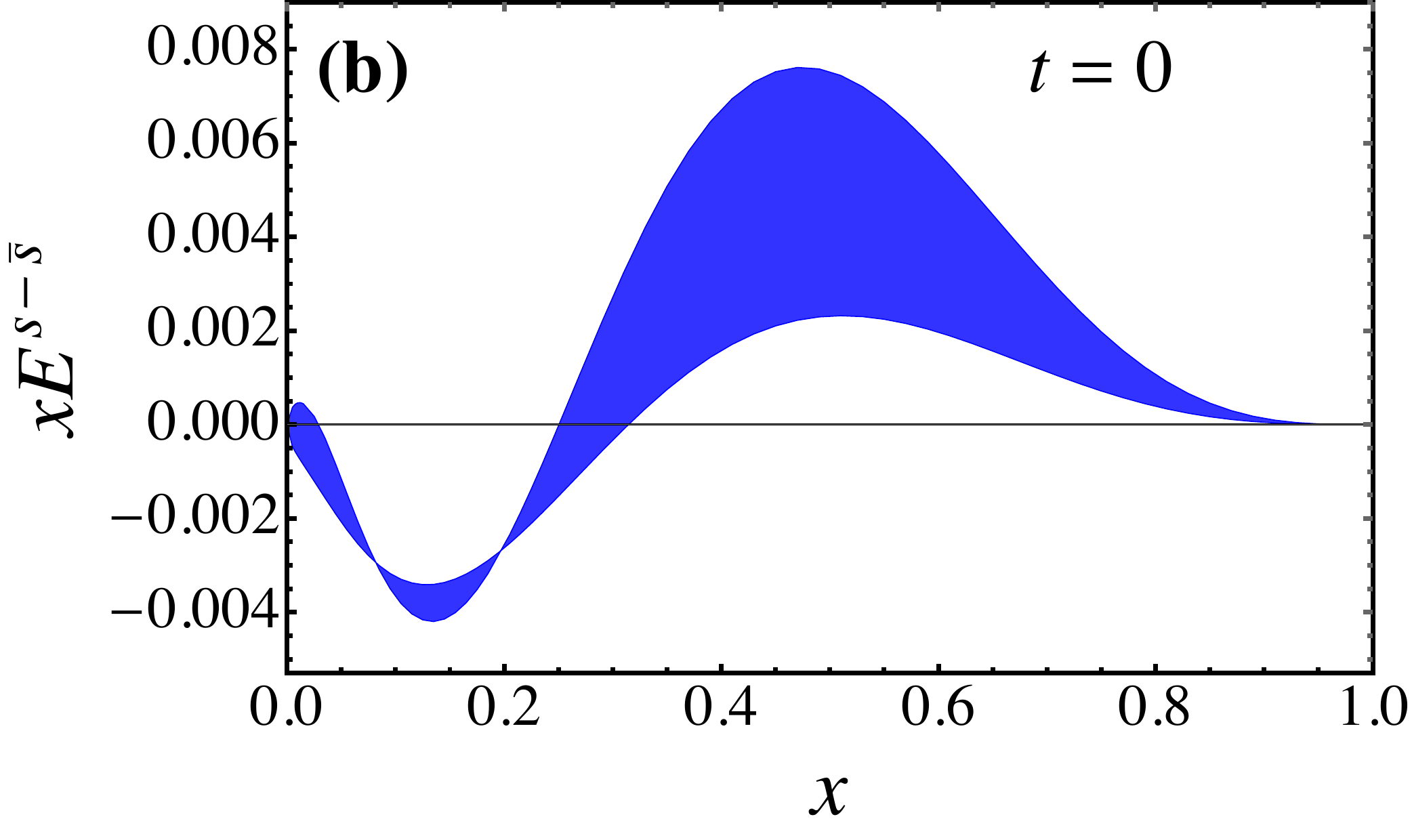} 
   \vspace{13pt}
\end{minipage}  
\\[-0.4cm]
\begin{minipage}[t]{.44\linewidth}
\hspace*{-0.1cm}\includegraphics[width=1.08\textwidth, height=5.5cm]{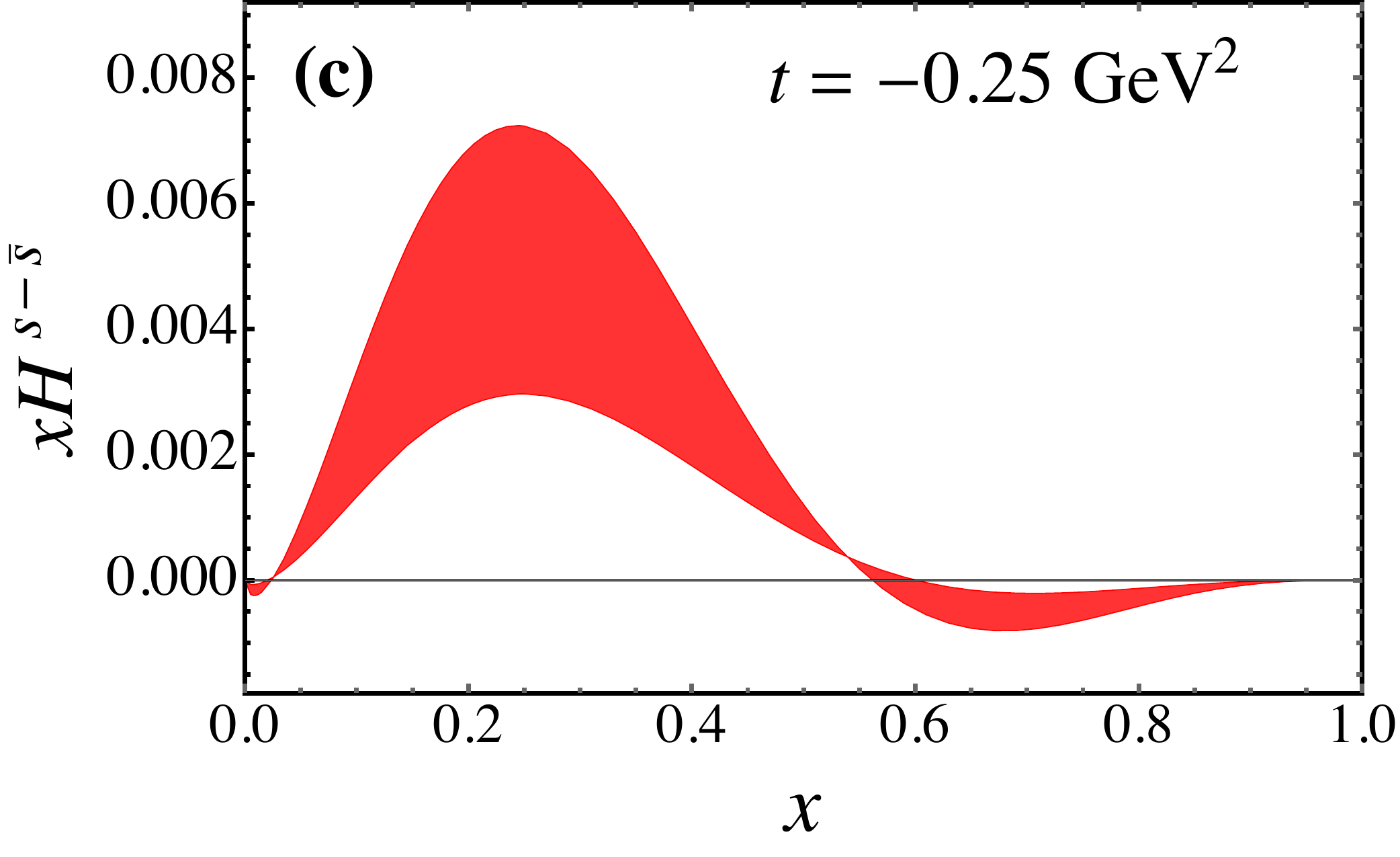}   
 \vspace{0pt}
\end{minipage}
\hfill
\begin{minipage}[t]{.45\linewidth}   
\hspace*{-1.05cm} \includegraphics[width=1.1\textwidth, height=5.5cm]{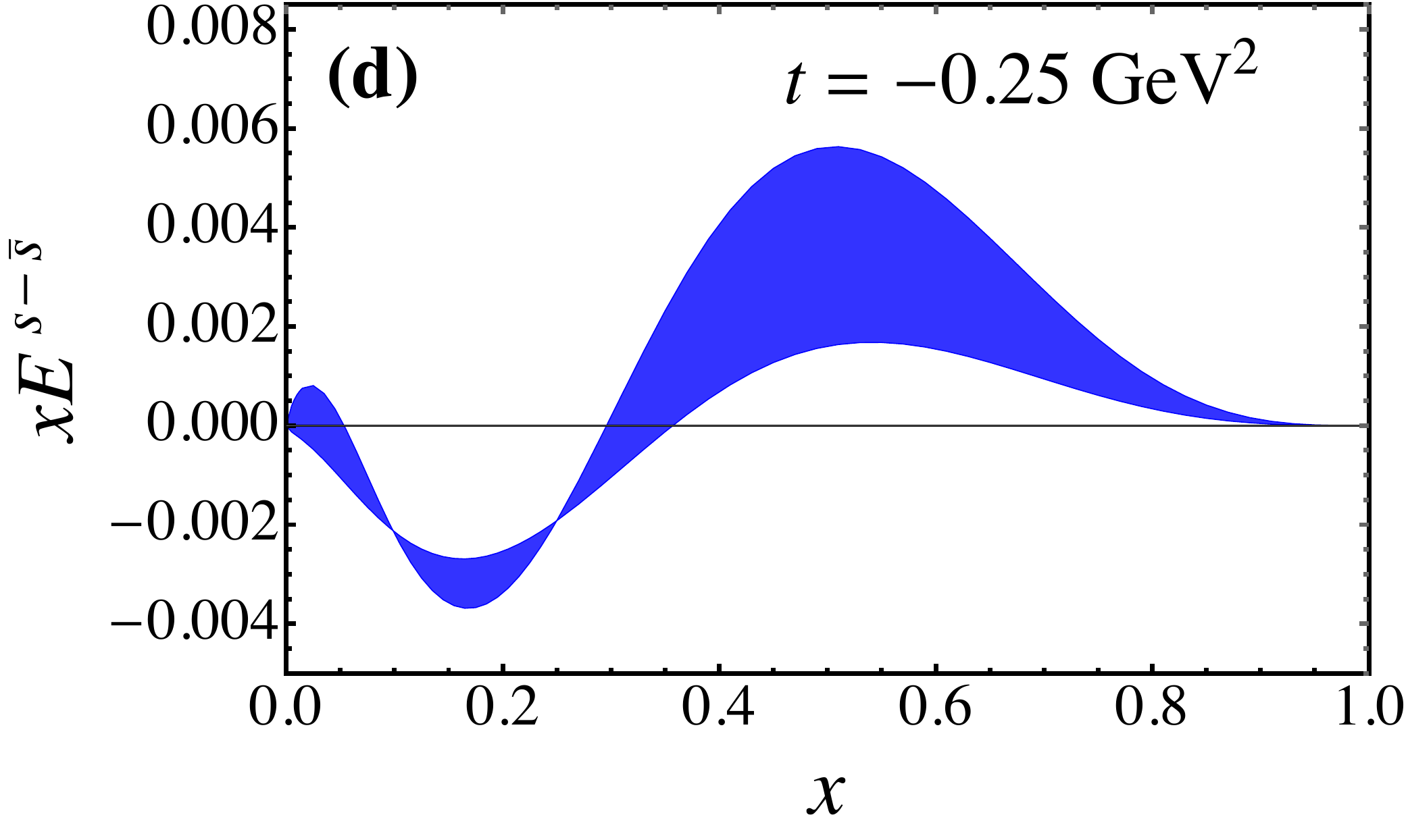}  
   \vspace{-8pt}
\end{minipage} 
\vspace*{-0.5cm}
\caption{Strange quark asymmetry for the $xH^{s-\bar{s}}$ (red bands) and $xE^{s-\bar{s}}$ (blue bands) GPDs versus $x$ at squared momentum transfers $t=0$ [{\bf (a)}, {\bf (b)}] and $-t = 0.25$~GeV$^2$ [{\bf (c)}, {\bf (d)}], with the bands corresponding to cutoff mass $\Lambda=1.0(1)$~GeV. The asymmetries are shown at the input scale $Q_0=1$~GeV, except for the electric asymmetry at $t=0$, which is compared with PDF parametrizations of $x(s-\bar{s})$ from JAM \cite{Cocuzza} (yellow band) and NNPDF \cite{Faura} (green band) evolved to $Q_0=m_c$. (Figure from Ref.~\cite{He5}.)}
\label{2ds-sbart0}
\end{figure}

The strange asymmetries $xH^{s-\bar{s}}$ and $xE^{s-\bar{s}}$ are shown in Fig.~\ref{2ds-sbart0}(a) and (b) at $t=0$ for $\Lambda=1.0(1)$~GeV.
At small $x$, the $xH^{s-\bar{s}}$ asymmetry is positive, but becomes negative at $x \gtrsim 0.5$.
The first moment of the strange asymmetry is computed to be
    $\int_0^1 \dd{x} xH^{s-\bar{s}}(x,0) = 0.0009^{(5)}_{(4)}$ 
for $\Lambda=1.0(1)$~GeV, which is comparable to recent estimates from Refs.~\cite{Bentz, Salamu2, XGWang}.
For $xE^{s-\bar{s}}$, the situation is opposite, with the asymmetry negative at small $x$ and positive for $x \gtrsim 0.3$.
The analogous integrated magnetic asymmetry is
    $\int_0^1 \dd{x} xE^{s-\bar{s}}(x,0) = 0.0009^{(12)}_{(8)}$,
while for the strange quark contribution to the proton's magnetic moment, one has
    $\int_0^1 \dd{x} E^{s-\bar{s}}(x,0) = \mu_s = -0.033^{(11)}_{(13)}$.
At $-t=0.25$~GeV$^2$ the strange asymmetry in Fig.~\ref{2ds-sbart0}(c) and (d) shows that $xE^{s-\bar{s}}$ is slightly smaller while $xH^{s-\bar{s}}$ is larger than the corresponding asymmetries at $t=0$.

\subsection{\it Transverse momentum dependent distributions}

In recent years the transverse partonic structure of hadrons has been the subject of growing theoretical and experimental investigations. 
The transverse momentum dependent parton distributions are of great interest, since they offer insight in the three-dimensional structure of hadrons \cite{Pisano}, complementary to that afforded through PDFs and GPDs.
At leading twist, there is a total of eight TMDs, among them the time-reversal odd Boer-Mulders function \cite{Boer} and the Sivers distribution \cite{Sivers}.
The Sivers function describes the asymmetric distribution of unpolarized quarks in a transversely polarized hadron, and has been studied in semi-inclusive DIS by the HERMES and COMPASS collaborations \cite{Airapetian:2009ae, Bradamante}.
Understanding this function is essential for explaining the single-spin asymmetries (SSAs) that were observed in semi-inclusive DIS experiments some time ago \cite{Sivers, Bravar, Airapetian:2001eg}.

In this section, we review the Sivers functions for the $\bar{u}$ and $\bar{d}$ distributions in the proton within effective field theory, as recently discussed in Ref.~\cite{He3}.
For a quark flavor $q$, according to the Trento convention, the unpolarized  $f^q_1(x,\bm{k}_\perp)$ and Sivers $f^q_{1T}(x,\bm{k}_\perp)$ TMD distributions can be written as \cite{Bacchetta:2004jz}
\begin{eqnarray}\label{Siversq} 
f_1^q(x,\bm{k}_\perp)
+ \frac{\epsilon^{ji}k_\perp^iS_\perp^j}{M_N}f_{1T}^{q}(x,\bm{k}_\perp)
= \frac12\int\!\frac{\dd\xi^-\dd^2\bm{\xi}_\perp}{(2\pi)^3}\,
 e^{-ixP^+\xi^-+i\bm{k}_\perp\!\cdot\, \bm{\xi}_\perp}\,
 \langle P,\bm{S}_\perp\mid\mathcal{O}^q\mid P,\bm{S}_\perp \rangle,
\end{eqnarray}
where $\bm{S}_\perp$ is the transverse spin vector of the proton. 
The gauge invariant bilocal operator $\mathcal{O}^q$ is defined as \cite{Ji:2002aa}
\begin{eqnarray}\label{inv}
\mathcal{O}^q=\bar{q}(\xi^-,\bm{\xi}_\perp)\mathcal{L}_{\xi_\perp}^{\dag}(\infty,\xi^-)\gamma^+
\mathcal{L}_0(\infty,0)q(0,0),
\end{eqnarray}
where the path-ordered light-cone color gauge link is given by
\begin{eqnarray}
\mathcal{L}_{\xi_\perp}^{\dag}(\infty,\xi^-)
= {\cal P} \exp\Big[ -ig_c\int_{\xi^-}^\infty\dd z^- A^+(z^-,\bm{\xi}_\perp \Big].
\end{eqnarray}

In analogy with the case of PDFs, the contribution to the antiquark TMD in the proton from pion loops can be written as a convolution of a hadronic TMD splitting function and a valence TMD in the pion,
\bea
\bar{q}(x,\bm{k}_\perp)
= \big[ f_\pi \otimes \bar{q}_\pi^v \big](x,\bm{k}_\perp) 
&\equiv& 
\int_0^1 \dd y 
\int_0^1 \dd z 
\int \dd^2\bm{k}_{\perp\pi}
\int \dd^2\bm{k}_\perp^{\rm in}\
\delta(x-yz)\,
\delta \big(\bm{k}_\perp^{\rm in}-(\bm{k}_\perp-z\bm{k}_{\perp\pi})\big)
~~~~
\nonumber \\ 
&& \hspace*{5.1cm} \times\,
f_\pi(y,\bm{k}_{\perp\pi})\,
\bar{q}_\pi^v(z,\bm{k}_\perp^{\rm in}),
\label{eq:convqbarT}
\eea
where $\bar{q}_\pi^{v}(z,\bm{k}_\perp^{\rm in})$ is the unpolarized valence quark TMD in the pion with intrinsic transverse momentum
$\bm{k}_\perp-z\bm{k}_{\perp\pi}$.
Integrating over $\bm{k}_\perp$ on both sides of Eq.~(\ref{eq:convqbarT}) will lead to the usual convolution form for PDFs, as in Eq.~(\ref{eq:conv}).
The splitting function $f_\pi(y,\bm{k}_{\perp\pi})$ is calculated from the matrix element of the twist-two hadronic operator, similar to that in Eq.~(\ref{eq:splitting}),
\be
f_\pi(y,\bm{k}_{\perp\pi})\,
\bar{u}(p)\gamma^+ u
= \int \dd^4 k\, \widetilde{\Gamma}(k)\,
\delta\Big(y-\frac{k^+}{p^+}\Big)\,
\delta(\bm{k}_\perp - \bm{k}_{\perp\pi}).
\ee

For the splitting function relevant to the Sivers distribution, the vector operator 
\be
{\cal O}^{\pi} =  i \left[ \pi^- \partial^+ \pi^+ - \pi^+ \partial^+ \pi^- \right]
\ee
makes no contribution.
To obtain a nonzero splitting function, the locally SU(2) invariant current is introduced by including a $\rho$-meson field explicitly \cite{He3},
\be
\label{gl}
\mathcal{O}^{\pi^+}_{\rm Sivers}
= g_{\rho\pi\pi} 
\Big[ \pi^-(y^-,\bm{y}_\perp)\, \partial^+\pi^+(0)
    - \pi^+(0)\, \partial^+ \pi^-(y^-,\bm{y}_\perp)
\Big] 
\left\{ \int_0^\infty \dd z^-\rho^{0+}(z^-,0)
      + \int_\infty^{y^-} \dd z^-\rho^{0+}(z^-,\bm{y}_\perp)
\right\}.
\ee
The splitting function for the Sivers distribution, or the Sivers distribution function of a pion in the proton, $f_{1T}^{\pi/p}(z,\bm{k}_{\perp\pi})$, can be written as
\begin{eqnarray}\label{Siverpi}
\frac{\epsilon^{ji}k_{\perp\pi}^iS_\perp^j}{M_N}
f_{1T}^{\pi/p}(z,\bm{k}_{\perp\pi})\,
= \frac12\int\,\frac{\dd y^-\dd^2\bm{y}_\perp}{(2\pi)^3}\,
e^{-i(z p^+ y^ - -\bm{y}_\perp\cdot\,\bm{k}_{\perp\pi})} 
\langle p,\bm{S}_\perp
\mid \mathcal{O}^\pi_{\rm Sivers} \mid 
p,\bm{S}_\perp\rangle.
\end{eqnarray}
To calculate the splitting function, in addition to the chiral effective Lagrangian, the interactions between baryons and $\rho$ mesons are necessary. 
The Lagrangians for the $NN\rho$, $\Delta \Delta \rho$ and $N\Delta \rho$ interactions are given by \cite{He3,Jenkins3,Geng,Jones3}
\begin{subequations}
\begin{eqnarray}\label{rD}
& &\hspace*{-0.8cm} 
\mathcal{L}_{\rho NN}
= -g\,\overline{N}\,
\Big( \gamma^\mu - \kappa_N\frac{\sigma^{\mu\nu}\partial_\nu}{2M_N}
\Big)\,
\bm{\rho}_\mu \cdot \bm{\tau}\, N,
\\
& &\hspace*{-0.8cm}
\mathcal{L}_{\rho\Delta\Delta}
= -g\,\overline{\Delta}_\alpha\,
\Big( \gamma^{\alpha\beta\mu}
    + g^{\alpha\beta} \kappa_\Delta \frac{\sigma^{\mu\nu}\partial_\nu}{2M_\Delta}
\Big)\,
\bm{\rho}_\mu \cdot \bm{\Sigma}\, \Delta_\beta,
\\
& &\hspace*{-0.8cm} 
\mathcal{L}_{\rho N\Delta}
= -i\frac{g\, G^M_{N\Delta}}{2M_N}\,
\overline{N} \gamma^\mu \gamma^5 
\big( \partial_\mu\bm{\rho}_\nu - \partial_\nu\bm{\rho}_\mu
\big)
\cdot\bm{I}\, \Delta^\nu
+ {\rm H.c.},
\end{eqnarray}
\end{subequations}
where $1+\kappa_\Delta = (3/5) (1+\kappa_N)$ and $G^M_{N\Delta} = (6\sqrt{2}/5) (1+\kappa_N)$ according to the quark model \cite{Brown}, and the value of the coupling $\kappa_N$ is taken as $\kappa_N=6.1(2)$~\cite{Mergell}.
The operators $\bm{\Sigma}$ and $\bm{I}$ are the isospin-3/2 and isospin transition matrices \cite{Haidenbauer}.

\begin{figure}[t]
\begin{center}
\includegraphics[scale=0.66]{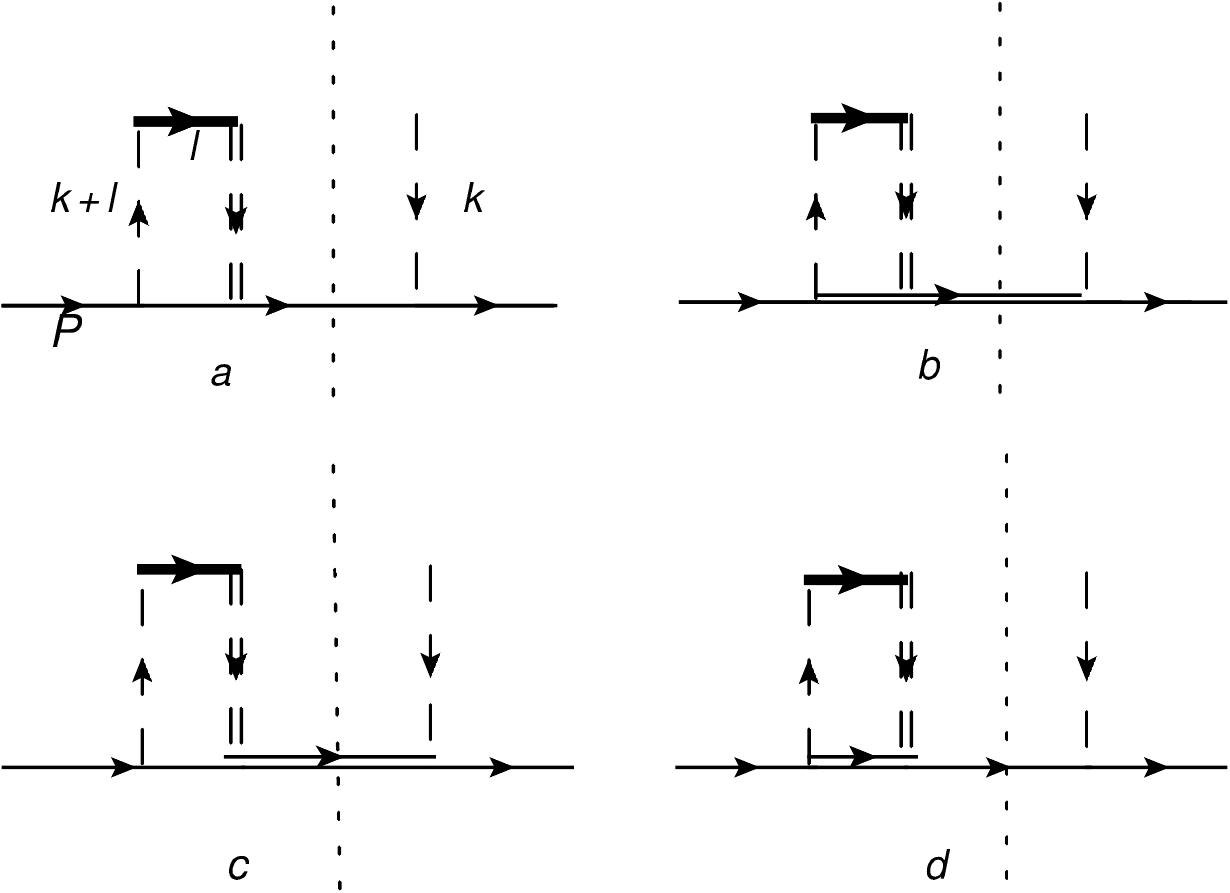}
\vspace*{-0.3cm}
\caption{Diagrams for the Sivers function for pseudoscalar mesons in the nucleon. The octet baryons (solid lines), pseudoscalar mesons (dashed lines), vector mesons (double-dashed lines) and decuplet baryons (double solid lines), as well as the eikonal propagator (thick solid lines) and the on-shell cut (dotted lines), are as indicated. (Figure from Ref.~\cite{He3}.)}
\label{fig:SiversFeynman}
\end{center}
\end{figure}

The leading loop diagrams are illustrated in Fig.~\ref{fig:SiversFeynman}.
For the intermediate octet baryons, the contribution to the Sivers splitting function $f_{1T}^{\pi/p}(z,\bm{k}_{\perp\pi})$ is written as
\begin{eqnarray}\label{eq:octetsivers}
\frac{\epsilon^{ji} k_{\perp\pi}^iS_\perp^j}{M_N}
f_{1T}^{\pi/p}(z,\bm{k}_{\perp\pi})
&=& \frac{ig^2 g_A^2}{4f^2}
\int\!\frac{\dd^4k}{(2\pi)^4}
\int\!\frac{\dd^4l}{(2\pi)^4}\,
\bar{u}(p,\bm{S}_\perp)
\slashed{k} \gamma^5 S_{\text{on}}(p-k)\,
V_\mu(l)\, S(p-k-l) \gamma^5   
\nonumber\\
& & \hspace*{-4cm} \times 
(\slashed{k}+\slashed{l})\,
u(p,\bm{S}_\perp)
\frac{(2k^++l^+)}{(l^++i\epsilon)}
S_\pi(k)\, S_\rho^{\mu+}(l)\, S_\pi(k+l)\,
\delta(k^+-zp^+)\,
\delta^{(2)}(\bm{k}_\perp-\bm{k}_{\perp\pi})
+ {\rm H.c.},
\end{eqnarray}
where $V_\mu(l)$ is the vertex of the interaction between nucleon and $\rho$ meson,
    $V_\mu(l) = \gamma_\mu + i\kappa_N \sigma_{\mu\nu} l^\nu/(2M_N)$. 
The functions $S$, $S_\pi$ and $S_\rho^{\mu+}$ are the nucleon, $\pi$ and $\rho$-meson propagators, respectively, while $S_{\text{on}}$ is the on-shell nucleon propagator factor, 
    $S_{\text{on}}(k) = 2\pi(\slashed{k} + M_N)\, \delta(k^2-M_N^2)$.
The imaginary part of the eikonal propagator $1/(l^+ + i\epsilon)$ gives the real Sivers distribution function of a pion in the nucleon.
The expressions for the other diagrams with decuplet intermediate states are similar but more complicated. 
With the derived splitting functions and valence $\bar{q}$ distribution in the $\pi$ \cite{Aicher}, the Sivers function for the $\bar{u}$ and $\bar{d}$ TMDs in the proton can be obtained.

\begin{figure}[h]
\begin{center}
\includegraphics[scale=0.4]{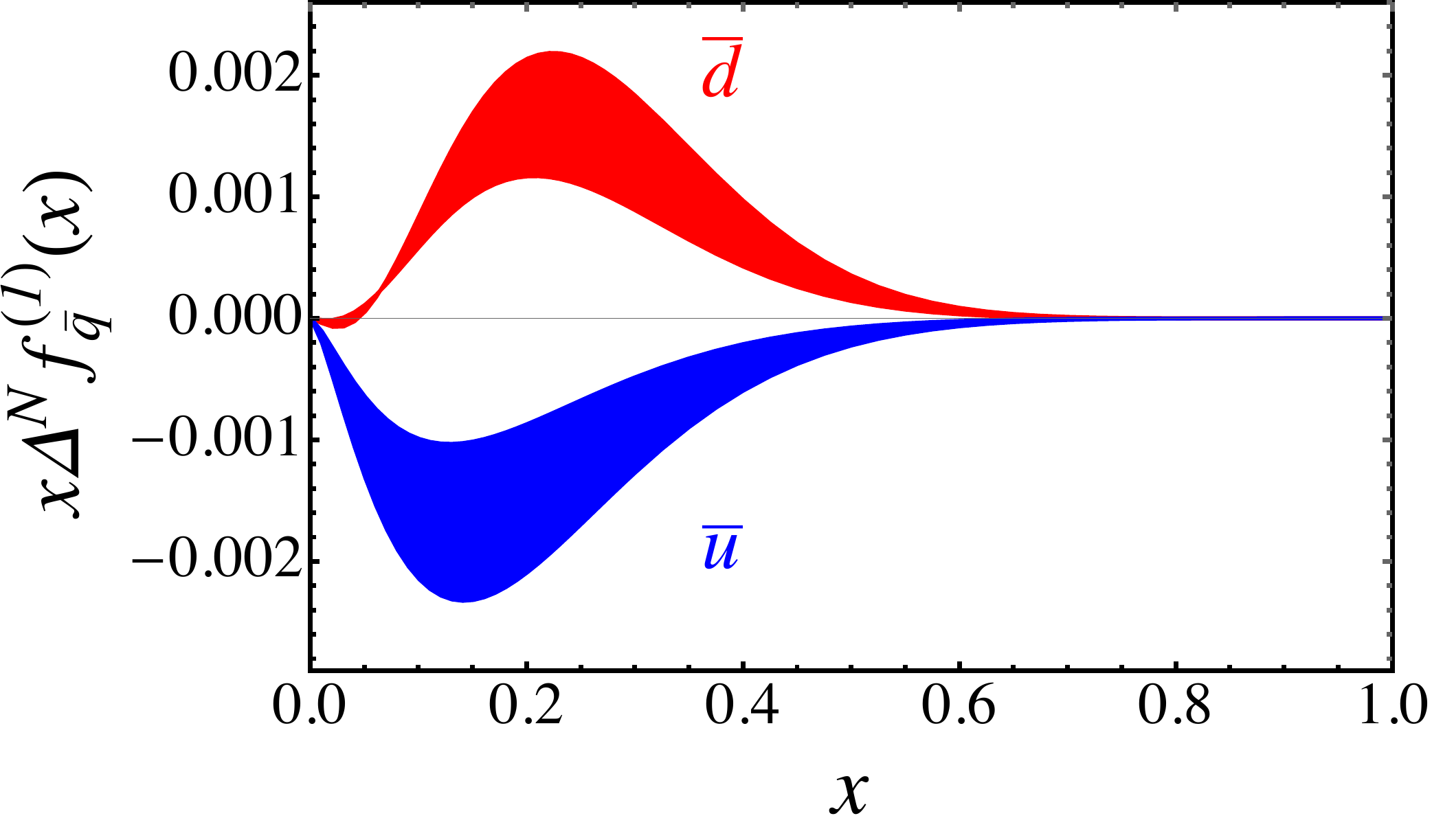}
\includegraphics[scale=0.4]{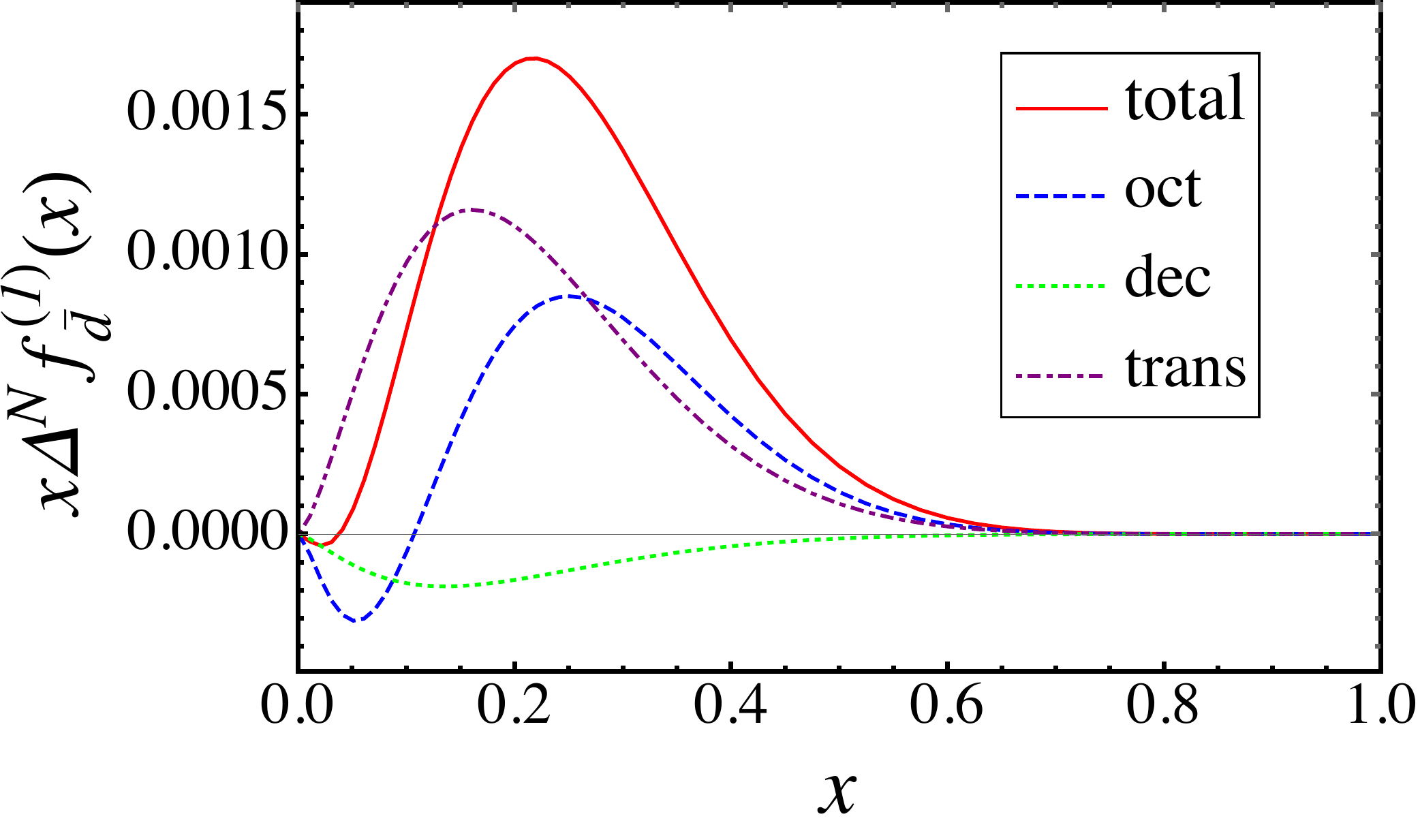}
\vspace*{-0.2cm}
\caption{(Left) First moment of the sea quark Sivers distribution $x\Delta^N f_{\bar{q}}^{(1)}(x)$ versus $x$ at $Q=0.63$~GeV for $\bar q=\bar d$ (red) and $\bar q=\bar u$ (blue), with 0.8~GeV $\leq \Lambda_\pi \leq 1.2$ GeV and $1.6$ GeV $\leq \Lambda_\rho \leq 2.0$ GeV. 
(Right) Contributions to $x\Delta^N f_{\bar{d}}^{(1)}(x)$ from intermediate states with baryon octet (blue dashed), decuplet (green dotted), octet-decuplet transition (purple dot-dashed) and total (red solid).
(Figure from Ref.~\cite{He3}.)}
\label{fig:Sivers}
\end{center}
\end{figure}

In Fig.~\ref{fig:Sivers} we show the first moment of the sea quark Sivers function, defined as
\be \label{eq:Siversmoment}
\Delta^Nf^{(1)}_{\bar{q}}(x)
\equiv \int\,\dd^2 \bm{k}_\perp\frac{-k_\perp^2}{2M_N^2}\,
f^{\bar{q}/p}_{1T}(x,\bm{k}_\perp),
\ee
for the $\bar d$ and $\bar u$ flavors.
For the $\bar{d}$ flavor in the proton, the first moment is positive, with a maximum value of $x\Delta^N f_{\bar{d}}^{(1)}(x) \approx 0.0008 - 0.0035$ at $x \sim 0.2$, which then decreases at large $x$ and vanishes beyond $x \approx 0.6$.
For the $\bar{u}$ in proton, $x\Delta^Nf_{\bar{u}}^{(1)}(x)$ is always negative, with maximum absolute value $\approx 0.0007 - 0.0037$ at $x \sim 0.15$, and similarly vanishing beyond $x \approx 0.6$.
As found in other phenomenological extractions, the value of sea quark Sivers function is rather small \cite{Bacchetta, Martin2}.

For $x\Delta^N f_{\bar{u}}^{(1)}(x)$, only the diagram in Fig.~\ref{fig:SiversFeynman}(b) makes a contribution, whereas all four diagrams in Fig.~\ref{fig:SiversFeynman} contribute to $x\Delta^Nf_{\bar{d}}^{(1)}(x)$.
To see the separate contributions explicitly, we plot in Fig.~\ref{fig:Sivers} the contributions to $x\Delta^Nf_{\bar{d}}^{(1)}(x)$ from different intermediate states. 
The contributions from intermediate octet and octet-decuplet transition are dominant, while the contribution from the decuplet intermediate state is very small. 
The decuplet intermediate state gives negative contributions to both $x\Delta^N f_{\bar{d}}^{(1)}(x)$ and $x\Delta^N f_{\bar{u}}^{(1)}(x)$, however, the $\bar{d}$ contribution is 9 times smaller than the $\bar{u}$ because of the smaller coupling constants for $\pi^+$ compared with the $\pi^-$ case.

\section{Summary and outlook}
\label{Sec.6}

In this article we presented a brief overview of chiral effective field theory with infrared, extended-on-mass-shell, and finite-range regularizations.
While EFT has been widely applied in hadronic physics, for the study of nucleon electromagnetic form factors, which are measured in elastic lepton-nucleon scattering, the theory is only valid at low momentum transfers.
This limited applicability of EFT restricts investigation of form factors at larger momentum transfer, as well as of PDFs and GPDs, which are extracted from high energy inclusive and exclusive reactions, respectively.
This problem motivates the introduction of the nonlocal chiral effective field theory, as one way to extend the study of physical observables to larger momentum transfers.

In the nonlocal Lagrangian, baryons are placed at spacetime coordinate $x$, while mesons and photons are located at coordinates $x+a$ and $x+b$.
To guarantee charge conservation and local gauge invariance, the path integral of the gauge field is introduced, and expansion of the gauge link leads to the generation of additional Feynman diagrams.
The ultraviolet regulator appears naturally in the loop integrals, as the Fourier transformation of the correlation function in the nonlocal Lagrangian.
The regulator makes the loop integrals convergent, and at the same time provides the momentum dependence of the form factors at tree level.

In the application of the nonlocal effective theory to the the study of elastic scattering observables, the nucleon electromagnetic form factors, strange form factors, light sea quark form factors, as well as the form factors of octet baryons were computed.
For PDFs, we reviewed the calculation of sea quark distribution functions, including the unpolarized distributions of $\bar{u}$, $\bar{d}$, $s$ and $\bar{s}$, and the distribution of longitudinally polarized strange quarks.
We further discussed the extension of the framework to GPDs, where we generalized to the nonforward sector the light sea quark and strange quark distributions.
To illustrate the application to TMDs, we considered the T-odd transverse Sivers momentum dependent distribution functions of $\bar{u}$ and $\bar{d}$ quarks.
Numerical results were presented for the electromagnetic moments, charge and magnetization radii, form factors, as well as the parton momentum fraction dependence of the nucleon PDFs and GPDs.

As future extensions of the work summarized in this report, one can study the applications of the nonlocal chiral effective theory to the full set of leading twist GPDs, including the four chiral-even (helicity-conserving) GPDs $H$, $E$, $\widetilde{H}$, $\widetilde{E}$ and four chiral-odd (helicity-flipping) GPDs $H_T$, $E_T$, $\widetilde{H}_T$, and $\widetilde{E}_T$, for both valence and sea quarks.
Similarly, the eight leading twist TMD distributions of the nucleon can be explored within this framework.
Beyond the nucleon, it would be of tremendous interest to consider the corresponding structures for the octet baryons, decuplet baryons and pseudoscalar mesons. 
Since the chiral Lagrangian already includes the octet and decuplet baryon degrees of freedom, the properties of these hadrons could be obtained from the existing theory without the introduction of additional parameters.
This would not only provide predictions for the properties of other hadrons that could be measured in experiments or in lattice QCD, but also offer a valuable check on the validity of the nonlocal theory itself.

Beyond hadronic observables, we also consider the intriguing possibility that the nonlocal behavior could be a general property for all interactions of physical particles.
As we discuss in the appendix below for the case of nonlocal QED, the discrepancy between the lepton Pauli form factors in the (local) SM and the corresponding nonlocal interaction could be visible at very large momentum transfer.
Since we do not know {\it a priori} the scale parameter $\Lambda$ for leptons, this is chosen to reproduce the anomalous magnetic moments of the electron and muon, and though the difference with the local values is much smaller than the current experimental uncertainties, the relative deviation from the SM is always large if the momentum transfer is high enough for any $\Lambda$.

Going one step further, we also discuss in the appendix a more general version of the nonlocal theory in which the free Lagrangian is also nonlocal.
The nonlocal Feynman propagator in this case corresponds to the solid quantization, where the $\delta^{(3)}(\bm{x}-\bm{y})$ function is replaced by a finite function $\Phi(\bm{x}-\bm{y})$.
An advantage of the general version of the nonlocal Lagrangian is that for distributions such as PDFs or GPDs, the $\delta$-function terms in the splitting functions could become smooth functions.
This may be a fruitful direction in which to pursue future phenomenolgical applications of chiral EFT in hadronic physics and beyond. \\

\section*{Acknowledgments}
\label{Sec.7}

We are grateful to Y.~Salamu, A.~W.~Thomas, and X.~G.~Wang for collaboration on the material presented in this review.
This work is supported by the NSFC under Grant No.~11975241, the DOE Contract No.~DE-AC05-06OR23177, under which Jefferson Science Associates, LLC operates Jefferson Lab and DOE Contract No.~DE-FG02-03ER41260. \\

\section[{Appendix: Generality of the nonlocal Lagrangian}]{Appendix: Generality of the nonlocal Lagrangian}
\label{Sec.8}

\subsection[{\it Nonlocal QED}]{\it Nonlocal QED}

In this review we have illustrated how the nonlocal chiral effective theory provides a systematic tool for the investigation of hadronic structure, including nucleon form factors, collinear PDFs, TMD distributions, and GPDs.
In contrast to traditional EFT, the nonlocal Lagrangian involves a correlation function $F(x-y)$, where the baryon is located at spacetime point $x$ and the meson or photon is at point~$y$. 
If $F(x-y)$ is chosen to be a $\delta$ function, $\delta(x-y)$, the nonlocal Lagrangian naturally reduces to the local one.
With the presence of the correlation function, there are no ultraviolet divergences appearing in the loop integrals.

In addition to being a useful method to deal with divergences in EFT, the nonlocal Lagrangian may also provide insights into general properties of physical interactions.
In other words, the nonlocal regulator could exist in fundamental interactions whatever divergences are present or not.
In this respect, we may also consider along similar lines the interaction between electrons and photons as being nonlocal.
In the classical scenario (at tree level), the nonlocal effect is certainly negligible at low momentum transfers.
On the other hand, for quantum fluctuations associated with loop diagrams, the internal photon can be sensitive to the structure of the physical particle since its momentum can be infinite.

It is informative, therefore, to study the Pauli form factors of leptons in the framework of nonlocal QED.
The anomalous magnetic moments of the electron and muon, $a_e$ and $a_\mu$, respectively, are among the most precisely determined observables in particle physics. 
The latest theoretical predictions for $a_e$ and $a_\mu$ in the Standard Model (SM) are 
    $a_e^{\text{SM}}=1159652182.032(720)\times10^{-12}$ \cite{Aoyama} 
and $a_\mu^{\text{SM}}=116591810(43)\times10^{-11}$ \cite{Aoyama2}.
The recent measurement of the muon anomalous magnetic moment in the E989 experiment at Fermilab finds 
\begin{equation}
\Delta a_\mu^{\text{FNAL}}
= a_\mu^{\text{FNAL}} - a_\mu^{\text{SM}} = 230(69) \times 10^{-11},
\end{equation}
which is a $3.3 \sigma$ discrepancy from the SM prediction \cite{Abi}.
Combined with the previous E821 experiment at BNL \cite{Bennett}, the result leads to a $4.2 \sigma$ discrepancy \cite{Abi}
\begin{equation}
\Delta a_\mu 
= a_\mu^{\text{FNAL}+\text{BNL}} - a_\mu^{\text{SM}} = 251(59) \times 10^{-11}.
\end{equation}
For the electron, the most accurate measurement of $a_e$ has been carried out by the Harvard group, and the discrepancy from the SM result was $2.4 \sigma$ \cite{Hanneke, Hanneke2},
\begin{equation}\label{eq:e1}
\Delta a_e
= a_e^{\text{exp}}-a_e^{\text{SM}} = -87(36) \times 10^{-14}.
\end{equation}
As SM predictions in general closely match other experimental information, the deviation in one of the most precisely measured quantities in particle physics remains a mystery, and has inspired considerable theoretical work.
The discrepancy between the SM results and experiments could be even larger for very large momentum transfers, where the structure of leptons could be visible. 
This motivates the investigation of the anomalous form factors of the electron and muon within the nonlocal QED, in analogy with from the nonlocal chiral effective theory \cite{He4}.

\begin{figure}[t]
\begin{center}
\includegraphics[scale=0.5]{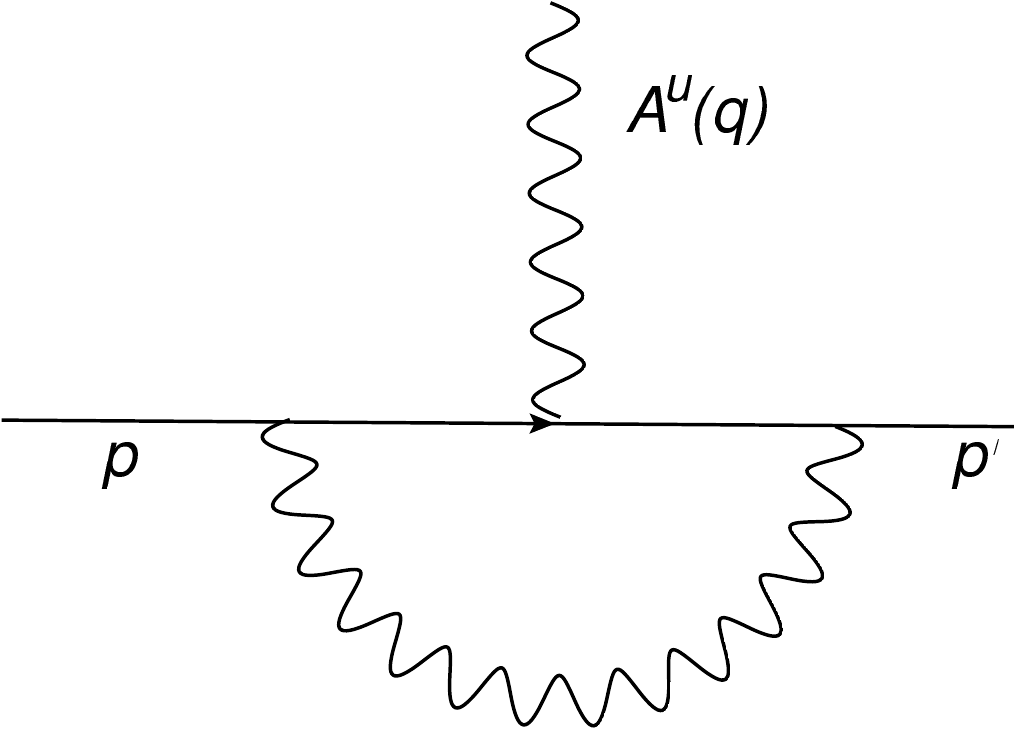}
\caption{Feynman diagram for the one-loop vertex correction to the lepton form factors.}
\label{fig:leptonFeynman}
\end{center}
\end{figure}

The nonlocal Lagrangian for QED can be written as
\begin{equation}\label{eq:Lqednl}
\mathcal{L}_{\rm QED}^{\rm nonloc}
= \bar{\psi}(x)(i\slashed{\partial}-m)\psi(x)
- \frac14 F_{\mu\nu}(x) F^{\mu\nu}(x)
- e\int\!\,\dd^4 a\, \bar\psi(x)\gamma^\mu \mathscr{A_\mu}(x+a)\psi(x)F(a),
\end{equation}
where the lepton field $\psi(x)$ with mass $m$ is located at spacetime point $x$ and the photon field $\mathscr{A_\mu}(x+a)$ is located at $x+a$.
The function $F(a)$ is the correlation function normalized as $\int \dd^4 a\, F(a)=1$, and for the case of a $\delta$ function reduces the nonlocal Lagrangian to the local one.
The nonlocal Lagrangian (\ref{eq:Lqednl}) is invariant under the gauge transformation
\begin{equation}
\psi(x)\to\,e^{i\alpha(x)}\psi(x),~~~~~~~~
\mathscr{A_\mu}(x)\to\,\mathscr{A_\mu}(x)-\frac1e\partial_\mu\,\alpha^\prime(x),
\end{equation}
where $\alpha(x) = \int\,\dd a\, \alpha^\prime(x+a)F(a)$. 
In contrast to the nonlocal Lagrangian for the chiral EFT, where the gauge link is introduced to guarantee local gauge invariance, in this case the gauge link is not necessary since the photon does not carry charge.

The one-loop Feynman diagram for the lepton form factors is shown in Fig.~\ref{fig:leptonFeynman}.
In this approximation, the vertex can be written as
\begin{eqnarray}\label{eq:loop}
\bar{u}(p')\Gamma^\mu(p',p) u(p)
= \bar{u}(p')\int\!\frac{\dd^4k}{(2\pi)^4}
\widetilde F(q^2) \widetilde F^2(k^2)
(ie\gamma^\nu) 
\frac{i}{\slashed{p'}-\slashed{k}-m}
\gamma^\mu
\frac{i}{\slashed{p}-\slashed{k}-m}
(ie\gamma^\rho)
\frac{-ig_{\nu\rho}}{k^2}u(p).
\end{eqnarray}
Similar to the case of the Dirac and Pauli form factors of the nucleon, one can obtain the form factors of the lepton as
\begin{eqnarray}\label{eq:F1lnonloc}
\hspace*{-0.6cm}
F_1^{\rm nonloc}(q^2)
&=& \frac{-ie^2\widetilde F(q^2)}{(4m^2-q^2)^2}
\int\!\frac{\dd^4 k}{(2\pi)^4}
\widetilde F^2(k^2) 
\bigg[
\frac{-24m^2 \big( (k\cdot p)^2 + (k\cdot p')^2 \big)
+ 8m^2 k^2 (4m^2-q^2)}
{\big( (p'-k)^2-m^2 \big) \big( (p-k)^2-m^2 \big) k^2}
\nonumber\\ 
&& \hspace*{-2.3cm} 
+ \frac{2(2m^2-q^2)(4m^2-q^2)^2-4(4m^2+2q^2)(k\cdot p)(k\cdot p') 
- 4 k\cdot (p+p') (8m^4-6m^2q^2+q^4)}
{\big( (p'-k)^2-m^2 \big) \big( (p-k)^2-m^2 \big) k^2}
\bigg]
\end{eqnarray}
and
\begin{eqnarray}\label{eq:F2lnonloc}
F_2^{\rm nonloc}(q^2)\!\!\!
&=& \!\!\!\frac{-ie^2\widetilde F(q^2) 8 m^2}{q^2(4m^2-q^2)^2}
\int\!\frac{\dd^4k}{(2\pi)^4}
\widetilde{F}^2(k^2)
\bigg[
\frac{(4m^2+2q^2) \big( (k\cdot p)^2+(k\cdot p')^2 \big) 
- 8(m^2-q^2)(k\cdot p)(k\cdot p')}
{\big( (p'-k)^2-m^2 \big) \big( (p-k)^2-m^2 \big) k^2}
\nonumber\\ 
&& +\, \frac{(q^4-4m^2q^2)(k\cdot p+k\cdot p'+k^2)}
{\big( (p'-k)^2-m^2 \big) \big( (p-k)^2-m^2 \big) k^2}
\bigg].
\end{eqnarray}
In Eqs.~(\ref{eq:F1lnonloc}) and (\ref{eq:F2lnonloc}) the momentum dependent vertices $\widetilde F(q)$ and $\widetilde F(k)$ appear.
For $\widetilde F(q)$, if the external momentum $q$ is much smaller than the scale of the lepton, the size of the lepton can be neglected and 
    $\widetilde F(q) \simeq 1$.
However, for $\widetilde F(k)$, the internal momentum $k$ varies from zero to infinity, and in this case the regulator is important as it renders the loop integrals for $F_1^{\rm nonloc}$ and $F_2^{\rm nonloc}$ convergent in the ultraviolet.

For the numerical calculation, as in the nonlocal chiral EFT case, the regulator $\widetilde F (k^2)$ is chosen to be a dipole form,
\begin{equation}
\widetilde F(k^2)=\frac{\Lambda^4}{(k^2-\Lambda^2)^2}.
\end{equation} 
The Pauli form factor at the one-loop level can then be obtained as
\be\label{eq:nonlocall}
F_2^{\rm nonloc}(Q^2)
= \frac{\alpha}{2\pi}\widetilde F(Q^2)
\int_0^1\,\dd x\int_0^{1-x}\dd y 
\frac{2\Lambda^{8}m^2(x+y)(1-x-y)^5}
{\big( Q^2xy+m^2(x+y)^2 \big) 
 \big( m^2(x+y)^2+Q^2xy+(1-x-y)\Lambda^2 \big)^4},
\ee
where $Q^2 \equiv -q^2$.
In the limit $\Lambda \to \infty$, the local result is recovered,
\begin{eqnarray}\label{eq:local}
F_2^{\rm local}(Q^2)
&=& \frac{\alpha}{2\pi}\int_0^1\,\dd x\int_0^{1-x}\dd y\frac{2m^2(1-x-y)(x+y)}{m^2(x+y)^2+Q^2xy},
\end{eqnarray}
which gives the well known anomalous magnetic moment at the one-loop level,     $F_2^{\rm local}(0) = \alpha/2\pi$.
In the nonlocal case, the anomalous magnetic moment is given by 
    $F_2^{\rm nonloc}(0) = (\alpha/2\pi)
        (1 - 8m^2/3\Lambda^2 
           - 65m^4/3\Lambda^4
           + O(m^6/\Lambda^6))$.
If $m \ll \Lambda$, the deviation of $F_2^{\rm nonloc}(0)$ from the SM is highly suppressed.
With the results from nonlocal and local QED, we can define their discrepancy as $\Delta F_2= F_2^{\rm local} - F_2^{\rm nonloc}$, and their relative discrepancy as $R =\Delta F_2 / F_2^{\rm local}$.
Note that the loop integrals for the Pauli form factors are ultraviolet convergent in both the local and nonlocal cases.
The regulator is naturally generated from the nonlocal Lagrangian with the naive idea that the interaction between the photon and lepton is not restricted to take place at one point. 
For the ultraviolet divergent integral in the local case, the regulator will render the integral convergent. 
For the integral which is convergent in the local case, the regulator also exists and will give deviations from the local result at large momentum transfer.

The analytic expression of the nonlocal Pauli form factor can be expanded in orders of the lepton mass $m$ when the momentum transfer and $\Lambda$ are $\gg m$.
At leading order, $F_2^{\rm nonloc}(Q^2)$ can be written as
\begin{eqnarray}\label{eq:nllead}
F_2^{\rm nonloc}(Q^2)
&=& \alpha m^2 
\left[ \frac{-\Lambda^4}{\pi Q^2 (Q^2+\Lambda^2)^2} 
\log \frac{m^2}{Q^2}
+ C(Q^2,\Lambda^2)
\right],
\end{eqnarray}
where the lepton mass independent function $C(Q^2,\Lambda^2)$ is given by
\begin{eqnarray}
C(Q^2,\Lambda^2)
&=& 
  \frac{\Lambda^4 (Q^4 + 3\Lambda^2 Q^2 - 6\Lambda^4) \log(\Lambda^2/Q^2)}
     {\pi Q^4 (Q^2-\Lambda^2) (Q^2+\Lambda^2)^2} 
-\frac{3 \Lambda^8 \log^2(\Lambda^2/Q^2)}{\pi Q^6 (Q^2+\Lambda^2)^2}
\nonumber\\
& &
-\, \frac{6 \Lambda^8 \text{Li}_2(1-\Lambda^2/Q^2)}
{\pi Q^6 (Q^2+\Lambda^2)^2}
- \frac{\pi \Lambda^8}{Q^6 (Q^2+\Lambda^2)^2}
+ \frac{3 \Lambda^4 (Q^2+4\Lambda^2)}{2 \pi Q^4 (Q^2+\Lambda^2)^2}.
\end{eqnarray}
The first term in the bracket of Eq.~(\ref{eq:nllead}) is the leading nonanalytic term.
For the local case with $\Lambda \to \infty$, one has $C(Q^2,\infty) = 0$ and $F_2^{\rm local}(Q^2)$ at leading order is given by
\begin{eqnarray}
F_2^{\rm local}(Q^2)
= -\frac{\alpha m^2}{\pi Q^2} \log\frac{m^2}{Q^2}.
\end{eqnarray}

\begin{figure}[tbh]
\begin{center}
	\begin{minipage}{8.5cm}
	\centering    
	\includegraphics[scale=0.4]{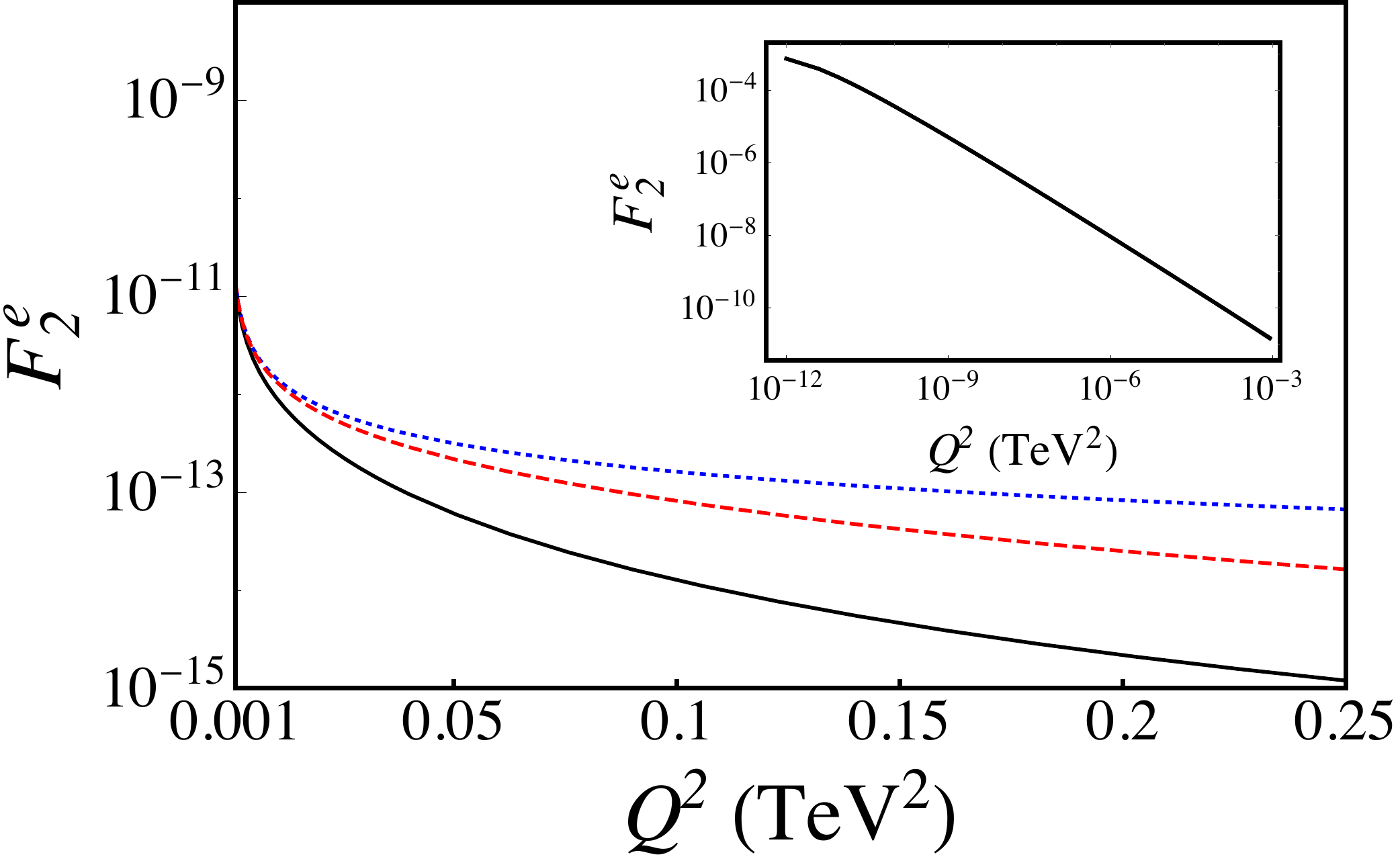} 
	\end{minipage}
	\begin{minipage}{8.5cm}
	\centering      
	\includegraphics[scale=0.4]{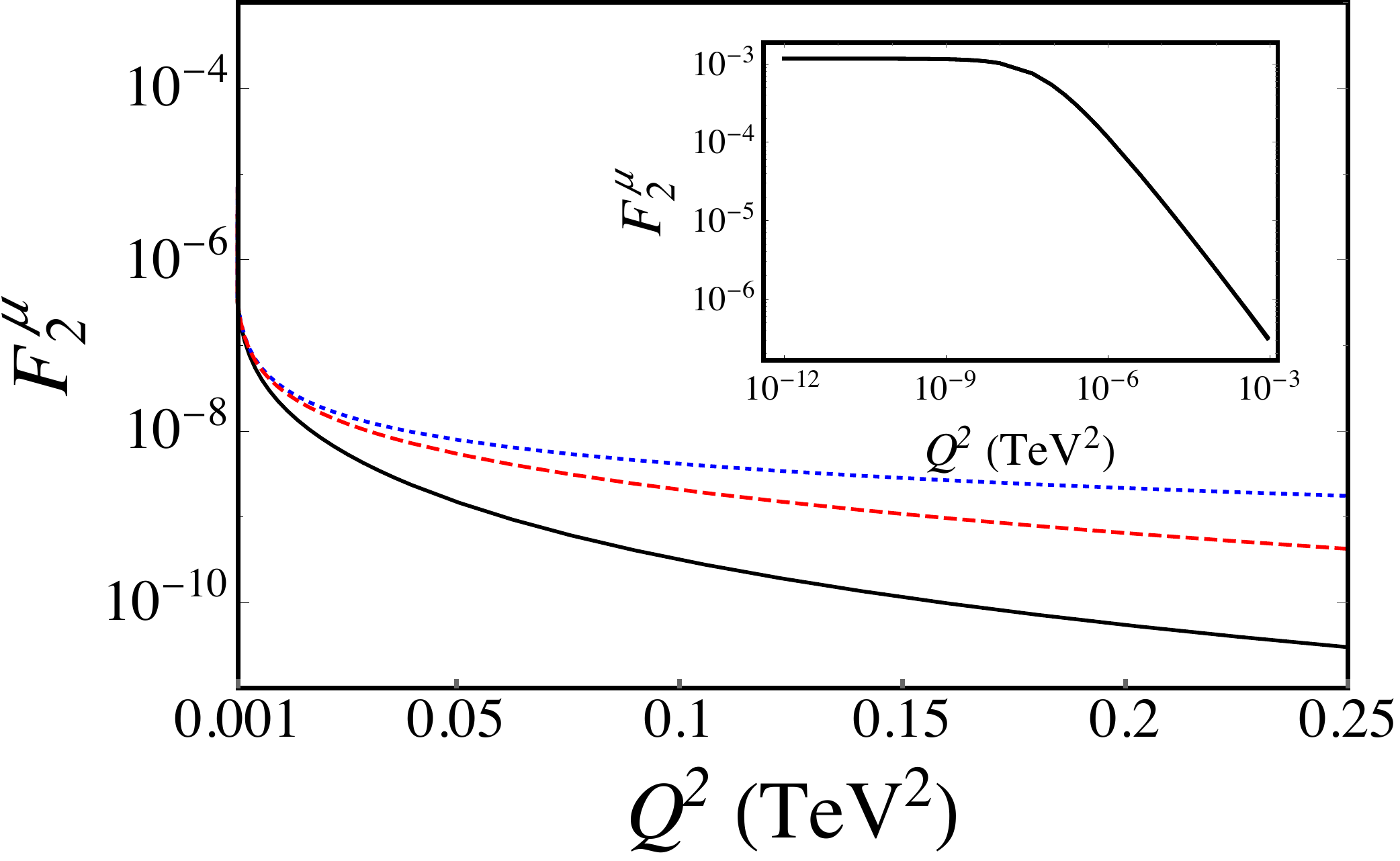} 
	\end{minipage}
\caption{Pauli form factor $F_2^l(Q^2)$ of the electron (left) and muon (right) versus momentum transfer squared $Q^2$ for $\Lambda=0.2$~TeV (solid), 0.5~TeV (dashed), and the local limit (dotted lines).
The insets show $F_2^l(Q^2)$ at low $Q^2$ up to $Q^2=0.001$~TeV$^2$.
(Figure from Ref.~\cite{He4}.)}
\label{fig:leptonF2}
\end{center}
\end{figure}

In the numerical calculation, the one free parameter, $\Lambda$, in the regulator needs to be determined.
For the nucleon case, $\Lambda$ is of the order of 1~GeV. 
For leptons, however, $\Lambda$ could be much larger given the much smaller lepton size.
Certainly, on the one hand, the smaller the value of $\Lambda$, the larger the deviation from the SM.
On the other hand, $\Lambda$ should be large enough to make the nonlocal results consistent with the experiments at the same level as the SM.
When $\Lambda=0.2$~TeV, the calculated $a_e^{\rm nonloc}$ in nonlocal QED is 0.00116171491307, which is a $2.0 \times 10^{-14}$ deviation from the corresponding SM value.
Considering the experimental accuracy $2.6 \times 10^{-13}$ \cite{PDG} and the discrepancy between experiment and the SM prediction of $8.7 \times 10^{-13}$, the choice of $\Lambda=0.2$~TeV seems reasonable.

For the muon, $a_\mu$ in nonlocal QED has some $8.6 \times 10^{-10}$ deviation from the local or SM case.
Comparing with $\Delta a_\mu = 2.5 \times 10^{-9}$, the value $\Lambda = 0.2$~TeV also appears reasonable to take for the muon.
In the numerical calculations, we therefore show the results with $\Lambda=0.2$~TeV and 0.5~TeV.
In principle, QED itself cannot determine the form of the regulator and the value of $\Lambda$, which can only be inferred from experimental data, especially at finite $Q^2$.
One can show, however, that whatever the value of $\Lambda$, the relative discrepancy between the local and nonlocal QED results is always significant if the momentum transfer $Q^2$ is large enough \cite{He4}.

The Pauli form factors of the electron and muon, $F_2^e(Q^2)$ and $F_2^\mu(Q^2)$, are shown in Fig.~\ref{fig:leptonF2} versus~$Q^2$.
The form factors of the leptons are seen to decrease rapidly with increasing $Q^2$ due to the small size of the lepton mass.
When $Q^2$ is small, the discrepancy between the nonlocal QED and the local SM QED is much smaller than the form factors themselves. 
For instance, at $Q^2=0$ the discrepancy $\Delta F_2^e$ is at least $10^{11}$ times smaller than the anomalous magnetic moments. 
With increasing $Q^2$, the discrepancy is more clearly visible when it becomes comparable with the form factors.
Since the mass of the muon is larger than that of the electron, its form factor drops more slowly than that for the electron.
Due to the nonlocal effect, the form factors in nonlocal QED are smaller than those in the SM at any value of~$Q^2$.

\begin{figure}[tb]
\begin{center}
\includegraphics[scale=0.45]{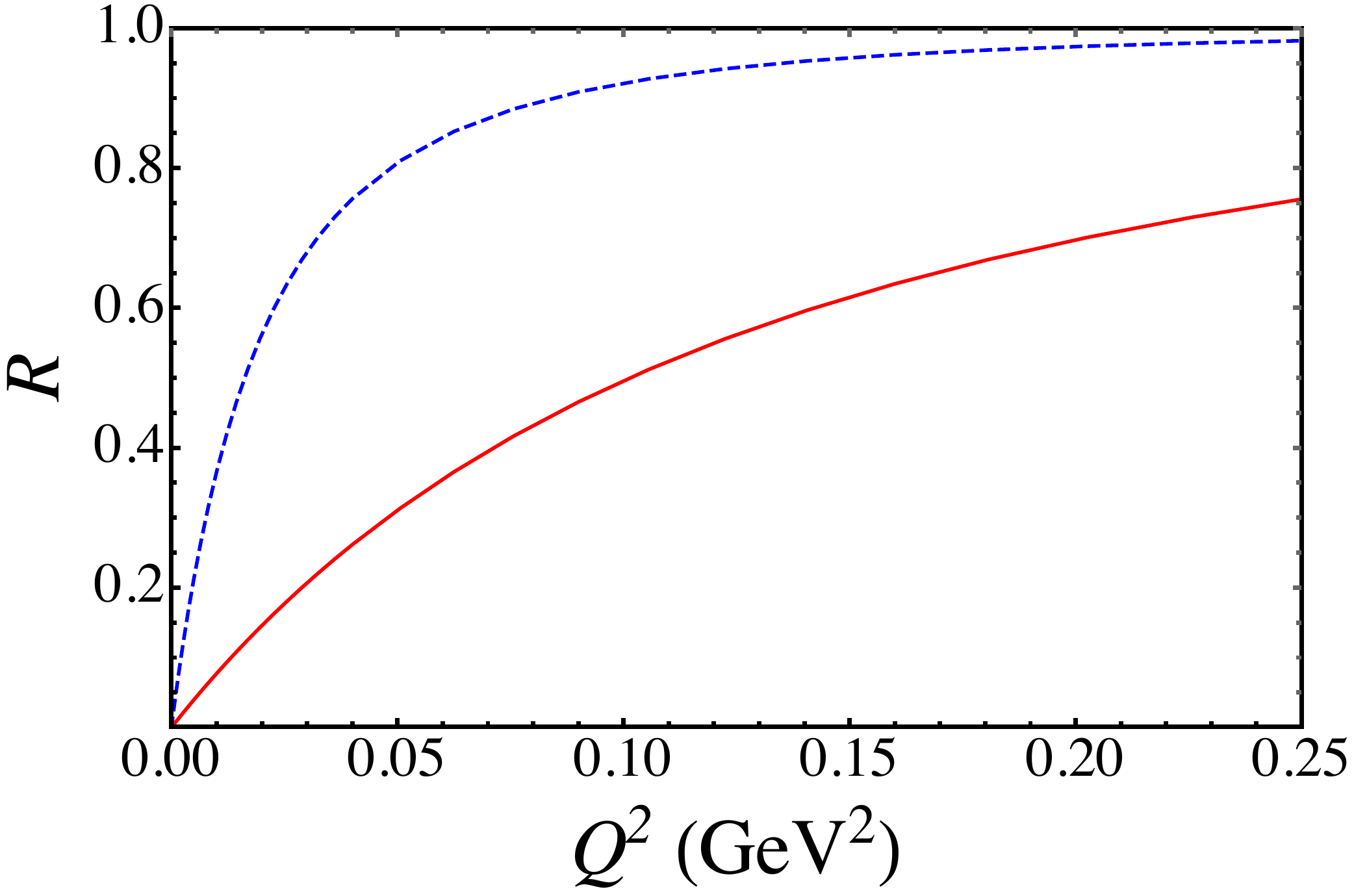}
\caption{Relative deviation $R(Q^2)$ for the electron and muon versus $Q^2$, for $\Lambda = 0.2$~TeV (dashed blue line) and 0.5~TeV (solid red line). (Figure from Ref.~\cite{He4}.)}
\label{fig:RF2}
\end{center}
\end{figure}

\begin{figure}[h]
\begin{center}
\includegraphics[scale=1.0]{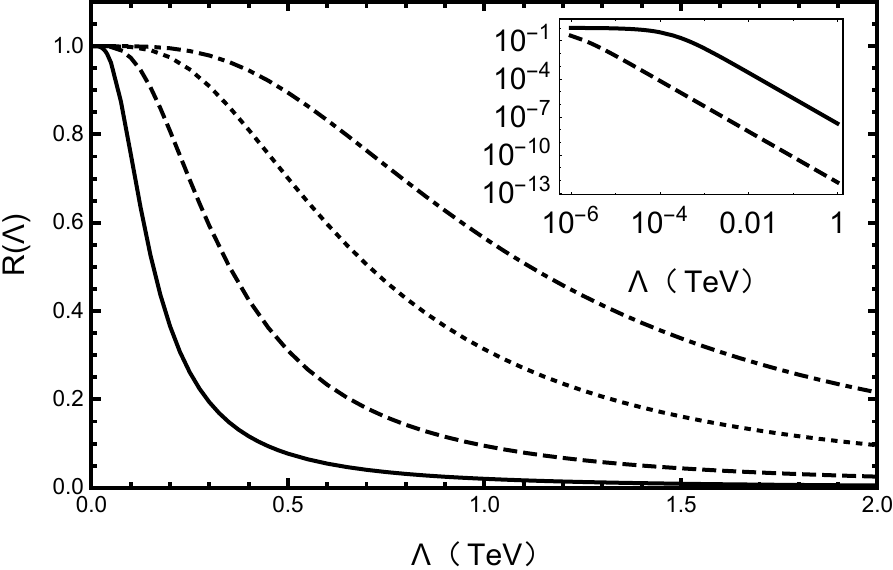}
\caption{Relative deviation $R(\Lambda)$ for the electron and muon versus $\Lambda$, for  $Q^2=0.01$ (solid line), 0.05 (dashed), 0.2 (dotted), and 0.5~TeV$^2$ (dot-dashed line). The inset is for $R$ at $Q^2=0$ for the muon (solid line) and electron (dashed line). (Figure from Ref.~\cite{He4}.)}
\label{diagrams}
\label{fig:RLambda}
\end{center}
\end{figure}

The relative deviation $R$, plotted in Fig.~\ref{fig:RF2} versus $Q^2$, clearly shows the discrepancy between the nonlocal QED and the SM for both the values of $\Lambda = 0.2$~TeV and 0.5~TeV. 
For a given $Q^2$, the discrepancy $\Delta F_2$ and the form factor $F_2$ are both much larger for the muon than for the electron.
However, the relative deviations $R$ are almost the same for the electron and muon.
At $Q^2=0$, the relative deviation $R$ is very small, of order $10^{-11}$ and $10^{-7}$ for the electron and muon, respectively, which is of the same order for the relative deviation of the SM from the experiments.
To confirm this discrepancy for the electron, the experiment should therefore be very accurate in order to attain an additional 10 effective digits.
The relative deviation $R$ increases with the increasing $Q^2$.
For example, at $Q^2=0.01$~TeV$^2$, $R \approx 0.37$ and 0.08 for $\Lambda=0.2$ and 0.5~TeV, respectively.
When $Q^2 \gtrsim 0.1$~TeV$^2$, $R$ is larger than 0.5 for both $\Lambda$ values.
For even larger $Q^2 \gtrsim 0.2$~TeV$^2$, $F_2^{\rm nonloc}$ could be one order of magnitude smaller than the SM value.

The relative deviation $R$ is also sensitive to the parameter $\Lambda$.
In Fig.~\ref{fig:RLambda}, the $\Lambda$ dependence of $R$ is plotted at different $Q^2$ values between $Q^2=0.01$ and 0.5~TeV$^2$.
For any given $Q^2$, $R$ clearly decreases with increasing $\Lambda$.
The relative deviations are very similar for electrons and muons, with the differences only visible on a logarithmic scale.
The exact value of $\Lambda$ can only be determined from future experiments.
Regardless of the specific value of $\Lambda$, the relative deviation $R$ is always significant as long as the momentum transfer $Q^2$ is sufficiently large.
Although the absolute value of the form factor is small, the relative deviation of nonlocal QED from the SM is very large.
Even if the experiment can only measure the form factor to one significant digit at finite $Q^2$, one may still be able to draw conclusions about physics beyond the SM.
Since the deviation is so large at finite $Q^2$, the conclusion is not changed by higher order QED corrections or hadronic effects, as these are typically less than one percent for both electrons and muons.

\subsection{\it Solid quantization}

In this report we have applied nonlocal EFT and nonlocal QED to the study of hadron and lepton properties, choosing dipole regulators for the correlation functions in the numerical calculations.
In this section, we will focus on the generality of the nonlocal Lagrangian without referring to a specific form for the correlation function. 
In the previous nonlocal Lagrangians, the fermion (baryon or lepton) is placed at spacetime coordinate $x$, while the boson (meson or photon) is at a different coordinate, $x+a$.
The free Lagrangian as well as the propagator is the same as the local one. Here, we make the nonlocal Lagrangian more general and assign each field to be at different spacetime points.
For illustration, we consider the U(1) gauge theory; however, it will be straightforward to generalize this to other types of interactions.

The most general nonlocal Lagrangian which is locally U(1) gauge invariant can be written as \cite{Wang3,Wang6}
\be \label{eq:NLU1}
{\cal L} = \int \dd^4 a\, \bar{\psi}\Big(x+\frac{a}{2}\Big)\,
e^{iI(x+a/2,x)} \gamma^\mu\, i (D_\mu + m)
e^{-iI(x-a/2,x)} \psi \Big(x-\frac{a}{2}\Big)\,
G_1(a),
\ee
where the covariant derivative is
\be
D_\mu = \partial_\mu - ig \int \dd^4 b\, \mathscr{A_\mu}(x+b)\, G_2(a,b),
\ee
and $I(y,x)$ is the path integral of the gauge field,
\be
I(y,x) 
= g \int_x^y \dd z^\mu \int \dd^4 b\, \mathscr{A_\mu}(z+b)\, G_2(a,b). 
\ee
The Lagrangian (\ref{eq:NLU1}) is invariant under the transformations
\begin{subequations}
\bea
\psi \left(x-\frac{a}{2} \right) 
&\to& \psi \Big(x-\frac{a}{2}\Big)'\,
=\, \exp\Big[ ig\theta\Big(x-\frac{a}{2}\Big) \Big]
\psi\Big(x-\frac{a}{2}\Big),
\\
\bar{\psi} \Big(x+\frac{a}{2}\Big) 
&\to& \bar{\psi} \Big(x+\frac{a}{2}\Big)'\,
=\, \exp\Big[ ig\theta\Big(x+\frac{a}{2}\Big)\Big]
\bar{\psi}\Big(x+\frac{a}{2}\Big),
\\
\mathscr{A_\mu}(x+b) 
&\to& \mathscr{A_\mu}'(x+b)\ =\, \mathscr{A_\mu}(x+b) + \partial_\mu \theta' (x+b),
\eea
\end{subequations}
where the functions $\theta'(x)$, $\theta(x)$ and $G_2(a,b)$ are constrained by the relation
\be
\theta(x) = \int \dd^4{b}\, \theta'(x+b) G_2(a,b).
\ee
In previous calculations, the simplification was made that $G_1(a) = \delta(a)$ and $G_2(a,b) = F(b)$.
As for the chiral EFT, the general version of the Lagrangian in Eq.~(\ref{eq:NLU1}) includes both the normal interaction from the minimal substitution and additional interactions from the expansion of gauge link.
The difference from the previous calculations is that the free Lagrangian is also nonlocal.

The free part of the nonlocal Lagrangian in Eq.~(\ref{eq:NLU1}) is 
\be \label{eq:NLfree}
{\cal L}_0(x) = \int \dd^4{a}\,
\bar{\psi}\Big(x+\frac{a}{2}\Big)
\gamma^\mu i (\partial_\mu+m)\,
\psi\Big(x -\frac{a}{2}\Big) G_1(a).
\ee
Using the translation operator to move the position to the same point, the free Lagrangian can be written as
\be \label{NLfree2}
{\cal L}_0(x) 
= \bar{\psi}(x)\gamma^\mu i (\partial_\mu+m)
\widetilde{G}_1(i\partial_\mu)\psi (x),
\ee
where $\widetilde{G}_1(i\partial_\mu)$ is the Fourier transform of $G_1(a)$,
$\widetilde{G}_1(i\partial_\mu) 
= \int \dd^4 a\, e^{ia \cdot i\partial}G_1(a)$.
In the nonlocal free Lagrangian, the fermion propagator is modified correspondingly as
\be \label{eq:NLpropagator}
S(x'-x) = \int \frac{\dd^4 p}{(2\pi)^4} 
\frac{ie^{ip\cdot(x'-x)}}
     {(\slashed p - m + i\epsilon)\, \widetilde{G}_1(p^2)}.
\ee
The propagator is related to the quantization of the field, and for a point particle one has
\be
\psi(x)
= \sum_{s=\pm} \int \frac{\dd^4 p}{(2\pi)^4} 
\delta(p^2 - m^2)
\big[ b^s_{\bm{p}}\, u_s(\bm{p})\, e^{-ip \cdot x}
    + d^{s\dag}_{\bm{p}}\, v_s(\bm{p})\, e^{ip \cdot x} 
\big],
\ee
where $b^s_{\bm{p}}$ ($d^s_{\bm{p}}$) and $d^{s\dag}_{\bm{p}}$ ($b^{s\dag}_{\bm{p}}$) are annihilation and creation operators satisfying
\be
\{b^s_{\bm{p}}, b_{\bm{p}}^{r\dag} \} 
= \{d^s_{\bm{p}}, d_{\bm{p}}^{r\dag} \} 
= (2\pi)^3\, 2\omega_{\bm{p}}\,
  \delta^{(3)}(\bm{p}-\bm{q})\, 
  \delta^{sr},
\ee
with $u_s(\bm{p})$ and $v_s(\bm{p})$ the Dirac spinors.
For a non-point particle with propagator in Eq.~(\ref{eq:NLpropagator}), the field can be written as
\be
\psi(x)
= \sum_{s=\pm} \int \frac{\dd^4 p}{(2\pi)^4}\,
H(p^2)
\big[ \alpha^s_p\, u_s(\bm{p})\, e^{-ip \cdot x}
    + \beta^{s \dag}_p\, v_s(\bm{p})\, e^{ip \cdot x} 
\big],
\ee
where $\alpha^s_p$ ($\beta^s_p$) and $\beta^{s\dag}_p$ ($\alpha^{s\dag}_p$) are the new annihilation and creation operators for the non-point particle. 
The creation and annihilation operators have the anticommutation relations
\bea
\{\alpha_p^s, \alpha_q^{r\dag} \} 
= \{\beta_p^s, \beta_q^{r\dag} \}
= (2\pi)^4\, \delta^{(4)}(p-q)\,
  \delta^{sr},
\eea
with all other anticommutators vanishing.
The corresponding anticommutation relation for the fermion field at equal time $t$ is
\be \label{eq:solid}
\{ \psi(\bm{x},t), \psi^\dag (\bm{y},t) \} 
= \int \frac{\dd^4 p}{(2\pi)^4}\,
H^2(p^2)\, 2p_0 \gamma^0\, e^{i \bm{p}\cdot (\bm{x}-\bm{y})}
= \int \frac{\dd^3 \bm{p}}{(2\pi)^3}\,
\gamma^0 \Psi(\bm{p})\,
e^{i \bm{p}\cdot (\bm{x}-\bm{y})}\,
\equiv\, \gamma^0\Phi(\bm{x}-\bm{y}),
\ee
where
\be
\Psi(\bm{p}) = \int \frac{\dd p_0}{\pi}\,H^2(p^2)\,p_0.
\ee
The key difference from the quantization condition for a point particle is that the $\delta(\bm{x}-\bm{y})$ is replaced by the function $\Phi(\bm{x}-\bm{y})$ in Eq.~(\ref{eq:solid}).

To derive the relations between $H(p^2)$ and $\widetilde{G}_1(p^2)$, we rewrite the fermion field as 
\bea
\psi(x)
&=& \sum_{s=\pm} \int \frac{\dd^4 p}{(2\pi)^4}
\int \dd M^2\, H(M^2)\,
\delta(p^2 - M^2)
\big[ \alpha^s_p\, u_s(\bm{p})\, e^{-ip \cdot x} 
    + \beta^{s \dag}_p\, v_s(\bm{p})\, e^{ip \cdot x} 
\big] 
\nonumber \\
&=& \sum_{s=\pm} \int \frac{\dd^3 \bm{p}}{(2\pi)^4\, 2\omega_M}
\int \dd M^2\, H(M^2)
\big[ \alpha^s_{\bm{p},\omega_M}\, u_s(\bm{p})\, 
        e^{i\bm{p} \cdot \bm{x}-i\omega_M t} 
    + \beta^{s \dag}_{\bm{p},\omega_M}\, v_s(\bm{p})\,
        e^{-i\bm{p} \cdot \bm{x} + i\omega_M t}
\big].
\eea
By definition, the Feynman propagator can then be written as
\bea \label{eq:propH}
S(x'-x) 
&=& \int \frac{\dd^3 \bm{p}}{4(2\pi)^4 \omega_M \omega_{M'}}
\int \dd M^2 \int \dd M'^2\, H(M^2) H(M'^2)\,
\delta(\omega_{M'} - \omega_{M}) 
\nonumber \\
&& \hspace*{2cm} \times
\big[ \theta(t'-t)(\slashed{p}+m)\, e^{-ip\cdot(x'-x)} 
    - \theta(t-t')(\slashed{p}-m)\, e^{ip\cdot(x'-x)}
\big]
\nonumber \\
&=& \int \frac{\dd^4 p}{(2\pi)^4} \int \frac{\dd M^2}{2\pi} \frac{iH^2(M^2)(\slashed{p}+m)}{p^2 - M^2 + i \epsilon}\,
e^{-ip\cdot (x'-x)}.
\eea
Comparing Eqs.(\ref{eq:NLpropagator}) and (\ref{eq:propH}), we then have \cite{Wang6}
\be
\int \frac{\dd M^2}{2\pi} \frac{H^2(M^2)}{p^2-M^2}
= \frac{1}{(p^2-m^2)\widetilde{G}_1(p^2)},
\ee
for general $\widetilde{G}_1(p^2)$, $H(p^2)$, or $G_2(a,b)$.

For a point particle, with
\be
H^2(p^2) = 2\pi\delta(p^2-m^2), ~~~~~
G_1(a)=\delta(a), ~~~~~
G_2(a,b) = \delta(b),
\ee
the solid quantization will revert back to the point quantization, and the nonlocal Lagrangian will reduce to the local one.
For Pauli-Villars type regularization, one would have
\be
H^2(p^2) = 2\pi \big[ \delta(p^2-m^2) - \delta(p^2-\Lambda^2) \big], ~~~~~
\widetilde{G}_1(p^2) = \frac{p^2-\Lambda^2}{m^2-\Lambda^2},~~~~~ G_2(a,b)=\delta(b).
\ee
For the nonlocal Lagrangian used in the calculation of the nucleon form factors and PDFs in this report, we have
\be
H^2(p^2) = 2\pi \delta(p^2-m^2),~~~~~
\widetilde{G}_1(p^2) = 1~\big[G_1(a)=\delta(a)\big],~~~~~
G_2(a,b) = F(b),
\ee
where $F(b)$ is the correlation function in the nonlocal Lagrangian.
For a non-point physical particle, the $\delta(p^2 - m^2)$ for the point case is replaced by the general function $H(p^2)$, and the creation and annihilation operators $b^s_{\bm{p}}$ ($d^s_{\bm{p}}$) and $d^{s\dag}_{\bm{p}}$ ($b^{s\dag}_{\bm{p}}$) are replaced by $\alpha^s_p$ ($\beta^s_p$) and $\beta^{s\dag}_p$ ($\alpha^{s\dag}_p$).

The solid quantization is easy to be understood in the nonrelativistic case. For example, for a non-point scalar field we can write \cite{Wang3,Wang6}
\be
\phi(\bm{x},t)
= \int \frac{\dd^3\bm{p}}{(2\pi)^3 2\omega_{\bm{p}}}\,
\psi(\bm{p})\,
\big[
  a_{\bm{p}}\, e^{i\bm{p}\cdot \bm{x}-i\omega_{\bm{p}} t}
+ a_{\bm{p}}^\dag\, e^{-i\bm{p}\cdot \bm{x}+i\omega_{\bm{p}} t}
\big],
\ee
where $\psi(\bm{p})$ is the wave function of the particle in momentum space.
The commutation relations of the scalar field in this case are 
\bea
&&[\phi(\bm{x},t),\phi(\bm{y},t)] = [\pi(\bm{x},t),\pi(\bm{y},t)] = 0, \nonumber \\
&&[\phi(\bm{x},t),\pi(\bm{y},t)] = i \Phi(\bm{x} - \bm{y}), 
\eea
where $\pi(\bm{x},t)$ is the conjugate field of $\phi(\bm{x},t)$ and
\be
\Phi(\bm{x} - \bm{y})
= \int \frac{\dd^3 \bm{p}}{(2\pi)^3}\,
\psi^2(\bm{p})\, e^{i \bm{p}\cdot (\bm{x} - \bm{y})}.
\ee
Again, the function $\delta^{(3)}(\bm{x} - \bm{y})$ in the point case is replaced by the function $\Phi(\bm{x} - \bm{y})$ in the case of solid quantization. \\ \\


\begin{thebibliography}{99}
\itemsep -2pt 
\bibitem{Hofstadter} R. Hofstadter and R. W. McAllister, \Journal{\PREV}{98}{217}{1955}
\bibitem{Janssens} T. Janssens, R. Hofstadter, E. B. Hughes, and M. R. Yearian, \Journal{\PREV}{142}{922} {1966}
\bibitem{Price} L. E. Price et al., \Journal{\PRD}{4}{45}{1971}
\bibitem{Rock} S. Rock et al., \Journal{\PRL}{49}{1139} {1982}
\bibitem{Lung} A. Lung et al., \Journal{\PRL}{70}{718} {1993}
\bibitem{Arnold} R. Arnold et al., \Journal{\PRL}{61} {806} {1988}
\bibitem{Berger} Ch. Berger et al., \Journal{\PLB}{35}{87}{1971}
\bibitem{Bruins} E. E. W. Bruins et al., \Journal{\PRL}{75}{21} {1995}
\bibitem{Bartel} W. Bartel et al., \Journal{\NPB}{58}{429}{1973}
\bibitem{Galster} S. Galster et al., \Journal{\NPB}{32}{221} {1971}
\bibitem{Simon} G. G. Simon, Ch. Schmitt, F. Borkowski, and V. H. Walther, \Journal{\NPA}{333}{381}{1980}
\bibitem{Bernauer2} J. Bernauer et al., \Journal{\PRC}{90}{015206}{2014}
\bibitem{Anklin2} H. Anklin et al., \Journal{\PLB}{428} {248} {1998}
\bibitem{Anklin1} H. Anklin et al., \Journal{\PLB}{336}{313}{1994}
\bibitem{Markowitz} P. Markowitz et al., \Journal{\PRC}{48}{5} {1993}
\bibitem{Christy} M. E. Christy et al., \Journal{\PRC}{70}{015206}{2004}
\bibitem{Qattan} I. A. Qattan et al., \Journal{\PRL}{94}{142301}{2005}
\bibitem{Lachniet} J. Lachniet et al., \Journal{\PRL}{102}{192001} {2009}
\bibitem{Sill} A. F. Sill et al., \Journal{\PRD}{48}{29}{1993}
\bibitem{Andivahis} L. Andivahis et al., \Journal{\PRD}{50}{5491}{1994}
\bibitem{Milbrath1} B. D. Milbrath et al., \Journal{\PRL}{80}{452} {1998}
\bibitem{Milbrath2} B. D. Milbrath et al., \Journal{\PRL}{82}{2221 (Erratum)} {1999}
\bibitem{Crawford} C. B. Crawford et al., \Journal{\PRL}{98}{052301} {2007}
\bibitem{Geis} E. Geis et al., \Journal{\PRL}{101} {042501} {2008}
\bibitem{Thompson} A. K. Thompson et al., \Journal{\PRL}{68}{2901} {1992}
\bibitem{Gao} H. Y. Gao et al., \Journal{\PRC}{50}{R546} {1994}
\bibitem{Pospischil} Th. Pospischil et al., {\em Eur. Phys. J.} A 12 (2001) 125
\bibitem{Ostrick} M. Ostrick et al., \Journal{\PRL} {83} {276} {1999}
\bibitem{Glazier} D. Glazier et al.,{\em Eur. Phys. J.} A 24 (2005) 101
\bibitem{Rohe} D. Rohe et al., \Journal{\PRL}{83}{4257} {1999}
\bibitem{Schlimme} B.  Schlimme, et al., \Journal{\PRL}{111}{132504} {2013}
\bibitem{Paolone} M. Paolone et al., \Journal{\PRL}{105}{072001} {2010}
\bibitem{Jones2} M. K. Jones et al., \Journal{\PRL}{84}{1398} {2000}
\bibitem{Gayou2} O. Gayou et al., \Journal{\PRL}{88}{092301} {2002}
\bibitem{Puckett2} A. Puckett et al., \Journal{\PRL}{104}{242301} {2010}
\bibitem{Warren} G. Warren et al., \Journal{\PRL}{92}{042301} {2004}
\bibitem{Riordan} S. Riordan et al., \Journal{\PRL}{105}{262302} {2010}
\bibitem{Anderson} B. Anderson et al., \Journal{\PRC}{75}{034003} {2007}
\bibitem{PDG} P. A. Zyla et al., {\em Prog. Theor. Exp. Phys.} 2020 (2020) 083C01
\bibitem{Xiong} W. Xiong et al., {\em Nature} 575 (2019) 147
\bibitem{Bezginov} N. Bezginov et al., {\em Science} 365 (2019) 1007
\bibitem{Epstein} Z. Epstein, G. Paz, and J. Roy, \Journal{\PRD}{90}{074027} {2014}
\bibitem{Lee} G. Lee, J. R. Arrington, and R. J. Hill, \Journal{\PRD}{92}{013013} {2015}
\bibitem{Lehmann} P. Lehmann, R. Taylor, and R. Wilson, \Journal{\PREV}{126} {1183}{1962}
\bibitem{Hand} L. Hand, D. Miller, and R. Wilson, \Journal{\RMP}{35}{335}{1963}
\bibitem{Murphy} J. Murphy, Y. Shin, and D. Skopik, \Journal{\PRC}{9}{2125}{1974}
\bibitem{Simon2} G. Simon, C. Schmitt, F. Borowski, and V. Walther, \Journal{\NPA}{333}{381}{1990}
\bibitem{Bernauer} J. Bernauer et al., \Journal{\PRL}{105}{242001} {2010}
\bibitem{Zhan2} X. Zhan et al., \Journal{\PLB}{705}{59} {2011}
\bibitem{Sick} I. Sick, \Journal{\PLB}{576}{62} {2003}
\bibitem{Sick2} I. Sick and D. Trautmann, \Journal{\PRC}{89}{012201} {2014}
\bibitem{Hill} R. J. Hill and G. Paz, \Journal{\PRD}{82}{113005} {2010}
\bibitem{Mohr} P. J. Mohr and B. N. Taylor, \Journal{\RMP}{77}{1} {2005}
\bibitem{Mohr2} P. J. Mohr, B. N. Taylor, and D. B. Newell, \Journal{\RMP}{80}{633} {2008}
\bibitem{Mohr3} P. J. Mohr, B. N. Taylor, and D. B. Newell, \Journal{\RMP}{84}{1527} {2012}
\bibitem{Pohl} R. Pohl et al., {\em Nature} 466 (2010) 213
\bibitem{Antognini} A. Antognini et al., {\em Science} 339 (2013) 417
\bibitem{Gasparian} A. Gasparian et al., Jefferson Lab Experiment 12-11-106
\bibitem{Mihovilovic} M. Mihovilovic et al., {\em EPJ Web Conf.} 72 (2014) 00017 
\bibitem{Gilman} R. Gilman et al., arXiv:1302.2160
\bibitem{Spayde} D. T. Spayde et al., \Journal{\PRL}{84}{1106}{2000}
\bibitem{Spayde2} D. T. Spayde et al., \Journal{\PLB}{583}{79}{2004}
\bibitem{Maas} F. E. Maas et al., \Journal{\PRL}{93}{022002}{2004}
\bibitem{Maas2} F. E. Maas et al., \Journal{\PRL}{94}{152001}{2005}
\bibitem{Armstrong} D. S. Armstrong et al., \Journal{\PRL}{95}{092001}{2005} \bibitem{Acha} A. Acha et al., \Journal{\PRL}{98}{032301}{2007}
\bibitem{Aniol} K. A. Aniol et al., \Journal{\PLB}{635}{275}{2006}
\bibitem{Aniol2} K. A. Aniol et al., \Journal{\PRC}{69}{065501}{2004} 
\bibitem{Young} R. D. Young, J. Roche, R. D. Carlini, and A. W. Thomas, \Journal{\PRL}{97}{102002}{2006} 
\bibitem{Baunack} S. Baunack et al., \Journal{\PRL}{102}{151803} {2009}
\bibitem{Jimenez} R. Gonzalez-Jimenez, J. A. Caballero, and T. W. Donnelly, \Journal{\PRD}{90}{033002}{2014} 
\bibitem{Ashman} J. Ashman et al., \Journal{\PLB}{206}{364} {1988}
\bibitem{Anthony} P. L. Anthony et al., \Journal{\PRD}{54}{6620} {1996}
\bibitem{Abe2} K. Abe et al., \Journal{\PRL}{79}{26} {1997}
\bibitem{Anthony2} P. L. Anthony et al., \Journal{\PLB}{493}{19} {2000}
\bibitem{Anthony5} P. L. Anthony et al., \Journal{\PLB}{553}{18} {2003}
\bibitem{Ackerstaff} K. Ackerstaff et al., \Journal{\PLB}{404}{383} {1997}
\bibitem{Airapetian0} A. Airapetian et al., \Journal{\PRD}{75}{012007} {2007}
\bibitem{Airapetian1} A. Airapetian et al., {\em Eur. Phys. J.} C 72 (2012) 1921
\bibitem{Adeva} B. Adeva et al., \Journal{\PRD}{58}{112001} {1998}
\bibitem{Adeva2} B. Adeva et al., \Journal{\PRD}{60}{072004} {1999}
\bibitem{Alexakhin} V. Y. Alexakhin et al., \Journal{\PLB}{647}{8} {2007}
\bibitem{Alekseev} M. G. Alekseev et al., \Journal{\PLB}{690}{466} {2010}
\bibitem{Prok2} Y. Prok et al., \Journal{\PRC}{90}{025212} {2014}
\bibitem{Fersch} R. Fersch et al., \Journal{\PRC}{96}{065208} {2017}
\bibitem{Parno} D. S. Parno et al., \Journal{\PLB}{744}{309} {2015}
\bibitem{Posik} M. Posik et al., \Journal{\PRL}{113}{022002} {2014}
\bibitem{Solvignon} P. Solvignon et al., \Journal{\PRC}{92}{015208} {2015}
\bibitem{Armstrong2} W. Armstrong et al., \Journal{\PRL}{122}{022002} {2019}
\bibitem{Adamczyk} L. Adamczyk et al., \Journal{\PRL}{113}{072301} {2014}
\bibitem{Adare} A. Adare et al., \Journal{\PRD}{93}{051103} {2016}
\bibitem{Adam} J. Adam et al., \Journal{\PRD}{100}{052005} {2019}
\bibitem{Florian2} D. de Florian, R. Sassot, M. Stratmann, and W. Vogelsang, \Journal{\PRL}{113}  {012001}{2014}
\bibitem{Hirai} M. Hirai and S. Kumano, \Journal{\NPB}{813}{106} {2009}
\bibitem{Blumlein} J. Blumlein and H. Bottcher, \Journal{\NPB}{841}{205} {2010}
\bibitem{Leader3} E. Leader, A. V. Sidorov, and D. B. Stamenov, \Journal{\PRD}{91}{054017} {2015}
\bibitem{Khanpour} H. Khanpour, S. T. Monfared, and S. A. Tehrani, \Journal{\PRD}{95}{074006} {2017}
\bibitem{Noceraet} E. R. Nocera et al., \Journal{\NPB}{887}{276} {2014}
\bibitem{Sato2} N. Sato, J. J. Ethier, W. Melnitchouk, M. Hirai, S. Kumano, and A. Accardi, \Journal{\PRD}{94}{114004} {2016}
\bibitem{Ethier} J. J. Ethier, N. Sato, and W. Melnitchouk, \Journal{\PRL}{119}{132001} {2017}
\bibitem{Fanourakis} G. Fanourakis et al., \Journal{\PRD}{21}{562}{1980}
\bibitem{Ahrens2} L. A. Ahrens et al., \Journal{\PLB}{202}{284}{1988}
\bibitem{Barish} S. J. Barish et al., \Journal{\PRD}{16}{3103}{1977}
\bibitem{Millerkl} K. L. Miller et al., \Journal{\PRD}{26}{537}{1982} 
\bibitem{Baker} N. J. Baker et al., \Journal{\PRD}{23}{2499}{1981}
\bibitem{Kitagaki2} T. Kitagaki et al., \Journal{\PRD}{42}{1331}{1990}
\bibitem{Armenise} N. Armenise et al., \Journal{\NPB}{152}{365}{1979}
\bibitem{Liesenfeld} A. Liesenfeld et al., \Journal{\PLB}{468}{20}{1999}
\bibitem{Bloom} E. D. Bloom et al, \Journal{\PRL}{30}{1186}{1973}
\bibitem{Brauel} P. Brauel et al, \Journal{\PLB}{45} {389} {1973}
\bibitem{Guerra2} A. del Guerra et al , \Journal{\NPB}{107}{65}{1976}
\bibitem{Joos} P. Joos et al , \Journal{\PLB}{62} {230}{1976}
\bibitem{Bernard} V. Bernard, L. Elouadrhiri, and U.-G. Meissner, {\em J. Phys.} G 28 (2002) R1
\bibitem{Aguilar-Arevalo} A. A. Aguilar-Arevalo et al., \Journal{\PRD}{82} {092005} {2010}
\bibitem{Golan} T. Golan, K. M. Graczyk, C. Juszczak, and J. T. Sobczyk, \Journal{\PRC}{88}{024612} {2013}
\bibitem{Camacho} C. M. Camacho et al., \Journal{\PRL}{97}{262002} {2006}
\bibitem{Defurne} M. Defurne et al., \Journal{\PRC} {92}{055202} {2015}
\bibitem{Airapetian} A. Airapetian et al., \Journal{\PRL}{87}{182001} {2001} 
\bibitem{Airapetian6} A. Airapetian et al., {\em JHEP} 06 (2010) 019
\bibitem{Airapetian7} A. Airapetian et al., \Journal{\PLB}{704}{15} {2011} 
\bibitem{Airapetian8} A. Airapetian et al., {\em JHEP} 07 (2012) 032
\bibitem{Adloff} C. Adloff et al., {\em Eur. Phys. J.} C 13 (2000) 371  
\bibitem{Aaron} F. D. Aaron et al., {\em JHEP} 05 (2010) 032
\bibitem{Chekanov} S. Chekanov et al., \Journal{\NPB}{718}{3} {2005}
\bibitem{Hadjidakis} C. Hadjidakis et al., \Journal{\PLB}{605}{256} {2005} 
\bibitem{Morrow} S. A. Morrow et al., {\em Eur. Phys. J.} A 39 (2009) 5  
\bibitem{Bedlinskiy} I. Bedlinskiy et al., \Journal{\PRL}{109}{112001} {2012}
\bibitem{Gao2} H. Gao et al., {\em Eur. Phys. J. Plus} 126 (2011) 2
\bibitem{Kubarovsky} V. Kubarovsky, {\em Nucl. Phys. Proc. Suppl.} 219 (2011) 118 
\bibitem{Armstrong3} W. Armstrong et al., arXiv:1708.00888 [nucl-ex]
\bibitem{d'Hose} N. d'Hose, E. Burtin, P. A. M. Guichon, and J. Marroncle, {\em Eur. Phys. J.} A 19S1 (2004) 47
\bibitem{Silva} L. Silva, {\em Few Body Syst.} 54 (2013) 1075
\bibitem{Kouznetsov} O. Kouznetsov, {\em Nucl. Part. Phys. Proc.} 270 (2016)  36 
\bibitem{Sawada} T. Sawada, W. C. Chang, S. Kumano, J. C. Peng, S. Sawada, and K. Tanaka, \Journal{\PRD}{93}{114034} {2016}
\bibitem{Kroll} P. Kroll, {\em JPS Conf. Proc.} 13 (2017) 010014
\bibitem{Accardi} A. Accardi et al., {\em Eur. Phys. J.} A 52 (2016) 268
\bibitem{Fernandez} J. L. Abelleira Fernandez et al., {\em J. Phys.} G 39 (2012) 075001
\bibitem{Airapetian10} A. Airapetian et al., \Journal{\PRL}{94}{012002} {2005}
\bibitem{Alexakhin2} V. Y. Alexakhin et al., \Journal{\PRL}{94}{202002} {2005}
\bibitem{Adolph} C. Adolph et al., \Journal{\PLB}{744}{250} {2015}
\bibitem{Adolph2} C. Adolph et al., \Journal{\PLB}{753}{406} {2016}
\bibitem{Qian} X. Qian et al., \Journal{\PRL}{107}{072003} {2011} 
\bibitem{Zhangy} Y. Zhang et al., \Journal{\PRC}{90}{055209} {2014} 
\bibitem{Huang} J. Huang et al., \Journal{\PRL}{108}{052001} {2012}
\bibitem{Yan} X. Yan et al., \Journal{\PRC}{95}{035209} {2017} 
\bibitem{Zhuly} L. Y. Zhu et al., \Journal{\PRL}{99}{082301} {2007}
\bibitem{Adamczyk2} L. Adamczyk et al., \Journal{\PRL}{116}{132301} {2016}
\bibitem{Meissner} S. Meissner, A. Metz, and M. Schlegel, {\em JHEP} 08 (2009) 056
\bibitem{Lorce} C. Lorcé and B. Pasquini, {\em JHEP} 09 (2013) 138
\bibitem{Lorce2} C. Lorcé and B. Pasquini, \Journal{\PRD}{84}{014015} {2011}
\bibitem{Arneodo} M. Arneodo et al., \Journal{\PRD}{50}{R1} {1994}
\bibitem{Ackerstaff2} K. Ackerstaff et al., \Journal{\PRL}{81}{5519} {1998}
\bibitem{Baldit} A. Baldit et al., \Journal{\PLB}{332}{244} {1994}
\bibitem{Hawker} E. A. Hawker et al., \Journal{\PRL}{80}{3715} {1998}
\bibitem{Towell} R. S. Towell et al., \Journal{\PRD}{64}{052002} {2001}
\bibitem{SeaQuest} J.~Dove et al., {\it Nature} 64 (2021) 052002
\bibitem{Zeller} G. P. Zeller et al., \Journal{\PRL}{88}{091802} {2002}
\bibitem{Bentz} W. Bentz, I. C. Clo\"{e}t, J. T. Londergan, and A. W. Thomas, \Journal{\PLB}{693}{462}{2010}
\bibitem{Barone} V. Barone, C. Pascaud, and F. Zomer, {\em Eur. Phys. J.} C 12 (2000) 243
\bibitem{Bazarko} A. O. Bazarko et al., {\em Z. Phys.} C 65 (1995) 189
\bibitem{Mason} D. Mason et al., \Journal{\PRL}{99}{192001} {2007}
\bibitem{Lilf} L. F. Li and T. P. Cheng, {\em Lect. Notes Phys.} 512 (1998) 115 
\bibitem{Chung} P. L. Chung and F. Coester, \Journal{\PRD}{44}{229} {1991}
\bibitem{Gross} F. Gross, G. Ramalho, and M. Pena, \Journal{\PRC}{77}{015202}{2008}
\bibitem{Gross2} F. Gross and P. Agbakpe, \Journal{\PRC}{73}{015203} {2006}
\bibitem{Gross3} F. Gross, G. Ramalho, and M. Pena, \Journal{\PRD}{85}{093005}{2012}
\bibitem{Isgur} N. Isgur and G. Karl, \Journal{\PRD}{18}{4187} {1978}
\bibitem{Capstick} S. Capstick and N. Isgur, \Journal{\PRD}{34}{2809} {1986}
\bibitem{Ferraris} M. Ferraris, M. M. Giannini, M. Pizzo, E. Santopinto, and L. Tiator, \Journal{\PLB}{364}{231} {1995}
\bibitem{Sanctis} M. De Sanctis, M. M. Giannini, L. Repetto, and E. Santopinto, \Journal{\PRC}{62}{025208} {2000}
\bibitem{Sanctis2} M. De Sanctis, M. M. Giannini, E. Santopinto, and A. Vassallo, \Journal{\PRC}{76} {062201} {2007}
\bibitem{Glozman} L. Y. Glozman, Z. Papp, W. Plessas, K. Varga, and R. F. Wagenbrunn, \Journal{\PRC}{57}{3406} {1998}
\bibitem{Glozman2} L. Y. Glozman, W. Plessas, K. Varga, and R. F. Wagenbrunn, \Journal{\PRD}{58}{094030} {1998}
\bibitem{Boffi} S. Boffi, L. Y. Glozman, W. Klink, W. Plessas, M. Radici, and R. F. Wagenbrunn, {\em Eur. Phys. J.} A 14 (2002) 17
\bibitem{Zhang} Z. Y. Zhang, Y. W. Yu, P. N. Shen, L. R. Dai, and A. Faessler \Journal{\NPA}{625}{59}{1997}
\bibitem{Dai} L. R. Dai, Z. Y. Zhang, Y. W. Yu, and P. Wang, \Journal{\NPA}{727}{321}{2003}
\bibitem{Miller2} G. A. Miller, A. W. Thomas, and S. Theberge, \Journal{\PLB}{91}{192} {1980}
\bibitem{Thomas} A. W. Thomas, S. Theberge, and G. A. Miller, \Journal{\PRD}{24}{216}{1981}
\bibitem{Lu} D. H. Lu, A. W. Thomas, and A. G. Williams, \Journal{\PRC}{57}{2628} {1998}
\bibitem{Oset} E. Oset, R. Tegen, and W. Weise, \Journal{\NPA}{426}{456}{1984}
\bibitem{Faessler} A. Faessler, T. Gutsche, V. E. Lyubovitskij, and K. Pumsaard, \Journal{\PRD}{73}{114021} {2006}
\bibitem{Lyubovitskij} V. E. Lyubovitskij, P. Wang, T. Gutsche, and A. Faessler, \Journal{\PRC}{66}{055204} {2002}
\bibitem{Faessler1} A. Faessler, T. Gutsche, B. R. Holstein, V. E. Lyubovitskij, D. Nicmorus, and K. Pumsaard, \Journal{\PRD}{74}{074010} {2006}
\bibitem{Obukhovsky} I. T. Obukhovsky, A. Faessler, T. Gutsche, and V. E. Lyubovitskij, \Journal{\PRD}{89}{014032} {2014}
\bibitem{Carroll} J. Carroll, D. B. Lichtenberg, and J. Franklin, \Journal{\PREV}{174}{1681}{1968}
\bibitem{Sanctis3} M. De Sanctis, J. Ferretti, E. Santopinto, and A. Vassallo, \Journal{\PRC}{84}{055201} {2011}
\bibitem{Ferretti} J. Ferretti, A. Vassallo, and E. Santopinto, \Journal{\PRC} {83}{065204} {2011}
\bibitem{Frank} M. R. Frank, B. K. Jennings, and G. A. Miller, \Journal{\PRC}{54}{920} {1996}
\bibitem{Schlumpf} F. Schlumpf, \Journal{\PRD}{47}{4114} {1993}
\bibitem{Miller} G. A. Miller and M. R. Frank, \Journal{\PRC}{65}{065205} {2002}
\bibitem{Pace} E. Pace, G. Salme, F. Cardarelli, and S. Simula, \Journal{\NPA}{666}{33} {2000}
\bibitem{Cloet} I. C. Clo\"{e}t and G. A. Miller, \Journal{\PRC}{86}{015208}{2012}
\bibitem{Chakrabarti} D. Chakrabarti, N. Kumar, T. Maji, and A. Mukherjee, {\em Eur. Phys. J. Plus} 135 (2020) 496
\bibitem{Zhang2} J. Zhang and B.-Q. Ma, \Journal{\PRC}{93}{065209} {2016}
\bibitem{CJ1999} H.-M. Choi and C.-R. Ji, \Journal{\PRD}{59}{074015}{1999} 
\bibitem{CJ2015} H.-M. Choi, C.-R. Ji, Z.Li and H.-Y. Ryu, \Journal{PRC}{92}{055203}{2015} 
\bibitem{Ivanov} M. A. Ivanov and V. E. Lyubovitskij, {\em Phys.  Lett.}  B 408 (1997) 435
\bibitem{Faessler2} A. Faessler, T. Gutsche, M. A. Ivanov, J. G. Korner, and V. E. Lyubovitskij, {\em Phys. Lett.} B 518 (2001) 55
\bibitem{Faessler3} A. Faessler, T. Gutsche, M. A. Ivanov, V. E. Lyubovitskij, and P. Wang, \Journal{\PRD}{68}{014011}{2003}
\bibitem{Klevansky} S. P. Klevansky, {\em Rev. Mod. Phys.} 64 (1992) 649
\bibitem{Weigel} H. Weigel, A. Abada, R. Alkofer, and H. Reinhardt, \Journal{\PLB}{353}{20}{1995}
\bibitem{Serrano} M. E. Carrillo-Serrano, W. Bentz, I. C. Clo\"{e}t, and A. W. Thomas, \Journal{\PLB}{759}{178}{2016}
\bibitem{Courtoy} A. Courtoy, S. Noguera, and S. Scopetta, {\em JHEP} 12 (2019) 045
\bibitem{Roberts} C. D. Roberts, {\em Prog. Part. Nucl. Phys.} 61 (2008) 50
\bibitem{Nicmorus} D. Nicmorus, G. Eichmann, and Reinhard Alkofer, \Journal{\PRD}{82}{114017}{2010}
\bibitem{Eichmann} G. Eichmann, \Journal{\PRD}{84}{2011}{014014}
\bibitem{Bednar} K. D. Bednar, I. C. Clo\"{e}t, and P. C. Tandy, \Journal{\PLB}{782}{675}{2018}
\bibitem{Iachello} F. Iachello, A. D. Jackson, and A. Lande, \Journal{\PLB}{43}{191}{1973}
\bibitem{Gari} M. F. Gari and W. Krumpelmann, \Journal{\PLB}{274}{159}{1992}
\bibitem{Williams} R. A. Williams and C. Puckett-Truman, \Journal{\PRC}{53}{1580}{1996}
\bibitem{Bijker} R. Bijker, {\em Eur. Phys. J.} A 32 (2007) 403
\bibitem{Faessler4} A. Faessler, M. I. Krivoruchenko, and B. V. Martemyanov, \Journal{\PRC}{82}{038201}{2010}
\bibitem{Christov} C. V. Christov et al., {\em Prog. Part. Nucl. Phys.} 37 (1996) 91
\bibitem{Penttinen} M. Penttinen, M.V. Polyakov, and K. Goeke, \Journal{\PRD}{62}{014024}{2000} 
\bibitem{Schweitzer} P. Schweitzer, M. Colli, and S. Boffi, \Journal{\PRD}{67}{114022}{2003} 
\bibitem{Ossmann} J. Ossmann, M. V. Polyakov, P. Schweitzer, D. Urbano, and K. Goeke, \Journal{\PRD}{71}{034011}{2005}
\bibitem{Wakamatsu} M. Wakamatsu and H. Tsujimoto, \Journal{\PRD}{71}{074001}{2005}
\bibitem{Goeke} K. Goeke et al., \Journal{\PRD} {75} {094021} {2007}
\bibitem{Abidin} Z. Abidin and C. E. Carlson, \Journal{\PRD}{79}{115003}{2005}
\bibitem{Brodsky3} S. J. Brodsky and G. F. de Teramond, \Journal{\PRD}{83}{036011}{2011}
\bibitem{Hashimoto} K. Hashimoto, T. Sakai, and S. Sugimoto, {\em Prog. Theor. Phys.} 120 (2008) 1093.
\bibitem{Traini} M. C. Traini, {\em Eur. Phys. J.} C 77 (2017) 246
\bibitem{Wilcox} W. Wilcox, T. Draper, and K.-F. Liu, \Journal{\PRD}{46}{1109}{1992}
\bibitem{Zanotti} J. M. Zanotti, D. B. Leinweber, A. G. Williams, and J. B. Zhang, {\em Nucl. Phys. B Proc. Suppl.} 129 (2004) 287
\bibitem{Gockeler} M. Gockeler et al., \Journal{\PRD}{71}{034508}{2005}
\bibitem{Sasaki} S. Sasaki and T. Yamazaki, \Journal{\PRD}{78}{014510}{2008}
\bibitem{Alexandrou} C. Alexandrou, G. Koutsou, J. W. Negele, and A. Tsapalis, \Journal{\PRD}{74}{034508}{2006}
\bibitem{Gockeler2} M. G\"{o}ckeler et al., {\em PoS} LATTICE2007 (2007) 161
\bibitem{Alexandrou2} C. Alexandrou et al., \Journal{\PRD}{88}{014509}{2013} 
\bibitem{Lin} H.-W. Lin, T. Blum, S. Ohta, S. Sasaki, and T. Yamazaki, \Journal{\PRD}{78}{014505} {2008}
\bibitem{Syritsyn} S. N. Syritsyn et al., \Journal{\PRD}{81}{034507}{2010}
\bibitem{Bhattacharya} T. Bhattacharya et al., \Journal{\PRD}{89}{094502}{2014}
\bibitem{Green} J. R. Green et al., \Journal{\PRD}{90}{074507}{2014}
\bibitem{Ishikawa} K.-I. Ishikawa et al., \Journal{\PRD}{98} {074510}{2018}
\bibitem{Alexandrou3} C. Alexandrou et al., \Journal{\PRD}{96}{034503}{2017}
\bibitem{Shintani} E. Shintani, K. Ishikawa, Y. Kuramashi, S. Sasaki, and T. Yamazaki, \Journal{\PRD}{99}{014510}{2019}, [Erratum: ibid 102 (2020) 019902] 
\bibitem{Jang} Y. Jang, R. Gupta, H.-W. Lin, B. Yoon, and T. Bhattacharya, \Journal{\PRD}{101}{014507} {2020}
\bibitem{Babich} R. Babich et al., \Journal{\PRD}{85}{054510}{2012}
\bibitem{Sufian} R. S. Sufian, Y. B. Yang, J. Liang, T. Draper,and K.-F. Liu, \Journal{\PRD}{96}{114504}{2017}
\bibitem{Alexandrou4} C. Alexandrou et al., \Journal{\PRD}{97}{094504}{2018}
\bibitem{Jang3} Y.-C. Jang, R. Gupta, B. Yoon, and T. Bhattacharya, \Journal{\PRL}{124}{072002}{2020}
\bibitem{Alexandrou5} C. Alexandrou et al., \Journal{\PRD}{96}{054507}{2017}
\bibitem{Green0} J. Green et al, \Journal{PRD}{95}{114502}{2017}
\bibitem{Horsley} R. Horsley et al., \Journal{\PLB}{732}{41}{2014}
\bibitem{Bali} G. Bali et al., \Journal{\PRD}{91}{054501}{2015}
\bibitem{Gupta} R. Gupta et al., \Journal{\PRD}{98}{034503}{2018}
\bibitem{Chang} C. C. Chang et al., {\em Nature} 558 (2018) 91
\bibitem{Green1} J. R. Green et al., \Journal{\PLB}{734}{290}{2014}
\bibitem{Alexandrou10} C. Alexandrou et al., \Journal{\PRD}{95}{114514} {2017}, [Erratum: ibid 96 (2017) 099906]
\bibitem{Bhattacharya2} T. Bhattacharya, V. Cirigliano, R. Gupta, H.-W. Lin, and B. Yoon, \Journal{\PRL}{115}{212002} {2015}
\bibitem{Green2} J. R. Green et al., \Journal{\PRD}{86}{114509} {2012}
\bibitem{Yoki} Y. Aoki et al., \Journal{\PRD}{82}{014501} {2010}
\bibitem{Abdel-Rehim} A. Abdel-Rehim et al., \Journal{\PRD}{92}{114513} {2015}, [Erratum: ibid 93 (2016) 039904]
\bibitem{Bali2} G. S. Bali et al., \Journal{\NPB}{866} {1}{2013} 
\bibitem{Alvarez-Ruso} L. Alvarez-Ruso, T. Ledwig, J. M. Camalich, and M. J. Vicente-Vacas, \Journal{\PRD}{88} {054507}{2013}
\bibitem{Alexandrou9} C. Alexandrou, V. Drach, K. Jansen, C. Kallidonis, and G. Koutsou, \Journal{\PRD}{90} {074501} {2014}
\bibitem{Alexandrou8} C. Alexandrou et al., \Journal{\PRL}{119}{142002}{2017}
\bibitem{Lin3} H.-W. Lin, R. Gupta, B. Yoon, Y.-C. Jang, and T. Bhattacharya, \Journal{\PRD}{98}
{094512} {2018}
\bibitem{Gupta2} R. Gupta et al., \Journal{\PRD}{98}{091501} {2018}
\bibitem{Izubuchi} T. Izubuchi, H. Ohki, and S. Syritsyn, {\em PoS} LATTICE2019 (2020) 290 
\bibitem{Lin2} H.-W. Lin, {\em Int. J. Mod. Phys.} A 35 (2020) 2030006
\bibitem{Gockeler3} M. G\"{o}ckeler et al., \Journal{\PRD}{71}{114511}{2005}
\bibitem{Ji} X. Ji, \Journal{\PRL}{110}{262002} {2013}
\bibitem{Radyushkin} A. V. Radyushkin, \Journal{\PLB}{767}{314} {2017}
\bibitem{Radyushkin2} A. V. Radyushkin, \Journal{\PRD}{96}{034025} {2017}
\bibitem{Ioffe} B. L. Ioffe, \Journal{\PLB}{30}{123} {1969}
\bibitem{Braun} V. Braun, P. Gornicki, and L. Mankiewicz, \Journal{\PRD}{51}{6036} {1995}
\bibitem{Lin4} H.-W. Lin, J.-W. Chen, S. D. Cohen, and X. Ji, \Journal{\PRD}{91}{054510} {2015}
\bibitem{Alexandrou11} C. Alexandrou et al., \Journal{\PRD}{92}{014502} {2015}
\bibitem{Chen2} J.-W. Chen et al., \Journal{\PRD}{97}{014505} {2018}
\bibitem{Chen} J.-W. Chen, S. D. Cohen, X. Ji, H.-W. Lin, and J.-H. Zhang, \Journal{\NPB}{911}{246}
{2016}
\bibitem{Alexandrou12} C. Alexandrou et al., \Journal{\PRD}{96}{014513} {2017}
\bibitem{Green3} J. Green, K. Jansen, and F. Steffens, \Journal{\PRL}{121}{022004} {2018}
\bibitem{Alexandrou13} C. Alexandrou et al., \Journal{\PRD}{98}{091503}{2018}
\bibitem{Weinberg} S. Weinberg, \Journal{\PRD}{13}{974}{1976}
\bibitem{Weinberg2} S. Weinberg, {\em Physica} A 96 (1979) 327
\bibitem{Gasser} J. Gasser and H. Leutwyler, {\em Ann. Phys.} 158 (1984) 142
\bibitem{Gasser2} J. Gasser and H. Leutwyler, \Journal{\NPB}{250}{465} {1985}
\bibitem{Jenkins} E. Jenkins and A. V. Manohar, \Journal{\PLB}{255}{558} {1991}
\bibitem{Jenkins2} E. Jenkins and A. V. Manohar, \Journal{\PLB}{259}{353} {1991}
\bibitem{Bernard2} V. Bernard, N. Kaiser, J. Kambor, and U.-G. Meissner, \Journal{\NPB}{388}{315} {1992}
\bibitem{Wang4} P. Wang, {\em Eur. Phys. J.} A 50 (2014) 172
\bibitem{Ellis} P. J. Ellis and H. B. Tang, \Journal{\PRC}{57}{3356} {1998}
\bibitem{Becher} T. Becher and H. Leutwyler, {\em Eur. Phys. J.} C 9 (1999) 643
\bibitem{Gegelia} J. Gegelia and G. Japaridze, \Journal{\PRD}{60}{114038} {1999}
\bibitem{Lutz} M. F. Lutz and E. E. Kolomeitsev, \Journal{\NPA}{700}{193} {2002}
\bibitem{Ellis2} P. J. Ellis and K. Torikoshi, \Journal{\PRC}{61}{015205} {2000}
\bibitem{Kubis} B. Kubis and U.-G. Meissner, {\em Eur. Phys. J.} C 18 (2001) 747
\bibitem{Zhusl} S. L. Zhu, G. Sacco, and M. J. Ramsey-Musolf, \Journal{\PRD}{66}{034021} {2002}
\bibitem{Bernard3} V. Bernard, T. R. Hemmert, and U.-G. Meissner, \Journal{\PRD}{67}{076008} {2003}
\bibitem{Bauer} T. Bauer, J. C. Bernauer, and S. Scherer \Journal{\PRC}{86}{065206} {2012}
\bibitem{Wein} P. Wein, P. C. Bruns, and A. Schaefer \Journal{\PRD}{89}{116002} {2014}
\bibitem{Borasoy} B. Borasoy and S. Wetzel, \Journal{\PRD}{63}{074019}{2001}
\bibitem{Fuchs} T. Fuchs, J. Gegelia, G. Japaridze, and S. Scherer, \Journal{\PRD}{68}{056005}{2003}
\bibitem{Schindler} M. R. Schindler, J. Gegelia, and S. Scherer, \Journal{\PLB}{586}{258}{2004}
\bibitem{Fuchs2} T. Fuchs, J. Gegelia, and S. Scherer, {\em J. Phys.} G 30 (2004) 1407
\bibitem{Kubis2} B. Kubis and U.-G. Meissner, \Journal{\NPA}{679}{698}{2001}
\bibitem{Young2} R. D. Young, D. B. Leinweber, A. W. Thomas, and S. V. Wright, \Journal{\PRD}{66}{094507}{2002}
\bibitem{Leinweber} D. B. Leinweber, A. W. Thomas, and R.D. Young, \Journal{\NPA}{755}{59}{2005}
\bibitem{Wang} P. Wang, D. B. Leinweber, A. W. Thomas, and R.D. Young, \Journal{\PRD}{75}{073012} {2007}
\bibitem{Wang2} P. Wang, D. B. Leinweber, A. W. Thomas, and R.D. Young, \Journal{\PRC}{79}{065202} {2009}
\bibitem{Hall} J. M. M. Hall, D. B. Leinweber, and R. D. Young, \Journal{\PRD}{89}{054511} {2014}
\bibitem{Li} H. Li, P. Wang, D. B. Leinweber, and A. W. Thomas, \Journal{\PRC}{93}{045203} {2016}
\bibitem{Shanahan} P.E. Shanahan, et al, \Journal{\PRD}{90}{034502} {2014}
\bibitem{He} F. He and P. Wang, \Journal{\PRD}{97}{036007} {2018}
\bibitem{He2} F. He and P. Wang, \Journal{\PRD}{98}{036007} {2018}
\bibitem{Salamu} Y. Salamu,  C.-R. Ji, W. Melnitchouk, A.W. Thomas, and P. Wang, \Journal{\PRD}{99} {014041} {2019}
\bibitem{Salamu2} Y. Salamu,  C.-R. Ji, W. Melnitchouk, A.W. Thomas, P. Wang, and X.G. Wang, \Journal{\PRD}{100} {094026}{2019}
\bibitem{He3} F. He and P. Wang, \Journal{\PRD}{100}{074032}{2019}
\bibitem{He4} F. He and P. Wang, {\em Eur. Phys. J. Plus} 135 (2020) 156
\bibitem{Scherer} S. Scherer, {\em Adv. Nucl. Phys.} 27 (2003) 277
\bibitem{Georgi} H. Georgi, Weak Interactions and Modern Particle Theory, Benjamin/Cummings, Menlo Park, 1984
\bibitem{Gasser3} J. Gasser, M.E. Sainio, and A. Svarc, \Journal{\NPB}{307}{779}{1988}
\bibitem{Fettes} N. Fettes, U.-G. Meissner, and S. Steininger, \Journal{\NPA}{640} {199}{1998}
\bibitem{Foldy} L. L. Foldy and S. A. Wouthuysen, \Journal{\PREV}{78}{29} {1950}
\bibitem{Tiburzi} B. C. Tiburzi and A. Walker-Loud, \Journal{\NPA} {764}{274}{2006}
\bibitem{Young3} R. D Young, D. B. Leinweber, and A. W. Thomas, {\em Prog. Part. Nucl. Phys.} 50 (2003) 399
\bibitem{Terning} J. Terning, \Journal{\PRD}{44} {887} {1991}
\bibitem{Yang2} M. Y. Yang and P. Wang, \Journal{\PRD}{102}{056024}{2020}
\bibitem{Anatomy} C.-R. Ji, W. Melnitchouk and A. W. Thomas, \Journal{\PRD}{88}{076005}{2013} 
\bibitem{Liu} J. Liu, R. D. McKeown, and M. J. Ramsey-Musolf, \Journal{\PRC}{76}{025202}{2007}
\bibitem{Jimenez2}R. González-Jiménez, J. A. Caballero, and T. W. Donnelly, {\em Phys. Rep.} 524 (2013) 1
\bibitem{Hemmert} T. R. Hemmert, B. Kubis, and U.-G. Meissner, \Journal{\PRC}{60}{045501} {1999}
\bibitem{Musolf} M. J. Musolf and H. Ito, \Journal{\PRC}{55}{3066} {1997}
\bibitem{Kubis3} B. Kubis, {\em  Eur. Phys. J.} A 24S2 (2005) 97
\bibitem{Yang} M. Y. Yang and P. Wang, {\em Chin. Phys.} C 44 (2020) 053101 
\bibitem{Leinweber2} D. B. Leinweber, \Journal{\PRD}{69}{014005} {2004}
\bibitem{Bernard4} C. W. Bernard and M. F. L. Golterman, \Journal{\PRD}{46}{853}{1992}
\bibitem{Wang5} P. Wang, D. B. Leinweber, and A. W. Thomas, \Journal{\PRD}{92}{034508}{2015}
\bibitem{Lin5} H.-W. Lin and K. Orginos, \Journal{\PRD}{79}{074507} {2009}
\bibitem{Shanahan2} P. Shanahan et al., \Journal{\PRD}{89}{074511} {2014}
\bibitem{Blin} A. Hiller Blin, \Journal{\PRD}{96}{093008}{2017}
\bibitem{Liu2} X. Liu, K. Khosonthongkee, A. Limphirat, and Y. Yan, {\em J. Phys.} G 41 (2014) 055008
\bibitem{Boinepalli} S. Boinepalli, D. B. Leinweber, A. G. Williams, J. M. Zanotti, and J. B. Zhang, \Journal{\PRD}{74}{093005}{2006}
\bibitem{Wangradii} P. Wang, D. B. Leinweber, A. W. Thomas, and R. D. Young, \Journal{\PRD}{79}{094001}{2009}
\bibitem{XGWang} X. G. Wang, C.-R. Ji, W. Melnitchouk, Y. Salamu, A. W. Thomas, and P. Wang, \Journal{\PRD}{94}{094035} {2016}
\bibitem{Chen3} J.-W. Chen and X. Ji, {\em Phys. Rev. Lett.} 87 (2001) 152002, [Erratum ibid 88 (2001) 249901]
\bibitem{Diehl} M. Diehl, T. Feldmann, R. Jakob, and P. Kroll, {\em Eur. Phys. J.} C 39 (2005) 1
\bibitem{Martin} A. D. Martin, R. G. Roberts, W. J. Stirling, and R. S. Thorne, {\em Eur. Phys. J.} C 4 (1998) 463
\bibitem{Leader4} E. Leader, A. V. Sidorov, and D. B. Stamenov, \Journal{\PRD}{75} {074027}{2007}
\bibitem{Burkardt} M. Burkardt, K. S. Hendricks, C.-R. Ji, W. Melnitchouk, and A. W. Thomas, \Journal{\PRD}{87}{056009} {2013}
\bibitem{Salamu3} Y. Salamu, C.-R. Ji, W. Melnitchouk, and P. Wang, \Journal{\PRL}{114}{122001} {2015}
\bibitem{XGWang2} X. G. Wang, C.-R. Ji, W. Melnitchouk, Y. Salamu, A.W. Thomas, and P. Wang, \Journal{\PLB}{762}{52} {2016}
\bibitem{Holtmann} H. Holtmann, A. Szczurek, and J. Speth, \Journal{\NPA}{596}{631} {1996}
\bibitem{Flauger} W. Flauger and F. Monnig, \Journal{\NPB}{109}{347} {1976}
\bibitem{Blobel} V. Blobel et al., \Journal{\NPB}{135}{379} {1978}
\bibitem{McKenney} J. R. McKenney, N. Sato, W. Melnitchouk, and C.-R. Ji, \Journal{\PRD}{93}{054011}{2016}
\bibitem{Barish2} S. J. Barish et al., \Journal{\PRD}{12}{1260} {1975}
\bibitem{Jaeger} K. Jaeger et al., \Journal{\PRD}{11}{1756} {1975}
\bibitem{Bockmann} K. Bockmann et al., \Journal{\NPB}{143}{395} {1978}
\bibitem{Alexandrou:2020sml} C. Alexandrou et al., \Journal{\PRD}{101}{094513} {2020}
\bibitem{Thomas:2008ga}  A. W. Thomas, \Journal{\PRL}{101}{102003} {2008}
\bibitem{Schreiber:1988uw} A. W. Schreiber and A. W. Thomas, \Journal{\PLB}{215}{141} {1988}
\bibitem{Myhrer:2007cf} F. Myhrer and A. W. Thomas, \Journal{\PLB}{663}{302} {2008}
\bibitem{Bass:2009ed} S. D. Bass and A. W. Thomas, \Journal{\PLB}{684}{216} {2010}
\bibitem{Yamaguchi:1989sx} T. Yamaguchi, K. Tsushima, Y. Kohyama, and K. Kubodera, \Journal{\NPA}{500}{429} {1989}
\bibitem{Lin:2018} H.-W. Lin et al., {\em Prog. Part. Nucl. Phys.} 100 (2018) 107
\bibitem{XGWang3} X. G. Wang, C.-R. Ji, W. Melnitchouk, Y. Salamu, A. W. Thomas, and P. Wang, \Journal{\PRD}{102}{116020}{2020}
\bibitem{He6} F. He, C.-R. Ji, W. Melnitchouk, Y. Salamu, A. W. Thomas, P. Wang, and X.G. Wang, \Journal{\PRD}{105}{094007}{2022}
\bibitem{NNPDF:2014} E. R. Nocera et al., \Journal{\NPB}{887}{276} {2014}
\bibitem{Hartland:2013} N. P. Hartland and E.R. Nocera, {\em Nucl. Phys. Proc. Suppl.} 234 (2013) 54
\bibitem{Brodsky:1997de} S. J. Brodsky, H.-C. Pauli, and S. S. Pinsky, {\em Phys. Rep.} 301 (1998) 299 
\bibitem{Ji1998} X. Ji, {\em J. Phys.} G 24 (1998) 1181
\bibitem{He5} F. He, C.-R. Ji, W. Melnitchouk, A. W. Thomas, and P. Wang, arXiv:2202.00266
\bibitem{Cocuzza} C. Cocuzza, W. Melnitchouk, A. Metz, and N. Sato, \Journal{\PRD}104{074031}{2021}
\bibitem{Faura} F. Faura, S. Iranipour, E. R. Nocera, J. Rojo, and M. Ubiali, {\em Eur. Phys. J.} C 80 (2020) 1168 
\bibitem{Pisano} S. Pisano, {\em EPJ Web Conf.} 85 (2015) 02033
\bibitem{Boer} D. B\"{o}er and P. J. Mulders, \Journal{\PRD}{57}{5780} {1998}
\bibitem{Sivers} D. W. Sivers, \Journal{\PRD}{41}{83} {1990}
\bibitem{Airapetian:2009ae} A. Airapetian et al., \Journal{\PRL}{103} {152002}{2009}
\bibitem{Bradamante} F. Bradamante, {\em J. Phys. Conf. Ser.} 938 (2017) 012004
\bibitem{Bravar} A. Bravar, {\em Nucl. Phys. Proc. Suppl.} 79 (1999) 520
\bibitem{Airapetian:2001eg} A. Airapetian et al., \Journal{\PRD}{64}{097101}
{2001}
\bibitem{Bacchetta:2004jz} A. Bacchetta, U. D’Alesio, M. Diehl, and C. A. Miller, \Journal{\PRD}{70}{117504}{2004}
\bibitem{Ji:2002aa} X. Ji and F. Yuan, \Journal{\PLB}{543}{66}{2002}
\bibitem{Jenkins3} E. E. Jenkins, \Journal{\NPB}{368} {190}{1992}
\bibitem{Geng} L. S. Geng, J. M. Camalich, and M. J. V. Vacas, \Journal{\PRD}{80}{034027}{2009}
\bibitem{Jones3} H. F. Jones and M. D. Scadron, {\em Annals Phys.} 81 (1973) 1
\bibitem{Brown} G. E. Brown and W. Weise, {\em Phys. Rep.} 22 (1975) 279
\bibitem{Mergell} P. Mergell, U.-G. Meissner, and D. Drechsel, {\em Nucl. Phys.} A 596 (1996) 367
\bibitem{Haidenbauer} J. Haidenbauer, S. Petschauer, N. Kaiser, U.-G. Meissner, and W. Weise, {\em Eur. Phys. J.} C 77 (2017) 760
\bibitem{Aicher} M. Aicher, A. Schafer, and W. Vogelsang, \Journal{\PRL}{105}{252003}{2010}
\bibitem{Bacchetta} A. Bacchetta and M. Radici, \Journal{\PRL}{107} {212001} {2011}
\bibitem{Martin2} A. Martin, F. Bradamante, and V. Barone, \Journal{\PRD}{95}{094024}{2017}
\bibitem{Aoyama} T. Aoyama, T. Kinoshita, and M. Nio, \Journal{\PRD}{97}{036001}{2018}
\bibitem{Aoyama2} T. Aoyama et al., {\em Phys. Rep.} 887 (2020) 1
\bibitem{Abi} B. Abi et al., \Journal{\PRL}{126} {141801} {2021}
\bibitem{Bennett} G. W. Bennett et al., \Journal{\PRD}{73}{072003}{2006}
\bibitem{Hanneke} D. Hanneke, S. Fogwell, and G. Gabrielse, \Journal{\PRL}{100}{120801}{2008}
\bibitem{Hanneke2} D. Hanneke, S. F. Hoogerheide, and G. Gabrielse, \Journal{\PRA}{83}{052122} {2011}
\bibitem{Wang3} P. Wang, {\em Chin. Phys.} C 35 (2011) 223
\bibitem{Wang6} P. Wang, {\em Can. J. Phys} 92 (2014) 25

\end{thebibliography}
\end{document}